\documentclass [11pt, oneside] {uwthesis}
\usepackage{amsmath}
\usepackage{amsfonts}
\usepackage{epsfig}
\usepackage{graphicx}

\newcommand{\phis}{$\{\phi_i\}$ }
\newcommand{\vom}{\vec{\Omega}}
\newcommand{\vla}{\vec{\lambda}}
\newcommand{\mcf}{\mathcal{F}}
\newcommand{\ptl}{\emph{pointlike} }
\newcommand{\ptln}{\emph{pointlike}}
\newcommand{\eopt}{E_{\mathrm{opt}}}
\newcommand{\egeo}{E_{\mathrm{geo}}}
\newcommand{\fermi}{\emph{Fermi}-LAT }
\newcommand{\nside}{N_{\mathrm{side}}}
\newcommand{\fwpsf}{f_{\mathrm{wpsf}}}
\newcommand{\band}{\emph{Band} }
\newcommand{\fluxunits}{ph cm$^{-2}$ s$^{-1}$ }
\newcommand{\dg}{^{\circ}}

\begin{document}
 
%

\prelimpages
 
%
%
\Title{Likelihood Methods for the Detection\\
	and Characterization of Gamma-ray Pulsars\\
  	with the \emph{Fermi} Large Area Telescope}
\Author{Matthew Kerr}
\Year{2010}
\Program{Department of Physics}
 \titlepage

%
%

\Chair{Thompson Burnett}{Professor}{Physics}

\Signature{Thompson Burnett}
\Signature{Scott Anderson}
\Signature{Miguel Morales}
\signaturepage

%

 

\doctoralquoteslip


%
%

\setcounter{page}{-1}
\abstract{%
The sensitivity of the {\it Large Area Telescope} (LAT) aboard the {\it Fermi Gamma-ray Space Telescope} allows detection of thousands of new $\gamma$-ray sources and detailed characterization of the spectra and variability of bright sources.  Unsurprisingly, this increased capability leads to increased complexity in data analysis.  Likelihood methods are ideal for connecting models with data, but the computational cost of folding the model input through the multi-scale instrument response function is appreciable.  Both interactive analysis and large projects---such as analysis of the full gamma-ray sky---can be prohibitive or impossible, reducing the scope of the science possible with the LAT.

To improve on this situation, we have developed \emph{pointlike}, a software package for fast maximum likelihood analysis of \fermi data.  It is interactive by design and its rapid evaluation of the likelihood facilitates exploratory and large-scale, all-sky analysis.  We detail its implementation and validate its performance on simulated data.  We demonstrate its capability for interactive analysis and present several all-sky analyses.  These include a search for new $\gamma$-ray sources and the selection of LAT sources with pulsar-like characteristics for targeted radio pulsation searches.  We conclude by developing sensitive periodicity tests incorporating spectral information obtained from \emph{pointlike}.
}
 
%
%
\tableofcontents
\listoffigures
\listoftables
 
%
%
\chapter*{Glossary}      
\addcontentsline{toc}{chapter}{Glossary}
\thispagestyle{plain}
\begin{glossary}
\item[LAT] Large Area Telescope, a pair-conversion telescope
\item[HE] high energy; a class of $\gamma$ rays with energies between about 100 MeV and 100 GeV; the passband of the \fermi
\item[GeV] $10^9$ electron Volts; in the text ``GeV'' is also used synonymously with ``HE'' for describing the \fermi passband, e.g., \fermi is a GeV telescope
\item[IRF] instrument response function; a statistical description of the measurement outcomes for an incident photon, including (a) the probability of an interaction and successful event reconstruction and (b) the distribution of measured variables about their true values
\item[PSF] point spread function; a probability density function describing the random spread in reconstructed position of photons incident from a point source
\item[TKR] the tracker, a section of \fermi comprising interleaved tungsten foils and silicon strip detectors
\item[CAL] the calorimeter, a section of \fermi comprising a hodoscopic array of CsI crystals
\item[ACD] anticoincidence detector; a component of \fermi comprising segments of scintillator designed to signal the passage of a charged particle into the detector
\item[FITS] Flexible Image Transport System, a standard for tabular data and images
\item[CALDB] a scheme for storing calibration information for instruments and telescopes
\item[FT1] a high-level \fermi data product in FITS format; columns give reconstructed quantities (e.g., energy, time) for events
\item[FT2] a high-level \fermi data product in FITS format; contains information about the spacecraft's orientation and position as a function of time
\item[l] Galactic longitude, a measure of azimuth with origin at Earth referenced to the Galactic center
\item[b] Galactic latitude, a measure of polar angle with origin at Earth referenced to Galactic center
\item[S/C] spacecraft, in particular the vehicle of the \fermi
\item[GTI] Good Time Interval, a tuple of a start and end time; events with reconstructed times lying within the interval pass some set of criteria to be ``good'' events
\item[ML] maximum likelihood; a statistical technique for estimating parameters of a distribution
\item[CT] conversion type, a classification for a reconstructed \fermi event indicating whether the event converted in a thin tungsten foil in the ``front'' of the TKR ($CT=0$) or in a thick tungsten foil in the ``back'' of the TKR ($CT=1$)
\item[SNR] signal-to-noise ratio; N.B. not supernova remnant!

\end{glossary}
 
%
%
\acknowledgments{ \vskip2pc
   {\narrower\noindent
It's said it takes a village to raise a child.  I don't know how many villages it takes to raise a grad student---in many ways still a child---but I've been lucky enough to have many.

My advisor, Toby Burnett, has made my course as a grad student remarkably straight and smooth.  He allowed me the freedom to choose and pursue my own projects while guiding their direction and providing insight on the research at hand.  He has gone above and beyond in supporting me, including allowing for travel to connect with LAT Collaboration members and other scientists in the field, an important part of joining the research community.

The science and performance of the Large Area Telescope is driven by the LAT Collaboration, and to them I owe much.  The members are happy to share advice, expertise, and time, and I learned much about the process of science by having the opportunity to regularly present results to a group of interested experts.  Within the collaboration, I owe thanks to a few individuals in particular:

\begin{itemize}
\item to David Smith, for having what turned out to be a rather important breakfast with me at Denny's in Cocoa Beach, Florida, where we were awaiting the launch of \emph{Fermi}.  He convinced me over greasy food that doing Galactic science with \emph{Fermi} was far too promising and exciting to miss out on, and this set the course of my graduate career.  Since then he has offered advice, friendship, and hospitality, and I am grateful for these and more.

\item to Eric Grove and Paul Ray for their enthusiasm about and support of the development of \ptln.

\item to Lucas Guillemot and Damien Parent for enjoyable collaboration.

\item to Tyrel Johnson, for hooking me on the Vela paper over beer in Cocoa Beach, for help in crunching all those spectra, and for hospitality.

\item to Seth Digel, for careful reading and valuable comments on many analyses and presentations.

\end{itemize}

More broadly, I owe a debt of gratitude to the scientists, engineers, and staff who made the design, construction, and launch of the \emph{Fermi} Gamma-ray Space Telescope a reality.

Thanks go to Fernando Camilo for a wonderfully productive collaboration and for teaching me everything I know about radio astronomy.  I am grateful to the radio community in general for their hard work in enabling so much of the pulsar science of the LAT.

At the University of Washington, I owe thanks to Eric Agol, Scott Anderson, Miguel Morales, Michael McCarthy, and Steve Sharpe for serving on my exam committee, and to Scott Anderson and Miguel Morales for additionally serving on my reading committee.

My classmates at the University of Washington are an excellent group of people and I thank them for the camaraderie and support over my years here.  Thanks especially go to my groupmates/officemates Marshall Roth and Eric Wallace, and to Jason Dexter, who has been all one could ask for in a friend.

My dear friends at home have held a place for me in their hearts despite years apart, for which I am profoundly grateful.  It means the world to be able to pick up right where we left off.

Most of all, I thank my family for making me who I am today and supporting me all these years.

   \par}
}

%
%
\dedication{\begin{center}For my parents\end{center}}

%

%
%

\textpages
 
\chapter{Overview}
\label{ch0}

In Chapter \ref{ch1}, we present an introduction to $\gamma$-ray astronomy, including the physics of $\gamma$-ray interaction in matter and the resulting constraints on $\gamma$-ray telescopes.  We review the major components of the \emph{Fermi} Large Area Telescope (LAT) and briefly describe the implementation of similar components in past experiments.  We conclude with an overview of some of the major sources of $\gamma$ rays.

In the next chapter, we detail the challenges of analysis of $\fermi$ data.  The primary consideration is the complex instrument response function (IRF), e.g. the power law dependence of angular resolution on energy.  We suggest the use of likelihood techniques which incorporate the full IRF {\it ab initio}.  We present the likelihood appropriate for photon-counting instruments and, by considering the IRF of the LAT in detail, develop a version suited to the LAT.  We note that the expressions require multi-dimensional integrals and are generally computationally expensive.

In Chapter \ref{ch3}, we discuss the implementation of \ptln, a package for maximum likelihood spectral analysis of \fermi data.  By adopting a binning scheme that scales with the detector resolution, we achieve considerable compression of the data at low energy.  Moreover, we make controlled approximations allowing for accurate but rapid evaluation of the likelihood.  We detail these approximations formally and show they adhere to a goal of $1\%$ accuracy.  By implementing a fast likelihood package, we open up new science through both interactive and large-scale analysis.

Chapter \ref{ch3_post} gives an overview of the validation of \ptl at the top level.  We verify that the code accurately reproduces Monte Carlo data from the model, i.e. that the approximations developed in Chapter \ref{ch3} are sufficiently accurate.  We then validate the most important functionality of \ptln, viz. estimating spectral parameters and parameter uncertainties, by performing fits of ensembles of simulated sources.

To demonstrate the potential of \ptln, we present in Chapter \ref{ch4} some actual \ptl analyses.  We work through an ``interactive'' analysis of sources in the Cygnus region, conveying the importance of an exploratory approach in identifying new and necessary components of the source model.  We develop the machinery for an all-sky analysis in which the spectra for all sources in the sky are determined consistently.  The product of this analysis---an excellent model of the GeV sky---enables other large-scale analyses, and we detail three: the construction of ``test statistic'' maps for the detection of new sources, the generation of useful visual representations of the data via kernel density estimation, and the selection of unidentified LAT with pulsar-like properties.  These pulsar-like sources have been targeted in radio pulsation searches, and we briefly present the results of two surveys.

We switch gears somewhat in Chapter \ref{ch5}.  We review statistics used for pulsation searches ($Z^2_m$ and $H_m$) and argue that their use of only time-domain information only is inadequate.  We modify the statistics to include a weight for each photon, and we use \ptl to calculate a weight giving the probability a photon originates from the source being tested.  By leveraging the additional information in the photon energy and position via the spectral analysis, the statistic becomes appreciably more resilient to Type I and Type II error.  We demonstrate the capabilities with an ensemble of Monte Carlo pulsars and show the weighted statistics improve the sensitivity by 50--100\%.

Finally, in the appendices, we present a derivation of the asymptotic distribution of the $H$ test\cite{dejager_1}.  This new result obviates the need for Monte Carlo calibration.  We also provide an extension of the methods presented in Chapter \ref{ch5} to sources with periods long compared to the time scale on which the LAT orientation changes, allowing sensitive searches for orbitally-modulated emission from binary systems.

 
\chapter{Gamma-ray Astronomy and the \emph{Fermi} Large Area Telescope}
\label{ch1}

Gamma-ray astronomy is the study of light in the limit of very few, very energetic photons.  The gamma-ray band extends from soft $\gamma$ rays with energies of 100 keV---see e.g. the instruments aboard the INTEGRAL\cite{integral} observatory---up to 100s of TeV, the domain of ground-based imaging air \v{C}erenkov telescopes, e.g., VERITAS\cite{veritas}, HESS\cite{hess}, and MAGIC\cite{magic}.  In between these two extremes, from $100$ MeV to $100$ GeV, lies the high-energy (HE) $\gamma$-ray band, which we shall also occasionally refer to as ``GeV $\gamma$ rays''.  To study this light is the purpose of the \emph{Fermi} Large Area Telescope and this work.

To order of magnitude, a blackbody spectrum peaking at $1$ MeV indicates a source temperature of $10^{10}$K.  Such temperatures occur in only the most extreme and ephemeral processes---e.g., core collapse supernovae\cite{ccsn_review}.  In general, persistent HE $\gamma$ rays are emitted through \emph{non-thermal} processes---inverse Compton scattering, Bremsstrahlung\footnote{We include radiation induced by magnetic fields---synchrotron and curvature---in this category.}, and $\pi^0$ decay\cite{longair}.  In \S \ref{ch1:sec:sources}, we consider some of the sources hosting such processes and shining in the GeV.

Before considering emission, however, we begin with the most important facet of $\gamma$-ray (or any!) astronomy: reception.  Earth's atmosphere is opaque to $\gamma$ rays, so GeV telescopes are necessarily balloon-borne or space-based.  HE source fluxes are generally low, and long integration times and reduced atmospheric background favor space telescopes (currently, AGILE\cite{agile} and the \emph{Fermi} Large Area Telescope).  In the following section, we give an overview of the relevant detector physics and resulting principles of operation of GeV telescopes.
 
\section{Gamma-ray Telescope Principles}
Just as the phenomenology of low-frequency radiation---diffraction, refraction, e.g.---determines the design of radio and optical telescopes, so the physics of high-energy particle interactions shapes the design of $\gamma$-ray telescopes.  Accordingly, we begin with an overview of the interactions of photons, particularly at high energy, in matter.

\subsection{$\gamma$ rays in Matter: Pair Production and Bremsstrahlung}

\begin{figure}
\begin{minipage}{6in}
\includegraphics[width=6in]{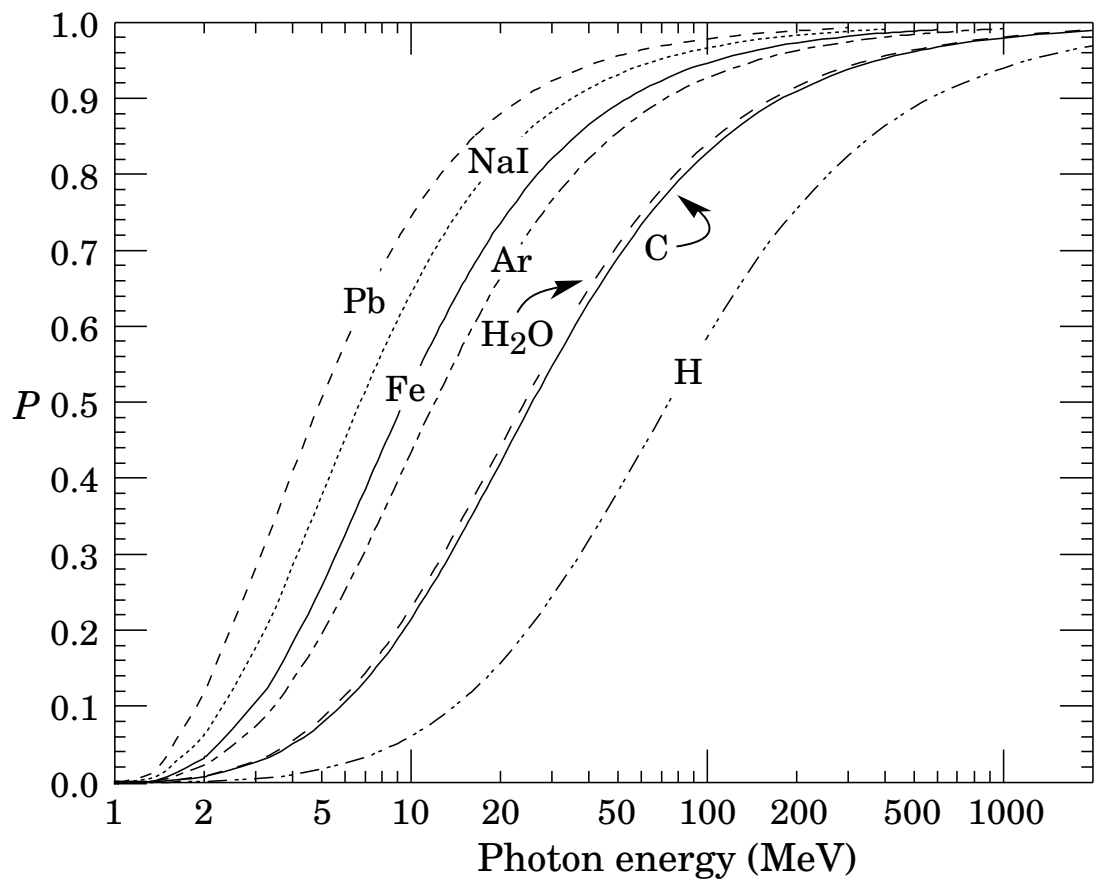}
\end{minipage}
\begingroup\renewcommand{\baselinestretch}{1.0}
\caption{The probability that the first interaction of an incident photon in the given material is pair production.  By 100 MeV, the cross section for photon interaction in heavy elements is dominated by pair production.  Reproduced from \cite{pdg}.}
\renewcommand{\baselinestretch}{1.5}\endgroup
\label{ch1_pair_prob_99}
\end{figure}

\begin{figure}
\center
\begin{minipage}{4in}
\includegraphics[width=4in]{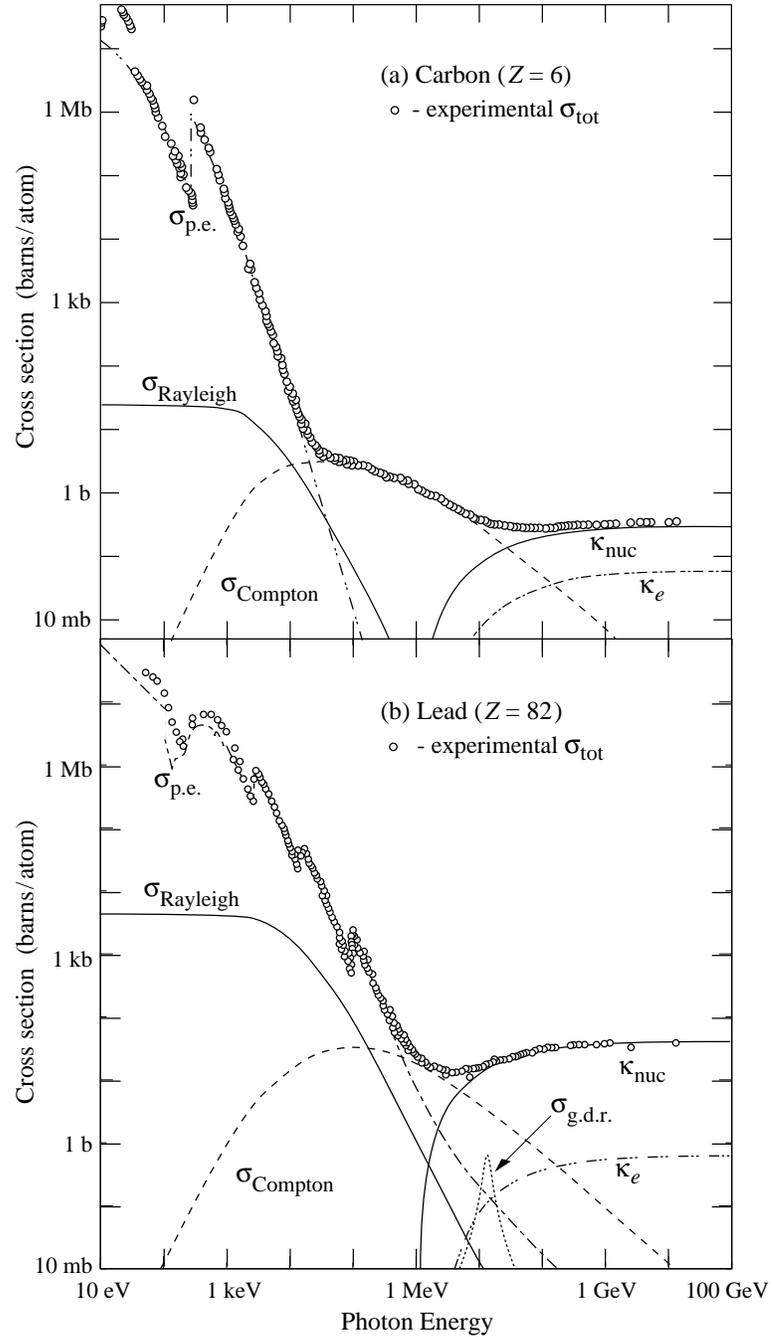}
\end{minipage}
\begingroup\renewcommand{\baselinestretch}{1.0}
\caption{The photon cross section in carbon (Z=6) and lead (Z=82).  The greater binding energy of atomic electrons in lead allows photoelectric resonances to X-ray energies, while Compton scattering is the dominant process in X-rays for carbon.  In both materials, pair production on nuclei becomes dominant at high energies, where the cross section is essentially flat.  The absolute value of the cross section is significantly higher for lead, as $\sigma \propto Z^2$.  Reproduced from \cite{pdg}.}
\renewcommand{\baselinestretch}{1.5}\endgroup
\label{ch1_sigma_both_06}
\end{figure}

At long wavelengths, light has a well-defined phase and interacts with matter as prescribed by classical electrodynamics\cite{jackson}.  At optical wavelengths, the energy associated with photons becomes comparable to atomic binding energies and the photoelectric effect is the dominant process.  For heavier elements, the binding energy of core electrons reaches X-ray energies, allowing for continuing resonances in the photoelectric cross section, while for lighter elements Compton scattering beings to dominate the cross section at energies of a few keV.  At 1 MeV (precisely, twice the electron mass), it becomes possible for a photon to produce an electron and a positron in the Coulomb field of a nucleus, the field absorbing the necessary four-momentum to ensure four-momentum conservation.  Stronger Coulomb fields provide for a larger cross section with a more rapid onset at threshold.  At sufficiently high energies (about 100 MeV for heavy elements), the pair-production cross section (a) dominates the total cross section (see Figure \ref{ch1_pair_prob_99}) and (b) is essentially flat (see Figure \ref{ch1_sigma_both_06}).

The flat cross section implies that the interaction rate is scale free, and there is effectively a mean free path for high energy photons\footnote{Equivalently, the energy loss rate is linear in the incident energy.}.  This quantity depends only on the material, and is commonly called the \emph{radiation length}\footnote{A radiation length is 7/9 of the photon mean free path\cite{pdg}.}, denoted $X_0$.  In this same high-energy r\'{e}gime, the dominant interaction for electrons is Bremsstrahlung via interaction with nuclear Coulomb fields.  (At the low-energy limit of this range, energy loss due to ionization of the material becomes important; this \emph{critical energy}\cite{pdg} is material-dependent but is roughly $10$ MeV for heavy elements.)  At first order, the energy loss rate for Bremsstrahlung is $dE/dx = -E/X_0$, i.e. electrons emit on average all but $1/e$ of their initial energy after traversing a distance $X_0$.

These conditions lead naturally to an electromagnetic shower in which an incident photon produces an $e^-$/$e^+$ pair, each member of which in turn radiates high energy photons, leading to an exponentially-growing cascade of electrons and photons.  As long as the next-generation particles remain well above the pair-production threshold (photons) or critical energy (electrons), the shower continues to develop exponentially.  Since the produced $e^-$/$e^+$ are highly relativistic, the pair-production cross section in the lab frame is sharply peaked along the initial photon trajectory.  The Bremsstrahlung is in turn beamed in a narrow cone about the electron trajectory.  The resulting well-contained showers allow inference about the properties of the incident photon.
 
An electron traveling in matter is subject to Coulomb scattering from surrounding atoms.  Unlike the significant momentum change from Bremsstrahlung spurred by close approach to a nuclear field, the interactions with distant charges are mere ``love taps'' perturbing the path of the electron.  For a sufficiently thick piece of material but small overall perturbations, this multiple Coulomb scattering is like a random walk, and the particle exits the material with an approximately Gaussian distribution about its initial direction.  The standard deviation of this distribution is given in the relativistic limit by\cite{pdg}
\begin{equation}
\label{eq:mult_scat}
\theta_0\approx8\dg\,\sqrt{X}\,\frac{100\ MeV}{E},
\end{equation}
with $X$ the length of material traversed measured in radiation lengths.

These basic physics considerations mean that a GeV telescope must be a full-fledged particle detector capable of tracking charged particles and measuring the energy of showers.  In the next section, we consider the components a typical GeV telescope needs to accomplish these tasks.

\subsection{Components of a GeV Telescope}
The nature of interaction of GeV photons with matter sets strong constraints on telescope design.  The \emph{only} available detection process is conversion of the incident photon into a pair, i.e., there is no focusing of $\gamma$ rays.  We have the following results:
\begin{itemize}
\item converting the bulk of incident photons requires sufficient radiation lengths of material;
\item maximizing the pair-production cross section (relative to, e.g., Compton scattering) and the electron critical energy requires a high-Z converter;
\item obtaining a good estimate of the photon trajectory requires measurement of the position of the first generation pair before it has suffered significant multiple scattering;
\item obtaining a precise measurement of the energy requires converting, containing, and measuring the bulk of the electromagnetic shower, i.e., many radiation lengths for conversion and additional material to measuring ionizing charged particles and radiation.
\end{itemize}

These constraints essentially design $\gamma$-ray telescopes for us, and nearly all successful GeV telescopes have used some form of the following components:
\begin{itemize}
\item A combination tracker/converter module.  To minimize multiple scattering and premature \emph{Bremsstrahlung}, thin, high-Z metal converter foils are interwoven with active material capable of tracking charged particles.  To increase the fraction of converted photons, multiple conversion foils and tracking layers may be stacked.
\item A calorimeter.  After having obtained position information, it is desirable to contain and measure as much of the shower as possible.  The calorimeter should bring high energy particles out of the radiation range (i.e., be many radiation lengths deep) and then measure the energy of the ionizing daughter particles.
\end{itemize}
Additional components require consideration of the telescope environment.  The charged particle number flux---from primary cosmic rays and secondary cosmic rays trapped in the earth's magnetosphere---is about from $10^3$ to $10^5$ that of the HE $\gamma$-ray flux, and this considerable background must be rejected.  Second, the earth is a bright source of $\gamma$ rays (see below and Chatper \ref{ch3}), and it is desirable to reject $\gamma$ rays entering the instrument from below.  Thus, we require two more components:
\begin{itemize}
\item An anticoincidence detector that signals the passage of a charged particle.
\item A mechanism to reject upward-going (zenith bound) $\gamma$ rays.
\end{itemize}

Having now identified the guiding physical principles and the major components of GeV telescopes, we consider the implementation of the most sensitive GeV telescope to date, the \emph{Fermi} Large Area Telescope.

\section{The \emph{Fermi} Large Area Telescope}
As we observed in the last section, a GeV telescope is naturally modular, and accordingly we organize this section around the components of the \emph{Fermi} Large Area Telescope (LAT).  To provide context and to acknowledge the significant scientific achievement of previous GeV experiments, we intersperse discussion of the LAT components with remarks on the implementation adopted in foregoing telescopes and the discoveries made possible by the technical advances.

\subsection{The Tracker}
\label{ch1:subsec:tkr}
The tracker/converter (TKR\cite{tkr}) is the heart of the LAT.  Enabled by advances in solid state particle tracking technology, it provides the sharpest view to date of the GeV sky.  To balance the tension between efficiency (governed by the total radiation lengths of converter material) and angular resolution (governed by multiple scattering in the converter material), the TKR is composed of 16 alternating layers of tungsten (Z=74) and tracking components\footnote{An additional two layers of tracking planes without conversion foils lie at the bottom of the TKR, for a total of 18 tracking layers.}.  The first 12 foils---referred to as the ``front'' of the TKR---are $0.028X_0$, while the final 4 foils---the ``back''---are $0.18X_0$.  Eq. \ref{eq:mult_scat} thus indicates that photons converting in the back layers suffer approximately twice the angular deviation of those converting in the front.  The total number of converted photons are roughly equally apportioned between the front and the back by the ratio of $X_0$ in the two TKR sections.

Following each tungsten layer is a silicon strip detector (SSD) with a pitch of $230$ microns.  The SSD comprises two planes with orthogonal strip orientations, allowing for a measurement of the transverse (to the detector plane) position of a charged particle.  The fine pitch and numerous layers require nearly one million readout channels\cite{lat_instrument}.

The TKR is not monolithic.  It is segmented into a $4$x$4$ array of 16 identical modules, each in turn made up of the 16 (18) conversion (tracking) layers described above.

Although using modern technology, the LAT TKR is also evolutionary.  The first $\gamma$-ray satellite telescope, OSO-3, had no tracking capabilities and instead relied on a rough collimation based on the anticoincidence veto, giving it a $\approx20\dg$ angular resolution.  Even this rough discrimination was sufficient to detect diffuse emission from the Milky Way\cite{oso3} (see below), but it was not until spark chambers were incorporated in SAS-2\cite{sas2} and COS-B\cite{cosb} that sufficient resolution and sensitivity to detect point sources was acquired.  The spark chamber reached its zenith in EGRET\cite{egret_intro,egret_calib}, where the 28 layers of tantalum-interleaved spark chambers (with 0.8mm wire spacing) achieved a high-energy resolution of $0.4\dg$.

\subsection{The Calorimeter}
To measure the energy from an incident photon, the LAT is equipped with a hodoscopic calorimeter (CAL\cite{calcal}).  As with the tracker, it is segmented into 16 autonomous modules.  (Together, the tracking and calorimetry elements and their electronics form a ``tower''.)  Each of these modules is formed from 96 $2.0\times2.7\times33.6$ cm cesium iodide (CsI) cyrstal scintillator bars arranged in an $8$ (depth, along instrument axis) by $12$ (breadth) array with the long dimension in the same plane as the TKR layers.  The orientation of the long dimension shifts by $90\dg$ in alternating layers.

The crystals serve both to continue to seed pair production and Brehmsstrahlung and, once the electromagnetic cascade has reached its maximum (particles are at the critical energy/ pair production threshold), the crystals scintillate as the main energy loss mechanism becomes ionization the CsI atoms.  Two photodiodes attached to each end of the crystal's long dimension allow for readout and a measurement of the transverse position of energy deposition by light asymmetry.

The total CAL is $8.4X_0$ on-axis\footnote{The CAL contains about 1800kg of CsI, constrained by the payload limit of the \emph{Fermi} launch vehicle.}, which depth becomes insufficient to fully contain showers initiated at high energies.  The segmentation allows characterization of the longitudinal profile of the shower and hence an estimate for the fraction of energy that escapes the CAL.  The segmentation also allows an accurate measurement of the centroid of energy deposition which serves as an anchor point in track reconstruction algorithm\cite{lat_instrument}.

While the depth of the CAL is similar to the calorimeter flown on EGRET, the greater transverse area of the CAL enables the wide field-of-view of the LAT.  Beyond facilitating track reconstruction and energy measurement, the imaging capabilities of the CAL also obviate the need for a dedicated time-of-flight coincidence system for rejecting upward-going $\gamma$ rays; EGRET, SAS-II, COS-B, and OSO-III all made use of such a system.

\subsection{The Anticoincidence Detector}
The charged particle background flux can exceed that of the desired $\gamma$-ray signal by up to $10^5$\cite{acd}.  While the TKR and CAL can reject hadronic showers with relatively high efficiency, cosmic ray electrons generate cascades very similar to those initiated by photons.  These too, of course, can be vetoed at the cost of efficiency by requiring ``empty'' TKR planes above the conversion point of the photon.  However, in the face of complex analysis and extremely high backgrounds, it makes much more sense to attempt to detect charged particles independently.  This function is provided by the Anticoincidence Detector (ACD), a plastic scintillator-based detector covering the top and sides of the LAT.

The ACD comprises 89 scintillator tiles, each with independent and redundant readout.  Spaces between tiles are filled with ribbon scintillators.  By segmenting the ACD, self-veto from ``backsplash''\footnote{``Backsplash'' denotes the phenomenon of soft X-rays produced in the electromagnetic cascade of a high-energy (above $\approx10$GeV) photon traveling back through the TKR and into the ACD.  Some of these X-rays will Compton scatter, causing an ACD trigger.  A segmented calorimeter allows for a fine-grained veto, as veto can be restricted to only events with a track extending to a triggered ACD tile.} is significantly decreased\cite{acd}.  The segmented ACD of the LAT is a significant improvement on those of previous experiments, which were largely monolithic and thus suffered from appreciable self-veto.  The sensitivity of EGRET, e.g., was essentially zero above 50 GeV\cite{acd}.

\subsection{Orbital Environment and Operating Modes}
The LAT was launched on June 11, 2008 from Cape Canaveral into an orbit with an altitude of $\approx565$ km inclined $\approx26\dg$ from the equator.  The orbit precesses with a period of about $53$ days.  The S/C typically operates in a ``sky survey'' mode in which with each orbit it rocks away from the zenith by $\pm D\dg$, alternately viewing the northern and southern orbital hemispheres.  The initial rock angle was $D=35\dg$ but has since increased to $D=50\dg$ to reduce the operating temperature of the S/C battery.

If the onboard software of the LAT or the GBM (Gamma-ray Burst Monitor) detects a signal consistent with a gamma-ray burst, the S/C can execute an Autonomous Repoint Request in which the sky survey mode is abandoned and the S/C maintains an attitude to keep the putative burst site near the center of the field-of-view\footnote{As with many telescopes, viewing a source directly on axis is a poor stategy since the ``cracks'' in the detector are then aligned with incident photon trajectories.}.  Since such pointing strategies allow the S/C axis to approach the limb of the earth, these periods are typically excised in standard analysis.

As noted in the previous section, the charged particle background for the LAT is considerable.  The typical trigger rate is $\approx2.2$kHz, but can increase by about a factor of 2 during some orbits as the earth's magnetosphere rotates in the plane of the S/C orbit\footnote{Background leakage is then also time-dependent on $<1$d timescales; care is required when searching for periodic signal from short-period binaries!}.  Reading out the LAT takes about $26$ $\mu$s, leading to deadtime of order $5\%$.  On the majority of orbits, the S/C encounters the ``South Atlantic Anomaly'', or SAA.  Within this region, the particle background is significantly higher than typical for the LAT's orbit, making nominal observations impossible.  When the S/C enters the SAA, the LAT ceases to trigger, and the time spent traversing the region is lost to normal science operations.  This episodic deadtime decreases the LAT duty cycle by about $10\%$ and leads to a modest north-south asymmetry in the exposure.

\section{Sources of Gamma Rays}
\label{ch1:sec:sources}
To complete the overview of $\gamma$-ray astronomy and \emph{Fermi}, we take a brief look at the classes of $\gamma$-ray sources filling the high-energy heavens.

\subsection{Diffuse Emission from Cosmic Rays}
Cosmic rays produce GeV radiation via three primary mechanisms\cite{diffuse1}.  At energies below and to the edge of the LAT passband, the emission is dominated by electronic processes, particularly Bremsstrahlung (induced by gas) and inverse Compton (IC) scattering of ambient light, e.g. the cosmic microwave background.  Coincident with the low end of the LAT passband, the contribution of hadronic processes turns on rapidly.  The inelastic scattering of high energy protons in gas nuclei produces $\pi^+$, $\pi^0$, and $\pi^-$ particles in roughly equal numbers, and the dominant decay process for $\pi^0$ ($98.8\%$ branching ratio\cite{pdg}) is to two photons.  The $\gamma$ spectrum depends on the $\pi^0$ spectrum, which in turn depends on the cosmic ray proton spectrum and the cross section for inelastic nucleon scattering.  This cross section turns on rapidly at the threshold for $\pi^0$ production\cite{gaisser,tune}, leading to the characteristic rapid rise of the observed diffuse $\gamma$ spectrum in the LAT band\cite{local_diffuse}.

At the high end of the LAT passband ($>100$ GeV), electronic processes can again become important via IC scattering of the interstellar radiation field.

In all of these processes, the $\gamma$-ray production depends on both the cosmic ray energy density and the ``target'' density---gas for Bremsstrahlung and radiation for IC scattering, leading to significant variations of the appearance of the HE sky as a function of energy, a topic we take up in a brief discussion of emission in our own Milky Way.

\subsubsection{The Milky Way}
As can be appreciated from the representation of the $\gamma$-ray sky in Figure \ref{ch4_allsky_im}, the dominant feature is diffuse emission associated with the structure of the Milky Way.  The intensity of the low latitude emission is a projection effect as the emissivity scales as the product of the cosmic ray density and the target density.  This sharp peak in intensity\cite{q2_diffuse} leads to difficulty in the analysis of point sources in the plane\cite{1fgl}.  Emission from higher latitude represents cosmic ray interactions in our local environment\cite{local_diffuse}.

\subsubsection{Starburst Galaxies}
If supernovae shocks provide the accelerators for cosmic rays, starburst galaxies\cite{starburst2}, with their intense patches of high star-formation rates, should be modestly bright $\gamma$-ray sources, as the high supernova rate and enhanced dust concentration (target material) lead to significantly elevated $\gamma$-ray production relative to, e.g., the Milky Way.  The detection by \fermi of two starburst Galaxies\cite{starburst2} confirms this idea, providing both a new opportunity for the study of cosmic ray production and insight into the observed extragalactic $\gamma$-ray background.

\subsubsection{The Earth}
As with much other emission from the Solar System (e.g., the moon and the quiescent sun\cite{sunmoon}, emission of $\gamma$ rays from Earth has its origin in cosmic rays\cite{fermi_limb}.  Due to its proximity, the limb of the earth is in fact the brightest source of $\gamma$ rays seen by Fermi, requiring care in separating its signal from that of celestial sources, see e.g. Figure \ref{ch3_plot2} and discussion in \S\ref{ch3:subsec:binning}.

\subsection{Active Galactic Nuclei}
The center of most galaxies are known to harbor massive $\geq10^6M_{sun}$ black holes.  While most are quiescent (e.g. our own Sgr A*), accretion at $\approx10\%$ of the Eddington rate onto these black holes can fuel powerful relativistic, collimated outflows known as jets.  These jets are often observed to have apparent superluminal velocities (e.g. \cite{hubble_m87}) implying the jet material has Lorentz factors of order $10$ and the motion is largely along the line of sight.  Such motion leads to a Doppler boost in both luminosity and apparent energy, making emission visible from cosmological distances.  GeV emission is produced through IC scattering of seed photons, which may be produced by synchtron emission by the electrons in the magnetic field of the jet or from sources external to the jet, e.g., the bright accretion disk\cite{blazar_jets}.  AGN bright in the GeV are known as blazars and represent the class of AGN with strong jets aligned very closely with the line of sight.  The class breaks down into flat-spectrum radio quasars and BL Lac objects, and Fermi sees roughly equal numbers of each\cite{1lac}.  An appreciable component of the observed isotropic diffuse emission\cite{isotropic_spec} is likely contributed by unresolved blazars.

\subsection{Pulsars}

Pulsars are rapidly-rotating, highly-magnetized ($10^8$ to $10^{14}$ gauss) neutron stars.  The intense currents induced lead to radiation emission over a broad energy range.  Although pulsars have a storied history in radio\cite{hewish68}, \fermi has essentially opened up the study of pulsars in GeV, e.g., with the detection of millisecond pulsars\cite{msp_pop}, the detection of tens of potentially radio-quiet pulsars\cite{blind_search_16,blind_search_8}, and thus we concentrate on their high-energy emission.

\begin{figure}
\begin{minipage}{6in}
\includegraphics[width=5.5in]{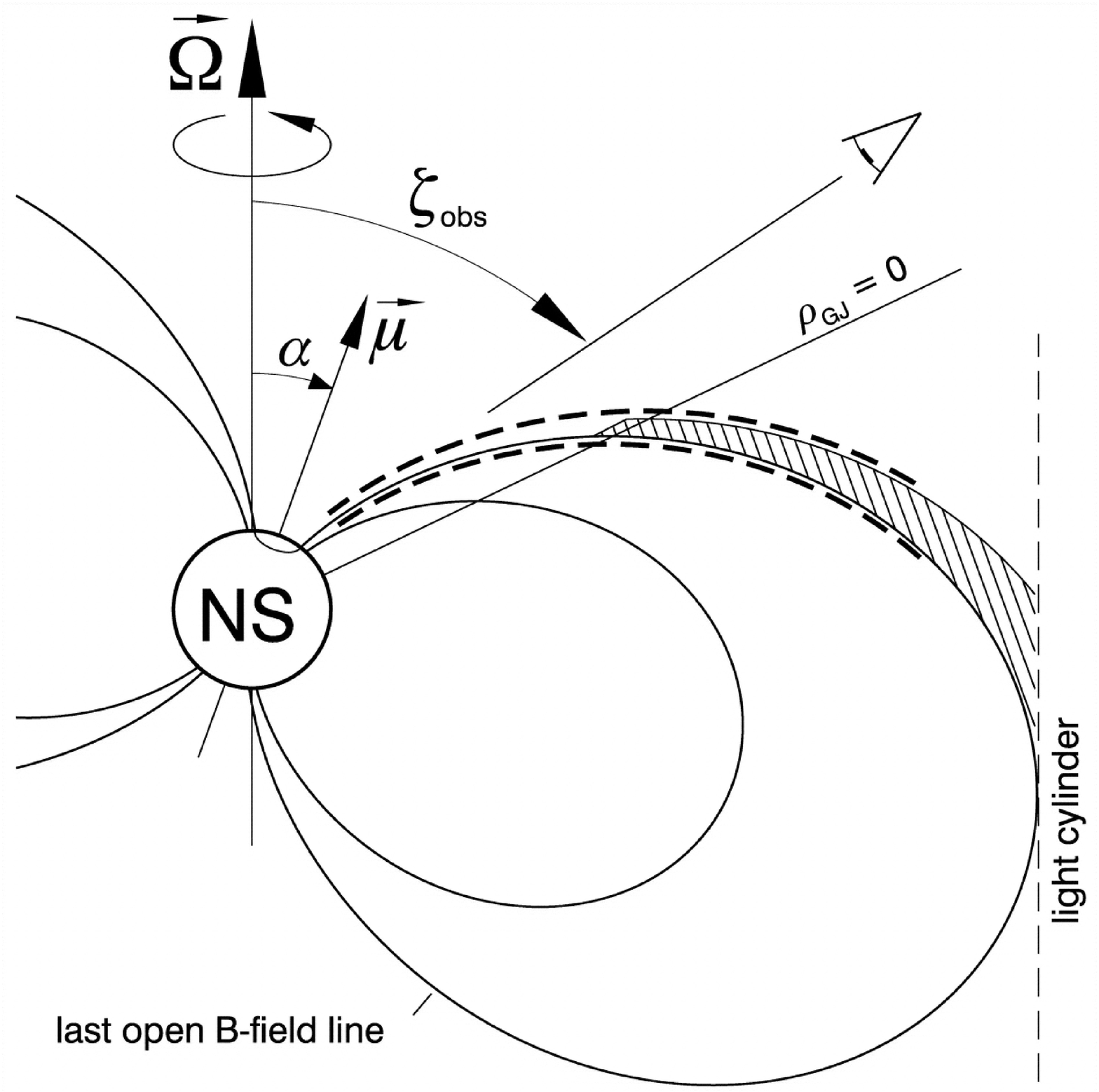}
\end{minipage}
\begingroup\renewcommand{\baselinestretch}{1.0}
\caption{The canonical representation of a pulsar magnetosphere, reproduced from \cite{dyks_rudak}.  The magnetic moment of the neutron star (NS), $\mathbf{\mu}$, is inclined to the NS spin axis, $\mathbf{\Omega}$, by an angle $\alpha$.  The field is assumed to be dominantly dipolar, though higher multipoles may be important\cite{xray_msp}.  The Goldreich-Julian equilibrium charge density\cite{gj} $\rho_{GJ}$ vanishes along the null charge surface.  The small patch on the NS surface at the foot of the open $\mathbf{B}$ field lines is the polar cap.  The slot gap (dashed lines) and outer gap (hatching) lie along the last closed (open) field line.  HE emission is beamed along $\mathbf{B}$ field lines, and the observer's inclination $\zeta$ determines which field lines are visible, strongly affecting the resulting light curve.}
\renewcommand{\baselinestretch}{1.5}\endgroup
\label{cartoon}
\end{figure}

In canonical emission models, the acceleration of charged particles to very high energies and subsequent production of $\gamma$ rays through synchrotron/curvature radiation or inverse Compton scattering relies on the ``gap'' paradigm.  A relative vacuum---a gap---develops in a magnetosphere otherwise filled with charged particles configured such that $\mathbf{E}\cdot\mathbf{B}\approx0$\cite{gj}.  The unshielded potential in the gap is sufficent to accelerate particles to ultrarelativistic energies, powering pair cascades which produce the observed GeV emission and allow current to flow.

Emission models are primarily distinguished by the site of the gap (Figure \ref{cartoon}).  Polar cap models\cite{pc82,pc96} place the gap immediately above the polar cap, the small patch on the neutron star surface defined by the footprint of the open magnetic field lines.  (A field line is open if it crosses the light cylinder (LC), the surface at which a corotating particle would attain light speed.)  Outer gap models\cite{chr,cr94,ry95,r96} invoke a gap at higher altitude, along the last closed field lines.  The gap extends from the null charge surface---where the corotating charge density $\propto\mathbf{\Omega}\cdot\mathbf{B}=0$ changes sign\cite{gj}---to the LC.  Slot gap models\cite{harding_slot} postulate narrow gaps encircling the PC and rising along the last closed field lines to the outer magnetosphere.  The two-pole caustic model\cite{dyks_rudak} is essentially a slot gap model with emission from a vanishingly thin surface along the last closed field lines.  The annular gap model of \cite{annular_gap} illuminates field lines along a thin annulus interior to the last closed lines.  Relativistic effects, such as aberration, become important in outer-magnetospheric emission and naturally furnish the characteristic cusped light curves of many HE pulsars\cite{chr,vela1}.

A second important consideration is the assumed structure of the magnetic field.  Most calculations employ the solution of Deutsch\cite{deutsch}, which posits an infinitely-conductive star and a vacuum exterior.  At the other extreme are force-free magnetospheres in which the star exterior is modeled as a highly-conductive fluid and numerical solutions are derived with magnetohydrodynamic codes\cite{spitkovsky06}.  The true solution must combine elements of both.  The two approaches differ most in the outer magnetosphere, where the co-rotation velocity becomes relativistic and the modification of field lines by plasma becomes important.  Since LAT results point to a primary emission site in the outer magnetosphere\cite{vela1,msp_pop}, this distinction becomes even more important.  In both models, the field geometry depends strongly on $\alpha$, the inclination of the dipole axis to the spin axis (Figure \ref{cartoon}).

\section{Summary}
The myriad sources of $\gamma$ rays---which we have only touched on here---offer opportunities to study some of the most extreme processes in the universe.  However, the physics of $\gamma$-ray interactions require non-traditional telescopes, viz. particle detectors.  By taking advantage of advances in solid state technology, the \fermi offers unprecedented angular resolution and sensitivity, particularly at energies $>10$ GeV via suppression of self-veto.  However, as we shall see in the following sections, the relevant physics still set fundamental limits on the capabilities of the detector.  In particular, multiple Coulomb scattering leads to an angular resolution that varies by more than two orders of magnitude over the LAT passband.  Dealing with this multi-scale response requires care, and in the next chapter we introduce the principle of maximum likelihood, leading to techniques optimally suited for analysis of \fermi data.

 
\chapter{Likelihood Analysis for High-energy Gamma-ray Telescopes}
\label{ch3_pre}

Analysis of data from \fermi and other HE gamma-ray telescopes (e.g., EGRET\cite{mattox}) requires techniques not normally found in analyses of data from other wavebands.  In optical astronomy, for instance, observations are typically made of a narrow range of wavelengths, and the angular resolution of the instrument is sufficient to resolve the astrophysical background.  To determine the flux from a star, e.g., one identifies the pixels in the CCD illuminated by the image of the star (aperture photometry), measures the collected charge, and divides by the telescope aperture, time of observation, and efficiency of the filter and CCD.  In practice, of course, these steps may be quite involved, but the point remains that \emph{physical} quantities, e.g., source intensities, can generally be extracted directly from the data.  In addition, time-dependent instrumental effects are typically small, as sources can be tracked with great precision, resulting in a stable instrument response over the integration.

\begin{figure}
\begin{minipage}{6in}
\includegraphics[width=6in]{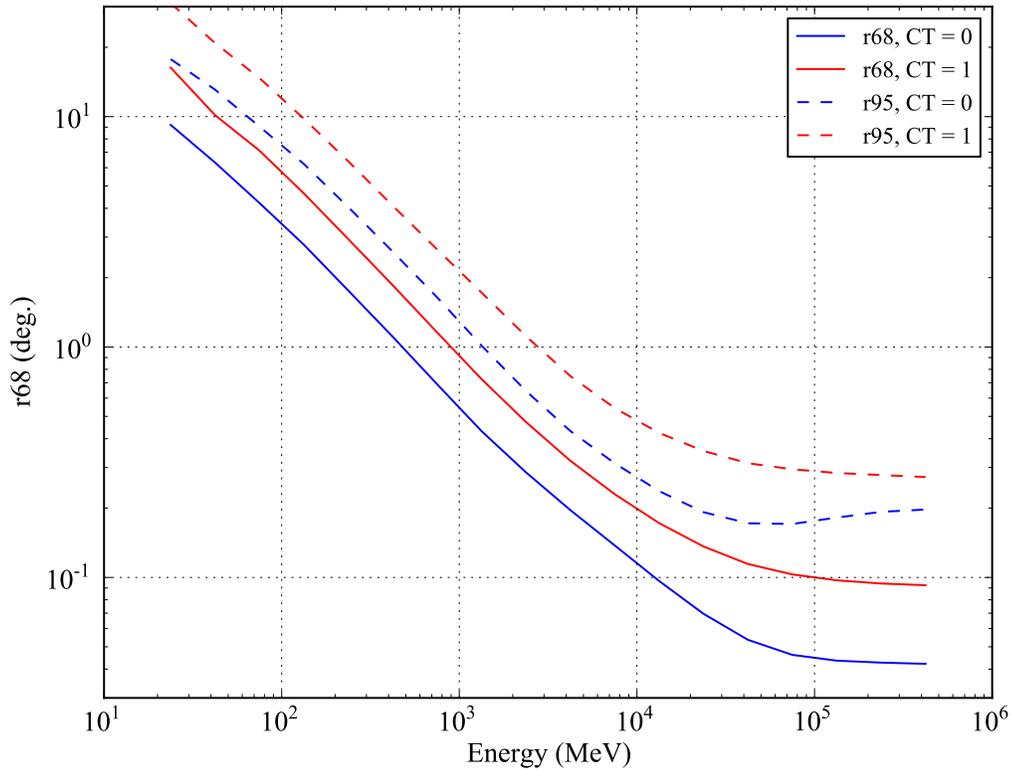}
\end{minipage}
\begingroup\renewcommand{\baselinestretch}{1.0}
\caption{The \fermi point-spread function as characterized by ``r68'' and ``r95'', the angular separation from a point source within which $68\%$ and $95\%$ of photons from the source lie.  The description here is for photons arriving close to the instrument's boresight; the angular resolution drops by up to $50\%$ at the edge of the field of view (see Figure \ref{ch3_plot1}).  As discussed in Chapter \ref{ch1}, the energy-dependence originates from multiple scattering of pair-produced electrons and positrons in the tungsten foils, and the asymptote is set by the pitch of the silicon strips in the tracking layers.}
\renewcommand{\baselinestretch}{1.5}\endgroup
\label{ch3_pre_plot1}
\end{figure}

Such methods fail for \fermi data for two reasons.  First, very few \fermi sources appear isolated.  Looking at Figure \ref{ch3_pre_plot1}, we note the absolute scale of the PSF at energies of order 0.1 GeV is about $5\dg$, and an immediate consequence is source confusion.  Suppose we want to extract the flux between 0.1 and 0.3 GeV for a particular source.  Except for the brightest blazars at high Galactic latitudes, there is generally no circle we can draw that is dominated by emission from the source of interest.  The situation is dire in the Galactic plane, as the projected density of point sources increases dramatically along with the diffuse background from cosmic rays.  At higher energies, the PSF scale drops to $<1^{\circ}$ and the situation improves.  By carefully calculating the integral of the PSF over an extraction radius, one could reliably extract fluxes for sources at high Galactic latitude by photon counting.  Sources in the Galactic plane, however, still suffer strong contamination.  In general, then, any technique that relies on simply apportioning photons to a source will deliver poor results.

Second, detecting and characterizing GeV sources may require integrations of a year or more.  The orientation of the instrument with respect to the source is constantly changing as the S/C orbits the earth, so there is no single ``number'' characterizing the instrument characteristic during the observation.  In essence, we are using a different telescope every few minutes.  Even if we could estimate accurately the number of photons we have observed from a particular source, we cannot straightforwardly convert it to a flux.

These considerations and others discussed below motivate likelihood techniques for the analysis of \fermi data.  Such techniques work in the ''forward-folded'' space, i.e., not with physical quantities but with instrumental quantities, and they naturally incorporate the complexity of the energy-dependent PSF and time-dependent observation.  Below, we introduce likelihood techniques in general and then specialize to GeV instruments and the LAT.

\section{Characterizing and Modeling Sources}
 
We characterize a source by its photon flux density.  (In the following, we neglect polarization, since it is very difficult for a pair-conversion telescope to measure the polarization of an incident beam.)  The particle flux density---the rate of photons incident per unit energy/area/time from a solid angle $d\Omega$ about the position $\vom$---completely encapsulates the properties of a source and is denoted $\mcf(E,t,\vom)$.  Its arguments are the observables associated with a photon, namely energy ($E$), time of arrival ($t$), and direction of origin ($\vec{\Omega}$), i.e., $-c\vec{p}/E$, with $\vec{p}$ the photon momentum vector.  Although the distinction is seldom of interest, $t$ is understood to be the retarded time.

Unlike infrared, optical, or even X-ray sources, the mechanisms that power HE gamma-ray sources almost universally produce broadband spectra, that is, spectra spanning multiple decades of energy and possessing no features and modest curvature.  Such spectra can be modeled with only a few parameters, and often a simple power law suffices:
\begin{equation}
\mcf(E,t,\vec{\Omega};N,\Gamma ,E_o) = N(t,\vom)\,\left(\frac{E}{E_0}\right)^{-\Gamma(t,\vom)}\,f(\vec{\Omega}).
\end{equation}
A simple extension, the exponential suppression of the flux above a cutoff energy,
\begin{equation}
\mcf(E,t,\vec{\Omega};N,\Gamma ,E_c,E_o) = N(t,\vom)\,\left(\frac{E}{E_0}\right)^{-\Gamma(t,\vom)}\exp\left(-\frac{E}{E_c(t,\vom)}\right)\,f(\vec{\Omega}),
\end{equation}
is adequate to characterize the emission from most $\gamma$-ray pulsars\cite{vela1}.  Here, $f(\vom)$ is a normalized function ($\int d\Omega\,f(\vom)=1$) describing the spatial morphology of the source.  We are mostly interested in point sources, i.e., those that cannot be spatially resolved and for which the spatial dependence is given by the Dirac delta function, $ f(\vom) = \delta(\vom - \vom_0)$, where $\vom_0$ is the true position of the point source.  In general, we will not specify a particular form of spectral model and instead label it with a set of parameters, $\vla$, and express the modeled flux density for a source as $\mcf(E,t,\vom;\vla)$.

\section{Principles of Maximum Likelihood}
\label{sec:like_principle}

With a model for the physical properties of sources in hand, the next step is to connect them to data and estimate the model parameters.  Below we outline the method of maximum likelihood (ML) parameter estimation which shall be our primary tool for this task.

Suppose an experiment is designed to measure a random vector $\vec{X}$, that the random variables admit a probability density function (pdf), and that the pdf is parameterized by some vector $\vla$.  We express the pdf as $f_{\vec{X}}(\vec{x};\vla)$\footnote{The notation for pdfs is discussed more fully in Chapter \ref{ch5}.}, meaning that the probability to measure a value near $\vec{x}$ is just $f_{\vec{X}}(\vec{x};\vla)\, d\vec{x}$.  Viewed this way---being given $\vla$---the probability is a function of $\vec{x}$, the possible outcomes of our experiment.  On the other hand, having performed the experiment and measured $\vec{x}$, we can view the same expression, $f_{\vec{X}}(\vec{x};\vla)$, as function of $\vla$, i.e., it characterizes the parameter values \emph{most likely} to have yielded the measured values.  When viewed as a function of the parameters, this quantity is called the \emph{likelihood} and we denote it $\mathcal{L}(\vla;\vec{x})$, although we often suppress the explicit dependence on $\vec{x}$.

The likelihood is useful for the common statistical/analysis task of determining estimators for the parameters of a distribution from data: natural estimators are those which maximize the likelihood, known as the \emph{maximum likelihood estimators}, or MLE.  In many cases, the MLE are \emph{consistent}, have an asymptotically normal distribution, and are efficient.  That is, with enough data, the estimators converge to their true values, the parameter uncertainty can be estimated from the shape of the likelihood, and the uncertainties are the smallest possible.

For large data sets, the likelihood is likely to be a very small number.  Since the logarithm of monotonic functions is again monotonic, we can instead maximize the ``log likelihood'', which is also often easier to compute.

We note in passing that, via Bayes' Theorem, the likelihood can be converted to the posterior probability by multiplication with the prior probability.  That is, since $f(\vec{x};\vla)f(\vla)=f(\vla;\vec{x})f(\vec{x})$, the posterior probability density, $f(\vla;\vec{x})$, is proportional to $\mathcal{L}(\vla;\vec{x})f(\vla)$.  Here, $f(\vla)$ is the \emph{prior probability}, or the distribution for the parameters we would expect in the absence or any data, or in light of previous experiments.  This interpretation allows one to interpret the statistical errors on the estimators as uncertainties on the true values, e.g., to construct credible intervals which have a $90\%$ chance of containing the true value of the parameter.

\section{The Observed Signal}

As outlined above, likelihood provides a connection between model parameters (as close to physical measurements as we can hope to get given the experimental constraints) and data.  Implicit (and perhaps hidden) in the simple notation we employed is the mapping of the source models onto the actual data.  That is, if we want to calculate the probability to observe some set of data (in the case of GeV instruments, some collection of photons) given some model, we must first figure out what that model looks like to the detector.  The first step is understanding how a signal from a source is processed by the instrument.

A GeV telescope is essentially a particle detector, and the fundamental data are events.  The detector, of course, is imperfect: it (a) fails to generate observables for all incident photons, and (b) introduces error into the observables for successfully detected photons.  To illustrate, suppose some number of photons from a steady, monoenergetic point source impinge upon the detector.  For the LAT, the tungsten foils in the tracker provide about $X_0=1.1$ radiation lengths\cite{lat_instrument}, and $\approx 40\%$ of photons at normal incidence\footnote{The apparent depth of a tungsten foil is inversely proportional to the cosine of the incidence angle.} will pass through the TKR without interacting\cite{tkr}.  These photons cannot be successfully reconstructed\footnote{In principle, events can be reconstructed using the hodoscopic calorimeter only, but this technique is not yet part of standard \fermi analysis.}.  Of the photons that interact in the TKR, some will fail to produce an identifiable signal\footnote{The cross section for pair production in tungsten is essentially flat in energy from a few hundred MeV to 100s of GeV\cite{pdg}, so the increase in effective area (Figure \ref{ch3_pre_plot2}) between 100 MeV and 1 GeV is primarily due to increasing performance of the event reconstruction algorithms.}.  And of course, for those that interact \emph{and} are successfully recognized as interacting photons, we will see a spread in the reconstructed energies and position.

These effects, both the \emph{efficiency} for successful reconstruction of an incident photon and the \emph{dispersion} of its true observables are characterized by the \emph{instrument response function}, or IRF.  In keeping with its two r\^oles, we explicitly factor the IRF into two pieces.  The instantaneous sensitivity, with units of area and denoted $\epsilon(E,t,\vec{\Omega})$, is the quantiy by which the flux density $\mathcal{F}$ should be multiplied to determine the instantaneous count rate in bin of width $dE\times dt\times d\Omega$.  The dispersion matrix, $P(E',t',\vec{\Omega'} | E,t,\vec{\Omega})$, gives the probability that a photon with true energy $E$, time of arrival $t$, and position $\vom$ will have a reconstructed energy $E'$, time of arrival $t'$, and position $\vom'$.  The dispersion matrix is a true probability density function and is normalized such that $\iiint dE\,d\Omega\,dt\, P(E',t',\vom' | E,t,\vom) = 1$.  The expected event rate in an infinitesimal bin centered on $E'$, $t'$, and $\vom'$ is then
\begin{equation}
\label{eq:source_rate}
r(E',\vom',t';\vla) = \iiint dE\,d\Omega\,dt\,\mcf(E,\vom,t;\vla)\,\epsilon(E,\vom,t)\,P(E',\vom',t'; E,\vom,t).
\end{equation}
The integral is over the support of $\mcf(E,\vom,t;\vla)$.

This quantity---the source flux as processed by the instrument---is what we need to calculate the likelihood.  Before we proceed, we first describe in some detail the \fermi IRF and make some simplifications to Eq. \ref{eq:source_rate}. 

\section{The Instrument Response Function of the Fermi-LAT}
\begin{figure}
\begin{minipage}{6in}
\includegraphics[width=6in]{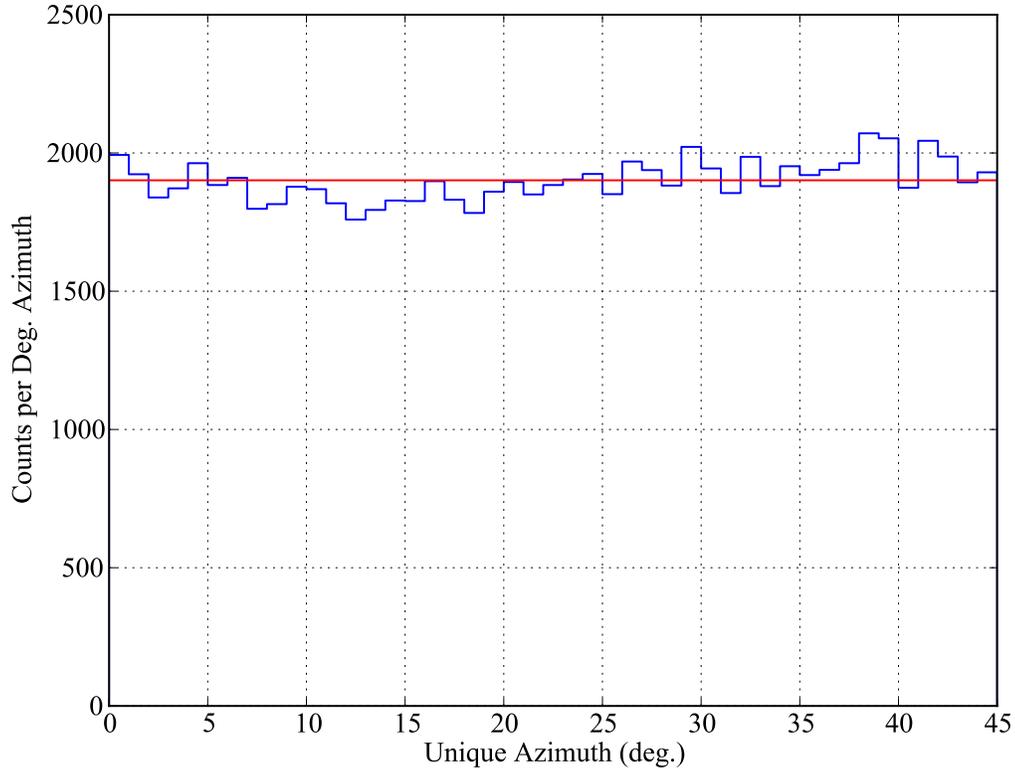}
\end{minipage}
\begingroup\renewcommand{\baselinestretch}{1.0}
\caption{The distribution in azimuth for a set of reconstructed events (P6 ``diffuse'' class) with 100 MeV $<$ E $<$ 100 GeV in a $10\dg$ aperture centered on (l,b) = (0,-45), integrated from Aug. 4 2008 to Jul. 18 2010.  Events with a reconstructed zenith angle $>105\dg$ have been discarded.  (See Figure \ref{ch3_plot2} which charts the zenith angle for this same data.)  The \fermi has twofold bilateral symmetry, so we can obtain the ``unique'' azimuth angles, i.e., those that truly probe the dependence on the shape of the detector, by $\phi_{unique}\equiv \phi \,\mathrm{mod}\,90\dg$ and, further, reflecting events with $\phi_{unique} > 45\dg$ to $90\dg - 45\dg$.  The red horizontal line indicates the mean, while the blue gives the reconstructed unique azimuth in $1\dg$ bins.  The deviation from the mean is quite modest.}
\renewcommand{\baselinestretch}{1.5}\endgroup
\label{ch3_pre_plot3}
\end{figure}

The calculation of the terms $\epsilon(E,\vom,t)$ and $P(E',\vom',t'; E,\vom,t)$ above requires some care for space-based gamma-ray telescopes such as the Fermi-LAT.  First, we note that the response of the LAT to an incident photon depends primarily on the geometry, viz. the angle between the photon's momentum vector and the instrument's boresight, which by convention defines the z-axis in the S/C frame, $\hat{z}$.  While this dependence is both on the polar angle, $\vom\cdot\hat{z}(t)\equiv\cos\theta(t)$ and the azimuth, for events not too far from the boresight, the azimuthal dependence is subsidiary to the polar dependence.  Furthermore, over long integrations\footnote{For transients such as gamma-ray bursts, only a short selection of data is used, and this approximation clearly fails.}, the azimuthal dependence tends to average out (see, e.g., Figure \ref{ch3_pre_plot3}).  Thus, we express the geometry solely in terms of the polar angle, which depends on the position and orientation of the S/C as it orbits.

\begin{figure}
\begin{minipage}{6in}
\includegraphics[width=6in]{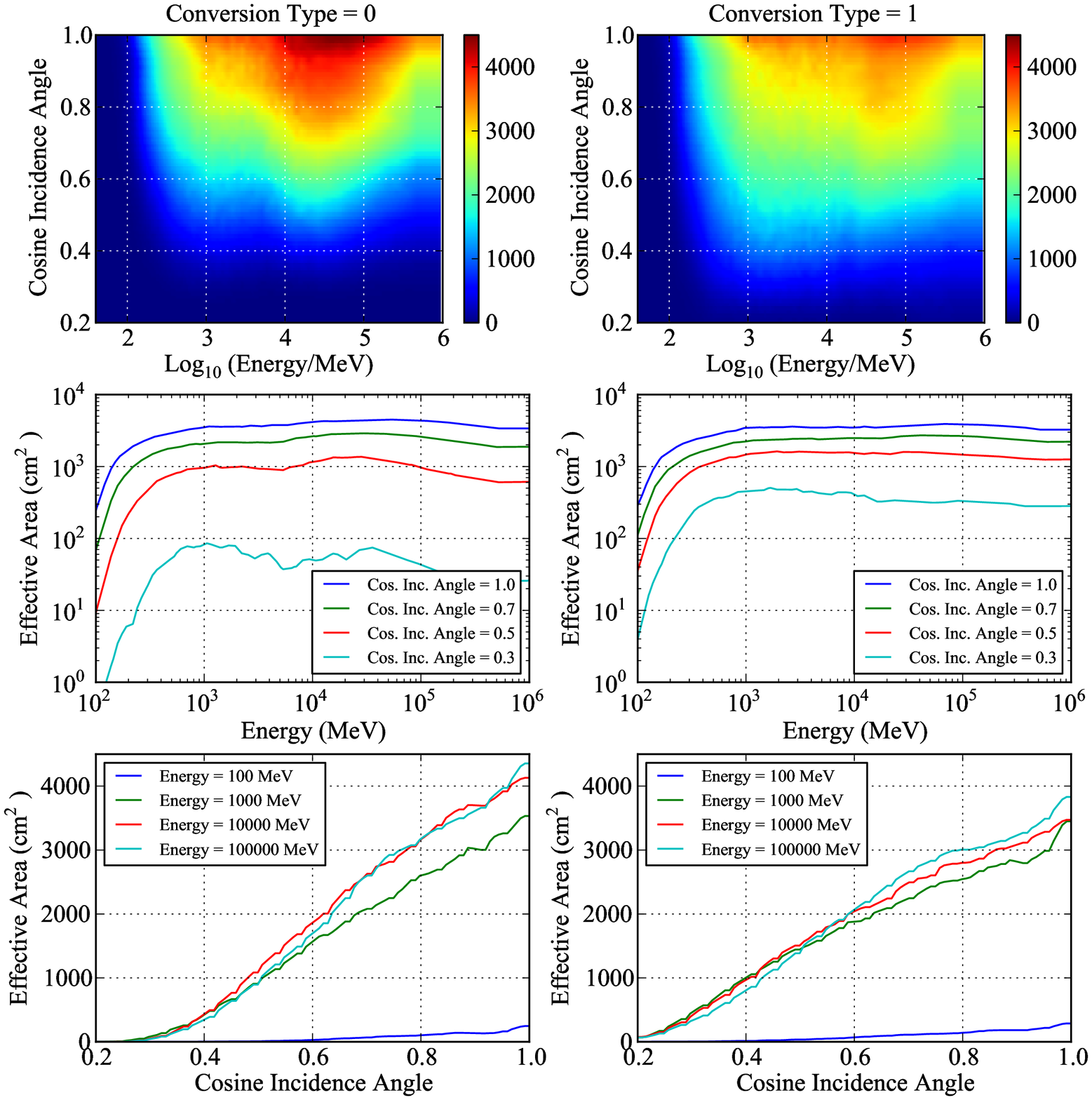}
\end{minipage}
\begingroup\renewcommand{\baselinestretch}{1.0}
\caption{The effective area for \emph{Fermi}-LAT as described by the \emph{P6\_v3\_diff} IRF.  The lefthand column shows the effective area of the front section of the detector (see \S\ref{ch1:subsec:tkr}) and the righthand column the back.  The top row shows the joint dependence of the effective area on energy and incidence angle.  The second row shows the dependence on energy for a set of incidence angles, while the third row shows the dependence on incidence angle for a set of energies.  We note that effective area is essentially zero below 100 MeV, turns on rapidly, and becomes essentially flat at 1 GeV.  This behavior is governed by the difficulty in reconstructing ``soft'' photons.}
\renewcommand{\baselinestretch}{1.5}\endgroup
\label{ch3_pre_plot2}
\end{figure}

We consider first the term $\epsilon(E,\vom,t)$, which is essentially the effective cross section of the detector at time $t$.  (As noted above, not all events can be reconstructed, so the effective area is always less than the geometric cross section.)  The term factors into the effective area, $A[E,\cos\theta(t)]$, with units of area, and a dimensionless efficiency, which we denote $e(E,t)$, a number between 0 and 1.  The effective area accounts for the ability of the instrument and software to successfully reconstruct a photon with energy $E$ incident at an angle $\theta$.  It gives \emph{to first order} the true collecting area of the detector.  We show the effective area for a particular version of the reconstruction algorithms, \emph{P6\_v3\_diff}, in Figure \ref{ch3_pre_plot2}.  The efficiency comprises two terms.  First, an energy-independent term depends on the trigger rate, as the instrument has a period of deadtime following each trigger while the electronics are read out.  Second, there is an energy-dependent effect resulting from ``pileup''.  For about 10\% of triggers, an additional charged background particle is interacting in the detector, degrading the reconstruction of the triggering particle and decreasing the efficiency.  The effect decreases with incident photon energy.  Since the trigger rate is dominated by charged particles, both of these effects depend on geomagnetic latitude and hence on time.

Next, we consider the term $P(E',\vom',t'; E,\vom,t)$.  We make the simplifying approximation that the three observables are statistically independent and write $P(E',\vom',t'; E,\vom,t) = f_{\mathrm{psf}}[\vom' ; \vom, \cos\theta(t), E]\times f_{\mathrm{edisp}}[E'; \cos\theta(t), E]\times f_{\mathrm{tdisp}}(t';E,\vom,t)$, with $f_{\mathrm{psf}}$ the point-spread function, $f_{\mathrm{edisp}}$ the energy dispersion function, and $f_{\mathrm{tdisp}}$ the time dispersion.  This assumption is certainly justified for $t'$.  The position (energy) reconstruction is dominated by the signal in the TKR (CAL), so to first order $E'$ and $\vom'$ are also independent.  We discuss each of these terms below.

\subsection{The Point-spread Function}
\label{ch3_pre:subsec:psf}
The point-spread function (PSF) lies at the heart of point source analysis.  The spatial distribution of photons on the sky is the single most important handle for disentangling confused sources.  Accordingly, we take some care in characterizing and (implementing) the PSF.

As with the effective area and energy dispersion, the PSF is determined by end-to-end Monte Carlo simulations of the detector.  The Monte Carlo events are binned in (true) energy and (true) incidence angle\footnote{The spectrum of incident events is uniform in logarithmic energy.}.  The resulting distribution of reconstructed incidence angle is fit with a parameterized analytic function.  These parameters describe the PSF over the instrument's phase space.

The PSF by definition is the probability density to observe an event in a small piece of solid angle $d\Omega$ centered on $\vom$ given that event originated from $\vom_0$.  That is, for some energy and incidence angle, suppressed here, the PSF is
\begin{equation}
\frac{dN(\vom)}{d\Omega} = \frac{dN(\vom)}{d\phi\, d\cos\theta} \approx \frac{1}{2\pi}\frac{dN(\cos\theta)}{d\cos\theta} \approx\frac{1}{\pi}\frac{dN(\theta^2)}{d\theta^2},
\end{equation}
where $\phi$ is the azimuthal angle and $\theta$ the polar angle measured with respect to the origin $\vom_0$, and where we have made the approximations of azimuthal symmetry and small angular deviations.  We see that the natural variable is $\theta^2$, the squared angular deviation, as isotropic sources ($dN/d\Omega$ constant) are flat in this quantity.  This motivates the definition of $u\equiv \theta^2/2\sigma^2$, in terms of which
\begin{equation}
\frac{dN(\vom)}{d\Omega} \approx\frac{1}{2\pi\,\sigma^2}\frac{dN(u)}{du}.
\end{equation}

\begin{figure}
\begin{minipage}{6in}
\includegraphics[width=6in]{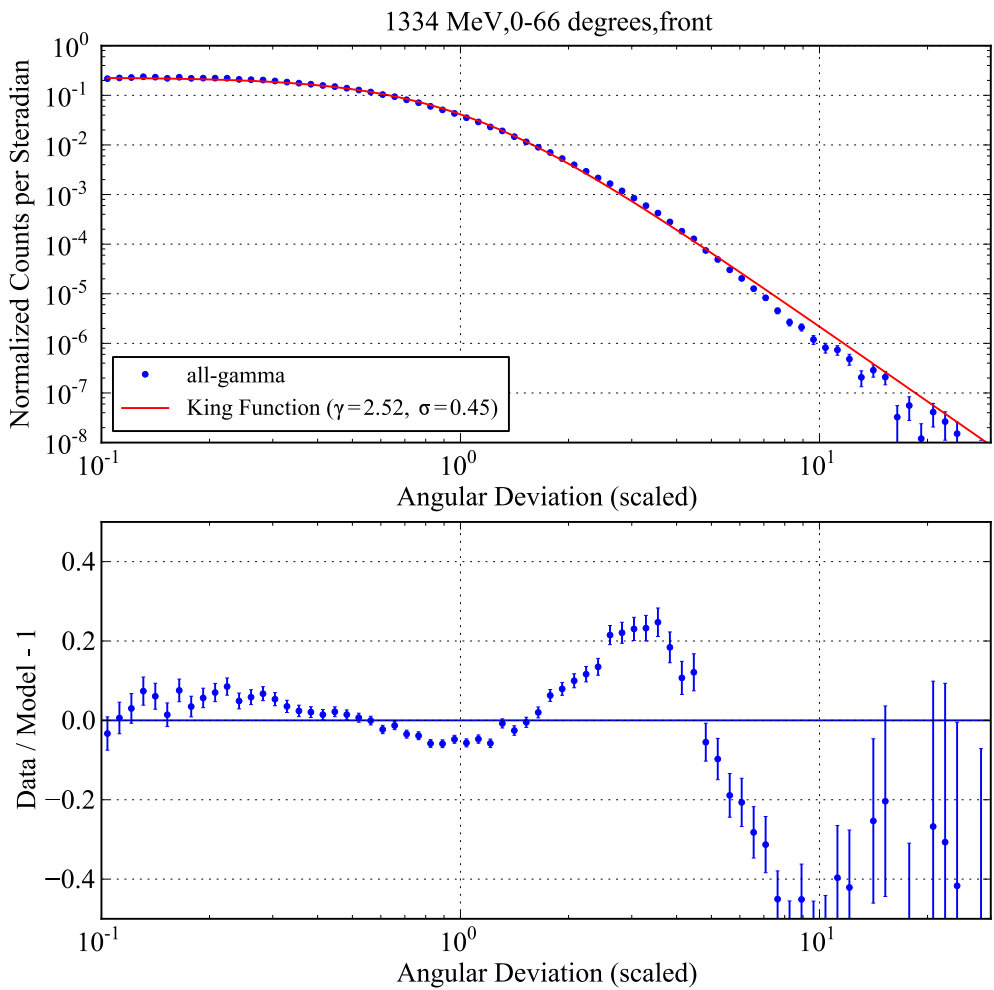}
\end{minipage}
\begingroup\renewcommand{\baselinestretch}{1.0}
\caption{Monte Carlo data from a detector simulation for a bin centered on 1334 MeV for front-converting events.  The events have been summed over incidence angle up to $66.4\dg$.  The behavior of the density function over many orders of magnitude is well-described by a single King function, i.e., that data show a sharp core and power law tails.  However, there is a slight break in the slope of the power law tail, and accordingly the King function cannot provide a perfect fit, leaving residuals $>10\%$.  In later versions of the IRF, this issue is ameliorated by the use of two King functions.}
\renewcommand{\baselinestretch}{1.5}\endgroup
\label{ch3_pre_plot4}
\end{figure}

While we could in principle use the Monte Carlo histograms directly as a representation of the PSF, a fitted analytic function is much more convenient; we thus need a function $dN(u)/du$ that adequately characterizes the instrument performance.  The functional form adopted for the PSF is the \emph{King function} which arises in characterization of many-body gravitational dynamics\cite{bandt}.  Mathematically, it is a precursor to the Gaussian distribution.  That is, if the King function is defined as 
\begin{equation}
f_k(u,\gamma)\equiv\left(1+\frac{u}{\gamma}\right)^{-\gamma},
\end{equation}
then since 
\begin{equation}
\lim_{\gamma\to\infty} \left(1+\frac{u}{\gamma}\right)^{-\gamma} = \lim_{\gamma\to\infty} f_k(u,\gamma) = \exp(-u),
\end{equation}
the King function can be seen as a ``finite $\gamma$'' version of the exponential.  Since $u\equiv \theta^2/2\sigma^2$, we see that the limiting form is a Gaussian in the angular deviation.  Thus, the smallness of $\gamma$ characterizes how far we are from the ``ideal'' Gaussian PSF, and $\sigma$ sets the overall angular scale.  (Note that $\gamma > 1$ for the function to be well-behaved.)  We see that, unlike the Gaussian, the King function has power law tails with slope $-\gamma$ for $u >> \gamma$, and is thus more suitable for characterizing a PSF with broad tails.  A King function fit to Monte Carlo data is shown in the top panel of Figure \ref{ch3_pre_plot4}.

Thus, for a given energy and incidence angle, the PSF is described by two parameters, $\sigma$ and $\gamma$.  In practice, these parameters are evaluted for bins in energy (4 bins per decade) and incidence angle (bins of width 0.1 in $\cos(\theta)$) and stored in a FITS table as a part of CALDB, the repository for instrument calibration information.  In order to retain the smooth energy dependence of the underlying behavior and to reduce the steepness of the parameter dependence on energy, the $\sigma$ parameter has a built-in energy dependence.  That is,
$\sigma \equiv \sigma_{\mathrm{CALDB}}\times f_{\sigma}(E)$, with
\begin{equation}
\label{eq:prescale_sigma}
f_{\sigma}(E)^2 \equiv \left(p_1 (E/100\,MeV)^{-p_2}\right)^2 + p_3^2,
\end{equation}
with $p_1$ and $p_3$ dependent on the conversion type of the event and $p_2=0.8$ in current versions of the IRF.  $p_3$, the high-energy asymptote, is governed by the pitch of the silicon strips in the tracker.  $\sigma_{\mathrm{CALDB}}$ is the parameter value stored in CALDB and is typically of order unity; $p_1$, $p_2$, and $p_3$ also appear in the CALDB.

The PSF must be normalized.  By integrating $f_k(u)$ from $0$ to $\infty$, we see $f_k(u)'\equiv(1-1/\gamma)f_k(u)$ is normalized to 1.  Thus, the final (King function) form of the PSF becomes
\begin{equation}
\label{eq:psf1}
\frac{dN}{d\Omega}\approx \frac{1}{2\pi\,\sigma^2}f_k'[u(\theta)] = \frac{1}{2\pi\,\sigma^2}\left(1-\frac{1}{\gamma}\right)
\left(1+\frac{\theta^2}{2\gamma\sigma^2}\right)^{-\gamma}
\end{equation}

In addition to determining the asymptotic behavior at large $u$, $\gamma$ also determines the transition point at $u=\gamma$.  This dual r\^{o}le for $\gamma$ sets a stringent relation between the tail steepness and the width of the core which is not always satisfied by the data (see Figure \ref{ch3_pre_plot4}).  E.g., improvements of the reconstruction algorithm delivered a PSF with an enhanced core and a ``broken'' power law tail.  A single King function is unable to model this more complex shape.  A new analytic function, the weighted sum of two King functions, on the other hand, adequately models the PSF.  Thus, each energy/$\cos(\theta)$ bin actually has 5 independent parameters, $\sigma_c$, $\sigma_t$, $\gamma_c$, $\gamma_t$, and $w$, i.e., the parameters for a core and a tail King function and the $w$, the relative weighting of the two.

\subsection{Energy Dispersion}
As with the reconstructed position, the reconstructed energy is subject to uncertainty.  At the highest energies, above about $100$ GeV, an appreciable fraction of the energy in the electromagnetic shower escapes the calorimeter, making an energy estimate difficult.  At lower incident energies, a significant fraction of the photon energy may be deposited in the TKR.  In between these two extremes, there is uncertainty due to shower particles missing the active elements of the calorimeter and the natural scatter induced when estimating the deposited energy from the crystal light yield.

\begin{figure}
\begin{minipage}{6in}
\includegraphics[width=6in]{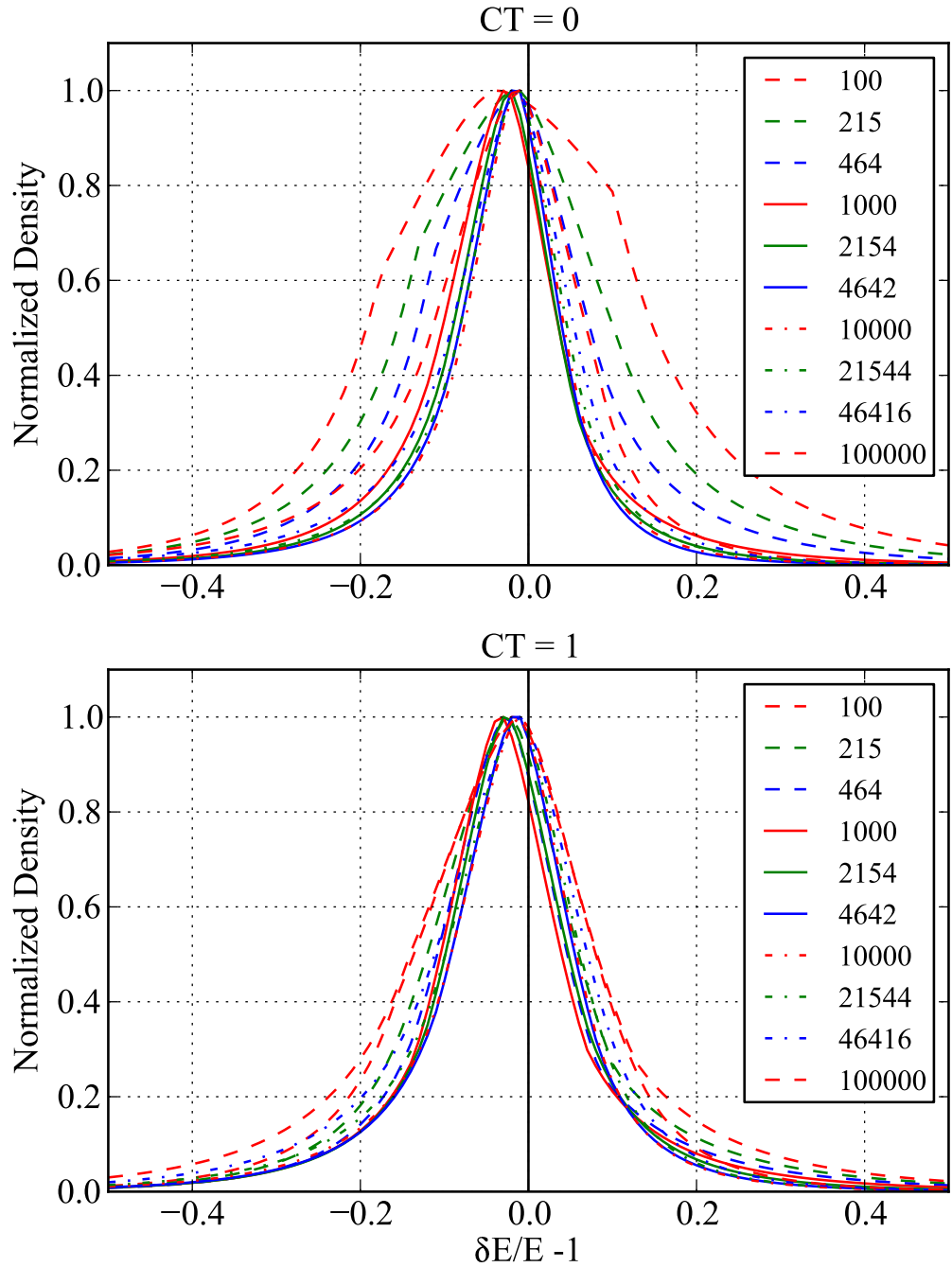}
\end{minipage}
\begingroup\renewcommand{\baselinestretch}{1.0}
\caption[]{The energy dispersion (observed/simulated - 1) for the \emph{P6\_v3\_diff} IRF for events incident on-axis.  The top (bottom) panel shows events converting in the front (back) of the TKR.  At low (order $100$ MeV) energies, the resolution is significantly better for back-converting events since the shower commences closer to the calorimeter, i.e. less energy is deposited in or escapes from the TKR.}
\renewcommand{\baselinestretch}{1.5}\endgroup
\label{ch3_pre_plot5}
\end{figure}

The magnitude of these effects are shown in Figure \ref{ch3_pre_plot5}, where it can be seen that the half-width at half-maximum (HWHM) at most energies is less than $10\%$ but with appreciable tails and rather wider HWHMs at low energy, about $15-20\%$.  This level of uncertainty is smaller than---but not \emph{much} smaller than---the width of the energy bins introduced in Chapter \ref{ch3}, which have a full width of about $30\%$ at 8 bins per decade.

In the following sections, we ignore the effects of energy dispersion by assuming it is negligible.  While we see here that this is not the case, we make this simplifying assumption because it drastically reduces the computational burden of spectral analysis.  (Including the effects of energy dispersion essentially requires an extra quadrature for all computed quantities.)  We expect this approximation may not be too bad since, unlike the PSF, the energy dispersion does not itself depend strongly on energy.  If our effective area were independent of energy, then for power law sources, the only effect of energy dispersion would be a slight softening of the measured slope and a slight increase in the measured flux.  Since the effective area rapidly rises for the first decade of the LAT passband, the energy dispersion does introduce some curvature in the measured spectra and causes \emph{hardening} of measured spectra and \emph{decrease} of measured flux, as we shall see in \S\ref{ch4:sub:energy_dispersion}.  However, the resulting bias on the measured parameters is found to be $<10\%$.

\subsection{Time Dispersion}
The LAT uses GPS for a time reference and measures the absolute time of photon arrival with an accuracy better than 1$\mu$s \cite{smith_timing,lat_instrument}.  The error is negligible for nearly all applications, so we approximate $f_{\mathrm{tdisp}}=\delta(t' - t)$.

\subsection{Approximate Observed Event Rate}
Given these approximations, the modeled observed event rate for a source becomes
\begin{equation}
\label{eq:diffuse_rate}
r(E',\vom',t';\vla) = \int d\Omega \, \mcf(E',t',\vom;\vla)\,A[E',\cos\theta(t')]\,e(E',t')\,f_{\mathrm{psf}}[\vom';\vom,\cos\theta(t'),E'].
\end{equation}
For a point source, we obtain
\begin{equation}
\label{eq:approx_psrate}
r(E',\vom',t';\vla,\vom_0) = \mcf(E',t';\vla) A[E',\cos\theta(t)]\,e(E',t')\,f_{\mathrm{psf}}[\vom' ; \vom_0, \cos\theta(t), E'].
\end{equation}
The simplification delivered for point sources by the accurate timing of the LAT and the neglect of energy dispersion is that we may express the observed rate without \emph{any} integration.  Diffuse sources, on the other hand, require a two-dimensional convolution to calculate the observed rate.

\section{Poisson Likelihood for the Fermi-LAT}

We now outline the formulation of the likelihood for the \fermi.  As with all particle detectors, \fermi is a counting experiment.  Events are binned by the observed quantities, particularly energy and position.  They may also be binned in time, or accumulated into a single bin spanning the length of the observation.  Thus, in the notation of \S \ref{sec:like_principle}, $\vec{X}$ is the set of counts observed in each bin, which we now relabel $\vec{N}$.

Each element of $\vec{N}$ is distributed according to a Poisson distribution with unknown mean, $r_i$.  We model this mean by integrating either Eq. \ref{eq:diffuse_rate} (diffuse sources) or Eq. \ref{eq:approx_psrate} (point sources) over the bin.  Since the components of $\vec{N}$, $N_i$, are mutually statistically independent, the probabily mass function for the data is the product of Poisson distributions with rates $r_i$.  We recall that for a Poisson distribution with mean $r$, the probability to observe $N$ counts is
\begin{equation}
\label{eq:poisson_distribution}
p(N;r) = \frac{r^N}{N!}\exp(-r).
\end{equation}
As noted in the introduction to this chapter, with the broad PSF of the LAT and the strong diffuse background present in the GeV sky, the rate for a bin of phase space involves contributions from multiple sources.  Thus, using the probability mass function in Eq. \ref{eq:poisson_distribution}, we write the (logarithm of the) binned likelihood for all selected data by summing over all $N_{\mathrm{bins}}$ bins and all $N_s$ sources:
\begin{align}
\label{eq:like1}
\log\mathcal{L}(\vla;\vec{N}) &= \sum_{i=1}^{N_{bins}} \left[ -\iiint\limits_{\mathrm{bin_i}} \sum_{j=1}^{N_s} r_j(E',\vec{\Omega}',t';\vla_j) + N_i\log \sum_{j=1}^{N_s}\iiint\limits_{\mathrm{bin}} r_j(E',\vec{\Omega}',t';\vla_j)\right] \\
&\equiv \sum_{i=1}^{N_{bins}} \left[ -\sum_{j=1}^{N_s} R_{ij} + N_i\log \sum_{j=1}^{N_s} R_{ij} \right]
\end{align}
Here, $N_i$ are the observed counts in the $i$th bin, and the triple integrals are over the energy, position, and time range specified for each bin.  In the second line, we have defined $R_{ij}$, the expected number of counts in the $i$th bin from the $j$th source.  We have discarded the $N!$ term from Eq. \ref{eq:poisson_distribution} since it is independent of the model parameters.  This is the form of the likelihood we shall find useful in the following chapters.

Typically, the bins are continguous, and we define the \emph{region of interest} (ROI) to be the portion of the total data (the volume of phase space) we have selected.    The likelihood then becomes
\begin{equation}
\label{eq:like2}
\log\mathcal{L}(\vla;\vec{N}) = -\iiint\limits_{\mathrm{ROI}} \sum_{j=1}^{N_s} r_j(E',\vec{\Omega}',t';\vla_j) + \sum_i^{N_{bins}} N_i\log \sum_{j=1}^{N_s}R_{ij}.
\end{equation}
This formulation facilitates the introduction of \emph{unbinned} Poisson likelihood, in which the bin widths are taken to be infinitesimal, such that only 0 or 1 counts can possibly be observed.  This formulation sacrifices no information to binning but can become prohibitive for large data sets.  The unbinned likelihood, with $\vec{N}$ now containing \emph{every} count, is given by,
\begin{equation}
\label{full_likelihood}
\log \mathcal{L}(\vla,\vec{N}) = - \iiint\limits_{\mathrm{ROI}} \sum_{j=1}^{N_s} r_j(E',\vec{\Omega}',t';\vla_j) + \sum_{i=1}^{N_{\mathrm{events}}} \log\sum_{j=1}^{N_s} r_j(E'_i,\vec{\Omega}'_i,t'_i;\vla_j),
\end{equation}
where $E'_i$ denotes the reconstructed energy of the $i$th event, etc.

\section{Summary}
We have motivated the use of likelihood principles in GeV astronomy by demonstrating that---at least for current telescopes---it is impossible to make reliable measurements of astrophysical quantities directly from the data.  Instead, we noted that GeV sources can be easily modeled with only a few parameters and suggested the method of maximum likelihood to estimate these parameters.  To calculate the likelihood, we fold the parameterized models through the instrument response function, thus naturally accounting for the complex IRF, source confusion, and the constantly-changing orientation of the instrument.  We outlined the features of the IRF and provided a convenient analytic approximation for the PSF in the form of the King function, and we argued that we could ignore the effects of energy dispersion at the cost of some bias to our measured parameters.  Finally, we provided an expression for the log likelihood appropriate for \fermi (Eq. \ref{eq:like1}) and for the expressions for the observed rates from diffuse sources (Eqs. \ref{eq:diffuse_rate} and \ref{eq:approx_psrate}) whose evaluation shall be our endeavor in the next chapter.

\chapter{The \emph{Pointlike} Package: Design and Implementation}
\label{ch3}

As outlined in Chapter \ref{ch3_pre}, likelihood analysis is the optimal method for characterizing $\gamma$-ray sources.  (We discuss in Chapter \ref{ch4} likelihood techniques for source \emph{detection} as well.)  Unfortunately, likelihood analysis is complex.  One must manage source models, descriptions of the instrument, and data.  Forward folding the models into ``data space'' can be computationally intensive.  Analysis for a single source (which, due to the PSF-broadening, entails analysis of many sources!) can be time consuming.  All-sky analysis requires large computer clusters to handle the computational burden.  However, by binning the data with scalable pixels and sacrificing some accuracy for time-saving approximations, this situation can be ameliorated.  In this section, we describe the \ptl software package, a collection of routines for performing maximum likelihood (ML) analysis of \fermi data, which aims for the canonical goal of ``making easy things easy and hard thing possible''.

At the highest level, \ptl is a collection of Python modules designed to be used both interactively (for analyzing a single source) and in batch mode (for analyzing many sources).  \ptl also relies internally on a large C++ codebase via SWIG\footnote{http://www.swig.org}, a tool providing wrapper code to access compiled libraries from interpreted languages.  The interface and organization are more properly left to documentation of the code, and in this chaper, we discuss the ``nuts and bolts'' of the implementation.

\section{\fermi Data}
For all the complexity of the instrument, the high-level \fermi data is simple, a list of meausured quantities from reconstructed events.  In the following section, we provide details about this list, our method for binning the events, and some necessary cuts that must be made to reject background.

\subsection{High-level Data Products}
Information about each event is telemetered from the S/C to a batch farm at SLAC for event reconstruction, in which the low-level hardware readouts are converted into estimates of physical quantities\cite{lat_instrument}: the energy, position, and time of arrival of the putative $\gamma$ ray.  There are two high-level data products.  The \emph{FT1} file is an event list in FITS format with columns containing reconstructed quantities (energy, R.A., Decl., time, etc.) for each event.  The \emph{FT2} FITS file tabulates the position and orientation of the S/C at 30-second intervals, information needed to construct the time-dependent IRF.
 
\subsection{Binning the Data: HEALPix}
\label{ch3:subsec:binning}
Binning \fermi data is not a trivial task.  Due to the energy-dependence of multiple scattering in the tracker, the LAT's angular resolution varies by nearly two orders of magnitude over its energy range (Figure \ref{ch3_pre_plot1}).  If one uses only a single size for position bins---e.g., the EGRET choice of $0.5^{\circ}\times0.5^{\circ}$\cite{mattox}---then one faces a tradeoff: lose information at high energies by using bins much larger than the ultimate instrumental resolution, or increase the computational burden by using bins much too small for the poor angular resolution at low energies.

We address this problem with HEALPix\cite{healpix}, a scheme for tessellating the sky with equal-area pixels\footnote{HEALPix also describes a particular projection used in demarcating the pixels.  We do not make explicit use of this feature.}.  The base tessellation comprises 12 pixels, and finer binnings are achieved by subdividing the base pixels; if there are $\nside$ subdivisions on a side of the base pixels, there are $12\times\nside^2$ pixels.  $\nside$ thus controls the granularity of the pixelization.  (A useful relation is that a pixel is approximately $60^{\circ}/\nside$ on a side.)  The finest binning possible is limited by computer architecture, as each pixel must be uniquely mappable to an integer.  We enforce $\nside\leq8192$\footnote{$8192$ is the largest power of two for which $12\times\nside^2 < 2^{31}$, the largest value of a signed integer available on 32-bit architecture.}, yielding a maximum resolution of about $30''$, more than adequate for the finest resolution PSF of the LAT. 

In addition to the finely-controlled pixel size, HEALPix facilitates sparse binning.  Due to the falling spectra of $\gamma$-ray sources and the flat effective area of \fermi, there are relatively few high energy photons and most pixels are empty.  By assigning each pixel a unique index, the data can take the form of a mapping of index to counts, and we need only store those indices with non-zero counts.

Data are initially binned into front-converting (those that pair produce in the forward section of the TKR with the thin tungsten foils) and back-converting (producing the first pair on a thick tungsten foil deep in the TKR) events.  This distinction is maintained throughout the \ptl package, and a distinct IRF is used for each class.  Data are next divided into uniform bins in logarithmic observed energy.  (Logarithmic energy is the natural choice for broadband spectra, which often look like power laws.)  By default, we use either 4 or 8 bins per decade, with the left edge of a bin at 100 MeV.  Such energy bins align with that used in CALDB for the PSF parameters and allow the use of the same PSF parameters over an entire bin width.

\begin{figure}
\begin{minipage}{6in}
\includegraphics[width=6in]{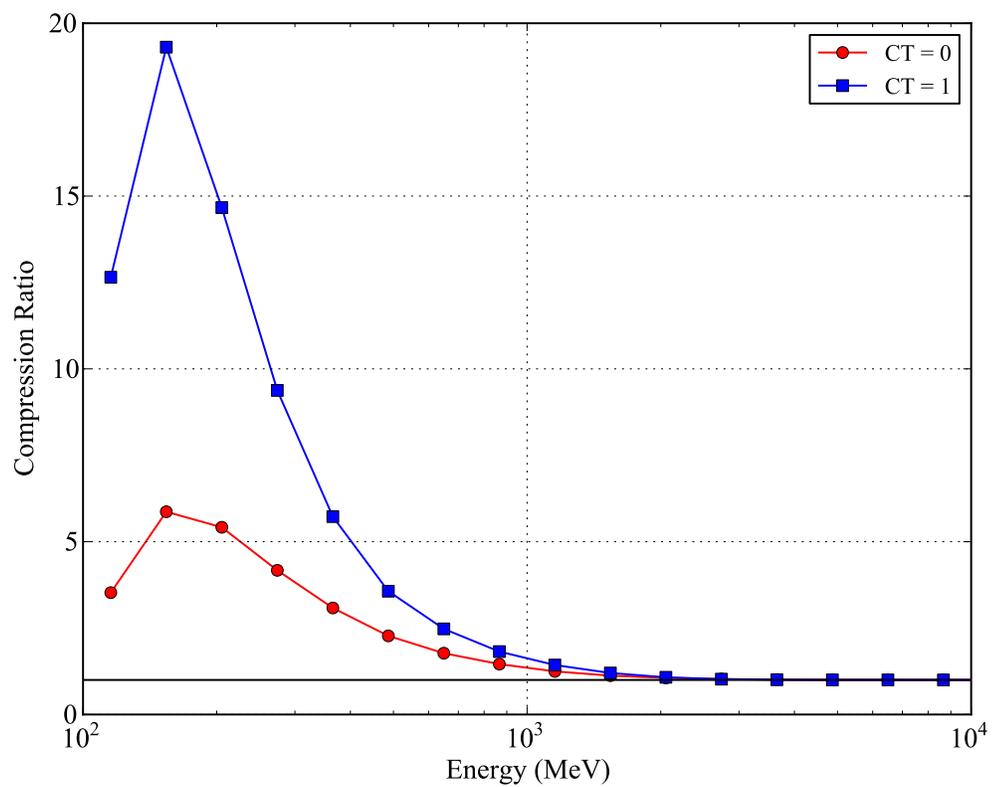}
\end{minipage}
\begingroup\renewcommand{\baselinestretch}{1.0}
\caption{The compression ratio of the data, i.e., the mean photons per pixel, for a given energy band.  The occupancy drops rapidly as the energy approaches 1 GeV due to the improving PSF and falling spectra of astrophysical sources.}
\renewcommand{\baselinestretch}{1.5}\endgroup
\label{ch3_plot6}
\end{figure}

For each energy bin (and conversion type), the $\nside$ parameter is chosen so that the pixels are somewhat smaller than the core of the PSF at that energy.  Precisely,
\begin{equation}
\label{eq:nside_scheme}
\nside = \frac{8192}{1 + 8192 / N_0 \times (E/100\,\mathrm{MeV})^{-0.8} \times \exp -(E/2000\,\mathrm{MeV})^2},
\end{equation}
with $N_0=86$ ($N_0=52$) for front (back) events.  This binning puts roughly 5 pixel widths within the $68$\% containment radius of the PSF at the given energy, up to about 1 GeV.  At 1 GeV, this binning rapidly asymptotes to $\nside=8192$.  At 1 GeV, the PSF-based pixels are sufficiently small and the data sufficiently sparse that the mean pixel occupancy drops to 1.  We gain no computational savings by using PSF-based pixels, and so use the smallest pixels possible.  Above 1 GeV, the \ptl binning becomes essentially unbinned in position.  (The binning in energy persists, of course, as this binning allows us to define a coarse-grained PSF.)  In Figure \ref{ch3_plot6}, we show the mean pixel occupancy for 18 months of data using the scheme in Eq. \ref{eq:nside_scheme}.  The compression ratio at low energies is large, especially for back-converting events, and will grow linearly with time.  (Essentially every pixel at low energies is occupied.)  For all energies, the total compression ratio is about 2.8.

The ``unbinned limit'' is relatively robust against loss of efficiency with increasingly large data sets.  The primary worry is if, by selecting very small pixel sizes, we prevent the mean occupancy of a particular band from increasing above 1, thus gaining performance.  However, the generally steep spectra of GeV sources and the sharpening PSF conspire to prevent this.  At 1 GeV, for instance, the scheme of Eq. \ref{eq:nside_scheme} gives $\nside= 644$ ($401$) for front (back).  These are approximately PSF-sized pixels, and there are 4.98 and 1.93 million pixels, respectively, for these $\nside$ settings, while for an 18-month data set, only 0.36 and 0.33 million of these pixels, respectively, are occupied.  Even in the extremely conservative limit that the number of occupied pixels grows with time, we must collect 9 years of data to reach a non-sparse pixelization for 1 GeV.  \emph{A fortiori} for higher energies with fewer photons and smaller PSF-based pixels.

\begin{figure}
\begin{minipage}{6in}
\includegraphics[width=6in]{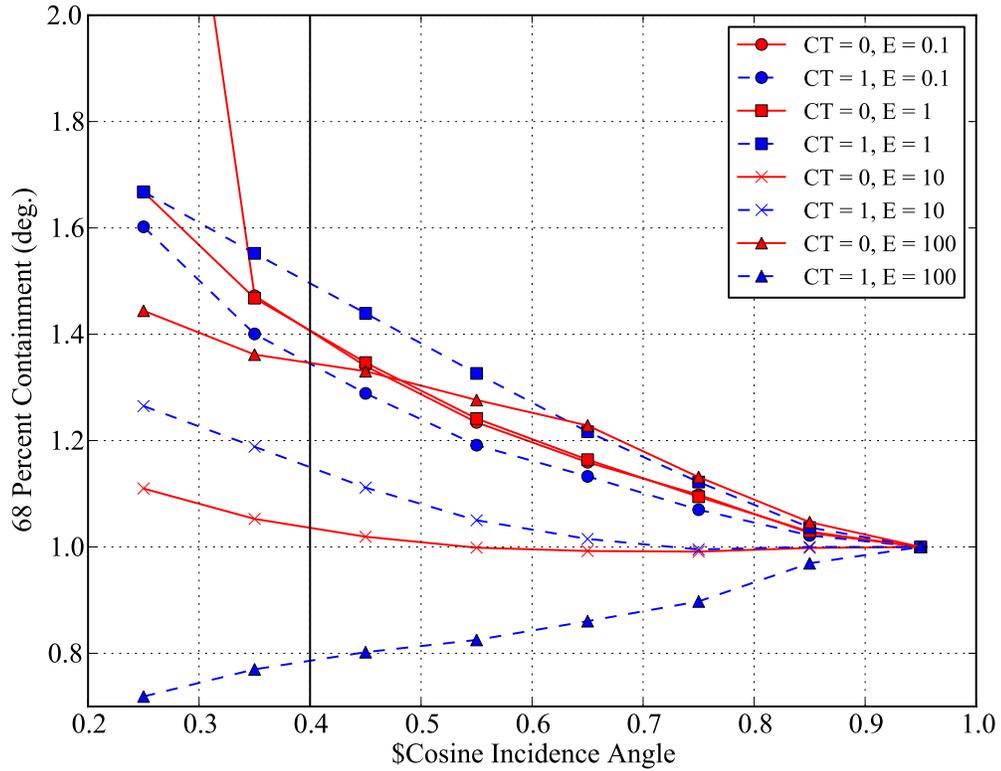}
\end{minipage}
\begingroup\renewcommand{\baselinestretch}{1.0}
\caption{The PSF---characterized by the radius in which 68\% events are enclosed---as a function of incidence angle, energy (in GeV), and conversion type (``CT'' in legend).  Generally, the angular resolution is best on-axis and degrades as the edge of the field-of-view is approached.  Using a larger containment radius, e.g., 95\%, as a metric yields nearly the same dependence on incidence angle and magnitude of shift.  The standard cut on reconstructed incidence angle for \ptl of $\cos(\theta)=0.4$, or $>66.4^{\circ}$, is indicated by the vertical black line.}
\renewcommand{\baselinestretch}{1.5}\endgroup
\label{ch3_plot1}
\end{figure}

During the binning process, we make additional cuts.  An important facet of the binned analysis is that the PSF not depend too much on incidence angle, so that a ``collective'' PSF for a bin makes sense.  This is generally true (see Figure \ref{ch3_plot1}); the variation of the PSF with incidence angle is on the order of $10\%$\footnote{Although the PSF containment radius increases to up to $50\%$ of the on-axis value, the effective area is approximately proportional to $\cos(\theta)$ and very few events are collected with this large PSF.}.  We typically reduce the variation even more by discarding events with a reconstructed incidence angle $>66.4\dg$.  Thus, a conservative choice of a pixel size appropriate to the on-axis PSF will remain efficient for all incidence angle. 

\begin{figure}
\begin{minipage}{6in}
\includegraphics[width=6in]{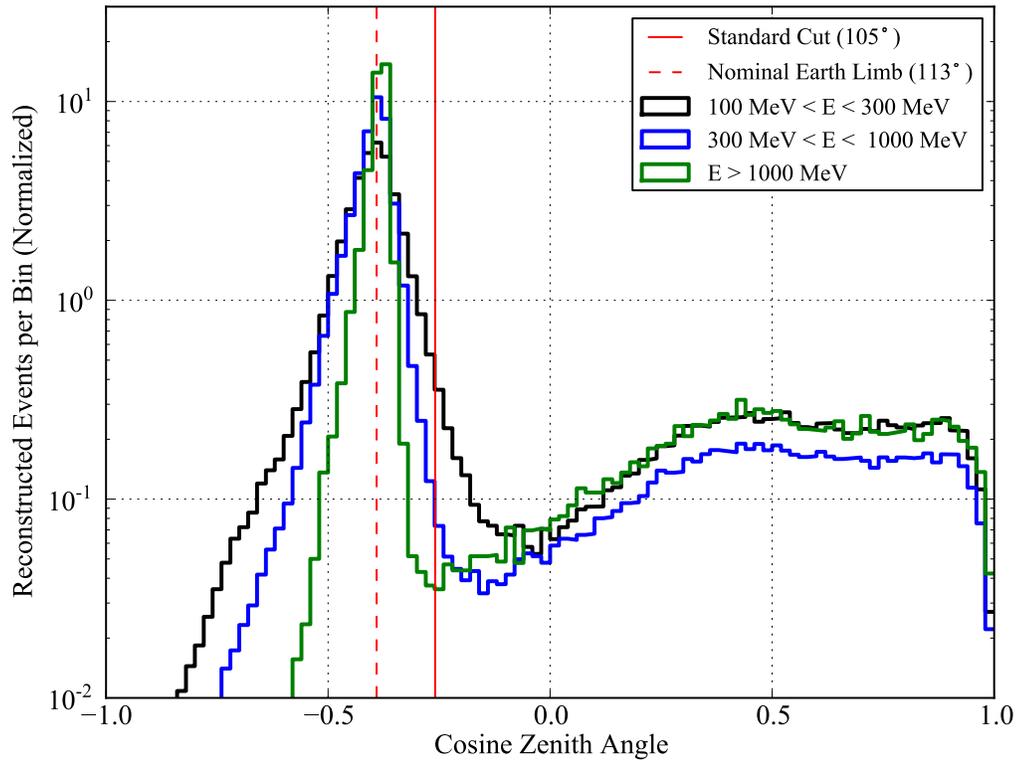}
\begingroup\renewcommand{\baselinestretch}{1.0}
\caption{Reconstructed events (P6 ``diffuse'' class) with 100 MeV $<$ E $<$ 100 GeV in a $10^{\circ}$ aperture centered on (l,b) = (0,-45), integrated from Aug. 4 2008 to Jul. 18 2010.  The binning is in the reconstructed zenith angle.  The effect of the PSF is clearly seen in the broadening of the distribution of limb photons about $\approx113\dg$.}
\renewcommand{\baselinestretch}{1.5}\endgroup
\label{ch3_plot2}
\end{minipage}
\end{figure}

Another cut is made to remove most of the $\gamma$ rays produced by cosmic ray interactions in Earth's upper atmosphere\cite{fermi_limb}.  Ultrarelativistic cosmic rays lead to a $\gamma$-ray cross-section in the Earth frame strongly peaked in the forward direction, and so from the point-of-view of the S/C, the intensity is approximately proportional to the total grammage of the atmosphere along the line-of-sight, i.e, the emission is dominated by the limb of the earth.  The polar angle of the limb as measured from the zenith is given by 
\begin{equation}
180^{\circ} - \arcsin\left(\frac{R_{\mathrm{earth}}}{R_{\mathrm{earth}}+R_{\mathrm{Fermi}}}\right),
\end{equation}
about $113^{\circ}$ from the zenith for Fermi's altitude of ~550km.  This number varies slightly (linear in the altitude) during the orbit, which is slightly elliptical, and decreases as the S/C slowly loses altitude to atmospheric drag.  Although the position of the limb is independent of the S/C orientation, more albedo photons are successfully reconstructed when more of the limb is in the field of view, i.e., when the S/C is at larger rocking angles.  At energies $>100$ MeV, the reconstructed zenith angle of albedo photons is generally $>105^{\circ}$ (see Figure \ref{ch3_plot2}); only events in the tails of the PSF at the lowest energies ``disperse'' more than $8^{\circ}$ from their true direction\footnote{In fact, a nontrivial---relative to \emph{astrophysical} source rates---number of limb photons with energies of order 100 MeV survive this cut.  The remaining photons contribute a non-isotropic foreground to astrophysical sources, and should be accounted for in background modeling.  [The contamination is increased with a larger rock angle.]} We therefore apply a cut on the reconstructed zenith angle $>105^{\circ}$.

Since the cuts on incidence and zenith angles use the reconstructed photon position, some events that would be cut if the true position were known are included, and others that would pass a cut in the limit of perfect reconstruction are discarded.  In addition to such contamination, a small bias results.  For instance, since the effective area increases approximately linearly with the cosine of the incidence angle, a cut on incidence angle will necessarily discard more events whose true incidence angle is less than the cut value.  Since the limb tends also to be on the edge of the FOV, similar remarks apply.  These effects are small, and we will ignore them.

Finally, we only allow events that lie within a \emph{Good Time Interval} (GTI). These GTI form an additional table in the \emph{FT1} file and are simply a list of tuples with the start and stop time of an interval during which we accept events.  By default, the GTI are determined by SAA passages.  (This cut is trivial since no data is taken during SAA passage.)  However, additional temporal cuts may be made based on the S/C position and orientation.  For instance, as the S/C rocks, it may overshoot its target rock angle and spend time at a larger rock angle with higher albedo background.  Similar excursions are caused when the S/C makes a pointed observation.  Such times can be excluded from the GTI.  Gamma-ray bursts (background for other sources) may also be so excluded.  Finally, when studying a particular source, in order to further reduce contamination from albedo events, times when the ROI is too close to the limb can be excluded.

\section{Point Sources Rates}

The data, whose treatment we have just discussed, and the source models (discussed in Chapter \ref{ch3_pre}) form the ``bookends'' of the likelihood analysis, and we have now to link the two by performing the integrals in Eq. \ref{eq:like1} and folding the intrinsic source rates through the instrument response function.  These integrals represent the primary computational (and infrastructural) difficulty in likelihood spectroscopy, and we devote some effort making time-saving approximations when possible while maintaining good overall accuracy.

There are two fundamental source classes: those that can be spatially resolved (diffuse sources) and those that cannot (point sources).  The former includes the Galactic diffuse background due to cosmic rays, the extragalactic background due to distant sources below detection threshold, nearby galaxies such as the Magellanic clouds, supernova remnants, and some pulsar wind nebulae.  The class of point sources includes the active nuclei of galaxies, pulsars, and small or distant remnants and nebulae.  Since diffuse sources require convolution with the PSF, the two classes are treated independently, and we discuss point sources and diffuse sources in turn.

We begin with Eq. \ref{eq:approx_psrate}, which we recall is the differential (in phase space) expected rate from a point source, neglecting energy dispersion.  To compare this to observations, we need to integrate it over each of the bins we have selected for the data.  We begin with integration over observed energy and time, i.e., we calculate the quantity
\begin{equation}
r_i(\vom') = \int dE' \int dt' \mcf(E',t';\vla) A[E',\cos\theta(t)]\,e(E',t')\,f_{\mathrm{psf}}[\vom' ; \vom_0, \cos\theta(t), E'],
\end{equation}
the expected rate per unit solid angle in the $i$th bin.  (We leave the particular energy and time abstract; the bounds of the integrals over energy and time are implicit and extend over the $i$th bin.)  For brevity, we suppress the prime notation in the sequel.

\subsection{Integral Over Time}
\label{ch3:subsec:time_integral}

In order to evaluate this expression quickly, we make approximations and assumptions that allow us to pre-compute some quantities and simplify the integral.  First, we assume that $\mcf(E,t;\vla) = \mcf(E;\vla)$, i.e., that the source is stationary during the time range specified by the bin.  Alternatively, we can view this assumption as measuring the time-averaged flux from the source\footnote{In fact, due to the time-dependent IRF, these statements are only equivalent to first order.}.  If we are interested in capturing the time dependence of the source, we simply use multiple time bins and proceed as before.

Next, we write $e(E,t) = e_0(t) + e_1(E,t)$.  $e_0$ is the dominant contribution to the efficiency, determined by the deadtime (trigger rate) and is energy independent.  $e_1$ is the second-order correction in which the reconstruction effectiveness changes based on the charged-particle background.  To a very good approximation, this correction depends linearly on the trigger rate, $e_0$, i.e., $e_1(E,t) = f_1(E)e_0(t) + f_2(E)$, and we can reconstruct the full quantity from the simple-to-measure $e_0$.

Now, we break up the integral over time into sub-integrals,
\begin{equation}
\int dt \rightarrow \sum_{i=1}^N \int_i dt,
\end{equation}
where each integral is over one of the (typically) 30-s intervals laid out in the FT2 file, and there are $N$ such intervals in the parent bin.  The sampling rate of the FT2 file is designed to be sufficiently high that the S/C orientation\footnote{This assumption is only marginally accurate when the S/C slews from its north-pointing attitude to its south-pointing attitude, but such a slew only occurs once per orbit, and $<10\%$ of the livetime is accumulated during a slew.} does not change much over the course of an interval.  Therefore, we evalute each sub-integral with a central-value approximation, and the time integral becomes
\begin{equation}
\sum_{i=1}^N \Delta t_i\,A[E,\cos\theta(t_i)]\,e(E,t_i)\,f_{\mathrm{psf}}[\vom ; \vom_0, \cos\theta(t_i), E],
\end{equation}
with $t_i$ the center of the $i$th FT2 bin and $\Delta\,t$ is the bin width, typically 30 seconds.

Next, we discretize $\cos\theta$, typically into 40 bins spanning from 0.2 to 1.0\footnote{The LAT field of view is $\cos\theta\geq 0.2$.}.  The time integral becomes
\begin{equation}
\sum_{j=1}^{N_{\cos}} \sum_{i=1}^N\Delta t_i\,\mathcal{I}_{\cos\theta_j}[\cos\theta(t_i)] A[E,\cos\theta(t_i)]\,e(E,t_i)\,f_{\mathrm{psf}}[\vom ; \vom_0, \cos\theta(t_i), E],
\end{equation}
with $\mathcal{I}_{\cos\theta_j}[\cos\theta(t_i)]$ the indicator function that evaluates as 1 if $\cos\theta(t_i)$ is in the $j$th $\cos\theta$ bin and 0 otherwise.  We define the livetime, $\tau_j(E)\equiv \sum_{i=1}^N \Delta t_i\,\mathcal{I}_{\cos\theta_j} e(E,t_i)$ and the exposure, $\epsilon_j(E) = \tau_j(E) \times A[E,\cos\theta_j]$, in terms of which the time integral becomes
\begin{equation}
\sum_{j=1}^{N_{\cos}} \epsilon_j(E)\, f_{\mathrm{psf}}[\vom ; \vom_0, \cos\theta_j, E],
\end{equation}
or in terms of the summed exposure,
\begin{equation}
\epsilon(E) \sum_{j=1}^{N_{\cos}} \epsilon_j(E)/\epsilon(E) f_{\mathrm{psf}}[\vom ; \vom_0, \cos\theta_j, E].
\end{equation}
We define the exposure-weighted PSF, $f_{\mathrm{wpsf}}(\vom ; \vom_0, E) = \sum_{j=1}^{N_{\cos}} \epsilon_j(E)/\epsilon(E) f_{\mathrm{psf}}[\vom ; \vom_0, \cos\theta_j, E]$, and the time integral becomes, finally,
\begin{equation}
\epsilon(E,\vom_0)\, f_{\mathrm{wpsf}}[\vom ; \vom_0, E].
\end{equation}
The immense benefit of all of this definition legerdemain is that the livetime calculation is entirely independent of the source.  We can pre-compute the livetime for some pixelization on the sky\footnote{$1\dg$ bins are typical} and have it available for all subsequent analyses using the same data.  The time integral is thus entirely eliminated in favor of a sum over $\approx40$ bins in incidence angle, a great improvement over the original sum over the S/C pointing history containing potentially millions of terms!

\subsection{Integral over Energy}
\label{ch3:subsec:energy_integral}
The bin rate is now
\begin{equation}
\label{eq:psrate_with_exposure}
r_j(\vom) = \int dE\, \mcf(E,\vla)\, \epsilon(E,\vom_0)\, f_{\mathrm{wpsf}}(\vom ; \vom_0, E).
\end{equation}
As it stands, the expression is still computationally burdensome, as we would need to evaluate the PSF at each data pixel over many energies.  To further simplify, we appeal to the Mean Value Theorem, according to which $\int dx\, f(x)\, g(x) = f(x') \int dx\, g(x)$, for some $x'$ within the bounds of integration.  Thus, if we determine the appropriate energy, $\eopt$, at which to evaluate $f_{\mathrm{wpsf}}$, we can remove the PSF from the energy integral, obtaining
\begin{equation}
\label{eq:eopt_psf}
r_j(\vom) = f_{\mathrm{wpsf}}(\vom ; \vom_0, \eopt) \int dE\, \mcf(E,\vla)\, \epsilon(E,\vom_0).
\end{equation}

Unfortunately, it is not possible to find a unique $\eopt$.  Although within a given energy bin, the energy dependence of the PSF is entirely given by Eq. \ref{eq:prescale_sigma}, the energy dependence of the King function depends on position, i.e., $\eopt$ is a function of $\theta=\arccos(\vom\cdot\vom_0)$.  Indeed, for small $\theta^2$, $f\propto\sigma^{-2}\propto E^{-1.6}$, while for large $\theta^2$, $f\propto\sigma^{-2(\gamma-1)}\approx\sigma^2\propto E^{1.6}$.

Rather than introducing an additional angular dependence through $\eopt$, we seek $\eopt$ that gives the best representation of the PSF over a wide range of source spectra and positions.  The natural metric for this is the log likelihood, Eq. \ref{eq:like1}.  We concentrate on a single source centered in the ROI and write the expected counts for a given band as
\begin{align}
&\int d\Omega\int dE\, \mcf(E,\vla)\, \epsilon(E,\vom_0)\, f_{\mathrm{wpsf}}(\vom ; \vom_0, E) \\
\approx \pi & \int_0^{\theta^2_{max}} d\theta^2 \int dE\,\mcf(E,\vla)\, \epsilon(E,\vom_0)\,f_{\mathrm{wpsf}}(\theta^2,E) \\
\equiv \pi &\int_0^{\theta^2_{max}} d\theta^2 \int dE\, N(E) f_{\mathrm{wpsf}}(\theta^2,E) \\
\equiv &\int_0^{\theta^2_{max}} d\theta^2\, g(\theta^2),
\end{align}
where the symmetry of the centered source allows the trivial azimuthal integral and we have made a small angle approximation.  We now define an approximate rate using the ``mean'' Mean Value Theorem,
\begin{align}
\pi &\int_0^{\theta^2_{max}} d\theta^2 \int dE\, N(E) f_{\mathrm{wpsf}}(\theta^2,E) \\
= &\pi\int_0^{\theta^2_{max}} d\theta^2 f_{\mathrm{wpsf}}(\theta^2,\eopt(\theta^2))\times \int dE\, N(E)\\
\approx &\pi\int_0^{\theta^2_{max}} d\theta^2 f_{\mathrm{wpsf}}(\theta^2,\eopt)\times \int dE\, N(E)\\
\equiv &\int_0^{\theta^2_{max}} d\theta^2 f(\theta^2,\eopt)
\end{align}
Now, the log likelihood for the energy band becomes
\begin{align}
\log\mathcal{L}_1 &= \int_0^{\theta^2_{max}} d\theta^2\, g(\theta^2) - \sum_{i=1}^{N} n_i \log \int_{\theta^2_{i1}}^{\theta^2_{i2}} d\theta^2 g(\theta^2) \\
& \rightarrow \int_0^{\theta^2_{max}} d\theta^2\, g(\theta^2)\times[1 - \log g(\theta^2)],
\end{align}
where in the second line we have replaced the observed counts with the expected rate and gone to the unbinned limit\footnote{The differential in the $\log$ term is not problematic, as it is multiplied by a second differential, and $\lim_{d\theta^2 \to 0} d\theta^2\log d\theta^2 = 0$.}.  In terms of the approximate rate, the log likelihood is
\begin{equation}
\log\mathcal{L}_2(\eopt) = \int_0^{\theta^2_{max}} d\theta^2\, N\,f(\theta^2,\eopt)\times[1 - \log N\,f(\theta^2,\eopt)],
\end{equation}
and now we simply solve the integral equation $\log\mathcal{L}_1=\log\mathcal{L}_2(\eopt)$ for $\eopt$.

\begin{figure}
\begin{minipage}{6in}
\includegraphics[width=6in]{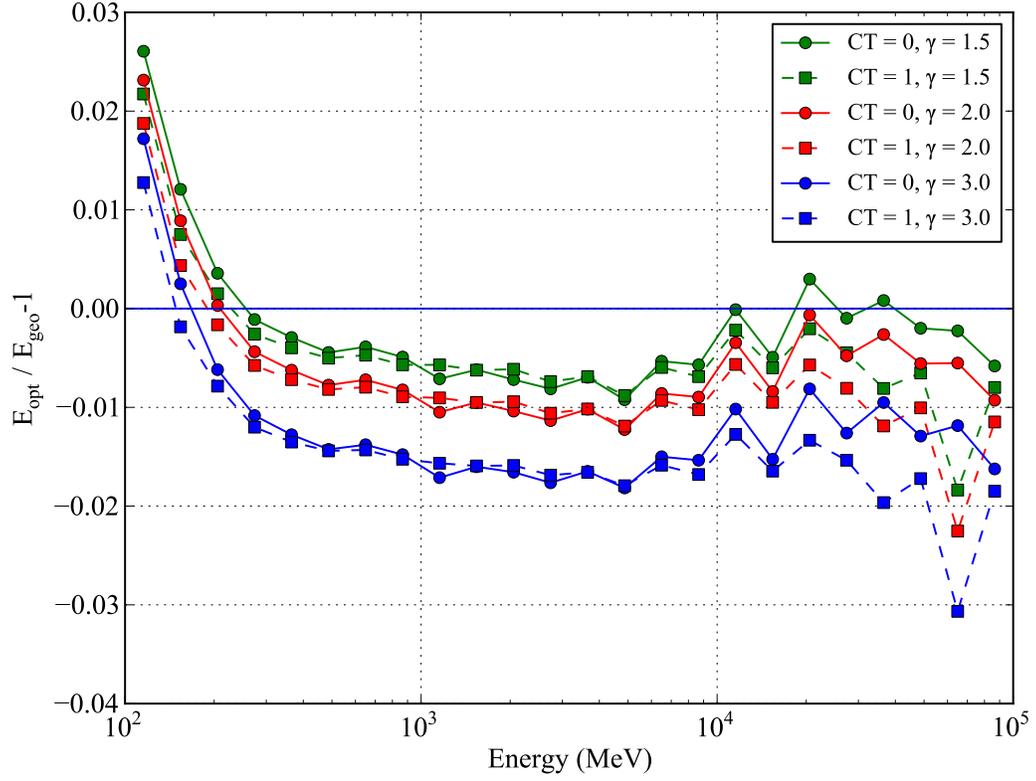}
\end{minipage}
\begingroup\renewcommand{\baselinestretch}{1.0}
\caption{Comparison of the geometric mean ($E_{\mathrm{geo}}\equiv \sqrt{E_{\mathrm{max}}\times E_{\mathrm{min}}}$) of an energy band with the optimal energy $E_{\mathrm{opt}}$ as estimated by the log likelihood method described in the text.  The binning is uniform in logarithmic energy with 8 bins per decade and separate bands for front- and back-converting events are shown.  The relative difference of $E_{\mathrm{opt}}$ from $E_{\mathrm{geo}}$ for three different power law source spectra with photon indices $\gamma = 1.5,\,2.0,\,3.0$ is shown.}
\renewcommand{\baselinestretch}{1.5}\endgroup
\label{plot8}
\end{figure}

In practice, $\eopt$ is within a few percent of the geometric mean, $E_{\mathrm{geo}}\equiv \sqrt{E_{\mathrm{max}}\times E_{\mathrm{min}}}$, of the energy band, and we obtain a sufficiently good solution for $\eopt$ with a single Newton-Raphson iteration,
\begin{equation}
\label{eq:eopt_ref}
\eopt \approx E_{\mathrm{geo}} + \left[\log\mathcal{L}_1-\log\mathcal{L}_2(E_{\mathrm{geo}})\right]\times \left[\frac{d\,\log\mathcal{L}_2(\eopt)}{d_{\eopt}}(E_{\mathrm{geo}})\right]^{-1}.
\end{equation}
We apply this procedure to obtain an estimate of $\eopt$ for each energy band (and event conversion type).  The results for a typical ROI (centered on the position of the Vela pulsar) are shown in Figure \ref{plot8}.  We see that, for a variety of source spectra, $E_{\mathrm{opt}}$ is within a few percent of $E_{\mathrm{geo}}$.  The largest discrepancy is at energies of a few hundred MeV.  Here, the effective area rises rapidly (Fig. \ref{ch3_pre_plot2}), meaning counts in the band are shifted towards higher energies, yielding a better effective PSF than $E_{\mathrm{geo}}$ implies.  The same effect causes the dependence on spectral index: softer sources emphasize lower energies, shifting $E_{\mathrm{opt}}$ lower.  The overall trend for an $\eopt$ lower than $E_{\mathrm{geo}}$ can then be understood to have its origin in the $1/E$ (uniform in logarithmic energy) spectrum used to determine the PSF.  In any case, at all energies, the $E_{\mathrm{opt}}$ for the different spectra are within a few percent of each other.  Therefore, we adopt the omnibus prescription of calculating $\eopt$ for a photon index ($\Gamma$) of 2.0.  We present empirical results showing the improvement of the PSF representation using the ``$\eopt$'' prescription in \S\ref{checking_band_shape}.

We now begin the final step.  We have the expected rate per unit solid angle, and we see from Eqs. \ref{eq:like1} and \ref{eq:like2} that we need to integrate this rate over both the entire ROI and over each data pixel (the terms on the right- and lefthand side of Eq. \ref{eq:like2} respsectively).  Recalling from the discussion on binning (\S\ref{ch3:subsec:binning}) that we choose $\nside$ such that our pixels are small compared to the PSF, we use a simple central value approximation to evaluate the righthand side terms.  The lefthand side terms, the integrals over the ROI for a given energy band, become
\begin{align}
\label{eq:overlap}
r(\vla) & = \left[\int d\Omega f_{\mathrm{wpsf}}(\vom ; \vom_0, E_{\mathrm{opt}})\right]\left[ \int dE\, \mcf(E,\vla)\, \epsilon(E,\vom_0)\right] \\
        &  \equiv \left[\int d\Omega f_{\mathrm{wpsf}}(\vom ; \vom_0, E_{\mathrm{opt}})\right] N(\vla) \\
        &  \equiv O(\eopt,\vom_0)\,  N(\vla).
\end{align}
Here, we have defined  $O(\eopt,\vom_0)$, the \emph{overlap integral}.  The overlap integral is simply the integral of the PSF over the ROI for a particular source.  It must be between 0 and 1, and it depends only on the source position, the PSF for the energy band as obtained through the method discussed above, and the radius of the ROI.  The payoff of the decoupling of the energy and position integrals we achieved by selecting $\eopt$ is evident when we consider the log likelihood for the energy band,
\begin{equation}
\label{eq:separated}
\mathcal{L}(\vla) = -\sum_{j}^{N_s}O(\eopt,\vom_j)\,N_j(\vla) + 
\sum_i^{N_{bins}} N_i \log \sum_j^{N_s}f_{\mathrm{wpsf}}(\vom_i ; \vom_j, \eopt) N_j(\vla),
\end{equation}
where we see that the dependence on $\vla$ comes only through the $N(\vla)$ terms.  Since only the $N(\vla)$ terms change during the likelihood maximization, we are free to pre-compute once and for all the overlap integrals \emph{and} the data-pixel contributions for each source.  We have already discussed the latter calculation; we now proceed with a description of the terms on the lefthand side of Eq. \ref{eq:separated}.

\subsubsection{Evaluating the Overlap Integrals}
While we make the central-value approximation to evaluate the PSF integral over the data pixels, the overlap integrals are more complex.  First, we assume that the ROI is circular.  Through this assumption and the azimuthal symmetry (for sufficiently long integrations) of the PSF, we can reduce the 2-d integral to a quadrature.  Let the ROI center be denoted $\vom_0$ and the source position be $\vom_s$.  By the symmetries of the ROI and PSF, the only relevant quantities are the arclength separating the source and the ROI center, denoted $\delta\equiv\arccos \vom_0\cdot\vom_s$, and the radius of the ROI, $R$.  In terms of these quantities,
\begin{align}
O(\eopt,\delta,R) &= \int\limits_{ROI} d\Omega\, f_{\mathrm{wpsf}}(\vom ; \vom_0, E_{\mathrm{opt}}) = \iint\limits_{ROI}d\phi\, \frac{d\theta^2}{2}\frac{1}{2\pi\sigma^2}\, f_{wk}\left(\frac{\theta^2}{2\pi\sigma^2}\right) \\
 & = \int\limits_{ROI}\frac{d\phi}{\pi}\int_{u_1(\phi)}^{u_2(\phi)} du\, f_{wk}(u) = \int\limits_{ROI}\frac{d\phi}{\pi}\, F_{wk}[u_2(\phi)] - F_{wk}[u_1(\phi)],
\end{align}
where we have introduced the explicit weighted-sum of King functions for the PSF.  (As discussed in \S\ref{ch3_pre:subsec:psf}, some versions of the LAT IRF express the PSF as the sum of two King functions.  The overlap integral is linear in the King function, so the derivation presented here carries over directly.  Likewise, $F_{wk}$ is an exposure-weighted sum, subject to the same linear arguments.)  $F_{wk}$ denotes the (analytic) cumulative distribution of the King function, and we have achieved reduction to quadrature.  All that remains is to determine the bounds of integration, i.e., $u_1(\phi)$, $u_2(\phi)$, and the bounds of the azimuthal integral.

For the case of a point source located within the ROI boundary, $\delta < R$, $u_1(\phi)=0$.  We place the x-axis on the diameter extending through the $\vom$ and the ROI center.  The integrand is clearly symmetric about the x-axis, so
\begin{equation}
O(\eopt,\delta<R,R) = \frac{2}{\pi} \int_{0}^{\pi} d\phi\, F_{wk}[u_2(\phi)].
\end{equation}
$u_2(\phi)$ can be determined from plane geometry\footnote{The assumption, of course, is that the ROI is sufficiently small that we may approximate the space as $\mathbb{R}^2$.}:
\begin{equation}
\sigma\sqrt{2u_2(\phi)} = \sqrt{R^2 - [\delta\sin(\phi)]^2} - \delta\cos(\phi),
\end{equation}
where $\sigma$ stands in for the $\sigma$ parameter for each relevant King function.

For an exterior source, $\delta > R$, both bounds in $u$ are non-trivial.  We again place the x-axis on the diameter joining $\vom_-$ and the ROI center and again note the reflection symmetry.  The $\phi$ integral extends from the x-axis to the tangent joining $\vom_0$ and the boundary of the ROI.  That is,
\begin{equation}
O(\eopt,\delta>R,R) = \frac{2}{\pi} \int_{0}^{\arccos(R/\delta)} d\phi\, F_{wk}[u_2(\phi)] - F_{wk}[u_1(\phi)].
\end{equation}
The bounds of integration are again determined from simple geometry:
\begin{align}
\sigma\sqrt{2u_1(\phi)}& = \delta\cos(\phi) - \sqrt{R^2 - [\delta\sin(\phi)]^2} \\
\sigma\sqrt{2u_2(\phi)}& = \delta\cos(\phi) + \sqrt{R^2 - [\delta\sin(\phi)]^2}
\end{align}

\subsubsection{Evaluating the Source Counts}
\begin{figure}
\begin{minipage}{6in}
\includegraphics[width=6in]{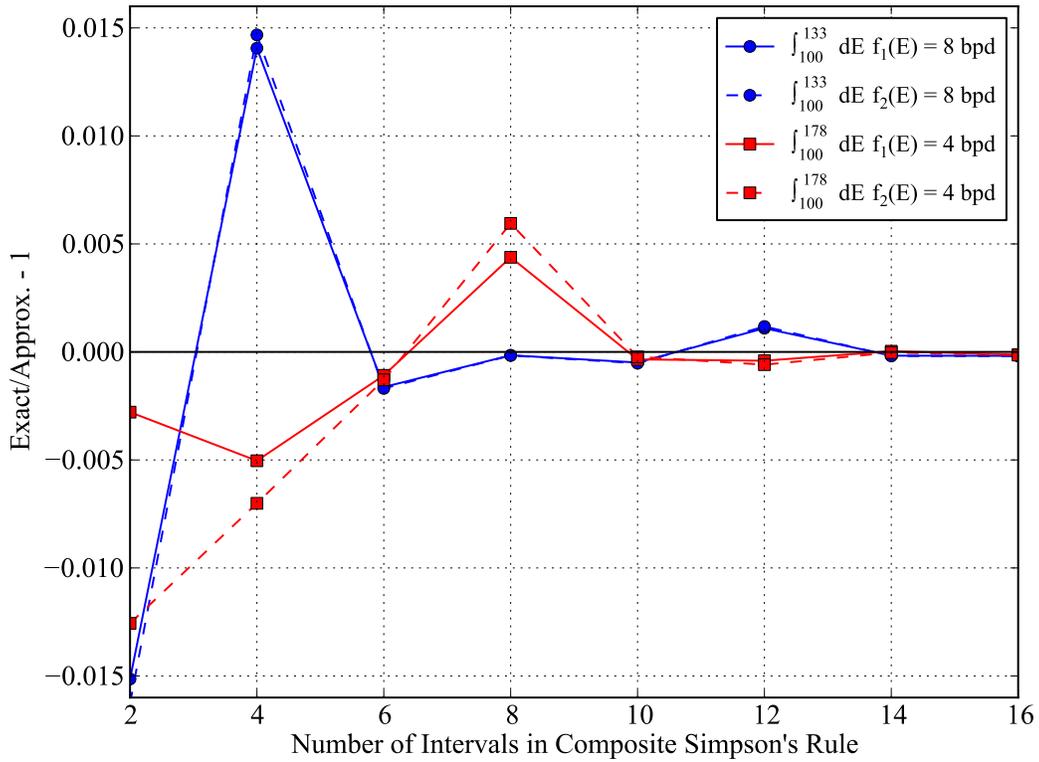}
\end{minipage}
\begingroup\renewcommand{\baselinestretch}{1.0}
\caption{Two power law spectra ($f_1(E) \propto E^{-1.5}$ and $f_2(E) \propto E^{-3.0}$) integrated over the most problematic energy band in a typical analysis, 100-133 MeV in 8 bin-per-decade mode and 100-178 MeV in 4 bin-per-decade mode.  The relative difference with the exact integral (as determined by numerical integration to machine precision) is shown as a function of the number of intervals used in the composite Simpson's rule.}
\renewcommand{\baselinestretch}{1.5}\endgroup
\label{plot3}
\end{figure}

Recall that we defined $N(\vla) = \int dE\, \mcf(E,\vla)\, \epsilon(E,\vom_0) \equiv \int dE\, f(E)$.  We evaluate this integral with a composite Simpson's rule quadrature over logarithmic energy.  If the energy range of the band runs from $e_a$ to $e_b$, then
\begin{equation}
\label{eq:simps_def}
N(\vla) \approx \frac{\log e_b/e_a}{3N_s} \left[ e_af(e_a) + e_bf(e_b) + \sum\limits_{i=1}^{N_s-1}\,(2 + 2\mathcal{I}_{\in 2\mathbb{Z}+1}(i))\times e_a\delta^{2i-1}f(e_a\delta^{2i-1})\right],
\end{equation}
where $\delta\equiv (e_b/e_a)^{1/N_s}$, $\mathcal{I}(i)$ is 1 (0) for odd (even) integers, and $N_s$ (an even integer) is the number of intervals into which we divide the band.  $N_s$ moderates the tradeoff between speed and accuracy.  In Figure \ref{plot3}, we explore the accuracy of the integral of a power law source over the (typically) lowest energy band in a \ptl analysis.  The effective area here is rapidly rising and bumpy (see Figure \ref{ch3_pre_plot2}), making this integral a good test case.  We see $N_s=8$ gives an error $<1$\%.  Since the evaluation of these integrals is not currently a serious bottleneck in the \ptl code, we adopt the conservative choice $N_s=16$.

\begin{figure}
\begin{minipage}{6in}
\includegraphics[width=6in]{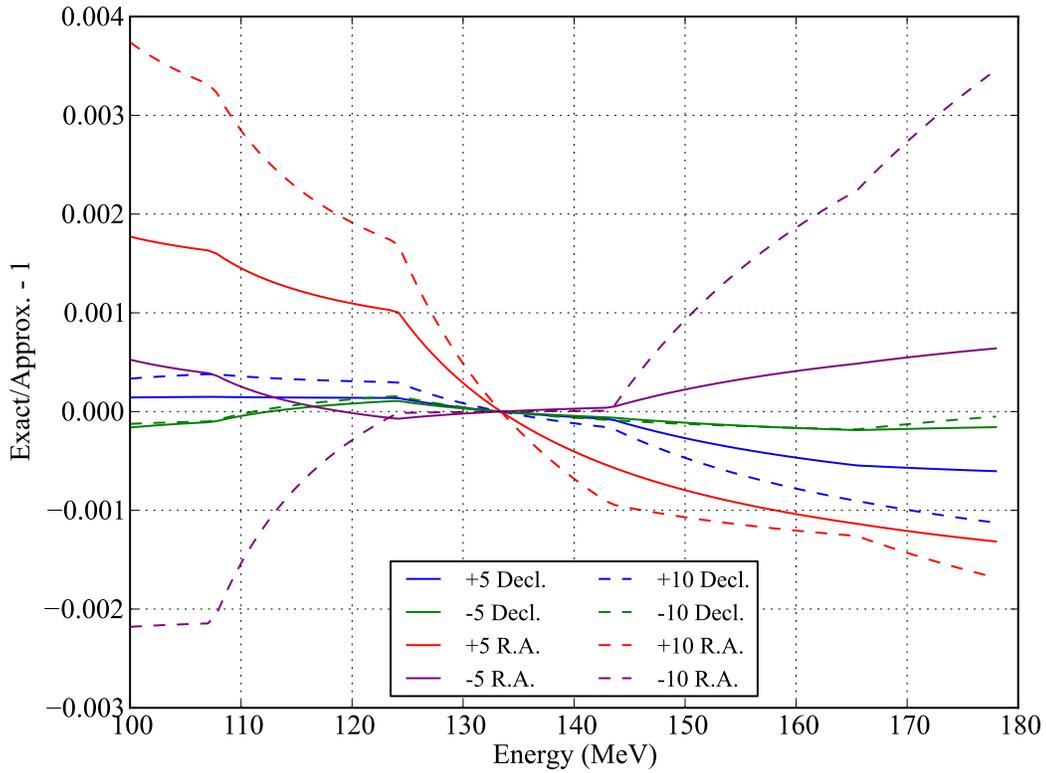}
\end{minipage}
\begingroup\renewcommand{\baselinestretch}{1.0}
\caption{The exposure variation over a typical ROI (here, located at the position of the Vela PSR.)  The approximated exposure assumes the same energy dependence as that at the ROI center and corrects only for the normalization at the geometric mean energy of the band.  The exact value accounts for the spatial variation at all energies.  Relative differences are shown for point sources at a modest ($5^{\circ}$) and extreme ($10^{\circ}$) separation from  the ROI center.  The band chosen, $100 < E/\mathrm{MeV} < 178$, has rapidly-varying effective area, making this a worst case.  The error is $<0.5\%$.}
\renewcommand{\baselinestretch}{1.5}\endgroup
\label{ch3_plot5}
\end{figure}

With a quadrature rule in hand, we can make one additional approximation for speed.  We cache the specified energies $E_i$ and factors (the 1s, 2s, and 4s in Eq. \ref{eq:simps_def}) for each band.  Further, since the exposure varies relatively slowly in position, over a typical ($10^{\circ}$) ROI we can approximate the exposure for a particular source by the exposure at the center, making a first order correction by normalizing to the exposure value at $E_{\mathrm{geo}}$.  We can then cache $w_{si}\equiv k_i\,E_i\,\epsilon(E_i,\vom_0)$, with $k_i$ the Simpson's rule factor, and evalute the integral, for an arbitrary source at position $\vom$, as
\begin{equation}
\label{eq:simps_fast}
\frac{\epsilon(\egeo,\vom)}{\epsilon(\egeo,\vom_0)} \sum_{i=1}^{N_s+1} w_{si} \mcf(E_i,\vla).
\end{equation}
Evaluating this expression is extremely fast, and the exposure approximation introduces negligible ($<<1\%$) error (see Figure \ref{ch3_plot5}).

\subsection{Point Source Summary}
Our goal in this section was to calculate the contribution of a point source to a particular energy band.  This task entailed evaluating the point source model (convolved with the IRF) both over the sparse data pixels and over an entire spatial aperture.  We made approximations that allowed us to define a single PSF for the energy band, $f_{wpsf}(\eopt)$, in turn allowing a factorization of the source rate integral into a spatial component and an energy component.  We reduced the spatial integral over the aperture to a quadrature, and we outlined a rapid method for performing the energy integral for each source.  Taken together, these approximations generally retain percent-level accuracy but greatly speed up computation of the contribution of point sources to the likelihood.

\section{Diffuse Sources Rates}
We now come to the calculation of the rates of diffuse sources, those that have resolvable angular extension in the LAT data.  We return to Eqs. \ref{eq:diffuse_rate} and \ref{eq:like1} which outline the integrations to be performed.  The argument parallels that for point sources.  We again begin by making the assumption that the diffuse sources are stationary\footnote{This is of course perfectly true for celestial diffuse sources.  Photons from the Earth's limb, and from irreducible charged particle background, have some time dependence based on the S/C precessional phase and the active rocking profile.  These effects are certainly second order.}.  As for point sources, we introduce the exposure and exposure-weighted PSF to arrive at the following expression for the expected rate per unit solid angle of a diffuse source for a given energy band:
\begin{equation}
\label{eq:diffuse_spatial_rate}
r(\vom') = \int dE \int d\Omega \, \mcf(E,\vom;\vla)\,\epsilon(E,\vom) \,f_{\mathrm{wpsf}}(\vom' ; \vom, E).
\end{equation}
This expression is similar to that for point sources, Eq. \ref{eq:psrate_with_exposure} but with an additional complication, the convolution of the diffuse source count rate per unit solid angle with the PSF.  In the following, we outline the evaluation of the integrals over energy and position.

The strategy differs slightly from point sources.  Typically, we deal with only a couple of diffuse sources, the Galactic diffuse background from cosmic rays and the (approximately) isotropic background from unresolved sources at cosmological distances, irreducible charged particle background, and contamination from the Earth's limb.  With only two sources compared to tens or even hundreds of point sources, we can be more lavish with computational resources.  Furthermore, all but the brightest point sources are utterly dominated by the diffuse background at low energy, particularly in the Galactic plane.  To avoid bias, it is important to model the diffuse sources with as much accuracy as possible.

\subsection{Energy Integral for Diffuse Sources}
As with point sources, we adopt a composite Simpson's rule for the integral over energy.  Adopting the same notation,
\begin{equation}
\label{eq:diffuse_energy_integral}
r(\vom') \approx \sum\limits_{i=1}^{N_s+1} w_i \int d\Omega \, \mcf(E_i,\vom;\vla)\,\epsilon(E_i,\vom) \,f_{\mathrm{wpsf}}(\vom' ; \vom, E_i) \equiv \sum\limits_{i=1}^{N_s+1} w_i\, r(\vom',E_i),
\end{equation}
where in this case $w_i$ is the product of the Simpson's rule factor and $E_i$.  The expression $\int d\Omega \, \mcf(E_i,t',\vom;\vla)\,\epsilon(E_i,\vom) \,f_{\mathrm{wpsf}}(\vom' ; \vom, E_i)$ is extremely costly to evaluate, so we are more cautious in selecting $N_s$.  Indeed, since only the low-energy bands, with their bumpy and rapid rise in the effective area, require care, we adopt $N_s=8$ for bands with $E < 200$MeV and $N_s=4$ for all remaining bands.

Unlike for point sources, we do not attempt to pull the PSF out of the energy integral for two important reasons.  First, in keeping with our remarks above, we want to evaluate the integral as accurately as is feasible.  Second, unlike point sources, the spatial morphology of diffuse sources can and does change with energy.  Such morphology shifts within an energy band are likely to be small but are in any case explicitly accounted for in our formulation here.

\subsection{Spatial Integral (Convolution) for Diffuse Sources}
We now have to evaluate
\begin{equation}
r(\vom',E) = \int d\Omega \, \mcf(E,\vom;\vla)\,\epsilon(E,\vom) \,f_{\mathrm{wpsf}}(\vom' ; \vom, E).
\end{equation}
The primary source of interest is the Galactic diffuse background.  The construction of the model is beyond the scope of this work, but uses methods similar to those employed in the analysis of \cite{q2_diffuse}.  It is typically structured as a \emph{mapcube}, a pixelized intensity map of the sky stored at a series of energy planes.  We construct the continuous function $\mcf(E,\vom;\vla)$ by first selecting the two image planes bounding $E$, determining the spatial intensity in the bounding image planes via bilinear interpolation of the four pixels closes to $\vom$, and finally logarithmically interpolating these two values in energy.  The isotropic background is typically tabulated, and we simply interpolate the table.  The isotropic model acquires spatial variation through multiplication with the exposure.  We write a general model as $\mathcal{D}(E,\vom)\equiv\mcf(E,\vom;\vla)\,\epsilon(E,\vom)$.  Since the PSF only depends on the \emph{difference} between the true and reconstructed positions, the spatial density is
\begin{equation}
r(\vom') = \int d\Omega \, \mathcal{D}(E,\vom)\,f_{\mathrm{wpsf}}(E,\vom'-\vom),
\end{equation}
readily seen to be a two-dimensional convolution.

Recall that a convolution of two functions, $f$ and $g$, is defined in one dimension as
\begin{equation}
(f*g)(x') = \int\limits_{-\infty}^{+\infty}dx\, f(x)\,g(x'-x).
\end{equation}
The convolution arises in probability theory as the probability density function for the sum of two statistically independent variables with probability density functions $f$ and $g$.  (The integral is over all combinations of values drawn from $f$ and $g$ whose sum has the correct value.)  Now, suppose $f$ or $g$ is concentrated near the origin.  The convolution is then well approximated with some cutoff:\begin{equation}
(f*g)(x') \approx \int\limits_{-X}^{+X}dx\, f(x)\,g(x'-x).
\end{equation}
We can evalute such an integral by sampling it at $N$ uniform points between $-X$ and $+X$ and applying some quadrature rule.  If we use $N$ points in the quadrature, then all of the information in the convolution can also be contained with $N$ points.  Thus, a full evaluation of the convolution over its support requires $\mathcal{O}(N^2)$ operations.

Similar arguments apply to the evaluation of a discrete Fourier transform (DFT) at its independent frequencies.  However, the use of the Fast Fourier Transform (FFT) algorithm\cite{fft} allows evaluation of all independent frequencies of a DFT in only $\mathcal{O}(N\log N)$ operations.  This development can be applied to convolutions via the Convolution Theorem:
\begin{equation}
\mcf(f*g) = \mcf(f) \times \mcf(g),
\end{equation}
i.e., the Fourier transform of the convolution of $f$ and $g$ is equal to the product of the Fourier transforms of $f$ and $g$.  The $N$ quadrature sampling points map to $N$ independent frequencies in a DFT.  Thus, by evaluating the FFTs of $f$ and $g$, multipyling, and taking the inverse FFT, we can calculate the convolution of $f$ and $g$ with $\mathcal{O}(N\log N)$ complexity.

We now apply these arguments to the evaluation the convolution of a diffuse source with the PSF.  To apply a 2D FFT, we need $\mathcal{D}(E,\vom)$ and $f_{\mathrm{wpsf}}(E,\vom)$ evaluated on a regular grid.  The diffuse map, of course, is a curved space, so in making such a projection we make an approximation.  However, since our grids are typically 10s of degrees on a side, a projection does not introduce too much distortion.  We assume for now that we are working on the equator in the Galactic coordinate system.  We use the plate carr\'{e}e projection such that the Cartesian coordinates of the projection $x$ and $y$ are simply $l$ and $b$.  (The carr\'{e}e projection is approximately equidistant near the equator; this is important, since the PSF depends only on the angular separation of two points.)  We cut off the integral to a square grid with a side length of $2\Delta$.  Then, the convolution becomes 
\begin{align}
r(l',b',E) &= \iint d\Omega \, \mathcal{D}(E,l,b)\,\fwpsf(E,l'-l,b'-b) \\
	     &\approx \int\limits_{-\Delta}^{+\Delta} dl\,\int\limits_{-\Delta}^{+\Delta}db\, \mathcal{D}(E,l,b)\, \fwpsf(E,(l'-l)^2+(b'-b)^2) \\
	     &\approx \sum\limits_{i=-N}^{i=+N}\sum\limits_{j=-N}^{j=+N}  \mathcal{D}(E,i\Delta ,j\Delta)\, \fwpsf(E,(l'-i\Delta)^2+(b'-j\Delta)^2)
\end{align}
The prescription is now clear: evaluate $\mathcal{D}$ and $f_{wpsf}$ on the $N\times N$ grid, perform a 2D FFT on each, multiply, and invert, and store the final $N\times N$ image plane containing the convolved diffuse model. By interpolating the pixels of this image, we can evaluate the convolved diffuse model for any $\vom$ within the grid, i.e., we can compute the rate in Eq. \ref{eq:diffuse_energy_integral} trivially.  We note, importantly, that we normalize $f_{wpsf}$ to the grid, with the net effect that ``diffuse photons are conserved''.  This adjustment inevitably leads to some edge effects, as at low energies some photons are scattered into/out of the ROI.  We thus choose a grid size to minimize these effects on the model in the ROI itself.  Currently, for each band, we calculate the $95\%$ containment radius of the PSF and require the grid be $2\times(R_{roi} + R_{95})$ on a side.

We note a few technical details.  First, we are effectively rebinning the model of the diffuse source onto the uniform grid.  Therefore, the pixel size of the uniform grid should be sufficiently small to avoid artifacts.  The \emph{gll\_iem\_v02} model\footnote{http://fermi.gsfc.nasa.gov/ssc/data/access/lat/BackgroundModels.html} of the Galactic diffuse uses $0.5^{\circ}$ pixels, so a sufficiently fine uniform grid size is $0.25^{\circ}$.  Second, the above derivation relies on a convolution at the equator, where the carr\'{e}e projection is relatively distortion free and, more importantly, equidistant.  In order to use this approach generally, we perform the following procedure for an ROI centered on $(l,b)$:
\begin{enumerate}
\item Construct a uniform grid centered at $(l,0)$.
\item Right-handedly rotate each point on the grid by $b^{\circ}$ about the vector pointing to $(l-90\dg,0)$ to get $(l_r,b_r)$.
\item Evaluate $\mathcal{D}(E,l_r,b_r)$ for each point on the rotated grid.
\item Perform the FFTs and convolution.
\end{enumerate}
To evaluate the convolved background model at a particular point, say $(l',b')$, the process is reversed: rotate by $-b^{\circ}$ about $(l-90\dg,0)$ and then linearly interpolate the grid values.  (Another benefit of using the plate car\'{e}e projection is the ease of indexing and interpolation, since the indices map directly to geographic coordinates with only arithmetic operations.)

\section{Calculating and Maximizing the Likelihood}

We have now described in some detail how the rates for point source and diffuse sources appearing in the likelihood, Eq. \ref{eq:like1}, are calculated in the \ptl framework.  Particularly, we have demonstrated how to calculate rates for \emph{bins in energy and event conversion type}.  We formalize this procedure in \ptl with the a \emph{band}, an object (in the computer science sense) that associates data and source rates for a particularly energy range and conversion type and performs a series of tasks to link the two and facilitate computation of the likelihood.

\subsection{The \band Object}

In terms of data, each \band maintains an array storing the positions ${\vom_i}$ and counts ${n_i}$ of the occupied pixels in the ROI that fall within its energy bounds. 

Next, the \band is responsible for pre-computing some quantities.  It causes $\eopt$ (Eq. \ref{eq:eopt_ref}) for its energy range/conversion type to be determined and, from this, determines and stores the proper (approximate) PSF $f_{wpsf}(\eopt)$ introduced in Eq. \ref{eq:eopt_psf}.  It determines and stores the Simpson's energies $E_i$ and Simpson's weights, $w_i$, from Eq. \ref{eq:simps_fast}.  For each point source, referenced by index $j$, a \band computes and stores, the overlap integral, $\mathcal{O}_j$ (Eq. \ref{eq:overlap}), and the exposure correction, $\alpha_j\equiv\epsilon(1\,\mathrm{GeV},\vom_j)/\epsilon(1\,\mathrm{GeV},\vom_0)$ appearing in Eq. \ref{eq:simps_fast}.  Using $f_{wpsf}$, it evaluates $f_{ij}$, the (normalized) contribution of the $j$th source to the $i$th pixel, i.e., $f_{ij}\equiv\fwpsf(\eopt,\vom_i;\vom_j)$, appearing in Eq. \ref{eq:separated}.  

To actually evaluate the log likelihood, we provide a $\vla$.  Let $\vla_j$ be the subspace of $\vla$ that parameterizes point source $j$.  The \band then evaluates the source rate using its cached Simpson's rule information:
\begin{align}
\label{eq:nsimps_band}
N_j(\vla_j) &\approx \alpha_j\,\int dE\, \mcf_j(E,\vla_j)\, \epsilon(E,\vom_0)
            \approx \alpha_j \sum\limits_{i=1}^{N_s+1} w_{si}\,\mcf_j(E_i,\vla_j) \\
            &\equiv \alpha_j \times \vec{w_s}\cdot \mcf_j(\vec{E},\vla_j).
\end{align}
Ignoring diffuse sources for the moment, the log likelihood for the \band is then
\begin{align}
\label{eq:loglike_band_ps}
\log \mathcal{L}(\vla) &= -\left[\sum\limits_{j=1}^{N_{\mathrm{sources}}} \mathcal{O}_j\,N_j(\vla_j)\right] + \left[\sum\limits_{i=1}^{N_{\mathrm{pix}}} n_i\, \log \sum\limits_{j=1}^{N_{\mathrm{sources}}} N_j(\vla_j)\,f_{ij} \right] \\
 &\equiv  -\vec{\mathcal{O}}\cdot \vec{N}(\vla) + \vec{n}\cdot \log\left[ \mathbf{f}\cdot\vec{N}(\vla)\right]
\end{align}
We have written the above equations for the source rate and log likelihood both as sums and as vectors to make clear (a) what is being done operationally and (b) what is being done internally.  It is easy to appreciate, especially in the vector notation, the computational savings gained by factoring the spatial and energy terms.  When we update $\vla$, all that is required to update the likelihood is a handful of vector arithmetic operations!  The vector notation is also schematic for how the computation is carried out in the Python source code.  Despite Python's interpreted nature, it is possible with \emph{numpy}\footnote{http://numpy.scipy.org} to run vectorized code at nearly the same speed as compiled code\footnote{It is also possible to extend the Python language with functions written in C, but \emph{numpy} typically suffices.}.  We thus strive to formulate vectorized implementations whenever possible.  (To avoid confusion, we note that, from left to right, the three inner products above are in ``source space'', in ``pixel space'', and in ``source space'', respectively.)  A second benefit to this notation is it makes determination of the gradient of the log likelihood, $d\log\mathcal{L}(\vla)/d\vla$, more tractable, as we shall see below.

\band objects also manage the likelihood calculation for diffuse sources in a manner similar to that described above.  Recall that, after constructing an image of the convolved diffuse model, we can evaluate the quantity
\begin{equation}
r(\vom',E) = \int d\Omega \, \mcf(E,\vom;\vla)\,\epsilon(E,\vom) \,f_{\mathrm{wpsf}}(\vom' ; \vom, E)
\end{equation}
for a given diffuse source by interpolating from the convolved model values evaluated on our uniform grid.  To integrate the rate over the ROI, we simply average all pixels on the uniform grid lying within the ROI, i.e.,
\begin{equation}
D(E,\vla)\equiv N(E,\vla)\int d\Omega'\,r(\vom',E) \approx \frac{\pi R_{ROI}}{\sum\limits_{i=0}^{N_{grid}}\mathcal{I}_{i\in ROI}}\sum\limits_{i=0}^{N_{grid}}\mathcal{I}_{i\in ROI}\, d_i,
\end{equation}
where $d_i$ is the the model value for the $i$th grid pixel and we use the indicator function $\mathcal{I}$ to select only grid pixels lying within a circle of radius $R_{ROI}$.  $N(\vla)$ is a scaling model which evaluates to $1$ for all energies at the ``default'' parameter values.  We integrate both $D(E)$ and $r(\vom',E)$ over the band using Simpson's rule analogously to the procedure for point sources and define
\begin{align}
D &\equiv \int dE \,D(E), \\
d(\vom,\vla)&\equiv \int dE\, r(\vom,E)  N(E,\vla) .
\end{align}
By using fixed Simpson's rule points, we are free to pre-compute and store the quantities $r(\vom,E)$ and $\int d\Omega'\,r(\vom',E)$, and updates from the likelihood will only require a sum with a few terms, viz. the Simpson's rule sum.  With these quantities, we now have the full \emph{Band} likelihood including point sources and diffuse sources:
\begin{align}
\label{eq:loglike_band}
\log \mathcal{L}(\vla) &= -\left[\sum\limits_{j=1}^{N_{\mathrm{ps}}} \mathcal{O}_j\,N_j(\vla_j)\right]
-\left[\sum\limits_{j=1}^{N_{\mathrm{ds}}} D_j(\vla)\right] +
\left[\sum\limits_{i=1}^{N_{\mathrm{pix}}} n_i\, \log \sum\limits_{j=1}^{N_{\mathrm{ps}}} N_j(\vla_j)\,f_{ij}
+ \sum\limits_{j=1}^{N_{\mathrm{ds}}} d_{ij}(\vla) \right] \\
 &\equiv  -\vec{\mathcal{O}}\cdot \vec{N}(\vla) -\left[\sum\limits_{j=1}^{N_{\mathrm{ds}}} D_j(\vla)\right] + \vec{n}\cdot \log\left[ \mathbf{f}\cdot\vec{N}(\vla) + \sum\limits_{j=1}^{N_{\mathrm{ds}}} \vec{d}(\vla)\right],
\end{align}
where $N_{ps}$ ($N_{ds}$) indicates the number of point (diffuse) sources.  We leave the sum over diffuse source explicit since there are typically only a few such sources modeled.

\subsection{Calculating the Likelihood Gradient}

We preface discussion on likelihood fitting with an important ingredient, an ``analytic'' calculation of the gradient---as opposed to finite differences---that provides significantly faster maximization of the log likelihood and determination of its curvature.  The calculation of $\partial\log \mathcal{L}(\vla)/\partial \lambda_i$ is complicated by the many-to-one mapping of parameters to sources.  We introduce the auxiliary matrix $\mathbf{M}$ defined by $M_{ij} = \partial N_j(\vla)/ \partial \lambda_j$.  This $N_{\mathrm{source}}\times N_{param}$ matrix will only have nonzero entries where the parameter in the denominator matches a parameter of the source in the numerator.  Then, ignoring diffuse sources,
\begin{equation}
\vec{\nabla}\log \mathcal{L}(\vla) = -\vec{\mathcal{O}}\cdot\mathbf{M} + \left[\frac{\vec{n}}{\mathbf{f}\cdot\vec{N}(\vla)}\right]\cdot\left[\mathbf{f}\cdot\mathbf{M}\right],
\end{equation}
where by $\vec{n}/(\mathbf{f}\cdot\vec{N}(\vla))$ we mean elementwise division of the numerator by the denominator.  We note that the denominator of this term has already been determined in the log likelihood calculation.  Thus, in practice, we first build up $\mathbf{M}$, requiring $N_{param}$ Simpson's rule quadratures and some bookkeeping, and the remainder of the gradient can be computed with a few matrix operations.  Terms arising from the diffuse sources can be incorporated analogously.  This method should be compared to $\geq2N_{param}$ likelihood evaluations required to estimate $\vec{\nabla}\log \mathcal{L}(\vla)$ using finite differences; the ``analytic'' approach is also free from numerical errors resulting from inappropriate step size choice.

\subsection{Spectral Models}
\label{ch3:subsec:spectral_models}
Up to now, we have been fairly abstract about $\mcf(E,\vla)$, the source flux.  A variety of spectral models are available in \emph{pointlike}, including those in most common use, a power law and a power law + exponential cutoff.  Using the computer science principle of inheritance, particular spectral models inherit from a single base class.  The base class manages all common tasks, while the subclasses implement the methods particular to the spectral model, e.g., the density, $dN/dE(E,\vla)$, and the gradient, $\partial(dN/dE(E,\vla))/\partial\vla$.

Parameters for spectral models are almost universally positive definite, and negative values are undefined or unphysical.  To prevent the fitting algorithm from attempting these disallowed values, we perform a logarithmic transformation of all parameters internally.  

\subsection{Maximizing the Likelihood and Estimating Errors}

The default fitter for \ptl is the implementation of the BFGS\cite{nocedal99} algorithm in \emph{scipy}\footnote{http://www.scipy.org} via \emph{fmin\_bfgs}.  This algorithm uses the computation scheme for the log likelihood and its gradient outlined above.  If a spectral model is included for which it is difficult to implement  $\partial\log \mathcal{L}(\vla)/\partial \lambda_i$, then the downhill simplex algorithm---which makes use of the log likelihood only---implemented in \emph{scipy} via \emph{fmin} is also available.

We estimate the errors from the information matrix, the inverse of the hessian of the log likelihood function.  That is, the covariance matrix is estimated by
\begin{align}
\sigma^2_{\lambda_i\lambda_j} &= \mathcal{H}^{-1}_{\lambda_i\lambda_j} \\
\mathcal{H}_{\lambda_i\lambda_j} &= -\frac{\partial^2\log\mathcal{L}}{\partial\lambda_i\partial\lambda_j}
\end{align}
If the analytic gradient is available, we employ a first-order finite difference scheme, while in the absence of a gradient we perform second-order finite differencing of the log likelihood function.  In both cases, we use an iterative process that attempts to find an ideal step size for estimation of the curvature, i.e., one for which the change in the log likelihood is of order unity.

\subsection{Source Localization}
While we have concentrated on spectral parameters, the source position is also a model parameter and we can find its MLE in a similar process.  Whereas the approximations we have made allow us to fit spectral parameters for point sources and diffuse sources without re-evaluating model values for pixels (e.g., $\mathbf{f}$ in Eq. \ref{eq:loglike_band}), this is manifestly not the case for source localization since we must obviously re-evaluate the row of $\mathbf{f}$ corresponding to the source being localized as its position is updated during ML fitting.  To speed up this process, we adopt an iterative process for source localization.  An initial spectral fit is performed, and then the spectral parameters are taken as given while the position of the source in question is varied to maximize a two-dimensional likelihood function (i.e., two angular coordinates).  The improved position can be used in a second spectral fit, and so forth until convergence.  The only drawback to this approach is a potential slight underestimate of the positional uncertainty since correlation with spectral parameters is not taken into account.  However, tests of the localization algorithm presented in Chapter \ref{ch3_post} indicate such an effect is negligible.

Let $\vec{b}\equiv \mathbf{f}\cdot\vec{N}(\vla) + \sum\limits_{j=1}^{N_{\mathrm{ds}}} \vec{d}(\vla) - \vec{f}_s\,N_s(\vla)$ (see Eq. \ref{eq:loglike_band}), i.e., the rate in each pixel for all sources under the best-fit spectral model with the rate for one point source, labeled ``s'', subtracted.  This quantity is constant when the position of ``s'' changes.  The log likelihood is then
\begin{equation}
\log \mathcal{L}(\vom) = -\mathcal{O}_s(\vom)\,N_s + \vec{n}\cdot\log\left[ \vec{b} + \vec{f}_s(\vom)\,N_s\right].
\end{equation}
To maximize the function, we fit a quadratic form to the log likelihood surface via least squares and iterate until the position change is sufficiently small.  The final quadratic form measures the likelihood surface and provides an estimate for the uncertainty in the two position parameters.

Finally, we note that this formulation is independent of the spectral model used for source ``s'', and while we typically employ one of the usual spectral models, e.g., a power law, we can also use a ``model-independent'' approach outlined below. 

\subsection{Model-independent Fits}
\label{ch3:subsec:model_independent}
In the previous sections, we have concentrated on the \emph{Band}.  Of course, when fitting a (broadband) spectral model, we accumulate the log likelihood contributions from each \emph{Band} to compute the total likelihood.  However, we may also be interested in the likelihood from a single band.  The typical use case is assessing the shape of the spectral energy density of a source, e.g. to look for deviations from the broadband model.  In this procedure, we model all sources initially with a broadband spectral model and maximize the likelihood to arrive at a good spectral model for the ROI.  We then freeze the parameters for all sources save source ``s''.  We assume that, \emph{within the band}, the source has a power law spectrum with $\Gamma=2.0$.  (Since the bands are narrow, there is little dependence on the particular slope chosen.)  Then, the log likelihood for the source flux \emph{within the band} is given by
\begin{equation}
\mathcal{L}(\mathcal{F}_s) = -\mathcal{O}_s\,\frac{N_s}{\mathcal{F}_{0s}}\,\mathcal{F}_s \vec{n}\cdot\log\left[ \vec{b} + \vec{f}_s(\vom)\,\frac{N_s}{\mathcal{F}_{0s}}\,\mathcal{F}_s\right], 
\end{equation}
with $\mathcal{F}_{0s}$ the flux calculated for source ``s'' in the initial spectral fit.  This single-dimension function is easily extremized by differentiating and finding the root of the resulting equation.  Thus, $\mathcal{F}_s$ can be rapidly estimated in each band, yielding a band-by-band estimate for the spectral energy density for the source.  These estimates can be used for localizing sources as outlined in the previous section, or for generating plots of the spectral energy density.  For the latter application, we typically use the joint likelihood for the \emph{Band} objects for front-converting and back-converting events at the same energy.

In this prescription, the only sensitivity of the final band-by-band fits to the Ansatz broadband model comes from the other sources.  Provided that model is not too far wrong---or that source ``s'' is not too bright---we expect the initial model to have little effect on the band-by-band estimates.

\section{Summary}
In this chapter, we laid out the major design principles and detailed the implementation of \emph{pointlike}, a full-featured maximum likelihood analysis package.  We first outlined our method of binning the data with pixels that scale with the PSF to achieve good performance at low energy and good accuracy at high energy.  We next began the ``heavy lifting'' of evaluating the source rates folded through the \fermi IRF by distinguishing point sources and diffuse sources.  For point sources, we defined an effective PSF for each band by estimating an optimal energy at which to evaluate the King function parameters, and we used this PSF to determine point source contributions to data pixels and in the overlap integrals, which we reduced to quadrature.  For diffuse sources, we described a method for evaluating the diffuse model on a locally flat grid and then applying the Convolution Theorem to quickly evaluate the convolution with a two-dimensional Fast Fourier Transform.  Finally, we described how these ingredients are unified in a \emph{Band} object to facilitate calculation of the likelihood and its gradient.

Along the way, we attempted to validate each piece of the calculation and verify that it met our goal of percent-level accuracy.  In the next chapter, we test the entire \ptl package for actual science tasks, viz. estimating the spectral and positional parameters for sources under a variety of conditions.
\chapter{The \emph{Pointlike} Package: Validation}
\label{ch3_post}

Although in Chapter \ref{ch3} the individual portions of \ptl have been considered and the efficacy of the various approximations tested, the most important check is a holistic one---the determination of spectral parameters from realistic data.  Although some bright $\gamma$-ray sources can be considered standard candles, \fermi is the first HE $\gamma$-ray observatory with good sensitivity above 1 GeV, and comparison to past experiments can only serve as a sanity check\cite{vela1}.  Since we are concerned with percent-level effects, we instead use data generated with Monte Carlo simulations using a method we outline below.

As described in previous sections, the LAT (as modeled) is entirely characterized by its IRF.  By combining a simulated or actual history of the S/C position and orientation (\emph{FT2} file) with the IRF and a source model, a realization of data from the sources can be created.  This task is implemented in the \emph{gtobssim} tool developed by the LAT Collaboration.  We make extensive use of it below.  Except where otherwise specified we use the \textit{P6\_v3\_diff} IRF for simulating and fitting validation data.  Additionally, except where otherwise noted, we use the true energy of the simulated photon, i.e, we turn off the effects of energy dispersion.

The majority of LAT sources are modeled as power laws.  While the brightest sources may require additional degrees of freedom for accurate representation, dimmer sources are adequately described, from a statistical standpoint, by power laws.  For the dimmest sources, even two degrees of freedom may be too much.  Thus, while we do test more complex models, especially power laws with exponential cutoffs given their strong connection with pulsars, we concentrate here on power law spectra.

We begin by simulating a diffuse background.  We adopt the model used for the 1FGL catalog analysis\cite{1fgl}, the \textit{gll\_iem\_ v02} mapcube for the Galactic diffuse and the \textit{isotropic\_iem\_v02} intensity tabulation\footnote{http://fermi.gsfc.nasa.gov/ssc/data/access/lat/BackgroundModels.html} for the isotropic background.  We simulate 1 year of data using the actual \emph{FT2} file describing the \fermi position and orientation during the calendar year 2009.  Simulating diffuse sources is a slow process, and 1 year of data occupies about 1 GB of disk space.  Therefore, for validation requiring multiple realizations of the same source configuration, we simulate photons in a $12^{\circ}$ disk (radius) about the position of the Vela pulsar, (R.A.,Decl.) = (128.8463, -45.1735).

Next, we simulate a point source with a power law spectrum, an integral flux from 100 MeV to 200 GeV of $10^{-5}$ \fluxunits and a photon index of either $1.5$, $2.0$, or $3.0$, spanning the typical spectrum of \fermi source spectra.  For convenience, we define $\mcf_X\equiv10^{-X}$\fluxunits and let $\mcf$ indicate a general flux.  A source with $\mcf=\mcf_5$ is comparable in brightness to the Vela pulsar, the brightest steady $\gamma$-ray source in the sky.  We are interested in the performance of \ptl over a range of source fluxes or, more importantly, signal-to-background ratios.  Strong sources expose inaccuracies when statistical fluctuations are unimportant, while weak sources test both statistical issues (e.g., error estimate) and bias from strong backgrounds.  Rather than simulate different data sets for each flux, we resample a subset of photons from the original $\mcf_5$ file using the following prescription:
\begin{enumerate}
\item Draw a Poisson random variable $N_{tar}$ from a distribution with rate $N_{tot}\mcf/\mcf_5$.
\item Select $N_{tar}$ photons uniformly from the original file set.
\end{enumerate}
Here, $N_{tot}$ is the number of photons in the original data set and $\mcf$ is the target flux.  Note that we make no cut on incidence or zenith angle.  

We simulate 20 base sets with $\mcf=\mcf_5$ each with an integration time of 1 year, and from these we can generate subsets for flux-dependent study of ensembles of sources.  But before we proceed to ensemble testing, we take advantage of the extreme precision allowed by summing data from many MC realizations to verify various aspects of the analysis outlined above. 

\section{Validation with Summed Data}

While the 20 independent realizations are valuable for ensemble testing, by combining them into a single data set we can test pieces of the \ptl framework to high precision.  When combined, since we use the exposure for only a year, we are essentially using a year of data for a source with a flux of $2\mcf_4$.  Likewise, the diffuse background is 20 times brighter than in actuality.  The resulting quantity of photons, a few million, allows measurements of deviations of the model and data to a few percent.  In the following sections, we examine our implementation of the point-spread function and convolution of diffuse sources.  We conclude with a comparison of the maximum likelihood parameters estimated from the summed data with their Monte Carlo truth values.

\subsection{Checking the Band PSF Shape}
\label{checking_band_shape}
We wish to verify two aspects of the approximate exposure-weighted PSF we defined for each energy band (see Eq. \ref{eq:eopt_psf}).  First, do we correctly characterize the \emph{shape} of the distribution of photons for a point source, and second, do we accurately calculate the overlap integral, Eq. \ref{eq:overlap}?  We begin with the PSF shape. 

First, recalling the azimuthal symmetry of the PSF, in each band we bin the photons in $\theta$, the angular separation of the reconstructed photon position and the point source position.  (Since we are using binned data, the reconstructed photon position is actually associated with a HEALPix pixel center, but this position differs negligibly from the actual position by design.)  To achieve an approximately equal number of photons in each band, we invert the cumulative distribution function furnished by the on-axis PSF and use the resulting quantiles such that, e.g., $\approx5\%$ of the photons are in each $\theta$ bin.  (We choose bins with equal counts, rather than bins with equal size in $\theta$ or $\theta^2$, so that the statistical weight of the bin on the results is easy to infer.)

In calculating the expected number of counts in a particular bin, we must account for the ``ragged binning'' of the HEALPixels.  To do this, we simply integrate the PSF over each $\theta$ bin numerically by evaluating the PSF at each HEALPixel, occupied or not, in the $\theta$ bin, and take the mean.  Since we are interested in the \emph{shape}, we normalize the sum of the predicted counts over all $\theta$ bins to the total observed counts.  Importantly, then, these results are \emph{independent} of how accurate the source model and estimate of the exposure are.

\begin{figure}
\begin{minipage}{6in}
\includegraphics[width=6in]{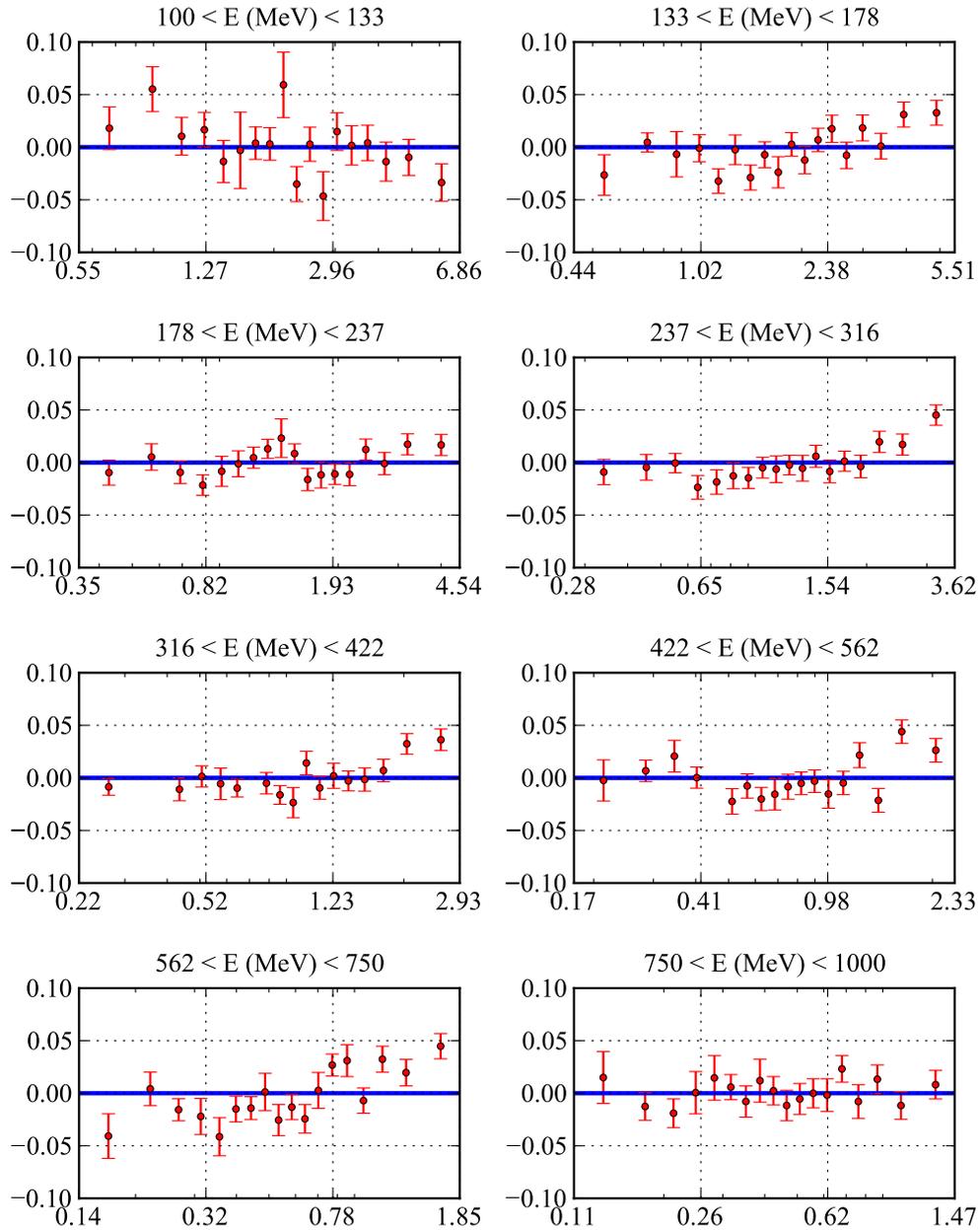}
\end{minipage}
\begingroup\renewcommand{\baselinestretch}{1.0}
\caption{The PSF profile for bands with front-converting events between 100 and 1000 MeV.  The x-axis gives the angular separation of the bin center from the point source position in degrees and is on a logarithmic scale.  The y-axis indicates the relative difference, observed less predicted, in percent.  These results are for a source with a power law spectrum of photon index $2.0$.}
\renewcommand{\baselinestretch}{1.5}\endgroup
\label{fig:ch3_post_plot9_pl_2.0_ct0_e100}
\end{figure}

\begin{figure}
\begin{minipage}{6in}
\includegraphics[width=6in]{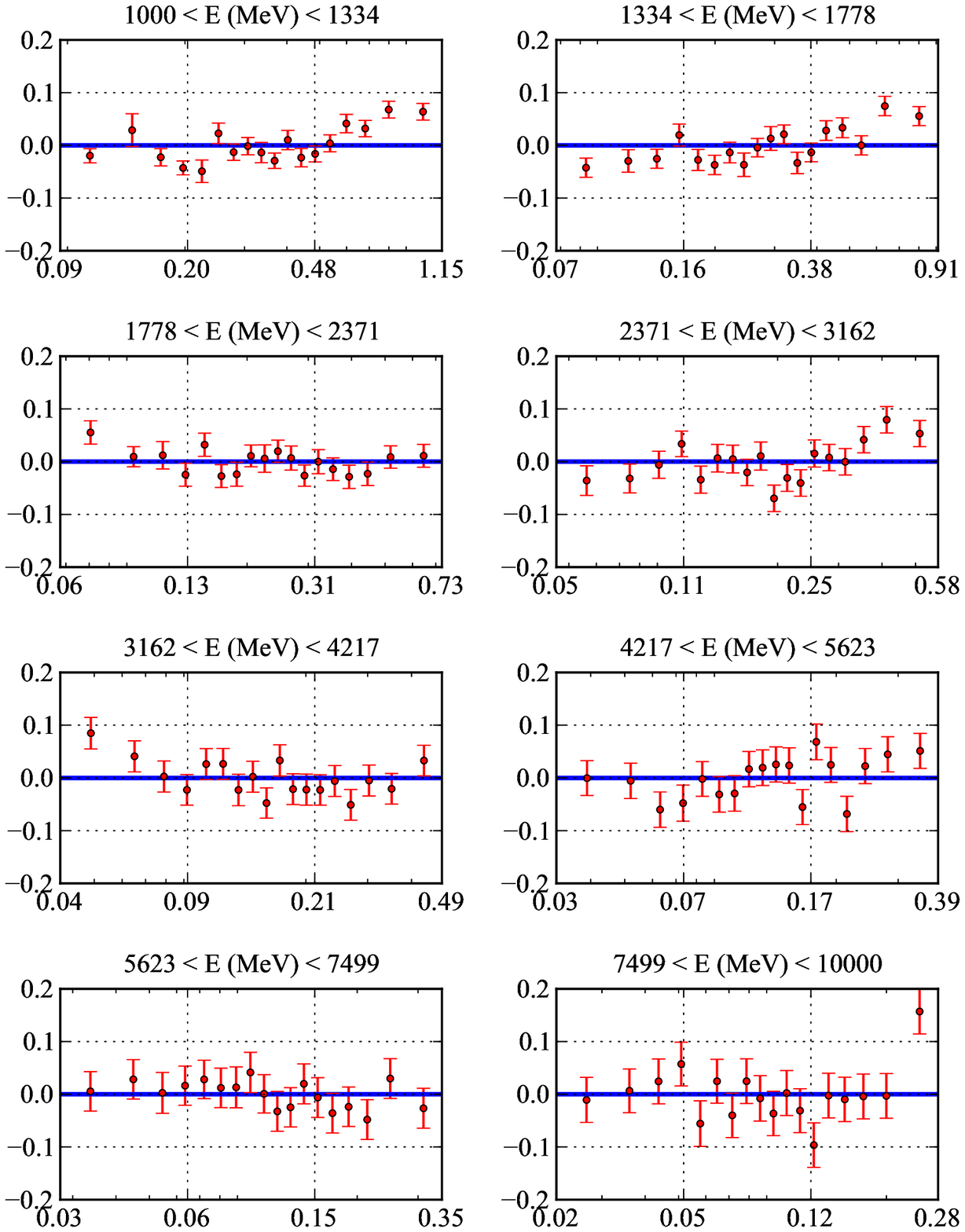}
\end{minipage}
\begingroup\renewcommand{\baselinestretch}{1.0}
\caption{As Figure \ref{fig:ch3_post_plot9_pl_2.0_ct0_e100}, for bands with front-converting events between 1000 and 10000 MeV.}
\renewcommand{\baselinestretch}{1.5}\endgroup
\label{fig:ch3_post_plot9_pl_2.0_ct0_e1000}
\end{figure}

\begin{figure}
\begin{minipage}{6in}
\includegraphics[width=6in]{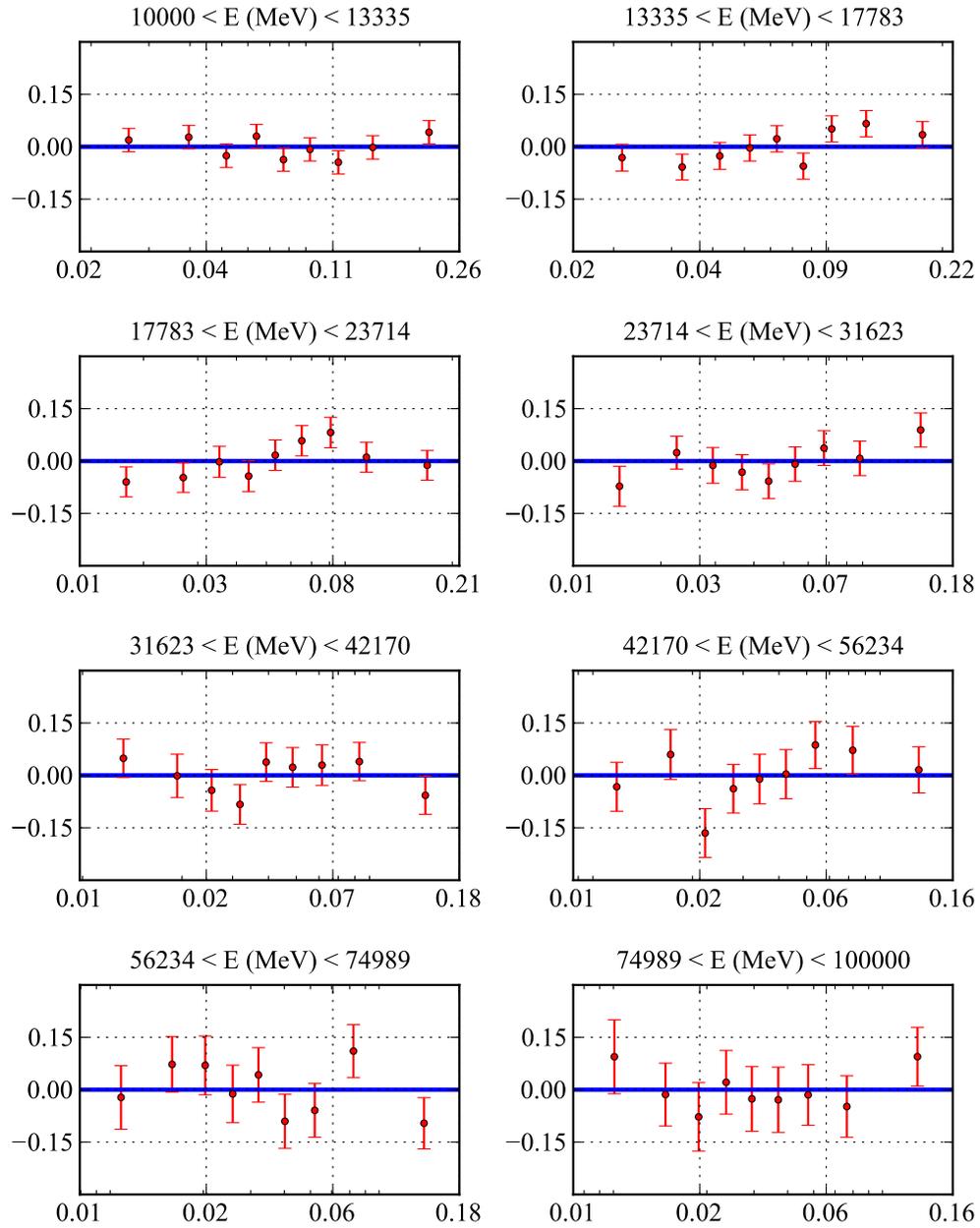}
\end{minipage}
\begingroup\renewcommand{\baselinestretch}{1.0}
\caption{As Figure \ref{fig:ch3_post_plot9_pl_2.0_ct0_e100}, for bands with front-converting events between 10000 and 100000 MeV.}
\renewcommand{\baselinestretch}{1.5}\endgroup
\label{fig:ch3_post_plot9_pl_2.0_ct0_e10000}
\end{figure}

The results for a power law source with $\Gamma=2.0$ are shown in Figs. \ref{fig:ch3_post_plot9_pl_2.0_ct0_e100}, \ref{fig:ch3_post_plot9_pl_2.0_ct0_e1000}, and \ref{fig:ch3_post_plot9_pl_2.0_ct0_e10000}.  These results, for front-converting events, are representative of results for $\Gamma=1.5$ and $\Gamma=3.0$ and for back-converting events.  Generally, the outcome is quite good.  For many bands, the residuals are perfectly flat to within the statistical error bars.  Others show a slight trend, i.e., the PSF is too narrow (broad), resulting in negative (positive) residuals in the core and positive (negative) residuals in the tail.  This is to be expected, as the band PSF is only approximate.

We can also investigate what gains we made by optimizing the energy at which we evaluate the scaled $\sigma$ parameters for our band PSF $\eopt$ (see \S\ref{ch3:subsec:energy_integral}) for our band PSF.  We compare in Table \ref{ch3:tab:eopt_comp} the log likelihoods for the three spectral shapes using both $\eopt$ as estimated by Eq. \ref{eq:eopt_ref} and the simple prescription of $\eopt = E_{geo}$, i.e., choosing the geometric mean energy as the optimal energy.  There is an appreciable increase in the log likelihood with $\eopt$ as estimated from Eq. \ref{eq:eopt_ref}.

\begin{table}
\center
\begin{tabular}{ l | c | c | c }
& $\Gamma = 1.5$ & $\Gamma = 2.0$ & $\Gamma = 3.0$ \\
\hline
$\Delta \log \mathcal{L}$ (Front) & 287 & 206 & 82 \\
$\Delta \log \mathcal{L}$ (Back)  & 539 & 378 & 156 \\  
\end{tabular}
\caption{The change in the log likelihood for 20 combined years of Monte Carlo data for three different spectral shapes when comparing band PSFs based on $E_{geo}$, the geometric mean energy for the band, and $E_{opt}$, the optimal energy estimated by Eq. \ref{eq:eopt_ref}.}
\label{ch3:tab:eopt_comp}
\end{table}

\subsection{Checking the Overlap Integrals}

Next, we compare the total number of counts predicted by the model to that actually observed in the entire ROI.  The accuracy of the predicted counts depends on the value of the exposure, the proper integration of the spectral model over the energy band, and the correct calculation of the PSF overlap.  The exposure is typically binned with $1\dg$ pixels, and varies sufficiently slowly that the resulting error is $<0.5\%$.  Likewise, we have shown in Figure \ref{plot3} that the integral of the spectral model introduces negligible error.  The primary source of error is then the calculation of the overlap integral, which in turn comprises two error sources.  Numerically, there is neglibible error, but since we use a single, approximate PSF for the band, the overlap integral will invariably depart from the true overlap.  Second, while we calculate the overlap integral assuming a circle aperture, the use of binned data results in an actual data set with a ``ragged'' edge.

\begin{figure}
\begin{minipage}{6in}
\includegraphics[width=6in]{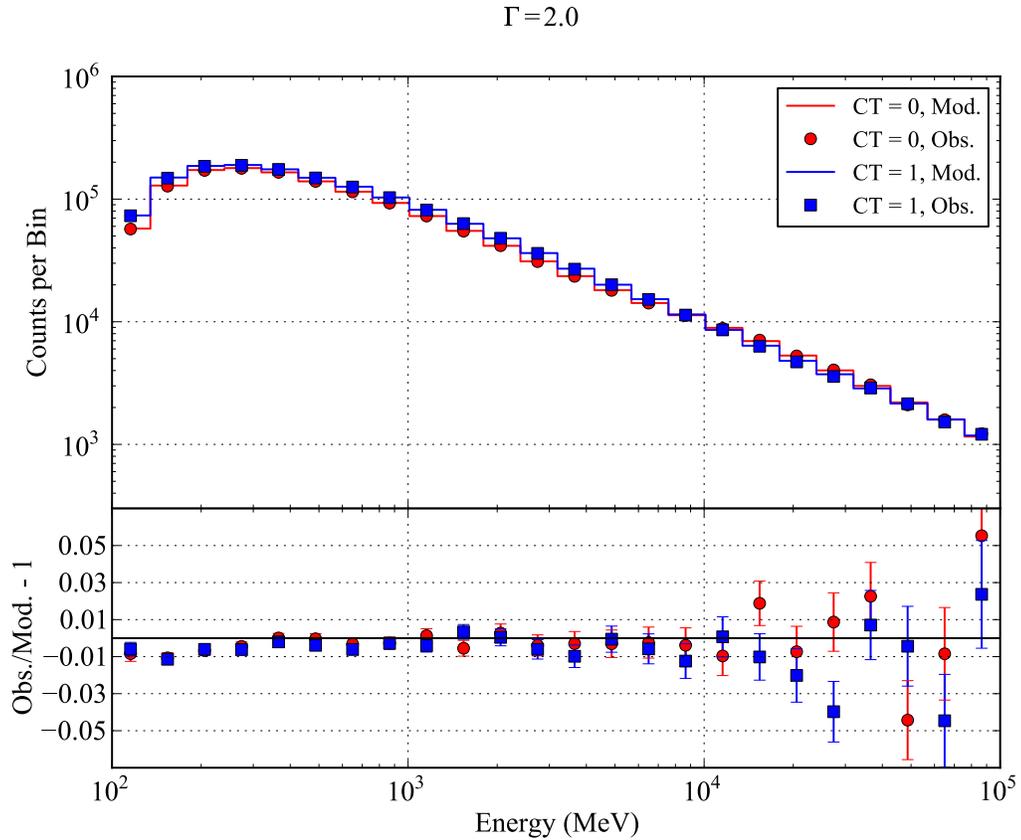}
\end{minipage}
\begingroup\renewcommand{\baselinestretch}{1.0}
\caption{Comparison of modeled and observed counts for a point source with a $\Gamma=2.0$ spectrum.  The histogram gives the model expectation and the data points the observed counts.  The residuals indicate that the accuracy is better than $1\%$ at all energies.  The minor but definite trend for an excess in the expected counts below 1 GeV can lead to a very small bias in the estimation of the photon index---see Table \ref{tab:valid_0}.  It is in this energy range the effects of PSF approximation and ``ragged'' edge effects are most pronounced.}
\renewcommand{\baselinestretch}{1.5}\endgroup
\label{plot12_2.0}
\end{figure}

\begin{figure}
\begin{minipage}{6in}
\includegraphics[width=6in]{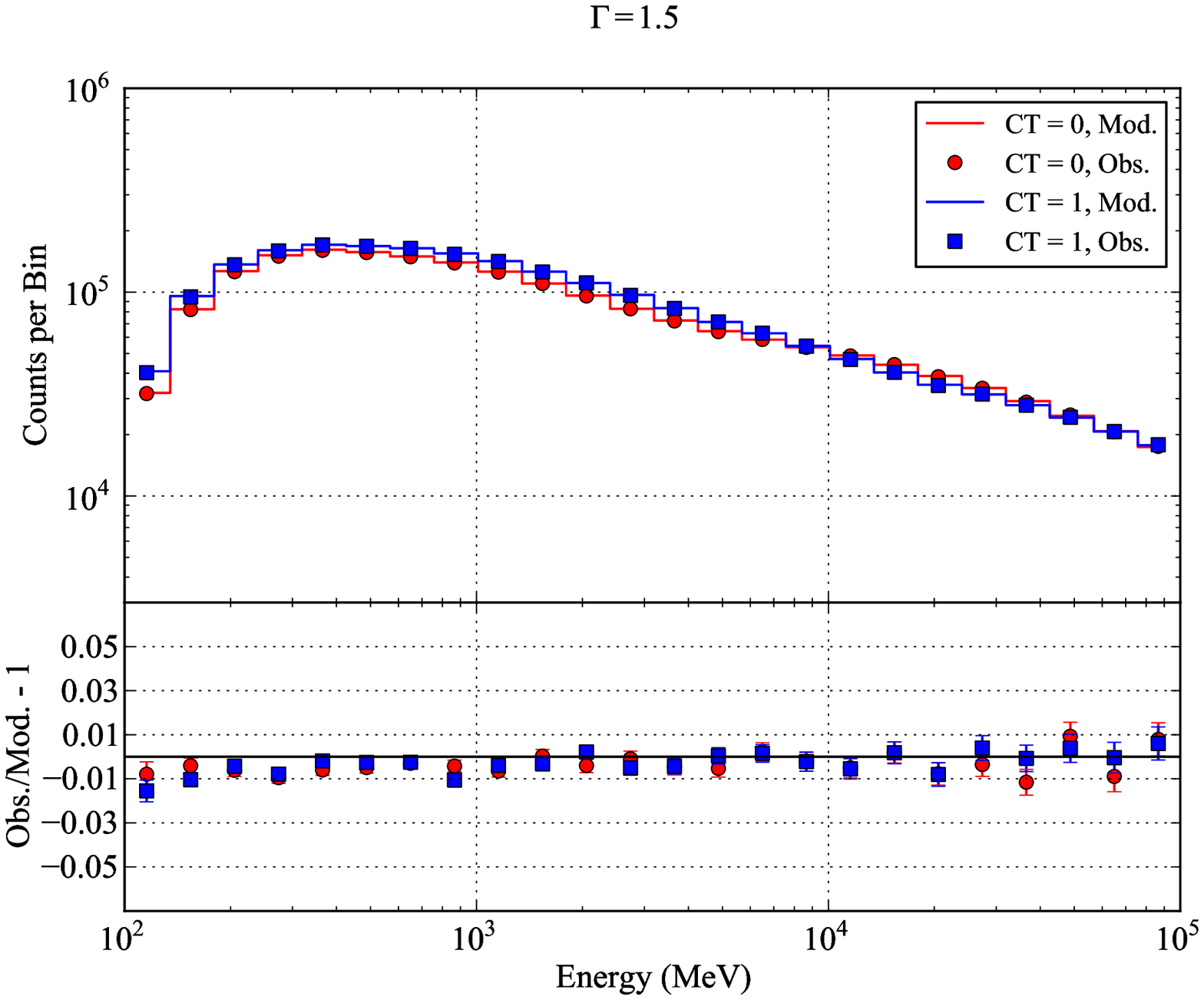}
\end{minipage}
\begingroup\renewcommand{\baselinestretch}{1.0}
\caption{Comparison of modeled and observed counts for a point source with a $\Gamma=1.5$ spectrum.  We include this model for its superior statistics at high energy, where it is clear the accuracy remains better than $1\%$.}
\renewcommand{\baselinestretch}{1.5}\endgroup
\label{plot12_1.5}
\end{figure}

We test these effects with the production code by again summing all 20 years of Monte Carlo data and comparing the observed and actual counts in a $10\dg$ ROI.  Since we are interested in the absolute accuracy of the model predictions, we do not renormalize the data or perform an initial spectral fit.  The results appear in Figures \ref{plot12_2.0} and \ref{plot12_1.5} and indicate overall accuracy of about $1\%$\footnote{The trend to overpredict the counts from $0.1$ to $1$ GeV is under investigation.}.

\subsection{Checking the Convolution}
The accuracy of the diffuse convolution is extremely important, particularly in the Galactic plane, because the the diffuse counts dwarf counts from all but the brightest point sources.  Small relative inaccuracies can become extremely statistically significant and bias point source fits.

\begin{figure}
\begin{minipage}{6in}
\includegraphics[width=6in]{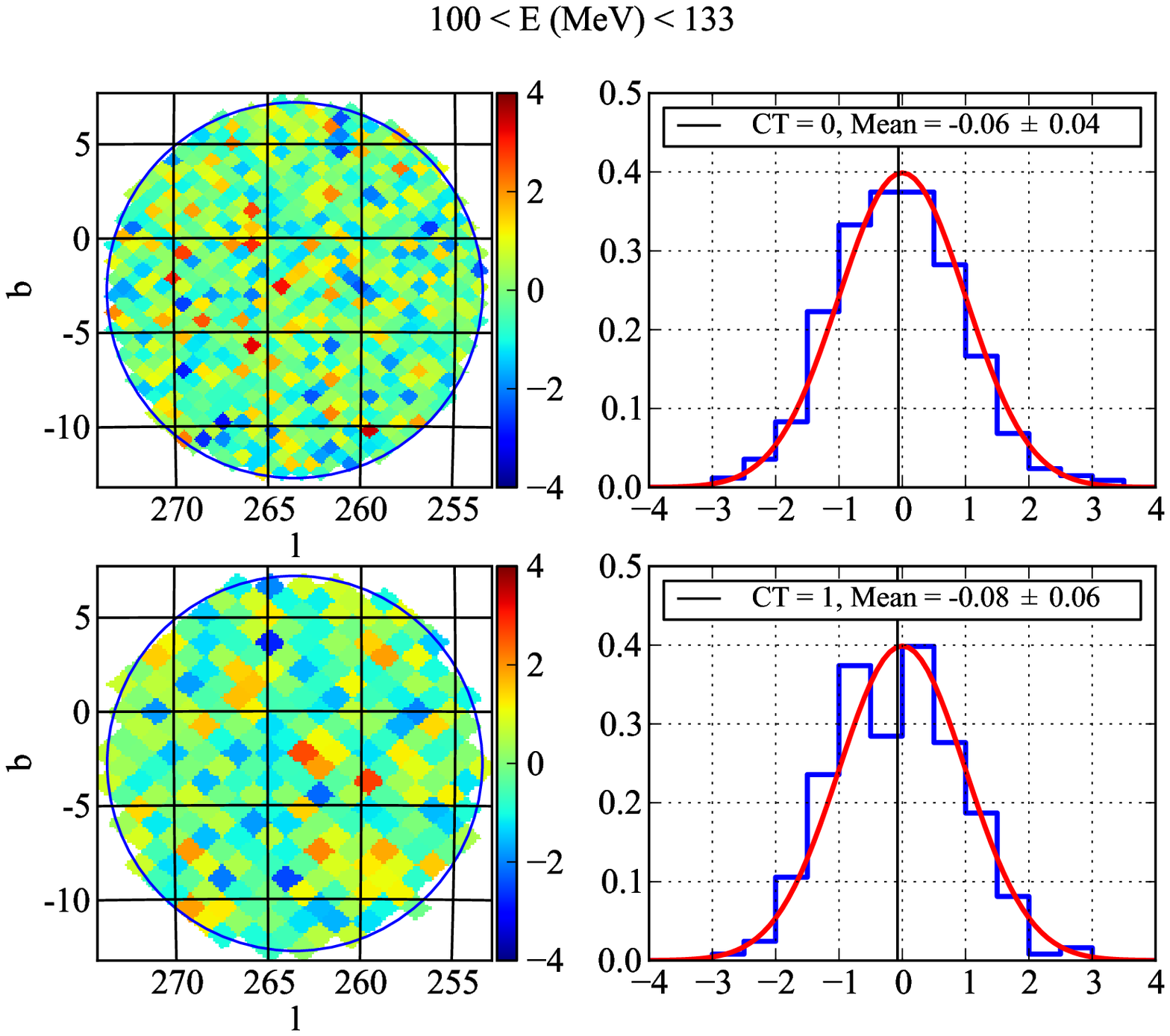}
\end{minipage}
\begingroup\renewcommand{\baselinestretch}{1.0}
\caption{The lefthand panels show the weighted residuals, $(N_{obs} - N_{mod})/\sqrt{N_{mod}}$, for the diffuse background (\emph{gll\_iem\_v02} + \emph{isotropic\_iem\_v02}) at low energy in zenithal equal-area (ZEA) projection.  The top (bottom) row shows front-converting (back-converting) events.  The ``ragged edge'' of the HEALPix data compared to the ROI boundary (blue circle) is clear at these low energies, particularly for the back events.  The righthand panels show a histogram of the weighted residuals.  There are many counts per pixel for these low energies (mean 102 (360) for front (back)), and the residuals should (and do) follow a normal distribution (shown in red).  The means of the distribution are consistent with 0, indicating no statistically significant bias.}
\renewcommand{\baselinestretch}{1.5}\endgroup
\label{ch3_plot15_00100}
\end{figure}

\begin{figure}
\begin{minipage}{6in}
\includegraphics[width=6in]{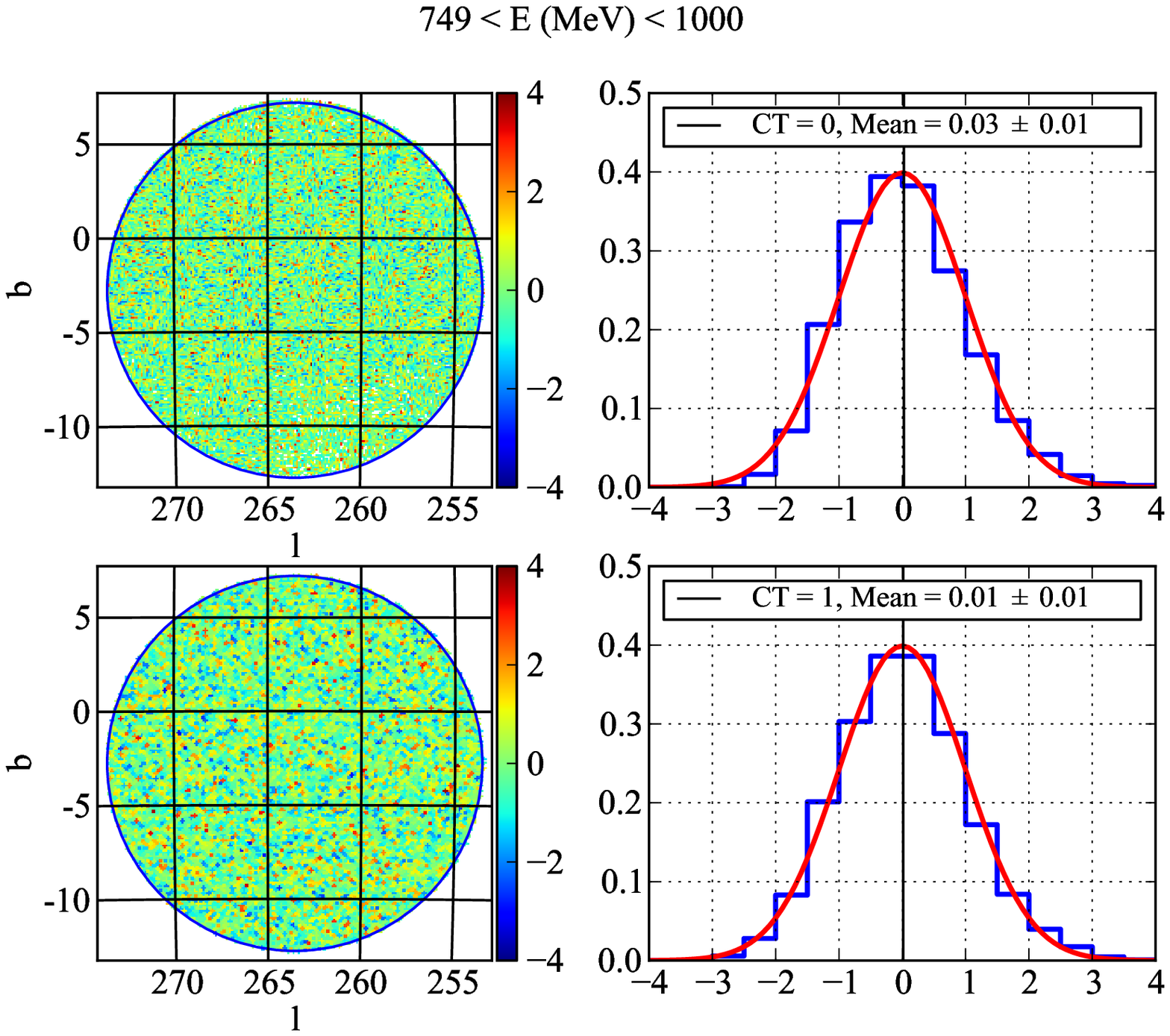}
\end{minipage}
\begingroup\renewcommand{\baselinestretch}{1.0}
\caption{Weighted diffuse residuals at moderate energies.  The spatial residuals are again featureless.  The count rate per pixel is at the limit of normal approximation (8 (24) for front (back)), and the residuals for the front-converting events clearly show a long right tail in keeping with the asymmetric Poisson distribution.  (The mean is accordingly slightly biased.)  The ``ragged'' edge is essentially eliminated by 1 GeV for nearly any reasonable ROI size.}
\renewcommand{\baselinestretch}{1.5}\endgroup
\label{ch3_plot15_00749}
\end{figure}

We first validate the spatial distribution, Eq. \ref{eq:diffuse_spatial_rate} as follows.  For each data pixel in the ROI, we evaluate Eq. \ref{eq:diffuse_spatial_rate}, as determined from our convolution scheme (including integration over energy) at the pixel center.  Since the pixels are small compared to the PSF, this is an accurate estimate of the pixel rate.  We then examine the weighted residuals, $(N_{obs} - N_{mod})/\sqrt{N_{mod}}$, as a function of position.  Representative results for two energy bands are presented in Figures \ref{ch3_plot15_00100} and \ref{ch3_plot15_00749} and are quite featureless, i.e., the model is correct in both shape and magnitude.

\begin{figure}
\begin{minipage}{6in}
\includegraphics[width=6in]{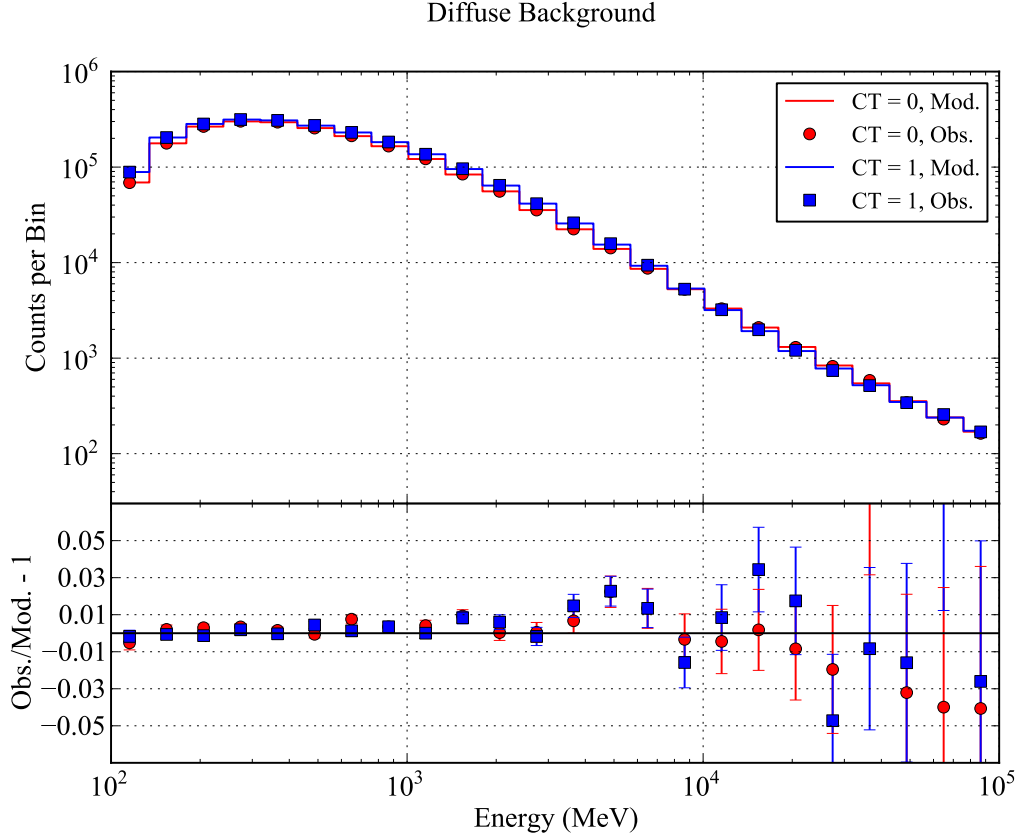}
\end{minipage}
\begingroup\renewcommand{\baselinestretch}{1.0}
\caption{The predicted and observed counts for the diffuse background (\emph{gll\_iem\_v02} + \emph{isotropic\_iem\_v02}) for all energies and conversion types.}
\renewcommand{\baselinestretch}{1.5}\endgroup
\label{ch3_plot14}
\end{figure}

At energies above 1 GeV, the pixelization becomes too small for this scheme to work.  However, above 1 GeV, the convolution is less important as the PSF scale is small compared with the model scale ($0.5\dg$ for \emph{gll\_iem\_v02}).  For the remainder of the validation, we simply present the integral of Eq \ref{eq:diffuse_spatial_rate} over the aperture in Figure \ref{ch3_plot14}.  The model predictions agree with the Monte Carlo data to about $1\%$ at all energies, although the poor statistics (from the relatively soft spectrum of the Galactic diffuse) at high energy preclude validation beyond about $5\%$.

\subsection{Checking the Maximum Likelihood Fit}
We test the absolute level of accuracy of the maximum likelihood estimates of the spectral parameters by performing maximum likelihood fits to the \emph{summed} data sets, i.e., by measuring the spectrum of a single point source with $\mcf = 2\mcf_4$.  Such an exercise indicates the magnitude of any bias for a point source in the limit of very small statistical errors.  Table \ref{tab:valid_0} shows the results for a point source alone, indicating better than percent-level accuracy.  Table \ref{tab:valid_1} shows results for the same sources in the presence of the diffuse background.  The additional degrees of freedom lead to a modestly increased bias in the flux measurement, but the accuracy is still about $1\%$; the accuracy of the photon index remains much better than $1\%$.

\begin{table}
\center
\begin{tabular}{ l | c | c | c }
      & $\Gamma=1.5$     & $\Gamma=2.0$     &   $\Gamma=3.0$ \\
\hline
  $100\times(\mcf_{sim}/\mcf_{fit}-1)$ (CT=0)     &-0.54       & 0.54    &   0.18 \\
  $100\times(\mcf_{sim}/\mcf_{fit}-1)$ (CT=1)     & 0.53       & 0.62    &   0.74 \\
  $100\times(\Gamma_{sim}/\Gamma_{fit}-1)$ (CT=0) & 0.07       & 0.11    &   0.02 \\
  $100\times(\Gamma_{sim}/\Gamma_{fit}-1)$ (CT=1) & 0.10       & 0.04    &  -0.01
\end{tabular}
\caption{The accuracy of maximum-likelihood estimated parameters for three point sources with a flux of $2\times10^{-4}$\fluxunits and three different power law photon indices.  The agreement is much better than $1\%$ for all three sources, and there is additionally good agreement between the estimates for front-converting events and back-converting events.  The general trend is for the source-as-fit to have a harder spectrum than simulated; this is in keeping with the slight excess of modeled events at low energy, as a hardened spectrum ameliorates this excess.}
\label{tab:valid_0}
\end{table}

\begin{table}
\center
\begin{tabular}{ l | c | c | c }
      & $\Gamma=1.5$     & $\Gamma=2.0$     &   $\Gamma=3.0$ \\
\hline
  $100\times(\mcf_{sim}/\mcf_{fit}-1)$ (CT=0)     & 0.84       & 1.01    &   0.87 \\
  $100\times(\mcf_{sim}/\mcf_{fit}-1)$ (CT=1)     & 0.92       & 1.26    &   1.62 \\
  $100\times(\Gamma_{sim}/\Gamma_{fit}-1)$ (CT=0) & 0.14       & 0.15    &  -0.03 \\
  $100\times(\Gamma_{sim}/\Gamma_{fit}-1)$ (CT=1) & 0.18       & 0.12    &  -0.07
\end{tabular}
\caption{As Figure \ref{tab:valid_0}, but with the Galactic and isotropic diffuse backgrounds added.  The agreement remains at the $1\%$ level.}
\label{tab:valid_1}
\end{table}

\subsection{Checking the Maximum Likelihood Fit: With Energy Dispersion}
\label{ch4:sub:energy_dispersion}

\begin{figure}
\begin{minipage}{6in}
\includegraphics[width=6in]{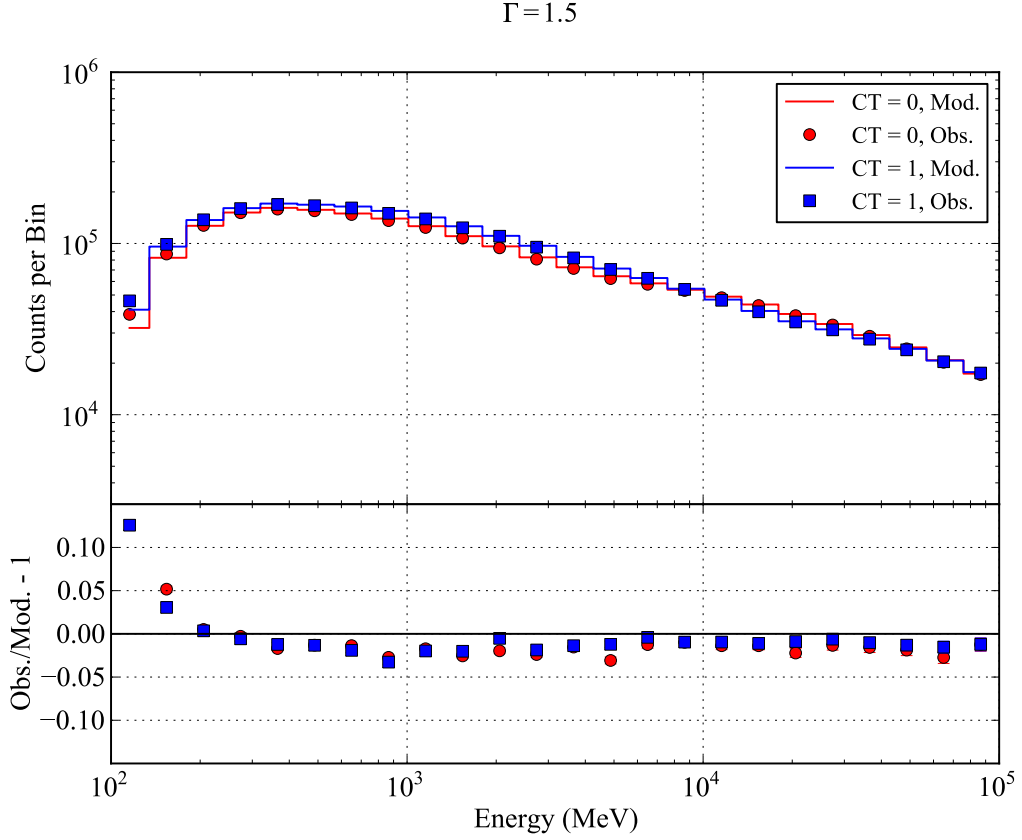}
\end{minipage}
\begingroup\renewcommand{\baselinestretch}{1.0}
\caption{As Figure \ref{plot12_2.0}, but for a power law source with photon index of 1.5 and allowing for energy dispersion in the data.  The model value is fixed at the Monte Carlo truth, so the residuals indicate how the photons have been redistributed from the true values.  The intrinsic spectrum of hard sources enhances the tendency for high energy photons to redistribute to low energies.}
\renewcommand{\baselinestretch}{1.5}\endgroup
\label{ch3_plot16_1.5}
\end{figure}

\begin{figure}
\begin{minipage}{6in}
\includegraphics[width=6in]{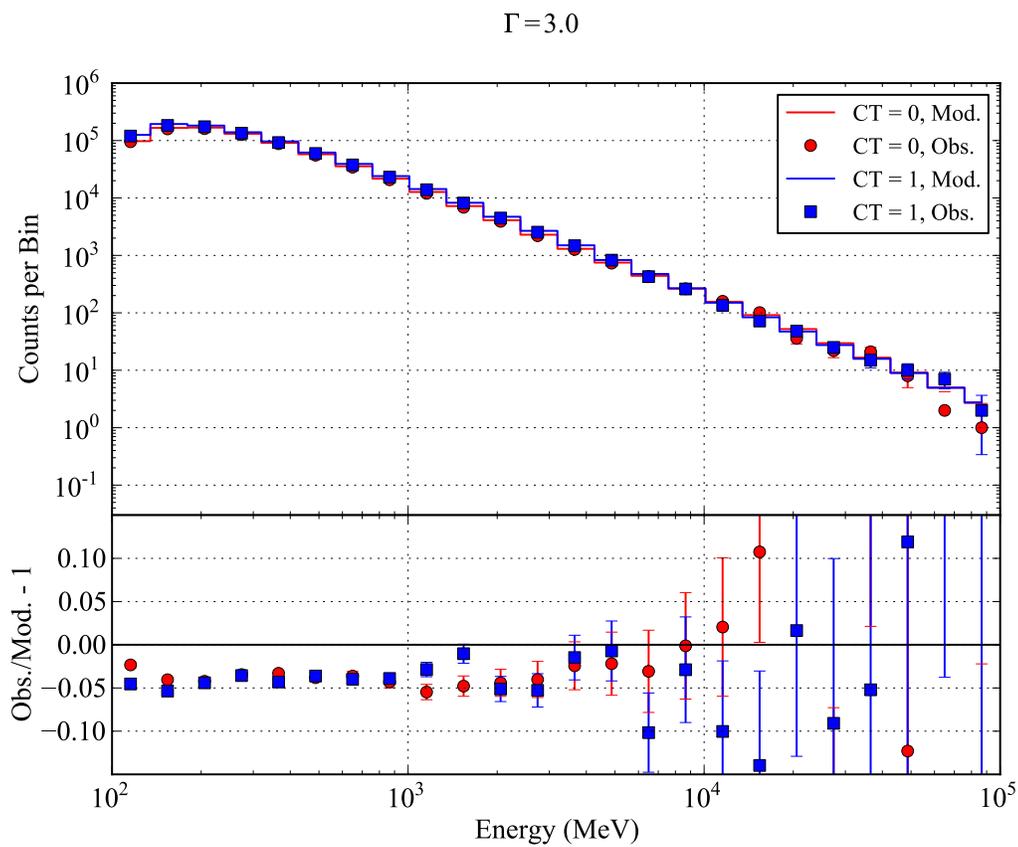}
\end{minipage}
\begingroup\renewcommand{\baselinestretch}{1.0}
\caption{As Figure \ref{ch3_plot16_1.5}, for a power law source with photon index of 3.0.  Soft sources show a decreased curvature at low energies, but the overall loss of photons from the passband is increased.}
\renewcommand{\baselinestretch}{1.5}\endgroup
\label{ch3_plot16_3.0}
\end{figure}

Previously, we have ignored the effects of energy dispersion by using data binned using the \emph{simulated} energy for each photon.  To illustrate and gauge the effect of energy dispersion, we repeat two of the above exercises with data binned using the \emph{reconstructed} energy.  In Figures \ref{ch3_plot16_1.5} and \ref{ch3_plot16_3.0}, the residuals---in which the model neglects energy dispersion but the data contains it---for two point sources with power law spectra appear.  It is clear that energy dispersion is not entirely negligible!  The residuals can be of order $10\%$, and the dispersion introduces slight curvature in the spectrum\footnote{The dominant mechanism is as follows: the effective area increases rapidly from $0.1$ to about $0.5$ GeV, while the energy dispersion is approximately symmetric in logarithmic energy, and there is a net ``migration'' of photons from high to low energy, causing the model to underpredict at low energy and overpredict at high energy.}.  However, it is clear that much of the curvature can be eliminated by restricting analysis to photons with reconstructed energies above 200 MeV.  Then---for power law sources, at least---the effect of energy dispersion is primarily a renormalization of the spectrum, decreasing the measured flux below its true value by a few percent.

While the residuals look somewhat dire, the parameter values for bright sources are not significantly affected.  Tables \ref{tab:valid_2} and \ref{tab:valid_3} show the absolute deviation of the measured parameters from the Monte Carlo truth without and with, respectively, the presence of a diffuse background.  The bias from neglect of energy dispersion is $<5\%$.

\begin{table}
\center
\begin{tabular}{ l | c | c | c }
      & $\Gamma=1.5$     & $\Gamma=2.0$     &   $\Gamma=3.0$ \\
\hline
  $100\times(\mcf_{sim}/\mcf_{fit}-1)$ (CT=0)     & 0.31       & 0.89     &   3.68 \\
  $100\times(\mcf_{sim}/\mcf_{fit}-1)$ (CT=1)     & 0.55       & 1.40     &   4.86 \\
  $100\times(\Gamma_{sim}/\Gamma_{fit}-1)$ (CT=0) & -0.53      & -0.64    &  -0.11 \\
  $100\times(\Gamma_{sim}/\Gamma_{fit}-1)$ (CT=1) & -0.31      & -0.41    &   0.25
\end{tabular}
\caption{As Table \ref{tab:valid_0}, but now including the effects of energy dispersion in the data.  No diffuse background is included.}
\label{tab:valid_2}
\end{table}

\begin{table}
\center
\begin{tabular}{ l | c | c | c }
      & $\Gamma=1.5$     & $\Gamma=2.0$     &   $\Gamma=3.0$ \\
\hline
  $100\times(\mcf_{sim}/\mcf_{fit}-1)$ (CT=0)     & 0.34      &-0.14     &  -1.86 \\
  $100\times(\mcf_{sim}/\mcf_{fit}-1)$ (CT=1)     & 0.06      & 0.56     &  -1.31 \\
  $100\times(\Gamma_{sim}/\Gamma_{fit}-1)$ (CT=0) &-0.06      &-0.23     &  -0.96 \\
  $100\times(\Gamma_{sim}/\Gamma_{fit}-1)$ (CT=1) & 0.08      &-0.27     &  -1.30
\end{tabular}
\caption{As Table \ref{tab:valid_2}, but now including a diffuse background.}
\label{tab:valid_3}
\end{table}

\section{Spectral Analysis of Ensembles: Verification of Central Values and Error Estimates}

The most important capability of \ptl is reliable and rapid extraction of spectral parameters and estimates for their statistical errors.  To test this, we perform a spectral analysis of 20 Monte Carlo realizations of the diffuse background and a point source at a given flux.  We plot the results for a series of decreasing point source fluxes, $\mcf_6$, $\mcf_7$, $\mcf_8$, and $5\mcf_9$.  The latter flux is below the detection threshold of the LAT for modest (1-2 years) integration baselines for sources near the Galactic plane, while the penultimate ($\mcf_8$) is marginal.  $\mcf_7$ is the flux of a typical Galactic source.  We show the results both in absolute units and in units in which the parameters become approximately bivariately normal.

\subsection{Spectral Parameters}
\begin{figure}
\begin{minipage}{6in}
\includegraphics[width=6in]{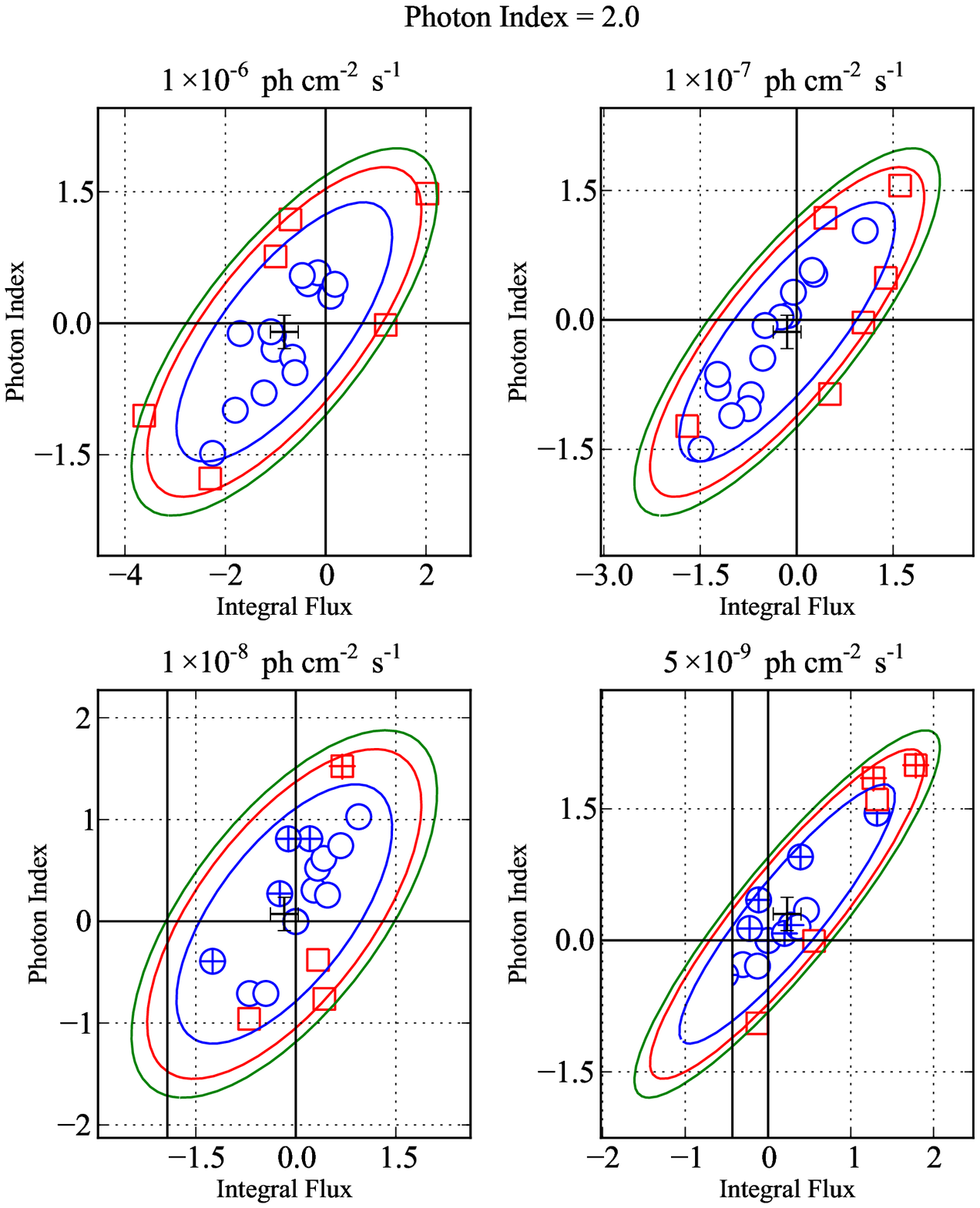}
\end{minipage}
\begingroup\renewcommand{\baselinestretch}{1.0}
\caption{Results for $\Gamma=2.0$ in error-weighted units.  The contours give the 68\%, 95\%, and 99\% confidence intervals as estimated from a Gaussian distribution using the sample covariance matrix.  Blue circles (red squares) are the 68\% (32\%) of the sample closest to (farthest from) the mean.  The single, black errorbar at the center of the contours gives the sample mean, and the length of the error bars gives the 1-d sample standard deviations divided by $\sqrt{N}$.  The black axis marks crossing at the origin represent the simulated (``true'') values.  Thus, a failure of the cross to overlap the origin indicates a systematic bias in the parameter estimate.  The position of the cross relative to the origin gives the magnitude of the bias in ``sigma'' units.
Sources with a test statistic $<10$ are indicated with a cross-filled symbol.}
\renewcommand{\baselinestretch}{1.5}\endgroup
\label{fig:plot13_pl_2.0_pull1}
\end{figure}

We consider first Fig. \ref{fig:plot13_pl_2.0_pull1} which shows a scatter plot of the estimated integral flux and photon index for a power law source with $\Gamma=2.0$ in error-weighted units: $x_{i}' = (x_i - \hat{\mu}_{x})/\sigma_i$, i.e., we have subtracted from each estimated parameter the sample mean and then divided by the estimated error for the parameter.  If the error estimates are accurate (implying the likelihood surface is approximately Gaussian), then the error-weighted estimates should be normally distributed.  Thus, we draw the confidence contours of a two-dimensional Gaussian (using the sample covariance matrix) on each figure to determine if the appropriate sample fraction lies within the appropriate contour.  In general, this is the case, indicating (a) the Gaussian approximation of the likelihood surface is a good one and (b) the likelihood surface is being accurately measured by the error estimation algorithms.

The sample mean has an error that scales as the square root of the sample size, say $1/\sqrt{N}$.  With sufficiently accurate estimates of the parameter errors (bright sources) and/or sufficiently numerous Monte Carlo realizations, we can begin to resolve systematic bias in the parameter estimates.  We adopted the former approach in the previous section and explore the latter approach here.  The magnitude of the bias, in units of the statistical error of a single source, can be read off of the figure by the position of the black cross with error bars.  For the bright sources ($\mcf=\mcf_6$), systematic bias is detectable but is at most comparable to the statistical error.  For dimmer sources, the statistical errors dominate.

\begin{figure}
\begin{minipage}{6in}
\includegraphics[width=6in]{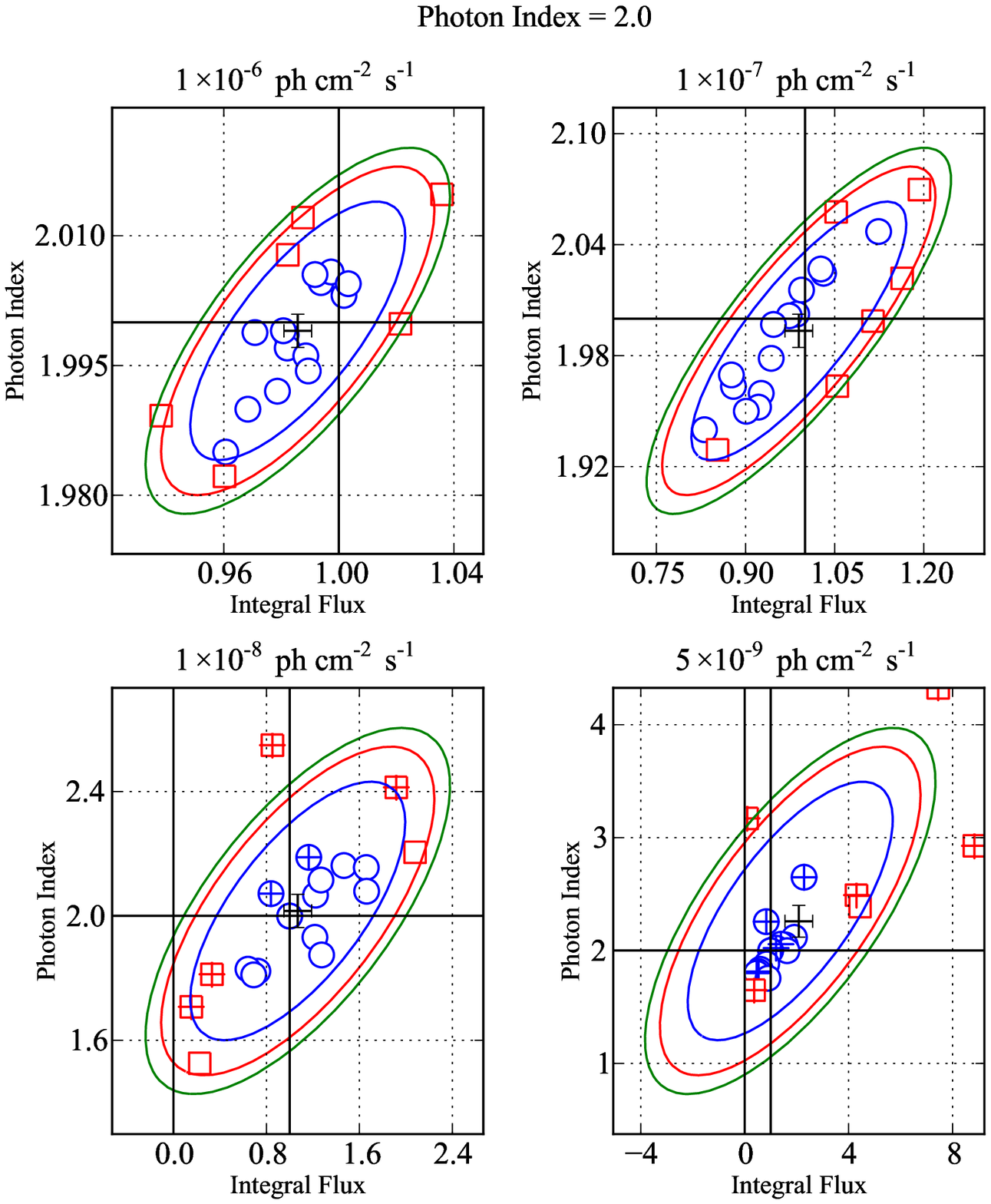}
\end{minipage}
\begingroup\renewcommand{\baselinestretch}{1.0}
\caption{As Figure \ref{fig:plot13_pl_2.0_pull1} in relative units.  The thick dashed line indicates zero flux.  Absolute outlying values are more apparent at low fluxes.}
\renewcommand{\baselinestretch}{1.5}\endgroup
\label{fig:plot13_pl_2.0_pull0}
\end{figure}

To gauge the absolute magnitude of both the statistical and systematic errors, we display the absolute parameter values in Fig. \ref{fig:plot13_pl_2.0_pull0}.  (Note we have still scaled the flux by the simulated value.)  Here, the position of the parameters relative to the contours are not meaningful and are only shown for reference.  In the case of $\Gamma=2.0$ and $\mcf=\mcf_6$, we see the absolute error on the flux is $<2\%$.

\subsection{Position Parameters}
\begin{figure}
\begin{minipage}{6in}
\includegraphics[width=6in]{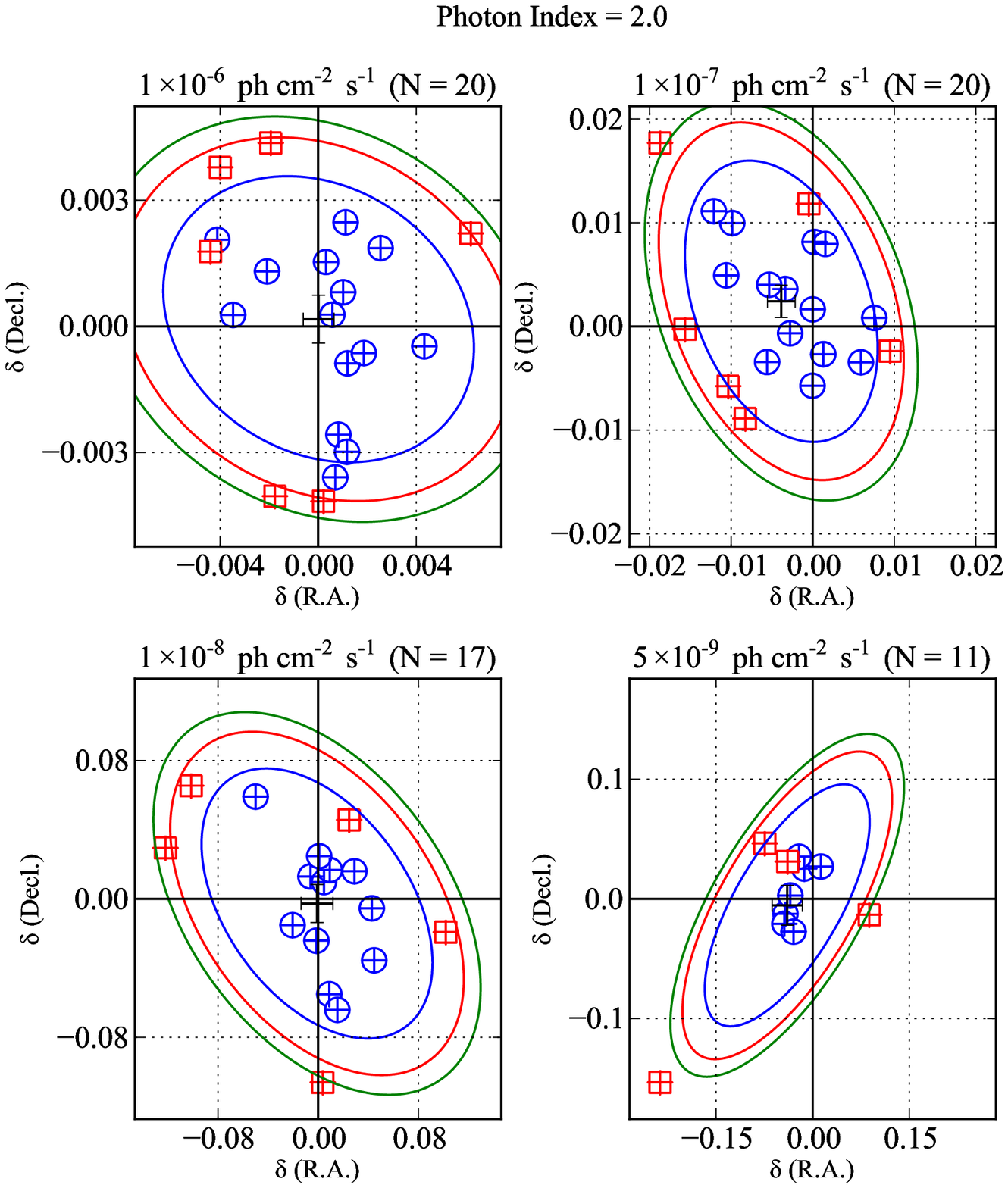}
\end{minipage}
\begingroup\renewcommand{\baselinestretch}{1.0}
\caption{The angular deviation in absolute units ($\dg$) for the position in the R.A. and Decl. directions.  The number of sources with successful position fits is indicated by $N$.}
\renewcommand{\baselinestretch}{1.5}\endgroup
\label{fig:plot17_pl_2.0_pull0}
\end{figure}

\begin{figure}
\begin{minipage}{6in}
\includegraphics[width=6in]{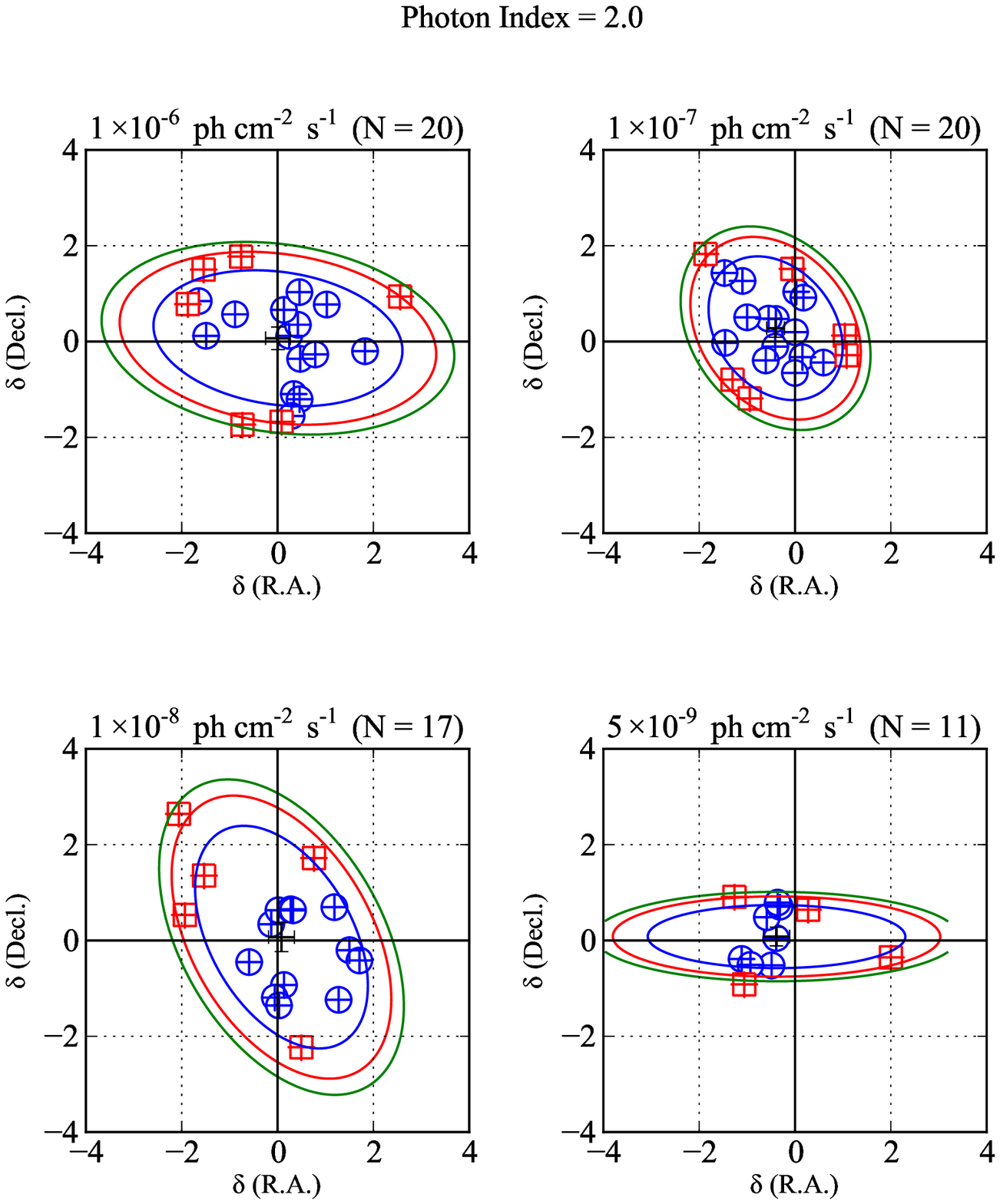}
\end{minipage}
\begingroup\renewcommand{\baselinestretch}{1.0}
\caption{As Figure \ref{fig:plot17_pl_2.0_pull0} but with error-weighted units.}
\renewcommand{\baselinestretch}{1.5}\endgroup
\label{fig:plot17_pl_2.0_pull1}
\end{figure}

To check the accuracy of the MLE for the source position and position uncertainty, we first performed a ML fit for the spectral parameters and followed with a ML fit for the position.  We did not iterate further.  The deviations in relative units and error-weighted units for Right Ascension and Declination are shown in Figures \ref{fig:plot17_pl_2.0_pull0} and \ref{fig:plot17_pl_2.0_pull1}, indicating that the mean of the position estimate is consistent with the simulated value and the the position uncertainty estimates are not too small, indicating only a small correlation between the position parameters and the spectral parameters.  Some of the low-flux ($\mathcal{F}\leq\mathcal{F}_8$) sources have a signal-to-noise ratio too low for robust localization.

By symmetry and the apparent insensitivity of the best-fit position to the precise value of the spectrum, we conclude that position estimates will be unaffected by energy dispersion.

\section{Summary}
By combining 20 years of simulated data for bright sources, we demonstrated that machinery used by \ptl to calculate expected rates for point sources and for diffuse sources is generally accurate to within a few percent of the expectation.  By performing spectral (and position) ML fits on ensembles of sources, we assessed the accuracy of the central values and uncertainty estimates reported by \ptl over 2.5 decades of source brightness, finding generally good agreement with systematic bias $<10\%$.
 
\chapter{Applications of \emph{Pointlike}}
\label{ch4}

Having detailed the inner workings of \ptl and validated the tool on a variety of test cases, we now outline some of the science results \ptl is capable of delivering.  We start with ``single-source'' analysis---by which we mean analysis of a relatively small region of the sky---to demonstrate the flexibility of \emph{pointlike}.  To showcase the speed of \emph{pointlike}, we then develop an all-sky analysis in which we determine spectra for every known GeV source.  Finally, we make use of this ``optimal'' model for other analysis tasks.

\section{Single-source Analysis}

To illustrate the use of \ptl in its ``single-source'' mode, we provide an example analysis of sources in the Cygnus region of the sky.  This region, rich in sources, has already delivered multiple discoveries, including the LAT's first detection of a new radio-loud pulsar\cite{j2021}, a particularly bright radio-quiet pulsar in the $\gamma$-Cygni supernova remnant\cite{blind_search_16}, the discovery of orbitally-modulated emission from Cygnus X-3\cite{cygx3}, and one of the first pulsars discovered in radio \emph{after} detection by \fermi\cite{two_pulsar_paper}.  This is clearly a region that rewards accurate analysis!

\subsection{Data Prepaparation}
The \emph{FT1} and \emph{FT2} files are obtained from a database maintained at SLAC\footnote{http://glast-ground.slac.stanford.edu/DataPortalAstroServer/, SLAC credentials required}.  Similar data products are available from the FSSC\footnote{http://fermi.gsfc.nasa.gov/ssc/data/access/lat/}.  For this exercise, we select data obtained between 4 August 2008 and 8 August 2010, processed using ``Pass 6'' reconstruction algorithms.  We restrict data to events with a reconstructed position lying with $15\dg$ of PSR J2021+3651 and with reconstructed energies between 100 MeV\footnote{To ameliorate effects of neglecting energy dispersion, we restrict \emph{analysis} to energies above 200 MeV.} and 100 Gev.  Finally, we use ``diffuse'' class events, a subset of events with a low background contamination.

Before ingestion into \ptln, we apply the Science Tool \emph{gtmktime}\footnote{http://fermi.gsfc.nasa.gov/ssc/data/analysis/scitools/help/gtmktime.txt} to the data.  The tool applies a general filter based on the S/C orientation and position.  We apply a filter that modifies the GTI in the \emph{FT1} file to exclude times when the S/C exceeds a $52\dg$ rocking angle and when the $15\dg$ ROI defined in the preceding paragraph approaches the horizon.  These excisions are designed to decrease contamination from Earth's limb and, since they use the \emph{known} S/C telemetry are essentially bias free.

Additional preparation, binning, and generation of secondary data products are performed within \ptln.  These procedures are discussed in Chapter \ref{ch3}, and we only provide some of the specifics here: data are binned with eight-bin-per-decade resolution, the livetime as a function of position is tabulated with $1\dg$ resolution, and we remove events with reconstructed zenith angles $>105\dg$ and reconstructed incidence angles $>66.4\dg$. 

\subsection{Source Model}
After preparing the data, we must construct a spectral model for the sources in the region.  Models for both the diffuse background and point sources are in a state of continual improvement as more data is acquired.  E.g., additional photons allow dimmer point sources to be detected, which in turn allow a better diffuse model to be generated, in turn allowing more point source detections, and so forth.  Systematic refinements independent of integration time also allow for model improvement.

\ptl is capable accepting a wide variety of input spectral models.  For instance, it can parse all published \fermi point source catalogs as well as many source lists internal to the LAT Collaboration.  It can ingest mapcubes representing a variety of diffuse emission.  However, rather than using the most recent internal models, we choose to make contact with the published literature and adopt models used in and resulting from the 1FGL\cite{1fgl} catalog analysis.  For diffuse models, we use the \textit{gll\_iem\_v02} mapcube for the Galactic diffuse and the \textit{isotropic\_iem\_v02} intensity tabulation\footnote{http://fermi.gsfc.nasa.gov/ssc/data/access/lat/BackgroundModels.html} for the isotropic background.  For point sources, we use the spectral models presented in the 1FGL catalog.

In general, there is a tradeoff between a statistically-optimal fit---a joint fit in which the parameters for all sources are allowed to vary---and the computational intractability of minimizing a high-dimensioned function with a large data set.  The typical approach is to select an ROI centered on the source of interest and allow sources within a fixed angular separation---say $5\dg$ or $10\dg$---to vary along with the source of interest.  Sources at larger radii remain fixed.  Due to the large tails of the PSF at low energy, sources lying outside the ROI must be included in the source model, and it is clearly impossible to fit these sources.  To estimate parameters for these sources, similar analyses must be performed for other regions of interest in what is essentially a bootstrap procedure.  We shall have more to say on this in devising an iterative, all-sky fit below.

In the following analysis, we model the 53 1FGL sources within $20\dg$ of the ROI center, and we maximize the likelihood with respect to the 17 1FGL sources within $8\dg$ of the ROI center as well as three parameters for the diffuse background, a power law scaling for the Galactic diffuse and a normalization for the isotropic diffuse.

\subsection{Broadband Spectroscopy}
\label{ch4:subsec:bba}
The first step in any analysis is to maximize the likelihood over the free sources.  This is done using the gradient fitter as described in Chapter \ref{ch3}.  During this process, spectral parameters and estimates are determined for all 17 free sources.  The process takes $1-2$ minutes.  As an example of the output, and to begin an example on the importance of correct modeling, we report the broadband spectral parameters for two sources, PSR J2021+3651\cite{j2021} and 1FGL J2015.7+3708, thought to be a blazar\cite{halpern_blazar}.

\begin{verbatim}
0 -- 1FGL J2021.0+3651 fitted with PowerLaw, e0=798
Norm      : (1 + 0.013 - 0.012) (avg = 0.013) 1.34e-010
Index     : (1 + 0.005 - 0.005) (avg = 0.005) 2.34
Ph. Flux  : (1 + 0.021 - 0.021) (avg = 0.021) 1.29e-006  (DERIVED)
En. Flux  : (1 + 0.012 - 0.012) (avg = 0.012) 7.62e-010  (DERIVED)

1 -- 1FGL J2015.7+3708 fitted with PowerLaw, e0=1051
Norm      : (1 + 0.043 - 0.041) (avg = 0.042) 1.47e-011
Index     : (1 + 0.014 - 0.014) (avg = 0.014) 2.46
Ph. Flux  : (1 + 0.073 - 0.068) (avg = 0.070) 3.27e-007  (DERIVED)
En. Flux  : (1 + 0.043 - 0.041) (avg = 0.042) 1.62e-010  (DERIVED)
\end{verbatim}

In this output, ``e0'' indicates the pivot energy, an estimate for the energy at which the flux density (``norm'') and photon index (``index'') are least correlated.  The integral photon flux ($>100$MeV,cm$^{-2}$ s$^{-1}$) and integral energy flux (erg cm$^{-2}$ s$^{-1}$) round out the parameters.  The relative error is reported, and since the parameters are transformed to log space internally, they are naturally two-sided when transformed back, but for bright sources become approximately symmetric.

The 1FGL models only provide power law parameters for the point sources.  Yet the two brightest sources in the region, PSRs J2021+4026\cite{blind_search_16} and J2021+3651, are known to have exponentially-suppressed spectra.  Further, 1FGL J2015.7+3708 is only $1.1\dg$ from PSR J2021+3651.  To correct this flaw in the model while simultaneously demonstrating the flexibility of \emph{pointlike}, we modify the sources interactively.  We first add a cutoff energy to J2021+4026 and re-maximize the likelihood, increasing its logarithm by 480.  We repeat the process for J2021+3651, improving the log likelihood by 296.  We now have:

\begin{verbatim}
0 -- 1FGL J2021.0+3651 fitted with ExpCutoff
Norm      : (1 + 0.026 - 0.026) (avg = 0.026) 2.01e-010
Index     : (1 + 0.020 - 0.020) (avg = 0.020) 1.74
Cutoff    : (1 + 0.065 - 0.061) (avg = 0.063) 2.93e+003
Ph. Flux  : (1 + 0.034 - 0.033) (avg = 0.033) 8.18e-007  (DERIVED)
En. Flux  : (1 + 0.018 - 0.017) (avg = 0.018) 5.44e-010  (DERIVED)

1 -- 1FGL J2015.7+3708 fitted with PowerLaw, e0=1051
Norm      : (1 + 0.040 - 0.038) (avg = 0.039) 1.6e-011
Index     : (1 + 0.015 - 0.014) (avg = 0.014) 2.62
Ph. Flux  : (1 + 0.071 - 0.067) (avg = 0.069) 4.67e-007  (DERIVED)
En. Flux  : (1 + 0.044 - 0.042) (avg = 0.043) 1.94e-010  (DERIVED)
\end{verbatim}

We note the reported spectrum for 1FGL J2015.7+3708 has softened significantly, $\Gamma=2.46\rightarrow\Gamma=2.62$, well outside of the statistical errors of $\pm0.04$, while the integral flux has increased by $>40\%$.  We take up again the (cautionary) tale of this source when we discuss model-independent spectroscopy, but first we give a brief example of localization with \emph{pointlike}.

\subsection{Source Localization}
\label{ch4:subsec:localization}
To demonstrate the localization capabilities of \emph{pointlike}, we again focus on the region near PSR J2021+3651.  Since the position of J2021+3651 is well-known from radio timing, its brightness in $\gamma$-rays allows a validation to high statistical precision of the ML best-fit position.  The results of localizing J2021+3651 (having first made the spectral modifications described in \S\ref{ch4:subsec:bba}, viz. adding cutoff energies to the two bright pulsars) appear in Table \ref{tab:psrj2021}.

\begin{table}
\begin{tabular}{l c c c c}
Method & R.A. & Decl. & r95 ($\dg$) & $\delta$/r95 \\
\hline
Timing & 305.2728 & 36.858 & --- & --- \\
BB, PL & 305.2533 & 36.850 & 0.010 & 1.5  \\
BB, PL+EC & 305.2543 & 36.848 & 0.010 & 1.5 \\
BF     & 305.2542 & 36.848 & 0.010 & 1.5
\end{tabular}
\caption{A comparison of the DC position obtained for PSR J2021+3651 with three methods to the (well-constrained) timing position\cite{j2021}.  The ``BB'' entries corresponding to using a broadband spectral model (respectively a simple power law and a power law with exponential cutoff), while ``BF'' refers to the model-independent method discussed in Chapter \ref{ch3}.  The three methods are self-consistent, but inconsistent with the radio position at high confidence.}
\label{tab:psrj2021}
\end{table}

In this iteration, it is clear that either the position is off or the uncertainty estimate is too small, as the best-fit position lies well outside of r95, the radius inside which we expect the ML position to lie in $95\%$ of realizations of the experiment.  In the time since the 1FGL catalog was collated, we have detected a new point source at (R.A., Decl.) $\approx$ (304.46, 36.5)\footnote{In fact, it was this exercise---and subsequent failure to arrive at a good localization---that led to the detection!}.  (We use this source as an example of the source detection algorithm in \S\ref{ch4:sec:tsmaps}.)  This source is $<0.8\dg$ from PSR J2021+3651, and we expect that failing to account for its emission may bias---albeit slightly---the position of the pulsar.  We therefore add a new point source to the model---a task which can be done interactively---at a trial position estimated from the TS map appearing in Figure \ref{ch4_cygnus_ts}.  We adopt a simple power law with Ansatz parameters ($\Gamma=2.0$ and an integral flux of $10^{-7}$\fluxunits) for the spectum, and we (a) re-maximize the likelihood, (b) re-localize the new source, and (c) re-maximize the likelihood.  The new source improves the log likelihood by 361, and after incorporating its emission in the spectral model, localization of J2021+3651 yields the results in Table \ref{tab:psrj2021_2}.  With the new model, the LAT position is consistent with the radio position, and with a precision of better than one arcminute!  As in the previous section, we see both the importance of correct modeling in likelihood techniques and the consequent benefit of interactive/exploratory analysis.

\begin{table}
\begin{tabular}{l c c c c}
Method & R.A. & Decl. & r95 ($\dg$) & $\delta$/r95 \\
\hline
Timing & 305.2728 & 36.858 & --- & --- \\
BB, PL+EC & 305.2685 & 36.859 & 0.010 & 0.83 \\
\end{tabular}
\caption{PSR J2021+3651, comparison of the DC position to the (well-constrained) timing position.  This fit includes a new, highly significant source, and the \fermi position now agrees with the timing position to a precision of $<1'$.}
\label{tab:psrj2021_2}
\end{table}

\subsection{Model-independent Spectroscopy}
\label{ch4:subsec:mis}
We introduced a method of generating (nearly) model-independent spectra in \S\ref{ch3:subsec:model_independent}.  By model-independent, we mean that we do not assume a broadband functional form for the source in question and we maximize a series of likelihoods to estimate the flux in each band.  However, we do assume an Ansatz model in the initial fit in order to obtain parameters for the other sources in the ROI in a consistent fashion.

\begin{figure}
\begin{minipage}{6in}
\includegraphics[width=6in]{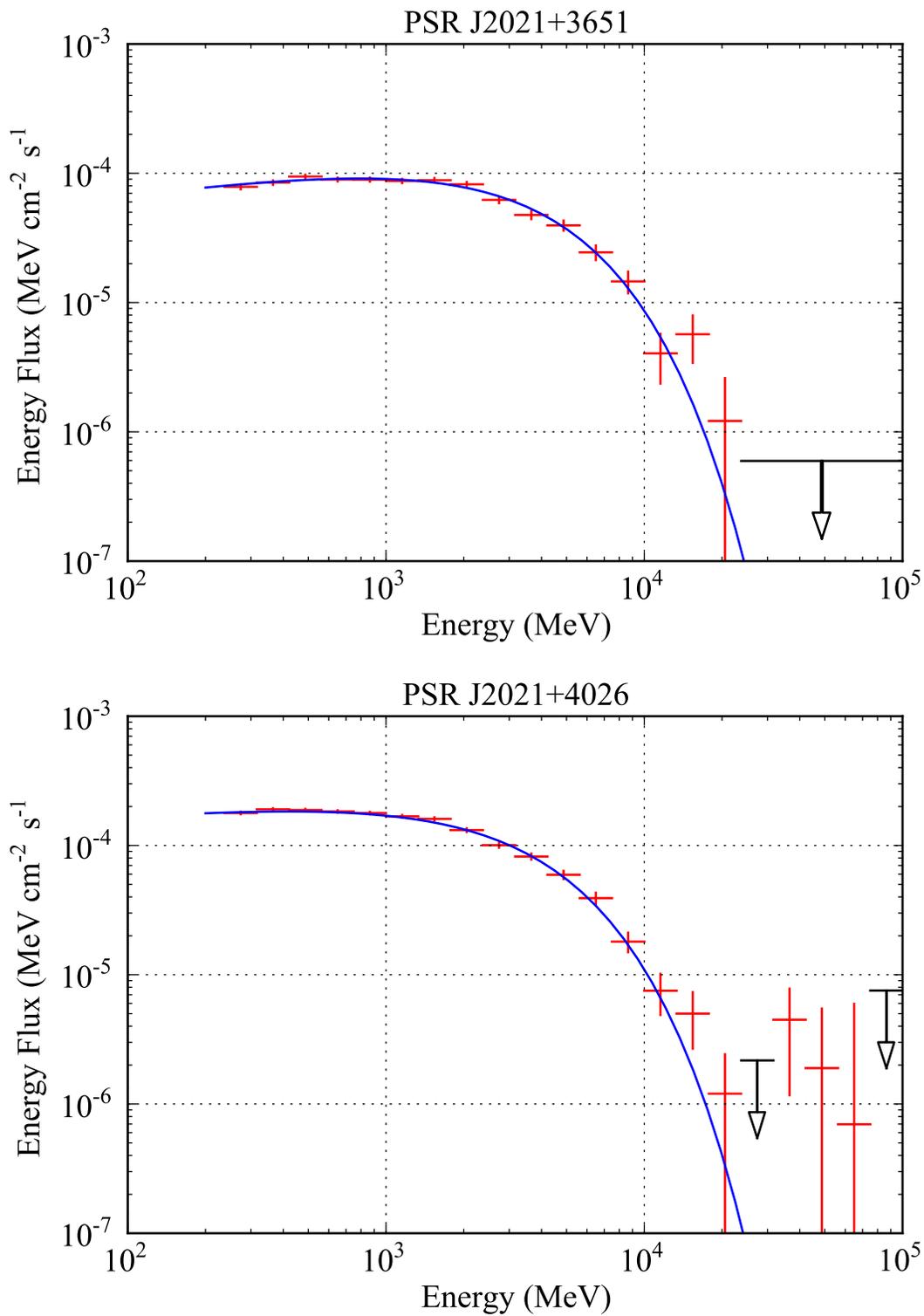}
\end{minipage}
\begingroup\renewcommand{\baselinestretch}{1.0}
\caption{Spectra for PSRs J2021+3651 and J2021+4026.}
\renewcommand{\baselinestretch}{1.5}\endgroup
\label{ch4_plot5_p2}
\end{figure}

We recall that the broadband spectral fit for 1FGL J2015.7+3708 was affected by the addition of a cutoff parameter to the nearby, bright PSR J2021+3651.  We can also examine what impact it has on the model-independent spectrum.  In Figure \ref{ch4_plot5_p2}, we show a series of spectra for 1FGL J2015.7+3708 as we progressively refine the model for the nearby sources as described above.  In the first panel, we see a strong suppression of the spectrum below $\approx 500$ MeV.  This is easily understood as the power law model for PSR J2021+3651 overpredicts the emission at low energy (the spectrum is concave down), and through the poor PSF at these energies the overprediction suppresses the flux estimates for neighboring sources.  When we model the pulsars correctly, with a cutoff, the low-energy behavior is much improved, as seen in the second panel of the figure.  Note that there is essentially no change above 1 GeV; the PSF is sufficiently good that the sources are resolved.

Looking closely at this new fit, we see there is tension between the simple power law model and the model-independent spectra.  Motivated by this evidence for spectral curvature, we can switch the model for 1FGL J2015.7+3708 to a log parabolic form:
\begin{equation}
\mathcal{F}(E) = N_0\left(\frac{E_b}{E}\right)^{\alpha + \beta\log(E_b/E)},
\end{equation}
which reduces to a power law for $\beta=0$.  After re-maximizing the likelihod, we see the new fit is in much better agreement with the model-independent points.

\begin{figure}
\begin{minipage}{6in}
\includegraphics[width=6in]{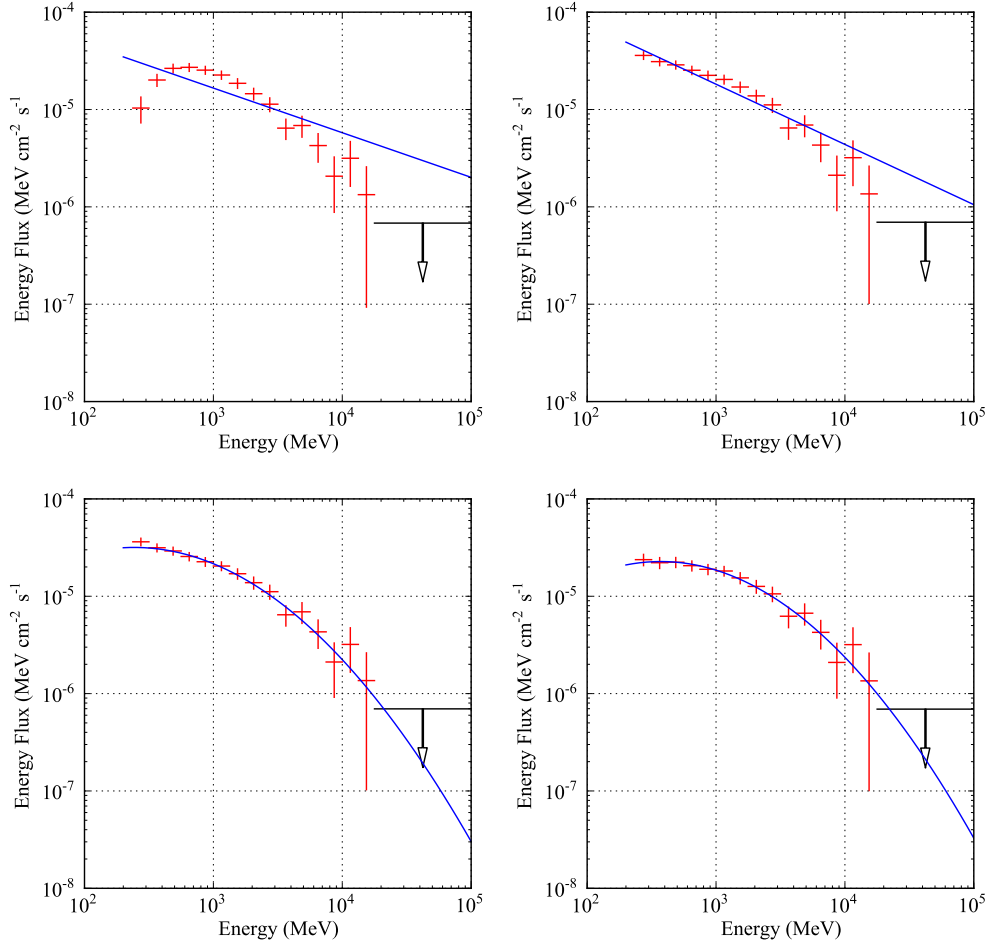}
\end{minipage}
\begingroup\renewcommand{\baselinestretch}{1.0}
\caption{The model-independent spectrum for 1FGL J2015.7+3708.  At upper left, the initial spectrum estimated with all sources fixed at 1FGL values.  At upper right, the spectrum after adding cutoff parameters to PSRs J2021+3651 and J2021+4026.  At lower left, no external sources have changed but we model 1FGL J2015.7+3708 with a log parabola (see text).  Finally, at lower right, we have added the new source discussed in the previous section.}
\renewcommand{\baselinestretch}{1.5}\endgroup
\label{ch4_plot5_p1}
\end{figure}

Finally, we note that the new source we introduced in the previous section is only $\approx0.8\dg$ from 1FGL J2015.7+3708 and, like J2021+3651 but to a much smaller extent, is likely to affect the low-energy spectrum of 1FGL J2015.7+3708.  In the final panel of Figure \ref{ch4_plot5_p1}, we show the spectrum after this source has been included in the model.  As expected, it ``soaks'' up some of the low-energy emission and the lowest-energy point of the 1FGL J2015.7+3708 spectrum drops accordingly.

For completeness, we also include the spectra of the two bright pulsars in Figure \ref{ch4_plot5_p2}.

\section{All-sky Analysis}
In discussion of single source (rather, single region) analysis, we mentioned a weakness of this approach: by selecting only a subset of the data, we cannot constrain all of the sources that contribute to the model for the data.  That is, since photons from sources from up to $10\dg$ outside the ROI can disperse into it, we must account for these sources in our model.  However, we cannot actually fit these sources since only low-energy photons contribute to the data.  We must have some \emph{a priori} model.  On the other hand, these same remarks apply to an ROI containing these external sources, and we have the makings of a ``chicken and egg'' scenario.

Obviously, the larger an ROI becomes, the less important are these ``edge effects'', particularly for a source in the center.  One extreme solution then is to include all data in the fit and attempt to jointly maximize the likelihood for \emph{all} sources in the sky.  At this juncture, such an approach is computationally infeasible.  A less extreme version of this solution would employ an ROI sufficiently large that these ``edge effects'' become less important.  However, using too large of an ROI brings its own drawbacks:
\begin{itemize}
\item the algorithms for convolution, likelihood evaluation, and other tasks have a complexity $\mathcal{O}(R_{ROI}^2)$ as well as additional overhead for large ROIs;
\item the plate car\'{e}e projection begins to show distortion for large ROI radius;
\item scaling parameters for the diffuse model become less effective for correcting local deficiencies.
\end{itemize}

Another approach is an iterative ``bootstrap'' analysis in which a set of ROIs are fit sequentially with the hope of convergence after a few iterations.  On the initial iteration(s), only the brighest sources---the most statistically independent and influential\footnote{\fermi sources span about four orders of magnitude in integral photon flux, from a handful of sources like the Vela pulsar with fluxes of order $10^{-5}$\fluxunits to myriad high-Galactic-latitude sources on the detection threshold with fluxes of order $10^{-9}$\fluxunits.  Sources in the top decade or so of flux utterly dominate the dimmer sources from a photon count standpoint, and (a) can be fit sufficiently without accounting for dimmer sources (b) must be sufficiently fit in order to fit---or even detect---dimmer sources.}---are fit, along with parameters for the diffuse background.  In later iterations, the flux threshold is relaxed.  With each iteration, the ``external'' background for an ROI in principle becomes better modeled, and the fits for sources within an ROI improve accordingly.

By combining these approaches---using an iterative, all-sky analysis (ASA) with the largest practical ROI radius---we can hope to obtain parameter estimates close to those that could be obtained from a true joint maximum likelihood fit.  Such an all-sky model, beyond the intrinsic interest of characterizing the GeV sky---is an ingredient {\it sine qua non} for more complex analyses to be discussed in the sequel.  Below, we outline a particular implementation using \emph{pointlike}.

\subsection{ROI Prescription}

In such an all-sky analysis, there are three initial decisions to make:
\begin{enumerate}
\item the geometry of the ROI (circular, square, other).  The infrastructure in \ptl is designed around a circular aperture.  This geometry maximizes symmetry (useful for calculating, e.g., PSF overlap integrals) and makes data extraction particularly simple.  However, it is impossible to tessellate the sky with circles, i.e., our ROIs must overlap, and we must then adopt a prescription for dealing with the same point source appearing in several ROIs as well as address a similar problem with the degrees of freedom of the diffuse background.  On the other hand, if we use a scheme like HEALPix to select data, we \emph{can} tessellate the sky, achieving statistically independent ROIs.
\item the size of the ROI.  As discussed above, the size of the ROI determines the magnitude of ``edge effects''.
\item the (possibly proper) subset of the ROI space within which point sources will be fit, i.e., the method for dealing with ROI overlaps.
\end{enumerate}

In light of these considerations, we adopt the following prescription:
\begin{enumerate}
\item Tessellate the sky using HEALPix pixels with $\nside=4$.  These pixels have an area of $(14.7\dg)^2$.  Although they are not regular in shape, in discussion we can approximate them as squares about $15\dg$ on a side.  There are 192 such pixels.
\item About the center of each pixel, extract data using a circular aperture with a radius of $R_{ROI}\equiv12\dg$.  The radius must satisfy the constraint that the pixel be entirely encircled, roughly that $R_{ROI}>30\sqrt{6/\pi}/\nside \approx 41\dg/\nside$.
\item Perform a maximum likelihood analysis in which point sources within the boundary of the HEALPix are allowed to vary while all sources lying in other HEALPix are kept fixed.
\end{enumerate}
These choices of $\nside$ and ROI radius represent a compromise between overall ROI size and the ratio of photons within the HEALPix pixel to the total number of photons, in this case $0.48$.  (A ratio of one gives statistically independent ROI.  Since each HEALPix has 7 or 8 neighbors, to first order, a given pair of pixels have $<10\%$ of their photons in common.)

Other choices are presented in Table \ref{tab:rad_hp_comp}.  Although an even larger ($15\dg$) ROI allows reduced redundancy and overlap, the background convolution begins to become problematic with an ROI of thise size.

\begin{table}
\center
\begin{tabular}{l l c}
$\nside$ & $R_{ROI}\ (\dg)$ & Fraction ROI within HEALPix \\
\hline
3 & 15 & 0.54 \\
4 & 12 & 0.48 \\ 
6 & 10 & 0.30
\end{tabular}
\caption{The fraction of the photons in a circular ROI with radius $R_{ROI}$ contained within a HEALPix (sharing the same center as the ROI) of size $4\pi/(12\nside^2)$ radians.}
\label{tab:rad_hp_comp}
\end{table}

\subsection{Iteration Prescription}

\begin{figure}
\begin{minipage}{6in}
\includegraphics[width=6in]{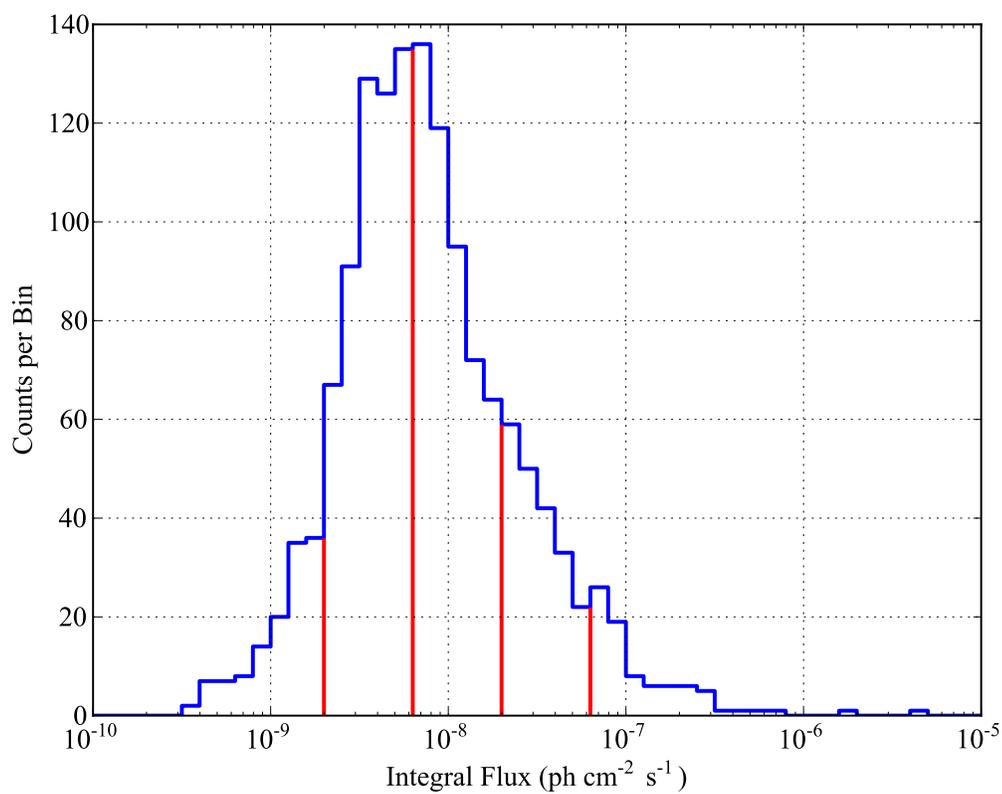}
\end{minipage}
\begingroup\renewcommand{\baselinestretch}{1.0}
\caption{The integral photon flux of 1FGL sources above 300 MeV.  The vertical lines indicate the flux thresholds in the iterative all-sky fit.}
\renewcommand{\baselinestretch}{1.5}\endgroup
\label{ch4_plot1}
\end{figure}

In order to allow the array of ROIs to ``relax'' to an approximately global ML solution, we proceed via iteration.  In each iteration, we allow all diffuse source parameters to vary, while only a fraction of point sources above a certain flux threshold are allowed to vary.  We find that the integral photon flux above $300$ MeV is a good indicator of source significance and we define the quantity $F_{300}\equiv \int\limits_{300}^{\infty} dE\,dN/dE$ for applying flux cuts.  This quantity for all sources in the 1FGL catalog is shown in Figure \ref{ch4_plot1}.  Concretely, the iteration strategy is:
\begin{enumerate}
\item Perform two fits in which only point sources with $F_{300}\geq10^{-7.2}$ are allowed to vary.  In Figure \ref{ch4_plot1}, this cut includes all sources to the right of the rightmost red horizontal line, essentially the hundred brightest point sources.  These sources are sufficiently strong to alter the diffuse background model parameters but are essentially independent of weaker sources.  These two iterations provide a ``baseline'' model with which to fit additional point sources.
\item Perform three iterations in which the flux threshold is lowered by one half of a decade each time.  These cuts are also indicated in Figure \ref{ch4_plot1}.
\item Perform three iterations in which the flux threshold is set to 0, i.e., all point source parameters are allowed to vary.
\end{enumerate}

Additionally, at each iteration, after performing an initial ML fit for the ROI, we check the free point source parameters for unphysical values, e.g., a photon flux consistent with zero or a photon index lower than $0.5$ or higher than $4.0$.  If we find such sources, we reset their parameters to a nominal value and refit the ROI.

Finally, we note that performing all 8 iterations for all ROIs requires on the order of $100$ CPU hours on modern machines, a small cost on a modest computing clusters.

\subsection{Tests for Convergence}
\begin{figure}
\begin{minipage}{6in}
\includegraphics[width=6in]{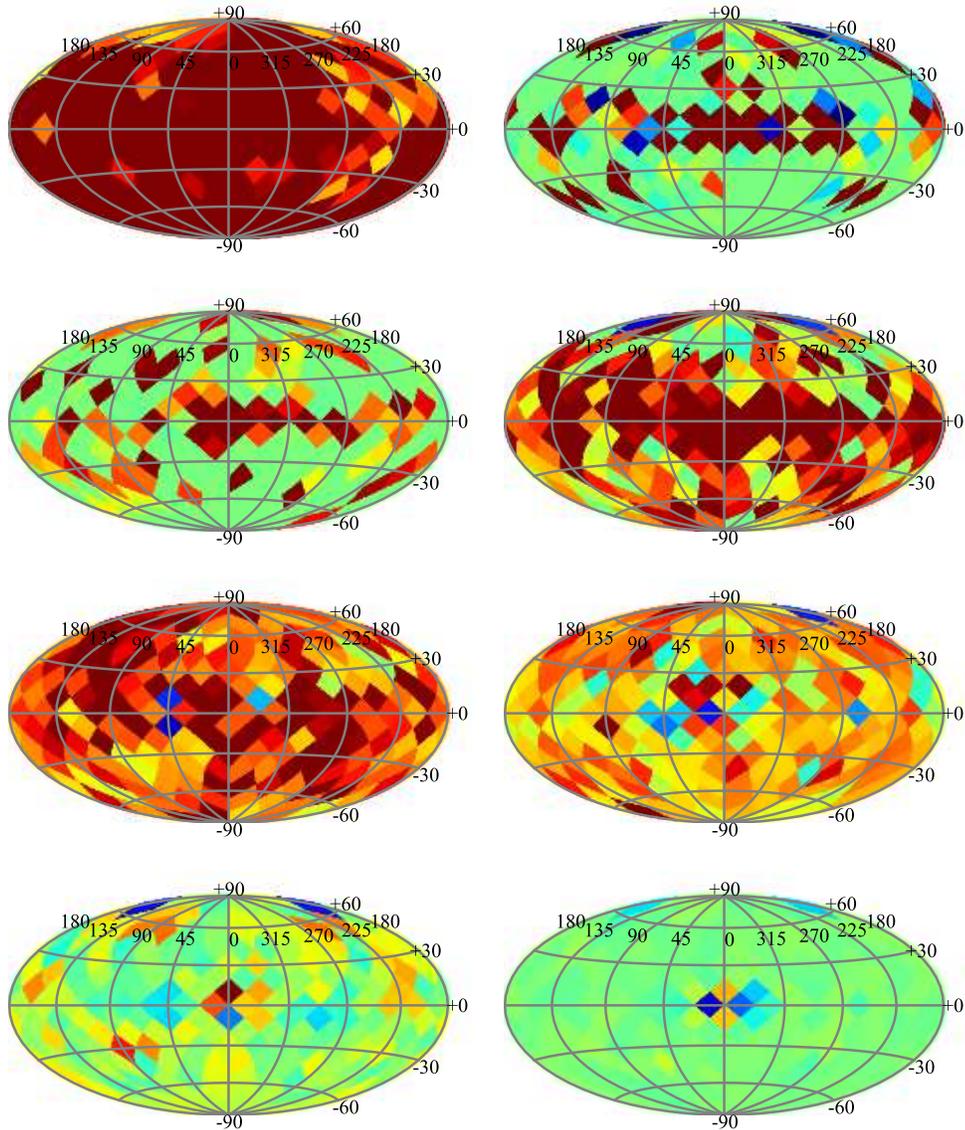}
\end{minipage}
\begingroup\renewcommand{\baselinestretch}{1.0}
\caption{The log likelihood changes in each ROI for each iteration.  The iterations proceed from left to right and top to bottom.  The precise quantity shown is $\mathrm{sign}(\Delta\log\mathcal{L})\times\sqrt{(\Delta\log\mathcal{L}}$, and the scale extends from $-5$ (dark blue) to $5$ (dark red).  For these large ROIs, a change in log likelihood of order a few is negligible.  In the upper lefthand corner---the initial iteration---the log likelihood shift is dominated by the initial fit of the parameters for the diffuse models.  The third through sixth panels show the log likelihood improving as more sources are allowed to vary.  The brightest sources are concentrated in the Galactic plane (selection and projection effects), so only by the fifth panel are a significant fraction of sources at all latitudes free.  The final three panels show the refinement with all point sources varying.  In the final panel, essentially all ROIs have ceased to improve.  Four pixels at the Galactic center continue to shift by order a few in an ``oscillation'' described in the main text.}
\renewcommand{\baselinestretch}{1.5}\endgroup
\label{ch4_plot3}
\end{figure}

To test the convergence of the above iteration scheme, we track the change in the log likelihood for each ROI from iteration to iteration.  As an example, we consider an ASA performed according to the standard iteration prescription using 18 months of data to which standard cuts\footnote{Incidence angle $\leq66.4\dg$, zenith angle $\leq105\dg$.} were applied.  The source model was taken to be that of the 1FGL analysis, i.e., for point sources the 1FGL catalog and for diffuse sources {\it gll\_iem\_v02} and {\it isotropic\_iem\_v02}.  The log likelihood changes for the eight iterations are shown in Figure \ref{ch4_plot3} and indicate the final iteration is well-converged.

Particularly for ROIs with relatively large overlaps with neighbors (Table \ref{tab:rad_hp_comp} e.g.), total convergence is rather difficult to achieve.  With each iteration, a pair of coupled ROIs will update their parameters to reflect the previous iterations background model, and this oscilllation can proceed for many iterations.  The magnitude of the log likelihood shift is typically of order unity, and we regard these shifts as negligible.  An approach that will in principle eliminate this issue uses the same HEALPix both for organizing free point sources \emph{and} extracting data, i.e., it abandons circular ROIs. 

\subsection{Results}
\begin{figure}
\begin{minipage}{6in}
\includegraphics[width=6in]{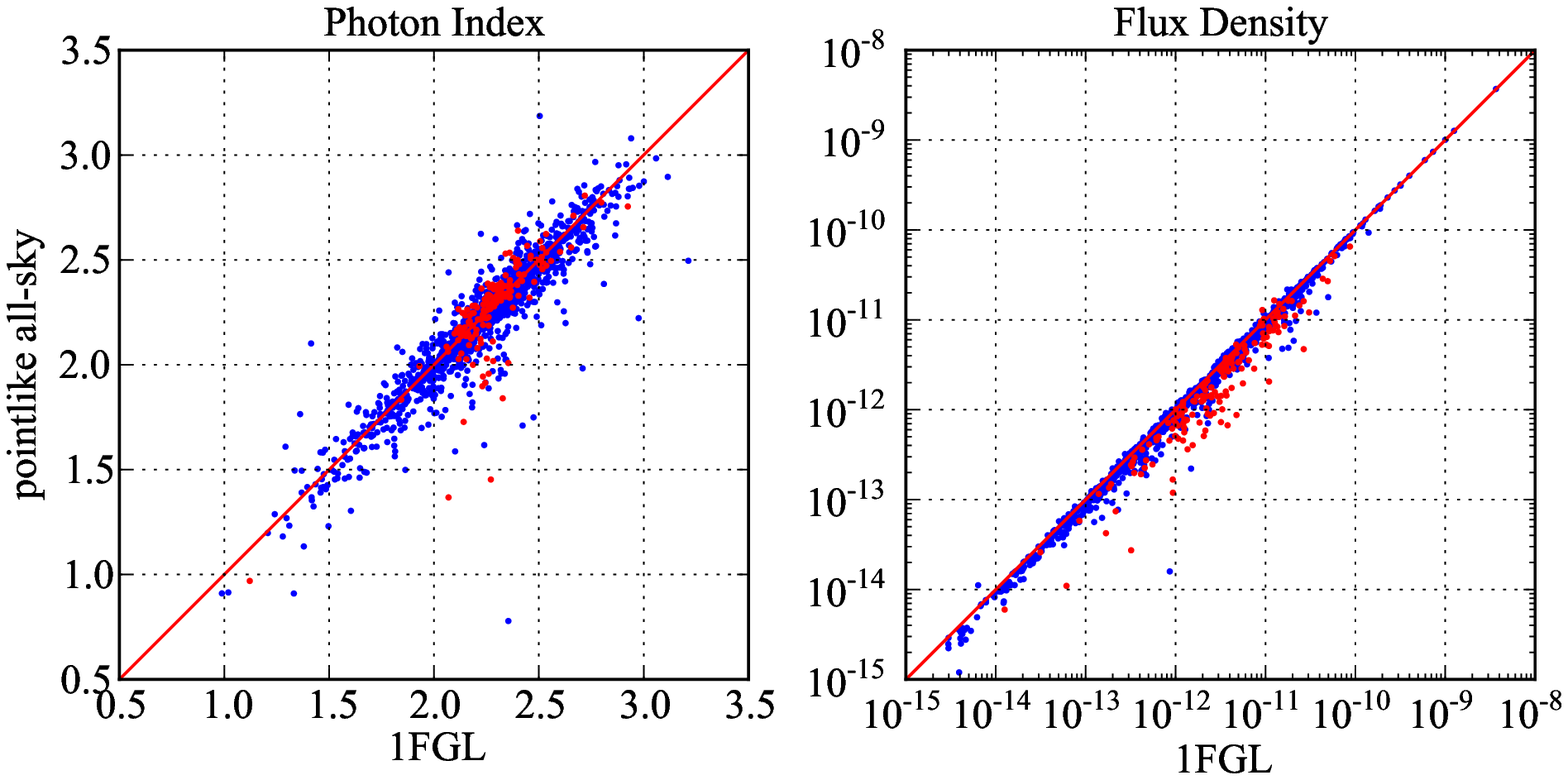}
\end{minipage}
\begingroup\renewcommand{\baselinestretch}{1.0}
\caption{A comparison of the ML estimates obtained by the \ptl ASA and the unbinned likelihood pipeline analysis of the 1FGL catalog.  All sources are fit with a power law.  The flux density is evaluated at the ``pivot energy'', an estimate of the energy at which the covariance of the photon index and flux density is minimized.  Sources in red correspond to 1FGL sources with a ``c'' appended, indicating they lie along the Galactic ridge and their spectral values should be taken with caution.  Many of these sources may be spurious.  A subset of these sources clearly departs from the main population for which \ptl and the 1FGL values are in excellent agreement.}
\renewcommand{\baselinestretch}{1.5}\endgroup
\label{ch4_plot4_scatter}
\end{figure}

\begin{figure}
\begin{minipage}{6in}
\includegraphics[width=6in]{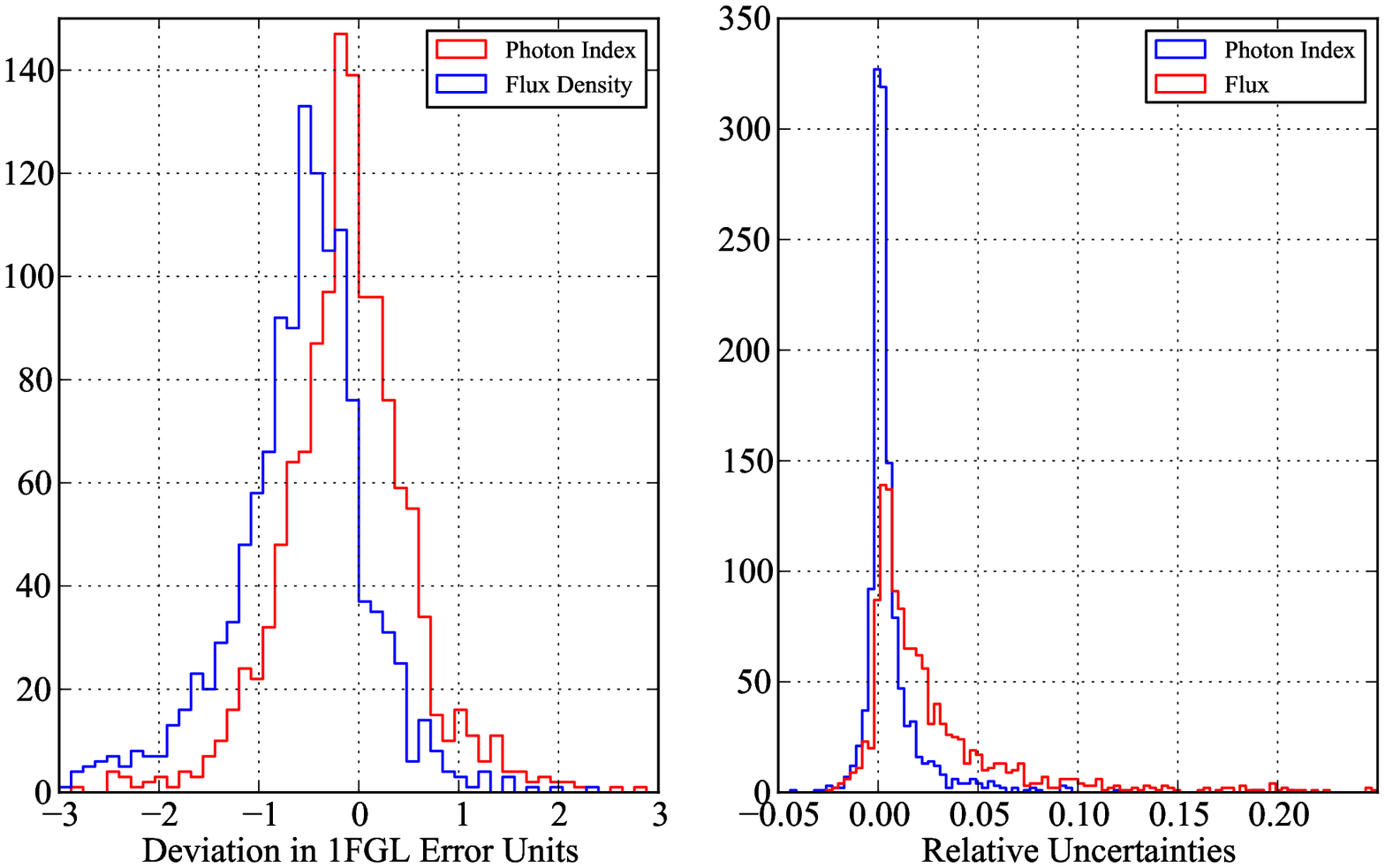}
\end{minipage}
\begingroup\renewcommand{\baselinestretch}{1.0}
\caption{In the lefthand panel, the difference in the ML parameter estimates scaled by the parameter error estimated by the 1FGL analysis.  The photon indices are in excellent agreement, while the flux densities shows a small trend toward underestimating the flux.  In the righthand panel, we compare the uncertainty estimated by each analysis.  This comparison is made by estimating the relative error for each method and then taking the difference (1FGL - \ptl ASA).  The error estimated by \ptl is generally in excellent agreement, with some flux error estimates exceeding the 1FGL estimate by up to $20\%$, an unimportant difference.  It is in any case not surprising that an unbinned method may deliver slightly smaller error estimates.}
\renewcommand{\baselinestretch}{1.5}\endgroup
\label{ch4_plot4_histo}
\end{figure}

As an example, we show a comparison of the parameters obtained with the \ptl ASA and those obtained by \emph{gtlike} as presented in the 1FGL catalog\cite{1fgl}.  We apply the iteration prescription to the same data set outlined in \cite{1fgl} but performing our own standard cuts (see previous section) and calculating the livetime with the \ptl package.\footnote{The livetime as calculated by the Science Tool application \emph{gtltcube} differs slightly from the \ptl application, as the livetime calculation in \ptl corrects for the cuts on zenith and incidence angle.}  In some sense, this exercise is only a sanity check, but we emphasize these results are obtained with an entirely independent toolchain.  We compare the central values obtained for the power law parameters for all 1451 1FGL sources in Figure \ref{ch4_plot4_scatter} and find good agreement. A detailed analysis in Figure \ref{ch4_plot4_histo} find good agreement between the error-weighted parameter estimate differences\footnote{The curious offset of the \ptl fluxes of about 1/2 an error unit is under investigation.} and the uncertainty estimates.

\subsection{Diffuse Source Studies}
\begin{figure}
\begin{minipage}{6in}
\includegraphics[width=6in]{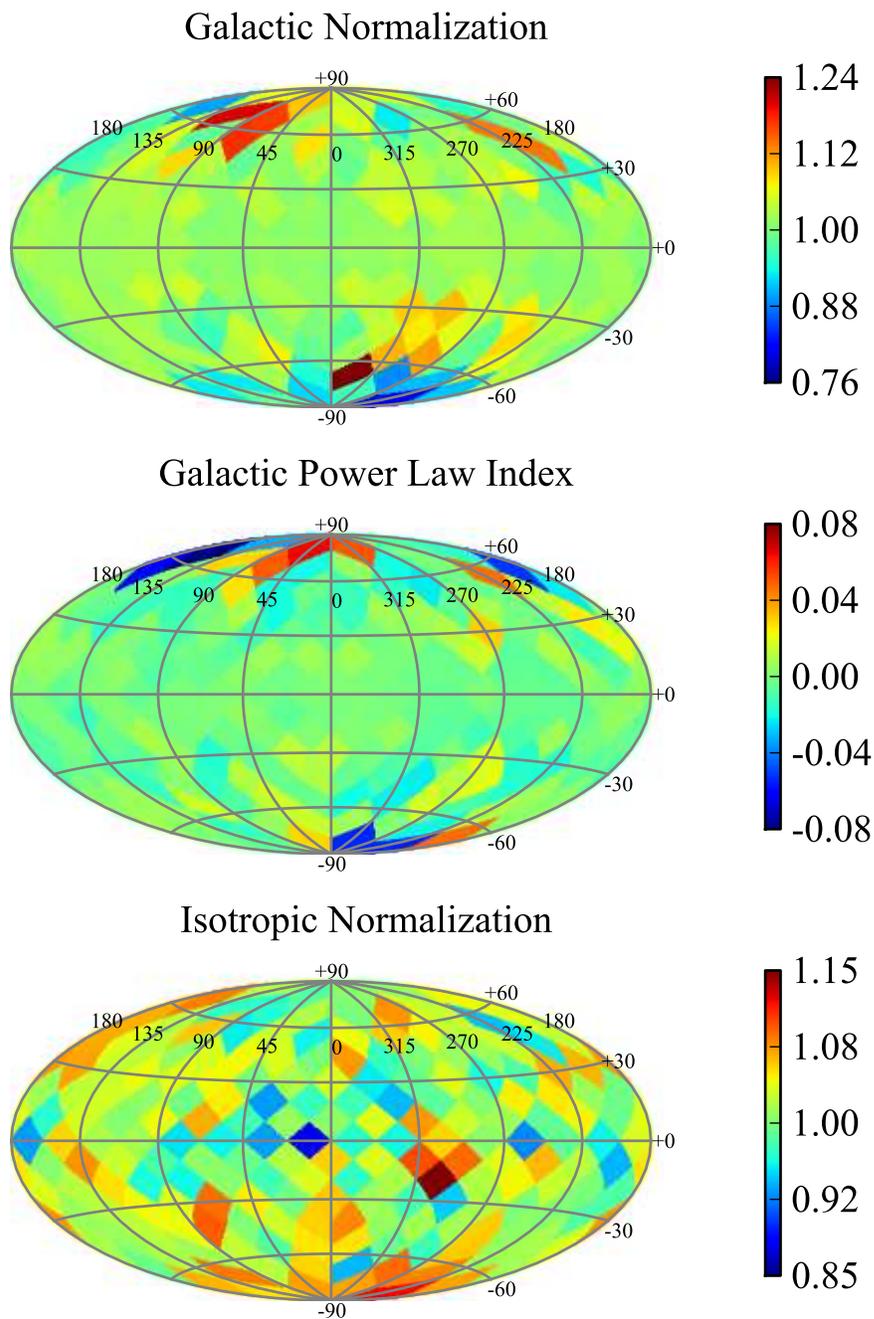}
\end{minipage}
\begingroup\renewcommand{\baselinestretch}{1.0}
\caption{The best-fit values of the parameters for the two scaling models discussed in the text.  The null value for the constants is 1, while for the photon index of the power law scaling the Galactic diffuse, 0.  The abundant photons in the Galactic plane strongly constrain the Galactic diffuse model, while at high latitudes the sparse photons allow large fluctuations.  The isotropic background is generally constrained to within $10\%$ independent of position on the sky.}
\renewcommand{\baselinestretch}{1.5}\endgroup
\label{ch4_plot2}
\end{figure}

As an additional validation of the treatment of diffuse sources presented in Chapter \ref{ch3}, we take advantage of the ASA infrastructure to obtain fits to diffuse scaling models and verify they are consistent with the input.  We used 18 months of data generated by \emph{gtobssim} for the {\it gll\_iem\_v02} Galactic diffuse and  {\it isotropic\_iem\_v02} isotropic diffuse sources and ran a single iteration of the ASA procedure.  The {\it gll\_iem\_v02} was scaled by a power law with free normalization and photon index, while the {\it isotropic\_iem\_v02} source was scaled only by a constant.  During a ML fit, these parameters vary due to statistical fluctuations, but the distribution means should be consistent with the input model.  The distribution of the best-fit parameters for each ROI is shown in Figure \ref{ch4_plot2}, from which it is clear that the fit parameters indeed agree with the input model.

\section{Sky Maps}
\label{ch4:sec:kde}

In this section, we deviate from likelihood analysis (but not from all-sky analysis!) to discuss depiction of \fermi data.  Constructing maps that give a good visual representation of the sky is challenging because (a) low energy events are significantly smeared by the PSF and (b) the emission is highly anisotropic, peaking strongly in the Galactic plane.  Constructing a simple counts map---a histogram---with a fixed pixel size is not optimal since we will either lose resolution in high-count regions or suffer from ``shot noise'' in low-count regions.

A well-known alternative to a histogram for visual representation of random data is \emph{kernel density estimation}.  In one dimension, e.g., the a kernel density estimator $\hat{f}$ for the probability density function $f$ given a set of data $\{x_i\}$, is
\begin{equation}
\hat{f}(x) = \frac{1}{N}\sum\limits_{i=1}^{N} g[(x-x_i)/b]
\end{equation}
where $g$ is the \emph{kernel}, some normalized density function, typically a gaussian, and $b$ is the \emph{bandwidth}, essentially a smoothing parameter.

An obvious extension to the current use case is adopting the PSF for the kernel, with a bandwidth then naturally parameterized by the reconstructed energy and event class.  As part of the all-sky analysis, we have already constructed for each ROI, in each energy band, an approximate, exposure-weighted PSF, as well as binned the data in a sparse format that eases the computation.  We can thus calculate a kernel density estimate for the photon density on the sky as
\begin{equation}
f(\vom) = \sum\limits_{i=1}^{12\nside^2} \sum\limits_{j=1}^{N_{bands}} \sum\limits_{k=1}^{N_{pix}} n_k\,f_{wpsf}(i,j,|\vom-\vom_k|),
\end{equation}
where we symbolically indicate the PSF dependence on energy and conversion type through the band subscript $j$ and where $n_k$ gives the observed counts in the $k$th pixel.  In practice, we would evaluate $f(\vom)$ over some grid corresponding to an image projection.

\begin{figure}
\begin{minipage}{6in}
\includegraphics[width=6in]{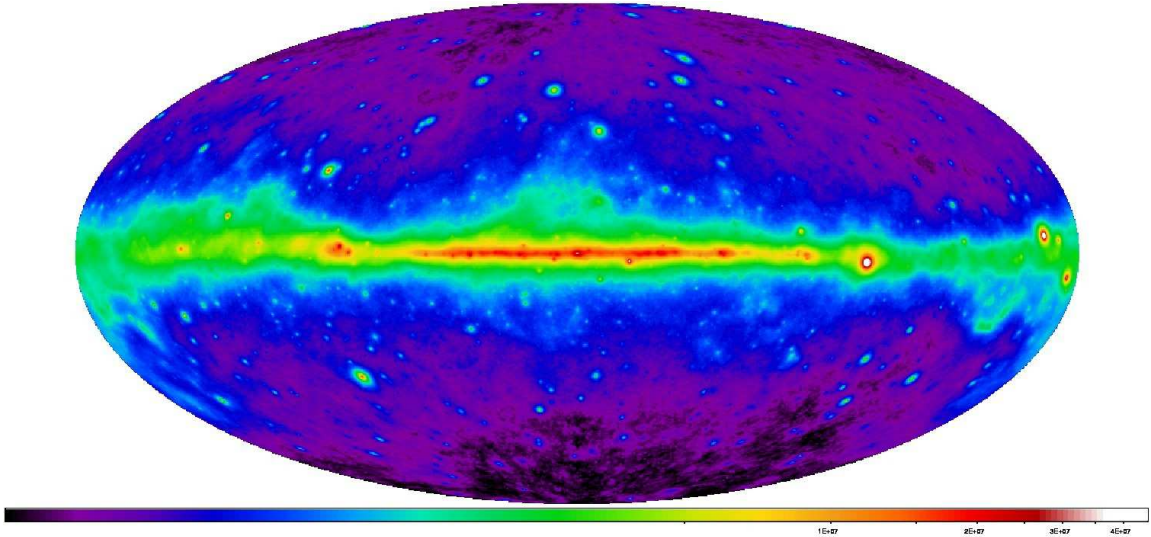}
\end{minipage}
\begingroup\renewcommand{\baselinestretch}{1.0}
\caption{An image (in Hammer-Aitoff projection of Galactic coordinates) of the GeV sky produced by kernel density estimation with the PSF with 18 months of data.  The units are somewhat arbitrary but can be roughly interpreted as counts per steradian.  All energies $>100$ MeV are included.}
\renewcommand{\baselinestretch}{1.5}\endgroup
\label{ch4_allsky_im}
\end{figure}

\begin{figure}
\begin{minipage}{6in}
\includegraphics[width=6in]{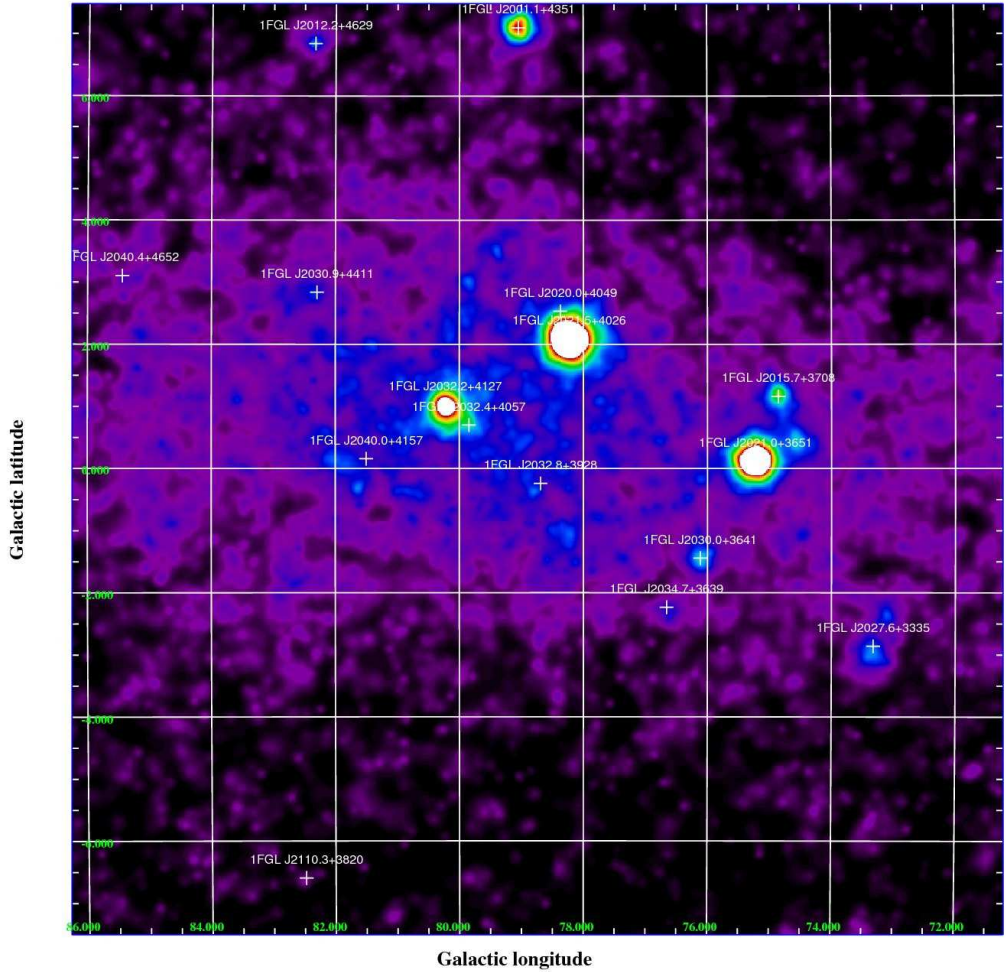}
\end{minipage}
\begingroup\renewcommand{\baselinestretch}{1.0}
\caption{An image (in zenithal equal-area projection of Galactic coordinates) of the GeV sky produced by kernel density estimation with the PSF with 18 months of data.  The image shows the Cygnus region and has overlaid the locations of sources in the 1FGL catalog.  In constructing this image, photon energies were restricted to $2200 < E/MeV < 10000$ ($4400 < E/MeV < 20000$) for events converting in the front (back) of the detector.  Due to the $\approx1/E$ dependence of multiple scattering and the factor of $\approx2$ difference in the radiation length of the front and back radiations, the two samples of events have approximately the same angular resolution. }
\renewcommand{\baselinestretch}{1.5}\endgroup
\label{ch4_cyg_hie_kde}
\end{figure}

We show such an image for the entire $\gamma$-ray sky in Figure \ref{ch4_allsky_im}.  An image restricted to the Cygnus region of the sky, and to relatively high-energy photons, is displayed in Figure \ref{ch4_cyg_hie_kde}.  It is useful to compare this image, particularly the evidence for the new source mentioned in the discussion of the spectral analysis of the Cygns region, with the series of TS maps in Figure \ref{ch4_cygnus_ts}.

\section{Source Finding with TS Maps}
\label{ch4:sec:tsmaps}

An important application of likelihood is searching for new sources.  By examining how the likelihood improves by expanding the model to include new sources, we can determine (in a statistically calibrated fashion, as we show below) whether the sources are required by the data.

\subsection{A Test Statistic for Source Detection}
The typical technique is a likelihood ratio test in which the ratio of the likelihood under the null hypothesis (no new source) is compared to the likelihood under the alternative hypothesis (there is a new source with some position/spectrum).  The likelihood ratio is a \emph{test statistic} (TS) and we reject the null hypothesis (claim detection of a new source) if the TS exceeds some threshold.

To ensure a small probability of Type I error (false positive), it is useful to know the distribution of the TS in the event the null hypothesis is true.  Then we immediately know our probability of Type I error is the tail probability of the distribution integrated from the threshold.  (If we test multiple, independent data sets, or check the same set for multiple sources, we must of course multiply this probability by the number of total trials.)  A celebrated result from Wilks\cite{wilks} gives the asymptotic (i.e., large sample) distribution for a certain class of likelihood ratios.  If the models for the null and alternative hypotheses are nested and have, respectively, $n_0$ and $n$ parameters, and if the null values do not fall on the boundary of the parameter space, and if the likelihood satisfies certain regularity conditions, then the likelihood ratio is distributed as $\chi^2_{n-n_0}$, i.e., chi-squared with degrees of freedom equal to the difference in the dimensions of the parameter spaces.  This result is extremely useful for testing the statistical significance of additional components in models \emph{provided the new component is not on the boundary of the parameter space in the null hypothesis.}

Unfortunately, the likelihood ratio test for source detection does not satisfy the criteria for Wilks' Theorem.  For example, suppose the likelihood for a data set in the presence of an additional point source at a fixed position with a fixed spectral shape but free flux parameter is calculated.  The models (with and without the new point source) are clearly nested, but the new parameter---the flux---is 0 in the null hypothesis.  Since flux is positive, this value lies on the boundary of the parameter space.

Fortunately, Chernoff\cite{chernoff_miracle} was able to extend the results of Wilks to show that, in cases where the alternative (null) model has a $n$ ($n-1$) dimensional parameter space, and the additional parameter is resticted to one side of an ($n-1$) dimensional plane---as is the case for the scenario outlined above---then the null distribution is an equal mixture of $\delta(0)$ and $\chi^2_1$.  It is easy to show then that the chance probability of observing a value in excess of $TS$ is given by the tail integral of a normal distribution from $\sqrt{TS}$ to $\infty$, and the confidence level can be quoted as ``$\sqrt{TS} \sigma$''.  Analysis of EGRET data, applying the same test for new sources, observed this distribution in Monte Carlo simulations\cite{mattox}.

\subsection{Generating a TS Map}

To generate a TS map, we begin with a single region, viz. one of the HEALPix/circular aperture pieces of the ASA.  After the ASA, we have a consistent model that should---if we have represented every source correctly---account for every photon in the GeV sky.  Next, we subdivide the HEALPix pixel into sub-pixels small enough that we can resolve new sources.  For $\nside=4$, we might choose to divide each side of the base HEALPix pixel into 150 segments, for $150^2$ sub-pixels approximatly $0.1\dg$ on a side.  (It will be recalled that this procedure is the typical method for producing a HEALPix pixel of finer resolution; this example grid has $\nside=600$.)

We regard the background model as fixed and we assume a power law model with $\Gamma=2.0$ for the putative point source.  (Other indices can be adopted.)  For the pixel at position $\vom_i$, the log likelihood for a given band (using the notation of Chapter \ref{ch3}) in the presence of a new source with flux $\mathcal{F}_s$ is
\begin{equation}
\mathcal{L}(\vom_i,\mcf_s) = -\mathcal{O}_s(\vom_i)\,N_s(\mcf_s) + \vec{n}\cdot\log\left[1 + N_s(\mcf_s)\frac{\vec{f}_s(\vom_i)}{\vec{b}_s}\right],
\end{equation}
where $\vec{b}$ represents the contribution of the background to each data pixel and $\vec{f}$ is the value of the PSF for a source centered at $\vom_i$ at each data pixel.  The likelihood is single-dimensional (by virtue of the fixed photon index) and can be easily maximized to determine the best-fit $\mcf_s$ (denoted $\hat{\mcf}_s$ for each $\vom_i$.  The TS is twice the quantity obtained by subtracting the null log likelihood ($\mcf_s=0$) from the log likelihood evaluated at $\hat{\mcf}_s$.  As defined above, $\mathcal{L}(\vom_i,0) = 0$, so
\begin{equation}
TS(\vom_i) = 2\times\mathcal{L}(\vom_i,\hat{\mcf}_s).
\end{equation}

We determine this quantity for each sub-pixel in each base ($\nside=4$) HEALPix pixel.  Since the background model is assumed to be fixed, the base pixels can be processed in parallel.  The output is the TS evaluated on a very fine grid, producing a TS map for the entire sky.  It is then a relatively simple matter to ``convert'' these TS values to new sources by selecting the highest, isolated TS excesses, incorporating them into the model, re-maximizing the likelihood via the ASA, crafting a new TS map, etc.

\begin{figure}
\begin{minipage}{6in}
\includegraphics[width=6in]{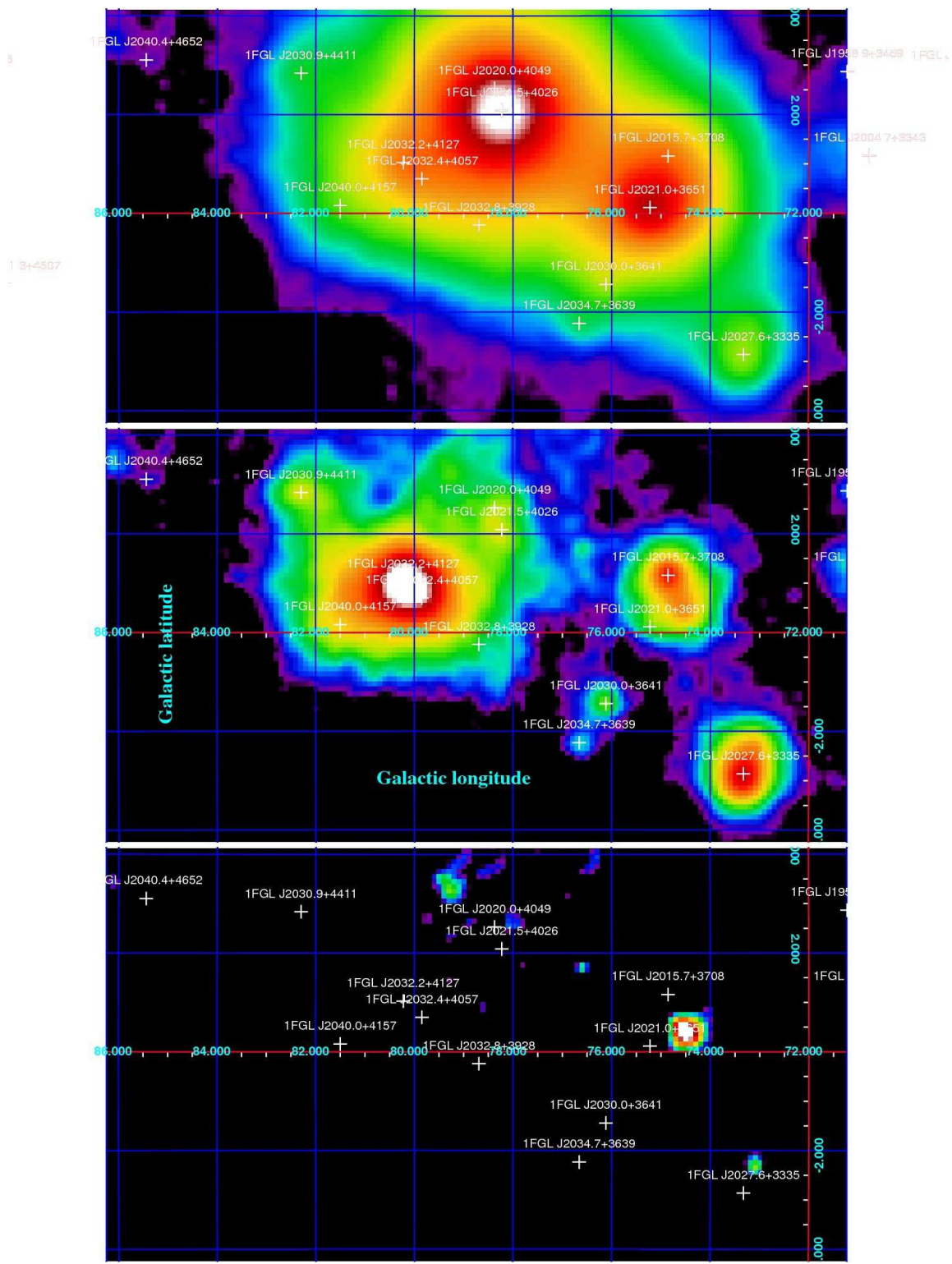}
\end{minipage}
\begingroup\renewcommand{\baselinestretch}{1.0}
\caption{A TS map constructed of the Cygnus region to demonstrate the source finding principle discussed in the text.  In the top panel, only the best-fit diffuse model (Galactic + isotropic) is used for a background model.  All point sources then contribute to the calculated TS, and the map is dominated by the exceedingly bright pulsars J2021+4026 and J2021+3651.  The range spanned by the color scale is 25--40,000.  In the second panel, the two bright pulsars are included in the background, and the color scale is now 25--2200, with the combination of Cygnus X-3 and PSR J2032+4127 setting the top end.  In the bottom panel, all 1FGL sources are included in the background model, and the scale is now 25--180.  The excess at $(l,b)\approx(74.5,0.5)$ is a new point source.}
\renewcommand{\baselinestretch}{1.5}\endgroup
\label{ch4_cygnus_ts}
\end{figure}

As an example of the efficacy of TS maps, we present a case example of searching for a source in the Cygnus region in Figure \ref{ch4_cygnus_ts}.

\section{Source Classification}
\label{ch4:sec:source_classification}

It has historically been the case that detection of $\gamma$-ray pulsations from neutron stars follows detection of radio pulsations\cite{sas2_vela,sas2_anticenter,j2021} or even X-ray pulsations\cite{geminga_xray_pulse,geminga_gray_pulse}.  This is due to the extreme paucity of photons collected from HE sources; long integration times are required, and even though pulsars are relatively stable rotators, the phase space in frequency, frequency derivative, etc. that must be searched has made direct pulsation detection infeasible.  Instead, radio observations of a particular pulsar are used to build a timing solution (see Chapter \ref{ch5}) that can be used to map the time of arrival of a detected $\gamma$ ray to the rotational phase of the neutron star.  With a stable timing solution, photons collected over weeks, months, or even years can be accumulated, building up profiles in phase that can be tested against a uniform distribution (Chapter \ref{ch5}).

With the improved angular resolution and effective area of \fermi, GeV telescopes have finally reached some parity.  Time-differencing techniques\cite{blind_search_technique} have led to the discovery of over 24 new pulsars\cite{blind_search_16,blind_search_8}, wrapping up two longstanding mysteries\cite{j1836p5925,cta1} on the way.  This number is comparable to the number of new $\gamma$-ray pulsars detected to date by \fermi using timing solutions\cite{1pcat}.

Besides direct detection of pulsations, ``$\gamma$-ray initiated'' detection of pulsars is opened by the sensitivity to \emph{unpulsed} sources.  The high yield so far of detected pulsars suggests many of the $>1000$ sources detected by \fermi so far\cite{1fgl} must harbor neutron stars emitting pulsed $\gamma$ rays.  There are many reasons why such sources may be difficult to detect with direct pulsations searches.  Many young pulsars exhibit stochastic timing noise\cite{timing_noise}, limiting the maximum integration period over which a detection may be obtained.  Even if the pulsar is stable, long integration times require fine-grained searches and the possible addition of additional search parameters (e.g., a second-order term in the Taylor expansion of phase-versus-time, i.e., a second derivative of frequency.)  Integration times comparable to a year also require a precise position to correct for the changing light travel time (the R\"{omer} delay) as the earth revolves about the sun.  Unfortunately, dim sources (requiring long integrations) are also difficult to localize and the problem becomes rapidly intractable.  Finally, millisecond pulsars are difficult to detect because of the extreme sensitivity to precise values of period and period derivative, and millisecond pulsars in binary systems are essentially undetectable by \fermi.

One approach, then, is to search for \emph{radio} pulsations from promising \fermi sources.  The dimmest known pulsars can be detected in hours with modern radio telescopes, solving most of the problems outlined above.  For this approach to fail (a) the source is not a pulsating neutron star (b) the position is not good (c) the pulsar radio beam is not aligned with the earth (d) the signal was temporarily suppressed by scintillation.  If pulsations \emph{are} detected, we are essentially guaranteed both a new radio pulsar \emph{and} a new $\gamma$-ray pulsar, since the chance coincidence of radio pulsars and LAT sources is quite low.

What constitutes a promising \fermi source?  A good candidate should
\begin{enumerate}
\item have a reliable position with an uncertainty comparable to or smaller than the main beam of the radiotelescope used to conduct the search.  For the 100-meter Green Bank telescope, this constraint ranges from a few arcminutes at $2$GHz frequencies to up to $0.3\dg$ (error radius) at $350$MHz.  Sources within the Galactic plane suffer some additional systematic errors (e.g. unmodeled neighboring point sources, mismodeled diffuse emission).
\item be unassociated with an extragalactic source.  By comparing the positions of \fermi sources with multiwavelength catalogs, probabilistic associations of $\gamma$-ray sources with counterparts can be made\cite{1fgl,1lac}.  If a source has a good probability to be associated with another source class (particularly active galactic nuclei (AGN), the largest class of \fermi sources), then it is sensible to pursue better candidates.
\item show no variability.
\item have a spectrum (a) like known pulsars and (b) unlike known AGN.  Essentially all pulsars detected to date have spectra consistent with a power law + exponential cutoff and have a constant flux when averaged over many periods\cite{1pcat}.  In particular, the observed cutoff energy (photon index) is generally $\leq5$ GeV ($\leq2$).  On the other hand, AGN have spectra broadly consistent with power laws.  Flat-spectrum radio quasars, a subclass of AGN, almost universally have $\Gamma>2$, and thus a candidate can be spectrally vetoed by observation of a falling spectral energy density from $0.1-1.0$ GeV.  BL Lac objects, a second subclass with harder spectra, can generally be excluded by observation of significant emission above 10 GeV.
\end{enumerate}

Thus, we seek to analyze \emph{all} known \fermi sources with the aim of (a) estimating the best possible positions and uncertainties (b) excluding known variable and known associated sources (c) fitting multiple spectra to assess the statistical strength of the exponential cutoff parameter (d) producing a spectral energy density plot to visually assess spectral features.  The all-sky analysis discussed above can handily achieve (a), (c), and (d).  We outline a specific analysis scheme along with radio surveys conducted to search the resulting candidates and present preliminary results.

\subsection{All-sky Analysis for Pulsar Candidate Identification}

\begin{figure}
\begin{minipage}{6in}
\includegraphics[width=6in]{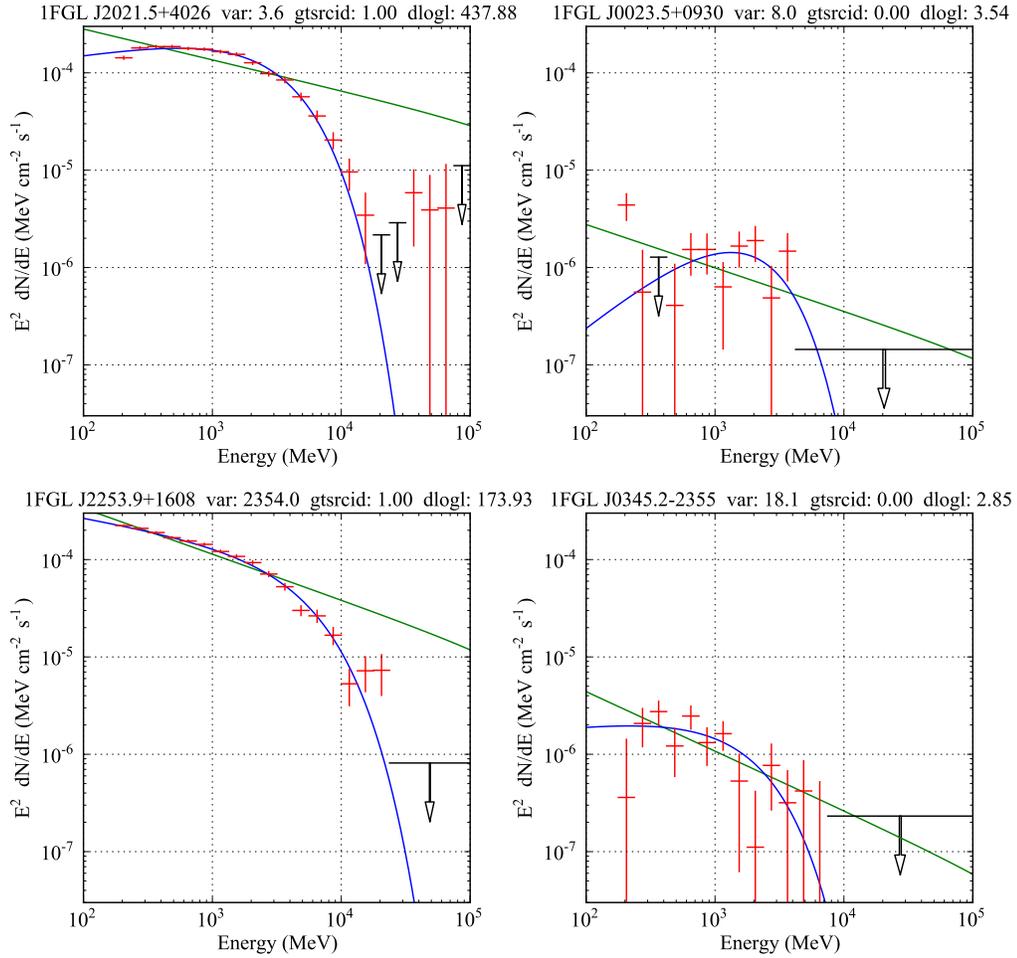}
\end{minipage}
\begingroup\renewcommand{\baselinestretch}{1.0}
\caption{A gallery of typical plots used for classifying unidentified sources as good pulsar candidates.  In the upper left panel, a known LAT PSR J2021.5+4026\cite{blind_search_16} gives an example of a typical pulsar spectrum: flat or rising in the $0.1-1$ GeV decade, a strong cutoff, and a low ``variability index''.  In the upper right panel, a much dimmer pulsar candidate (at which location radio pulsations were subsequently detected), which shows similar features scaled down by two orders of magnitude.  At bottom left is the quasar 3C 454.3\cite{3c_cat} exemplifying the features of typical AGN emission: a falling spectrum in the $0.1-1$ decade and extreme variability.  At bottom right, a comparable (unidentified) source showing similar features scaled down by two orders of magnitude in flux.  It has modest variability and would not be considered a good candidate due to this and its falling spectrum at low energy. }
\renewcommand{\baselinestretch}{1.5}\endgroup
\label{ch4_plot6}
\end{figure}

We performed an all-sky analysis as described above (in a more preliminary version) using 11 months of \fermi data and an internal (to the LAT Collaboration) version of the source list that became the 1FGL catalog.  (The 1FGL catalog was based on the same 11-month dataset employed here.)  To supplement the analysis, we made use of the association and variability studies performed for the catalog preparation.  For each source, we fit a power law spectrum, then added a cutoff, allowing a likelihood ratio test of the significance of the cutoff.  We generated a spectral energy density plot for each source to assess its shape qualitatively.  An example of these seds---which include also the other pieces of information save position---appears in Figure \ref{ch4_plot6}.

\subsection{Observations}

\subsubsection{Green Bank Telescope, 350 MHz}
A Fermi Cycle 2 Guest Investigator\footnote{http://fermi.gsfc.nasa.gov/ssc/proposals/cycle2/} proposal (P.I. Mallory Roberts) to observe approximately 50 unidentified Fermi sources at high Galactic latitudes with the 350 MHz receiver of the 100-m Green Bank Telescope (GBT) was accepted in 2009.  Since the spectrum of pulsars in radio is typically a falling power law, observations at low frequency are more sensitive provided there are not significant disperserive effects from the interstellar medium (ISM).  By restricting candidates to high Galactic latitudes, ISM effects are minimized.

The ASA preparation and classification scheme was used to draw up a candidate list.  An initial pass rejected any candidates at low ($<-5\dg$ or $>5\dg$) Galactic latitude and with Decl. $<-40\dg$, essentially the lowest latitude visible from GBT.  Since the full-width at half-maximum (FWHM) of the 350 MHz beam at GBT is about $0.6\dg$\footnote{To order of magnitude, the FWHM is given by the diffraction limit, FHWM $\approx\lambda/D\approx0.5\dg$, with $D\equiv100$m.  The actual beam depends on the exact geometry of the receiver and the blazing of the dish surface.}, no cut was made on the positional uncertainty candidate sources as virtually all sources in the 1FGL catalog have a $95\%$ error radius $<0.3\dg$.  For sources satisfying the position consitraints, plots such as those presented in Figure \ref{ch4_plot6} were generated for the remaining sources.  By considering spectral shape, cutoff strength, association with known blazars, and variability (at time scales of $\approx1$ month), sources were classified as 0, 1, 2, or 3, viz. unacceptable, poor, good, and excellent.

Observations were carried out using the GUPPI\cite{guppi} backend with the 350 MHz receiver.  During the campaign, most sources classifed as ``2'' or ``3'' were visible and observed.  In total, 47 sources were observed with a typical integration time of 32 minutes.  The collected data requires considerable computation to fully search for pulsations\footnote{E.g., many millisecond pulsars will be found in binary systems, requiring an ``acceleration'' search in which a linear correction for gravitational acceleration of a given magnitude is made.} and have not yet been fully analyzed.  However, the first 215 seconds of each observation have been fully searched, and five new MSPs have been detected.  A brief summary of their properties is given in Table \ref{ch4:tab:gbt350}\footnote{After a poster presented by P. Bangale at the 2010 meeting of the High-energy Astrophysics Division of the AAS.}.

\begin{table}
\begin{tabular}{l c c c}
Name & NS Period (ms) & Orbital Period (hr) & Companion Mass ($M_{sun}$) \\
\hline
J0023+09 & 3.05 &  3.3        &0.016\\
J0340+41 & 3.30 &  Isolated   &N/A\\
J1302-32 & 3.77 &  $>$24        &N/A\\
J1810+17 & 1.66 &  4.9        &0.089\\
J2215+51 & 2.61 &  4.2        &0.22
\end{tabular}
\caption{The properties of the 5 millisecond pulsars detected to date in the GBT 350 MHz survey.  The binary J1302-32 awaits further timing to constrain the orbital parameters.  PSRs J0023+09 and J1810+17 are likely of the ``Black Widow'' class\cite{bwp}, while PSR J2215+51 presents eclipses.}
\label{ch4:tab:gbt350}
\end{table}

\subsubsection{Parkes Observatory, 1400 MHz}
A second serendipitous\footnote{Observations were carried out during scheduled but unutilized maintenance time.} survey was conducted with the 1.4 GHz receiver on the 64-m dish at the Parkes Observatory (P.I. Fernando Camilo).  The source list was prepared as described in the previous section, {\it mutatis mutandis}, e.g. restricting the candidates to have Decl. $<-40\dg$.  A total of 14 sources were observed with integration times ranging from 1 to 2 hours.  Five new millisecond pulsars, listed in Table \ref{ch4:tab:parkes}, were detected.

\begin{table}
\center
\begin{tabular}{l c}
Name & Classification \\
\hline
J0100-64 & Bi \\
J1514-49 & Bi \\
J1658-53 & Iso \\
J1747-40 & Iso \\
J1902-51 & Bi
\end{tabular}
\caption{The five millisecond pulsars discovered in a serendipitous Parkes survey.}
\label{ch4:tab:parkes}
\end{table}

\section{Summary}

In this chapter, we demonstrated the applicability of \ptl to both a restricted analysis (in the sense of a small region with a subset of sources) and to all-sky analysis (ASA).  For the former, we carried out a spectral analysis in the Cygnus region, obtaining estimates of the parameters of broadband spectral parameters (\S\ref{ch4:subsec:bba}), source positions (\S\ref{ch4:subsec:localization}), and model-independent estimates of source spectra (\S\ref{ch4:subsec:mis}).  An exploratory approach was crucial to the success of this analysis as we observed the interplay of the bright source PSR J2021+3651 with two dim, neighboring sources.

Next, we outlined an approach for ASA in which we can obtain a good estimate for the spectral parameters (and positions) of all (known) sources in the sky.  The method adopted---organizing sources by HEALPix while extracting data by cone---was a compromise between manageable data sets and optimal estimates for the parameters.  We demonstrated that the iterative approach converged, while comparison with the 1FGL catalog indicated good agreement between the two methods, i.e., a valid solution.

Finally, we made use of the model estimated from the all-sky analysis (ASA) for additional applications: we constructed TS maps for source finding using the ASA spectral model as background; we generated a kernel density estimator map of the sky; and we used the ASA pipeline to perform a custom analysis of known sources to identify good pulsar candidates.  In the next chapter, we shall make use of ASA-estimated spectra in building more sensitive tests for periodicity detection.
 
\chapter{Probability-weighted Statistics and Pulsed Sensitivity}
\label{ch5}

We discussed in \S \ref{ch4:sec:source_classification} the difficulties involved in detecting pulsed signals with HE instruments.  There, we focused on how the improved sensitivity of the LAT enabled detection of pulsation using $\gamma$ rays only, or through an initial detection of a point source with subsequent pulsation detection in a targeted radio search.  Here, we focus on the ``other side of the coin''---since the LAT boasts much improved sensitivity, by using timing solutions, we should be able to detect periodic emission from many known sources.

And this has turned out to be the case, e.g., the first detection of orbitally modulated HE $\gamma$ rays from LS I +61$^{\circ}$ 303 \cite{lsi61303}, the first detection of pulsed $\gamma$ rays from a millisecond pulsar \cite{j0030}, and the first new young pulsar detected using a radio timing solution\cite{j2021}.

The pulsars detected so far have been sufficiently bright that relatively simple techniques have sufficed to discern pulsation.  We discuss the details fully below, but in essence a set of photons is extracted from the data, the arrival times are converted to phase using the timing solution, and a statistical test for uniformity is performed.  For the same reason that aperture photometry is inadequate for spectral analysis, so too is this procedure suboptimal for time-domain analysis: any given aperture that one selects will either be strongly contaminated by other sources or be selected so stringently that most of the signal is removed.  In either case, the test results will depend strongly on the cut.

For spectroscopy, we were able to tackle the problem of the multi-scale PSF and source confusion by using likelihood to account for the IRF.  However, while we could model nearly any spectrum well with a simple model with only a few parameters, this is not the case for time-domain analysis.  Gamma-ray light curves tend to be complex, and more importantly, since we are searching for periodicity, we do not know {\it ab initio} what the light curve should look like.  Therefore, we do not pursue a full likelihood analysis here but shall instead adopt a hybrid approach in which we combine the results of spectral analysis with simple pulsation tests.

In further distinction to our approach to spectroscopy, we adopt an \emph{unbinned} time-domain analysis.  In this way we lose no information and---more importantly---incur no bias from an arbitrary choice of binning\footnote{Pulsar light curves can be quite sharply peaked, e.g. P2 of PSR J0030+0451\cite{j0030}, or relatively broad, e.g. PSR J1836+5925\cite{j1836p5925}.  Choosing the ``wrong'' bin size can drasticaly decrease sensitivity to a class of light curves.}.  Since we will be using the results of a previous spectral analysis and don't have to worry about jointly fitting sources, we can use a relatively small ($\leq2\dg$) ROI.  This makes unbinned analysis practical from a computational standpoint.

Specifically, we shall study a class of unbinned statistics composed of sums of sines and cosines of the observed phases.  Next, we show how these statistics can be altered to \emph{weighted} sums of sines and cosines.  By using the results of a ML spectral analysis, we can calculate the probability that each photon originated from the source we are testing for pulsation, and this probability is an excellent choice of weight.  We go on to show that the probability-weighted versions of the statistics drastically reduce the probability of Type I and Type II error through tests on ensembles of simulated pulsars.  The machinery that we develop in this section can be easily applied to determine the pulsed sensitivity of the LAT.

Before proceeding to periodicity searches, we discuss the necessary first step of converting the observables---photon times of arrival---into phase of the object of interest.

\section{Time-to-phase Mapping}
A preliminary step in any periodicity search is to map the event times to the appropriate phase, e.g., the rotational phase of a neutron star or the orbital phase of a binary.  For instance, the reconstructed time associated with a \fermi event is recorded in \emph{Mission Elapsed Time}, essentially TAI (International Atomic Time) with an origin of January 1, 2001.  In order to connect with sources at astronomical distances, the proper reference is time measured at the Solar System barycenter, which eliminates dependence on the (time-dependent) configuration of the Solar System.  Using either an exact ephemeris for some astronomical source, or even simply a guess for the period, the barycentered times can be converted to the phase of the distant object.  Thus, to convert a time in MET to phase, we must:
\begin{enumerate}
\item correct for light travel time, e.g., to the center of the earth, requiring knowledge of the S/C position at the given MET.  Small general relativistic effects may also be included.
\item correct for the earth's position as well as other Solar System effects (e.g., if the direction of the source passes near the Sun at the given MET, it suffers a Shapiro delay).  The unit of time at the Solar System barycenter is also slightly different from TAI.  Light travel time and Doppler shifting from the earth's revolution about the sun are included.
\item correct to the distant object rotational phase.  If the object is a member of a binary system, the correction to the distant system barycenter is nontrivial, and must account for Doppler shift due to orbital acceleration, Shapiro delays, and light travel time delays.
\end{enumerate}
For \fermi data, there exists a Science Tool, \emph{gtbary}, to convert MET to TDB (Barycentric Dynamical Time) or to geocentric time, i.e., time measured at the center of the earth.  A plugin for \emph{tempo2}\footnote{http://www.atnf.csiro.au/research/pulsar/tempo2/} will also perform this transformation as well as fold the photons on standard-form ephemerides for pulsars.

For the remainder of this section, the exact procedure we use will be irrelevant, and we will simply assume some black box has produced from a set of event times, $\{t_i\}$, a set of phases, $\{\phi_i\}$.  Our convention is to identify all values of phase that are congruent modulo 1, so a sweep of $\phi$ from 0 to 1 defines a complete cycle.

\section{Statistics for Pulsation Searches}

Suppose that in the null hypothesis---emission from the source in question is not periodic---the \phis have the cumulative distribution function (cdf) $F_{\Phi}(\phi)$.  That is, the probability to observe the random variable (rv) $\Phi$ with a value less than or equal to $\phi$ is $F_{\Phi}(\phi)$.  If the distribution admits a probability density function (pdf), $d/d\phi F_{\Phi}(\phi)$, we denote this function $f_{\Phi}(\phi)$.  We follow this notation for distributions throughout.  Often, the \phis in the null case will be uniformly distributed, $\Phi\sim U[0,1) \rightarrow F_{\Phi}(\phi)=\phi$.  However, the null distribution will \emph{not} remain uniform if the instrument response varies on timescales comparable to the period of interest.  Many of the results discussed below only hold under a uniform null distribution.  Departure from uniformity is thus a serious complication, though we present two avenues of attack in Appendix \ref{appC}.

The alternative hypothesis specifies another distribution, say $F'_{\Phi}(\phi)$.  Ideally, this distribution is known \textit{a priori}; in practice, we must estimate it, that is, provide a morphology for the light curve.  We concentrate here on the class of tests---a subclass of those considered by Beran\cite{beran1}---that represent the light curve using a Fourier expansion estimated from the data, i.e., the true (unknown) light curve is
\begin{equation}
f(\phi) = 1 + \sum\limits_{k=1}^{\infty} \alpha_k \cos(2\pi k\phi) + \beta_k \sin(2\pi k\phi),
\end{equation}
and we will attempt to estimate the coefficients.  In practice, we will truncate this sum at some finite harmonic.  (We eschew the complex notation for Fourier series here to avoid the complication of complex rvs.  However, for brevity in the sequel, we will refer to the $\alpha_k$ and the $\beta_k$ collectively as the $\psi_k$.)  We define
\begin{eqnarray}
c_k(\phi) &=& \cos(2\pi k\,\phi);\\
s_k(\phi) &=& \sin(2\pi k\,\phi).
\end{eqnarray}
For brevity we will refer to these variables collectively as $G_k$.  Since $\Phi$ is a rv, so are the $G_k$, although the $G_k$ and are clearly mutually dependent.  If $\Phi\sim U[0,1)$, the distribution function for these variables is simple and independent of $k$: $f_{G_k}(x) = [\pi\sqrt{-x^2}]^{-1}$.  We consider the case of nontrivial distributions for $\Phi$ in Appendix \ref{appC}.  We note in passing the characteristic function for the distribution $f(x) = [\pi\sqrt{1-x^2}]^{-1}$
\begin{equation}
\phi_X(t) = <\exp itx> = \int\limits_{-1}^{1} \frac{dx \exp(itx)}{\pi \sqrt{1-x^2}} = \frac{2}{\pi}\int\limits_{0}^{1}\frac{\cos(tx)}{\sqrt{1-x^2}}
= J_0(t),
\end{equation}
where we have appealed to symmetry in the integrand and the definition of the Bessel functions of the first kind.  It is easier to calculate the moments directly than to differentiate the above expression, and we have that the $n$th absolute moment is given by
\begin{equation}
\label{eq:gk_moments}
<|x^a|> = \frac{2}{\pi}\int_0^1 dx\, \frac{x^a}{\sqrt{1-x^2}} = \frac{1}{\pi}\beta(\frac{a+1}{2},\frac{1}{2}) = \frac{1}{\sqrt{\pi}}\frac{\Gamma(\frac{a+1}{2})}{\Gamma(\frac{a+2}{2})},
\end{equation}
where $\beta(x,y)$ is the Euler beta funciton and $\Gamma(x)$ is the usual Euler gamma function.

Using the new variables, we define the empirical trigonometric moments, 
\begin{equation}
\alpha_k = \frac{1}{n} \sum_{i=1}^{n} c_{ki};\ \ \beta_k = \frac{1}{n}  \sum_{i=1}^{n} s_{ki},
\end{equation}
where e.g. $c_{ki} = \cos(2\pi\,\phi_i)$.  These quantities are essentially the Monte Carlo estimators for the Fourier coefficients for the $k$th harmonic.  From the expression above for the absolute moments (or simply noting that $G_k\in[-1,1]$ regardless of the distribution), the mean ($\mu$) and standard deviation ($\sigma$) are finite and the central limit theorem (CLT) applies to the $\psi_k$.  That is,
\begin{equation}
\lim_{n\to\infty} \sqrt{n}(\psi_k - \mu_{G_k})/\sigma_{G_k} \sim \mathcal{N},
\end{equation}
where $\mathcal{N}$ denotes the standard normal distribution.  If $\Phi\sim U[0,1)$, $\mu_{G_k}=0$ and $\sigma_{G_k}=1/\sqrt{2}$ (Eq. \ref{eq:gk_moments}), then
\begin{equation}
\lim_{n\to\infty} \sqrt{2n}\ \psi_k \sim \mathcal{N}.
\end{equation}

The third moment of $G_{k}$ is $4/3\pi$, so the Berry-Esseen inequality gives the rate of convergence as $\propto 1/\sqrt{n}$.  Specifically, let $\psi_k(n)$ denote one of the statistics above for a sample size of $n$.  Then
\begin{equation}
\sup\limits_x|F_{\psi_k(n)}(x) - \Phi(x)| \leq \frac{C\,8\,\sqrt{2}}{3\,\pi\,\sqrt{n}} \approx \frac{0.847}{\sqrt{n}},
\end{equation}
i.e., the difference between the cumulative distribution of the statistic and the cdf of the normal distribution, $\Phi(x)$, is absolutely bounded.  (Here, we have used $C=0.7056$, the upper bound determined Shevtsova \cite{be_bound}.)  In practice, it is not this rate of convergence that will be the limiting step in later statistics we develop.  However, before moving on, we mention that an exact expression for $F_{\psi_k(n)}(x)$ may be obtained by inverting the characteristic function noted earlier.  Since $\psi_k$ is simply a linear combination of $G_k$, the characteristic function for $\psi_k$ is given by
\begin{equation}
\phi_{\Psi_k(n)}(t) = J_0\left(\frac{t}{n}\right)^n.
\end{equation}
Alternatively, if an approximate result, but more accurate than the CLT, is desired, one can use the known moments in an Edgeworth series.

The quantities $\psi_k$ are used in several statistical tests in the literature; see \cite{dejager_1} for a brief review.  In particular, invariance under rotations of phase become manifest when using these variables.  For instance, if we redefine our zero of phase $\phi\rightarrow\phi + \delta$, then
\begin{equation}
\left( \begin{array}{c} \alpha_k' \\ \beta_k' \end{array} \right) = \begin{bmatrix} \cos(2\pi k\,\delta) & -\sin(2\pi k\,\delta) \\ \sin(2\pi k\,\delta) & \phantom{-}\cos(2\pi k\,\delta) \end{bmatrix} \times \left( \begin{array}{c} \alpha_k \\ \beta_k \end{array} \right),
\end{equation}
i.e., a phase shift amounts to a rotation in $\alpha/\beta$ space for each harmonic, so any statistic involving only functions of the magnitude of these vectors, $\alpha_k^2 + \beta_k^2$, will be independent of $\delta$.  The simplest such statistic, a sum of $\psi_k^2$ for the first $m$ harmonics, has a long history in high-energy gamma-ray observations, and we introduce it below.

\section{The $Z^2_m$ Test}

The $Z^2_m$ statistic, defined as
\begin{equation}
\label{eqn:ztest}
Z^2_m = 2n \sum_{k=1}^{m} \alpha_k^2 + \beta_k^2,
\end{equation}
has been the workhorse of searches for $\gamma$-ray pulsars for years.  A $Z^2_2$ test was used in a search for pulsations in COS-B data using timing solutions for $145$ radio pulsars \cite{cos-b_pulse_search}.  A similar search of EGRET data\cite{egret_pulse_search} used $Z^2_m$ tests with 1, 2, and 10 harmonics, the $H$ test (see below), and the ``$Z^2_{2+4}$'' test which is defined as above but with summation restricted to the 2nd and 4th harmonics.  It forms an integral part of the $H$ test, and continues to see use in analysis of \fermi data, especially for sources where approximately sinusoidal modulation is expected.

From the discussion above, in the case of uniformly-distributed phases, $\sqrt{2n}\,\psi_k$ asymptotically follows the standard normal distribution.  $Z^2_m$ is thus, asymptotically, distributed as the sum of the squares of $2m$ standard normal variates.  As is well known, the square of a normal variate is distributed as $\chi^2_1$, i.e., chi-square with one degree of freedom, and the sum of $2m$ squares of \emph{independent}, normally distributed variates is distributed as $\chi^2_{2m}$, chi-square with $2m$ degrees of freedom.  Hence, if and only if these $2m$ $\psi_k$ are statistically independent, then $Z^2_m$ is asymptotically distributed as $\chi^2$ with $2m$ degrees of freedom.  In addition to this simple null distribution, the test is powerful\cite{dejager_1}, quick to compute, and has the already-discussed property of invariance under phase translation.  It underpins much of the statistical framework we will develop, and so we take some time to assess its properties.

\subsection{Validity of Asymptotic Calibration}
First, there is the point of the asymptotic ``independency'' of the $\psi_k$.  It is clear that, for only a few phases, these variables are highly dependent.  Indeed, for a single observation, $\psi_k$ can be inverted to find find the original rv, $\phi$ (up to uncertainty from congruence modulo 1), and hence any other $\psi_k$.  However, for the uniform null distribution, this ability to infer the phase from the sum of sines and cosines breaks down with increasing sample size.  E.g., if we have observed 10 phases and we have $\alpha_0=10$, then we know that $\phi=0$ for each phase.  However, $\alpha_0=10$ will almost surely not be observed, whereas $\alpha_0\approx0$ is quite likely, and there are many combinations of 10 phases that can yield it, i.e., it tells us very little about the constituent phases.  As $N$ increases, the tails are further suppressed, and consequently information about the $\{\phi\}$ and hence the distribution of $\psi_i$ conditioned on the value of $\psi_j$.   It is in this sense and limit that the $\psi_k$ become ``independent''.  

On the other hand, for some null distributions, it is impossible for the $\psi_k$ to become independent, even as $N\rightarrow\infty$.  A clear albeit unrealistic example is $f(\phi)=\delta(\phi-\phi_0)$, in which case $\psi_k$ remain perfectly correlated.  In a more realistic case, any peaking in phase of the null distribution will lead to long-surviving mutual depedence of the $\psi_k$.  Thus, while knowledge of the null distribution will allow us to define $\psi_k$ such that the marginal distributions are still asymptotically normal (i.e., the CLT applies), it will not be the case that the sum of the $\psi_k$ becomes chi-squre distributed.  More details can be found in Appendix \ref{appC}.

\begin{figure}
\begin{minipage}{6in}
\includegraphics[width=6in]{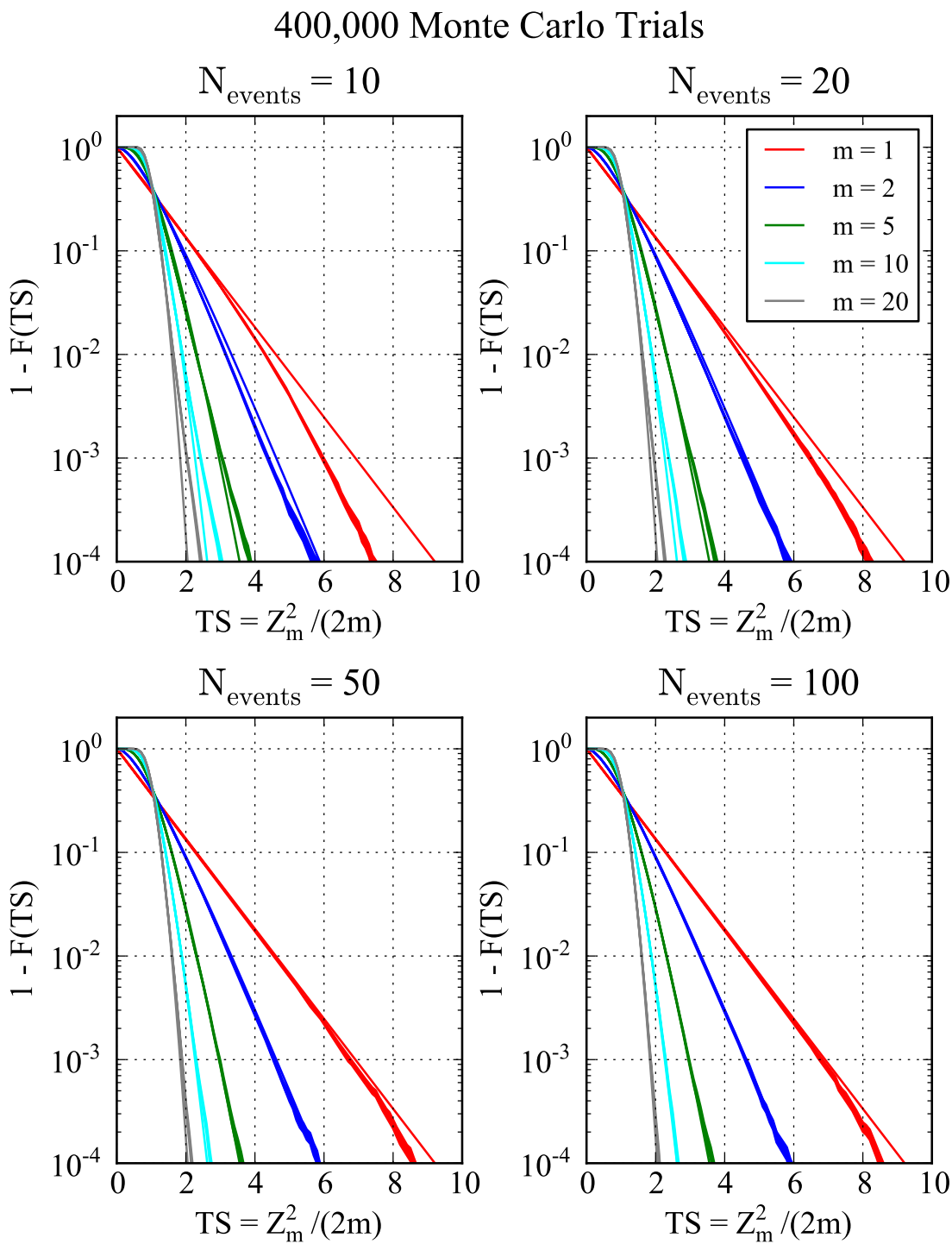}
\end{minipage}
\begingroup\renewcommand{\baselinestretch}{1.0}
\caption{The observed distribution for the $Z^2_m$ statistic for various sample sizes and a collection of maximum harmonics.  The asymptotic calibration is shown as a solid line, while the solid band gives the empirical distribution function of the Monte Carlo realizations.  The width indicates the statistical uncertainty estimated as $\sqrt{N(>=TS)}$, i.e., the square root of the number of counts with a TS greater than or equal to the current TS.
}
\renewcommand{\baselinestretch}{1.5}\endgroup
\label{ch5_plot1}
\end{figure}

Related to the question of mutual independence of the $\psi_k$ is the rate of convergence to the asymptotic $\chi^2$ distribution.  A Monte Carlo study of the converge as a function of both sample size and maximum harmonic ($m$) is shown in Figure \ref{ch5_plot1}.  Evidently, a sample size of about $50$ phases are required for robust significance estimation at the $3\sigma$ level.  Convergence appears to improve for higher values of $m$.  Interestingly, for low photon counts, low values of $m$ \emph{underestimate} the significance, leading to an increased probability of making a Type II error.  In contrast, high values of $m$ \emph{overestimate} the significance, increasing the chance of making a Type I error.

\subsection{Modification for Event-weighting}
A set of \phis will typically contain background events.  As discussed above, for most high-energy instruments, background events make up a significant fraction of the data selected, and may even dominate the signal.  Since the significance of a detection scales with the signal-to-noise ratio (SNR), we may try to apply stringent cuts to the data to try to enrich the signal.  We don't \textit{a priori} know the optimal cuts, so we must often choose between sub-optimal cuts or tuning the cuts with the data.  The former dilutes the power of the test, while the latter increases the risk of a Type I error if we fail to re-calibrate the test statistic, typically with Monte Carlo.

An alternate approach to enriching the SNR is event-weighting: we apply no stringent cuts to the data but instead assign each event a weight.  Events associated with the source are weighted heavily, those with the background lightly.  Below, we discuss a means of estimating the probability that the event in question was caused by detection of a source photon and we show that using these probabilities as weights is a powerful technique.

To include the weights, we define the weighted empirical trigonometric moments,
\begin{equation}
\nonumber\alpha_{wk} = \frac{1}{n} \sum_{i=1}^{n} w_i\,c_{ki};\ \ \beta_{wk} = \frac{1}{n}  \sum_{i=1}^{n} w_i\,s_{ki},
\end{equation}
where $w_i$ is the weight assigned to each event.  (Here, the $c_{ki}$ and $s_{ki}$ can either be derived directly from the phases or from the cumulative distribution in the case of non-uniformity as outlined in Appendix \ref{appC}.)  The weights may themselves be random variables (if determined from some model fit to the data), or they may be simple scalars.  If the former, the distribution of the $\{w_i\}$ must possess finite mean and variance since $w\in[0,1]$.  In the null hypothesis, the weights and the phases are independent rvs, so the moments of the products are simply the product of the moments, and it can be shown that the Lyupanov Condition for the CLT to hold will apply for any choice of $F_{\Phi}$ and any such weights distribution.  On the other hand, if the weights are taken as simple, non-random numbers, $\psi_k$ is now simply a different linear combination of $g_k$.  The characteristic function for $\psi_k$ becomes
\begin{equation}
\phi_{\Psi_k}(t) = \prod\limits_{i=1}^{n} J_0(2w_i\sqrt{n}t) \approx \prod\limits_{i=1}^{n} \left[1-\frac{w_i^2\,t^2}{2\,n}\right] \approx \left[1-\left(\sum\limits_{i=1}^{n}{w_i^2}\right)\frac{t^2}{2}\right].
\end{equation}
For this to match on to the characteristic function of a normal distribution, i.e., for the CLT to again apply, we must simply include $\sum\limits_{i=1}^{n}{w_i^2}$ in a new definition of $\psi_k$, i.e.,
\begin{equation}
\lim_{n\to\infty} \sqrt{2\,n}\left(\sum\limits_{i=1}^{n}{w_i^2}\right)^{\frac{-1}{2}}\, \psi_k \sim \mathcal{N}.
\end{equation}
Practically, the distribution function for the weights is unknown, so we would replace the distribution moments with the sample moments, i.e., we would follow the above prescription regardless of how we interpret the weights.

The weighted $\psi_k$ are now asymptotically normally distributed.  Their asymptotic mutual independence also remains, in the null case, as the distribution of weights and phases are independent.  Thus, the sum of the weighted $\eta_k$ will again be $\chi^2_{2k}$.  With this result, we define the weighted $Z^2_m$ test statistic, $Z^2_{mw}$,
\begin{equation}
\label{eq:weighted_z2m}
Z^2_{2mw} \equiv 2\,\left(\sum\limits_{i=1}^{n}{w_i^2}\right)^{\frac{-1}{2}} \sum_{k=1}^{m} \alpha_{wk}^2 + \beta_{wk}^2.
\end{equation}

\begin{figure}
\begin{minipage}{6in}
\includegraphics[width=6in]{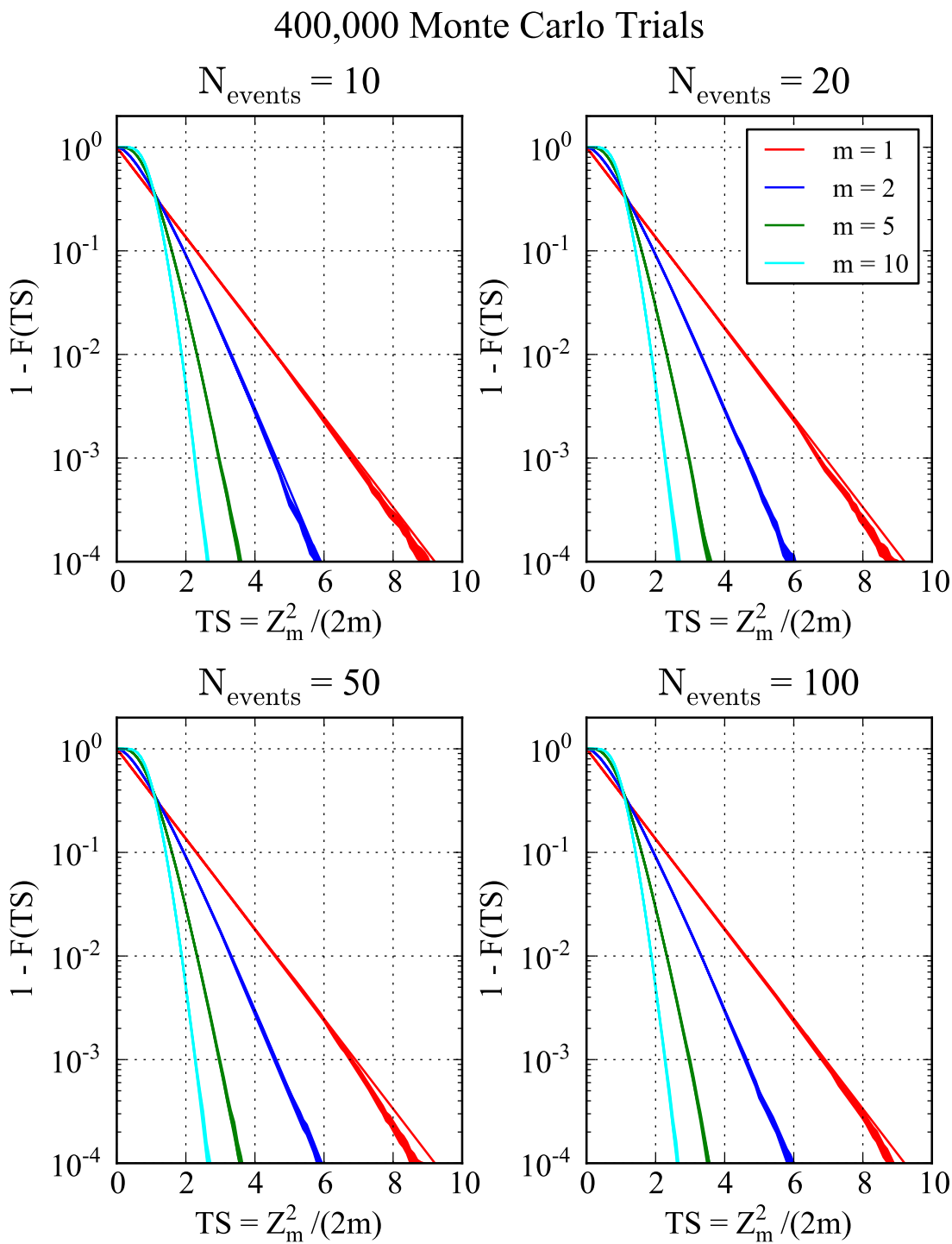}
\end{minipage}
\begingroup\renewcommand{\baselinestretch}{1.0}
\caption{The observed (Monte Carlo) and predicted (asymptotic) distribution for the \emph{weighted} test statistic $Z^2_{mw}$.  The procedure is identical to that described for Figure \ref{ch5_plot1} save for the addition of weights and the omission of the $m=20$ case for decreased computational burden.  The addition of weights is shown to make no difference in agreement with the predicted distribution; only insufficient sample size can leads to a discrepancy.}
\renewcommand{\baselinestretch}{1.5}\endgroup
\label{ch5_plot3}
\end{figure}

To verify the distribution, we generated a set of weights representative of those observed in actual data.  Precisely, we drew a signal, $s_i$, from a $\chi^2_2$ distribution and a background, $b_i$, from a $\chi^2_{50}$, and calculated $w_i=s_i/(s_i+b_i)$.  We incorporated the weights as in Eq. \ref{eq:weighted_z2m} and performed many Monte Carlo trials.  The results, showing perfect equivalence between $Z^2_m$ and $Z^2_{mw}$, appear in Figure. \ref{ch5_plot3}.

\section{ The H Test }

While the $Z^2_m$ test is unbinned in phase we still must choose the number of harmonics to include in the Fourier decomposition.  Choosing $m$ too small will result in a loss of power against sharply-peaked light curves, while $m$ too large will lose power against broad, sinusoidal light curves.  To ameliorate this problem, de Jager et al. (\cite{dejager_1}) proposed the H test, which seeks to estimate the optimal $m$ from the data.  They defined
\begin{equation}
\label{eqn:htest}
H_m = \max\left[Z^2_i - c\times(i-1)\right], 1 \leq i \leq m
\end{equation}
and specifically recommended $m=20$ and $c=4$ as an omnibus test.  They provided a Monte Carlo calibration of the tail probability and, in a recent paper\cite{dejager_2}, increased the Monte Carlo statistics, giving an estimate $1-F_{H_{20}}(h)\approx\exp(-0.4\times h)$ good to $h\approx70$.  We derive the analytic, asymptotic calibration for all values of $m$, $c$, and $h$ in Appendix \ref{appB} and use this calibration anywhere a conversion from $h$ to chance of Type I error (i.e., ``$\sigma$'') is needed\footnote{Throughout this section, we use a two-tailed convention when converting to ``$\sigma$'' units, e.g., $2\sigma$ implies a chance probability of $5\%$.}.

In connection with the preceeding material, we note that, since the calibration of the H-test relies only on the asymptotic $\chi^2_{2m}$ calibration and independence of the underlying $\eta_k$ variables, any transformation leaving these properties unchanged (e.g. weighting) also leaves the $Z^2_m$ and H test calibrations unchanged.  Thus, we define a \emph{weighted H-test statistic}
\begin{equation}
H_{mw} \equiv \max\left[Z^2_{iw} - c\times(i-1)\right], 1 \leq i \leq m.
\end{equation}
In the sequel, we adopt the original values for $c$ and $m$ unless otherwise noted, i.e.,
\begin{equation}
H_{w} \equiv \max\left[Z^2_{iw} - 4\times(i-1)\right], 1 \leq i \leq 20
\end{equation}
and likewise for $H$ (unweighted).

\section{Probability-weighted Statistics: Pulsars}

With the definitions and derivations above, we are now ready to implement \emph{probability-weighted} statistics for pulsation searches.  By probability-weighted, we mean we seek to weight each phase (or rather, cosine or sine thereof) with some number that expresses a genuine probability that the photon associated with the phase originated from the (possibly pulsed) source.  In constructing this quantity, we must first choose between an instantaneous or a time-averaged probability.  Allowing for time dependence provides a more precise treatment of time-varying backgrounds, but it requires a complicated bookkeeping effort.  However, for periodicity searches for periods on timescales comparable to or longer than the timescale for exposure variations, this bookkeeping is crucial.  Time dependence also allows for \emph{phase-dependence}, but we must assume a form for the light curve in order to calculate the weights, and this is more appropriate for a likelihood approach.

Since we are concentrating on pulsars, we thus choose to compute the time-averaged probability.  (See Appendix \ref{appC} for a treatment of searches for periods $>>1$s.)  To define a time-averaged probability, we refer to Eq. \ref{eq:psrate_with_exposure} and write down a slightly generalized version here, the expected rate from the $j$th source contributing to the ROI in an infinitesimal slice of observed energy and position
\begin{equation}
r_j(E,\vom) = \mcf(E,\vla)\, \epsilon(E,\vom_0)\, f_{\mathrm{wpsf}}(\vom ; \vom_0, E).
\end{equation}
Here, we recall that $\epsilon(E,\vom_0)$ is the exposure integrated over the pointing history of the S/C, $f_{\mathrm{wpsf}}(\vom ; \vom_0, E)$ is the incidence-angle-averaged PSF, and $\mcf(E,\vla)$ gives the intrinsic flux density of the source.  In this expression, $E$ and $\vom$ are the \emph{measured} energy and position for a particular photon; this expression is unbinned in all quantities save time.

We now assume we have performed a likelihood analysis as outlined in Chapter \ref{ch4} and have estimates for $\vla_j$, the parameters describing the spectral model for the $j$th source.  Then, for a photon observed near $E$ and $\vom$, the probability that it originated with the $j$th source is simply
\begin{equation}
w_j \equiv \frac{r_j(E,\vom,\vla_j)}{\sum\limits_{i=1}^{N_s} r_i(E,\vom,\vla_i)},
\end{equation}
where $N_s$ is the number of sources contributing to the ROI.  $w_j$ clearly satisfies the requirements of a probability: $w_j\in[0,1]$ and $\sum\limits_{i=1}^{N_s}w_i=1$.

Having defined our approach, we can now make contact with previous efforts to improve periodicity searches.  As we shall discuss in more detail in \S \ref{ch5:subsec:type1}, applying any statistical test to phases only with an omnibus extraction scheme---e.g., all photons within a fixed radius, or within a PSF-dependent cone---is not at all optimal.  At least two methods for improving on this basic scheme have been attemped.  Brown et al. proposed constructing ``photon probability'' maps\cite{clemson}, essentially kernel density estimators for the sky constructed in a fashion analagous to that outlined in \S \ref{ch4:sec:kde}.  By using these quantities in light curves constructed with COS-B data, they obtained an increase in the SNR ratio for the Crab pulsar.  This method was profitably employed in searches for periodicity in EGRET data\cite{ramanamurthy,maura}.

This scheme---essentially probability weighting with a weight depending only on the reconstructed photon position---is similar to our method but does not adequately account for varying SNRs.  To see this, consider searching for pulsations from a very bright (very dim) source.  At all energies of interest, the aperture over which the PSF is appreciably different from $0$ will have a background component, particularly in the Galactic plane.  If, in a given energy band, the source is much brighter (dimmer) than the background component,  we want to include (exclude) events lying the tails of the PSF.  This ``choice'' is naturally made by using probability weights derived from a full spectral model in which the relative strength of sources enters directly into the weight calculation.

A second approach was outlined by {{\"O}zel} and {Mayer-Ha{\ss}elwander}\cite{hans} who proposed a method for constructing an optimal aperture of arbitrary shape in position-energy space based on the \emph{expected} counts, i.e., a tentative spectral model for the candidate pulsar folded through the IRF.  By adopting an approximate spectral model, the authors effectively estimated the SNR in each of the three energy bands they considered, overcoming partially the deficiency of PSF-weighting.  However, their prescription was insensitive to light curve morphology, i.e., their algorithm returns the same aperture for a given position/spectrum, but the optimal aperture certainly depends on the light curve.  As we shall show below, incorporating weights directly into the statistical test leads to a near-invariance in choice of aperture.

We now proceed with a comparison of weighted and un-weighted versions of the $Z^2_m$ and $H$ test.

\subsection{Performance: Type 1 Errors}
\label{ch5:subsec:type1}

The primary interest in constructing the weighted versions of pulsation search statistics (and the weights) is, of course, to craft tests that find more (real) pulsars.  In statistical language, we want to minimize false positives (Type I Error) and false negatives (Type II error).

Type I error stems from two sources.  First, there is the chance of a fluctuation in the test statistic sufficiently large to pass our pre-set threshold for rejection of the null hypothesis, i.e., claiming detection of a pulsar.  As long as we understand the null distribution of the test statistic, this particular source of error is easy to control: we simply determine in advance our overall tolerance to false positives and set the TS threshold accordingly.  We must be cautious about applying the aysmptotic calibration of the null distribution to small sample sizes; in such cases, it is always a good idea to verify the chance probability with a Monte Carlo simulation.

A more insidious source is related to the nature of LAT data.  As discussed in Chapter \ref{ch3}, the PSF is a strong function of energy, and at energies of a few GeV and below, there is strong source confusion, particularly in the Galactic plane.  Simply going to higher energies for the better SNR is, however, not feasible for pulsar as the spectra are exponentially supressed above a few GeV.  Thus, the sensitivity to pulsations can depend quite strongly on the particular selection criteria for the data, e.g., the energy-dependent extraction radius.  Without knowing these cuts in advance, one is tempted to tune the cuts using the test statistic itself, a famous source of bias.  To compensate, one should correct with a trials factor equal to the number of cuts tested.  Alternatively, the use of omnibus cuts will result in a sub-optimal SNR, increasing the chance of a false negative.  Finally, there is the issue of diminished sample size---and concomitant departure from the asymptotic calibration---with very stringent cuts.

\begin{figure}
\begin{minipage}{6in}
\includegraphics[width=6in]{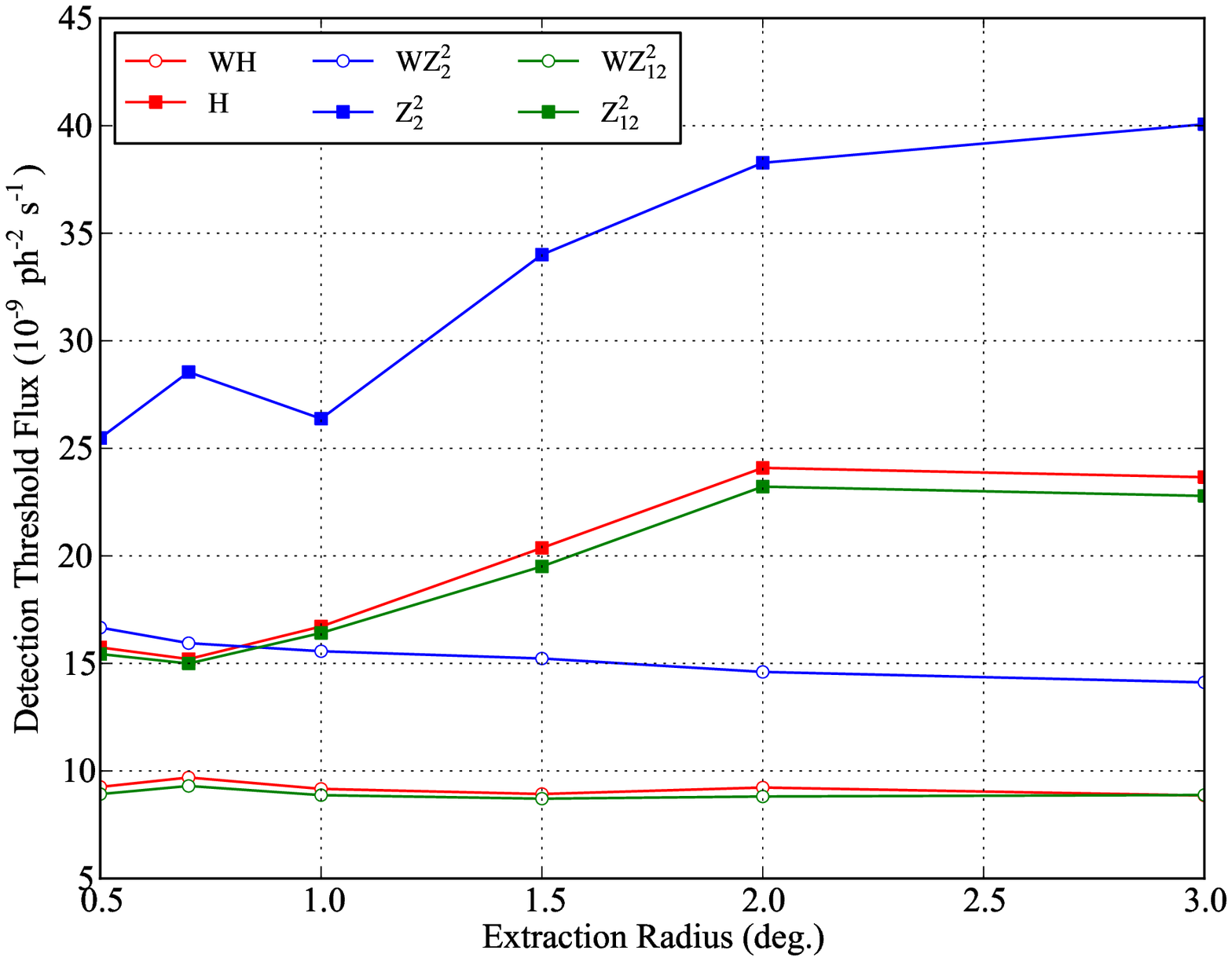}
\end{minipage}
\begingroup\renewcommand{\baselinestretch}{1.0}
\caption{A comparison of the weighted and un-weighted versions of three statistics.  Photons with energies above 200 MeV are selected according to the extraction radius on the x-axis.  The light curve in this case is a Gaussian with $\sigma=0.03$, so the H test and high-harmonic Z test are much more sensitive than the $Z_2$ tests.  The salient feature is the relative independence of the detection flux threshold on extraction radius for the weighted methods.  The sensitivity for the standard tests drops by about 50\% as we increase the aperture size.  The results indicate that little is gained for weighted methods with an extraction radius larger than $2^{\circ}$.}
\renewcommand{\baselinestretch}{1.5}\endgroup
\label{ch5_plot5}
\end{figure}

We expect that \emph{probability weighted statistics} will be all but immune to the issue of aperture selection.  High energy photons falling in the core of the PSF will naturally receive a weight close to 1, i.e., we include all photons that would be accepted by a stringent cut on SNR\footnote{Since the weighted statistics are in some sense normalized, the actual proximity to the maximum value of 1 is irrelevant.  A more precise statement: ``photons falling in the core of the PSF at high energies will have significantly higher than average weights''.}.  However, we also include a large set of photons from lower energies that receive modest weights but undoubtedly still contain pulsed signal.  Finally, photons falling in the tails of the PSF at any energy receive a very low---essentially zero---weight.  Unweighted, these low-energy, distant photons would swamp the signal.  Weighted, they are neglibible.  We can then with impunity select a single, large region of interest and be guaranteed good performance.  Practically, no signal is gained beyond $2$ to $3^{\circ}$ save for the very strongest pulsars, which are of course already known; see Figure \ref{ch5_plot5}.  Since we never tune the data selection, we incur no undue risk of Type 1 error.  In this regard, probability weighting significantly increases the performance of the $Z^2_m$ and $H$ statistics.

\begin{figure}
\begin{minipage}{6in}
\includegraphics[width=6in]{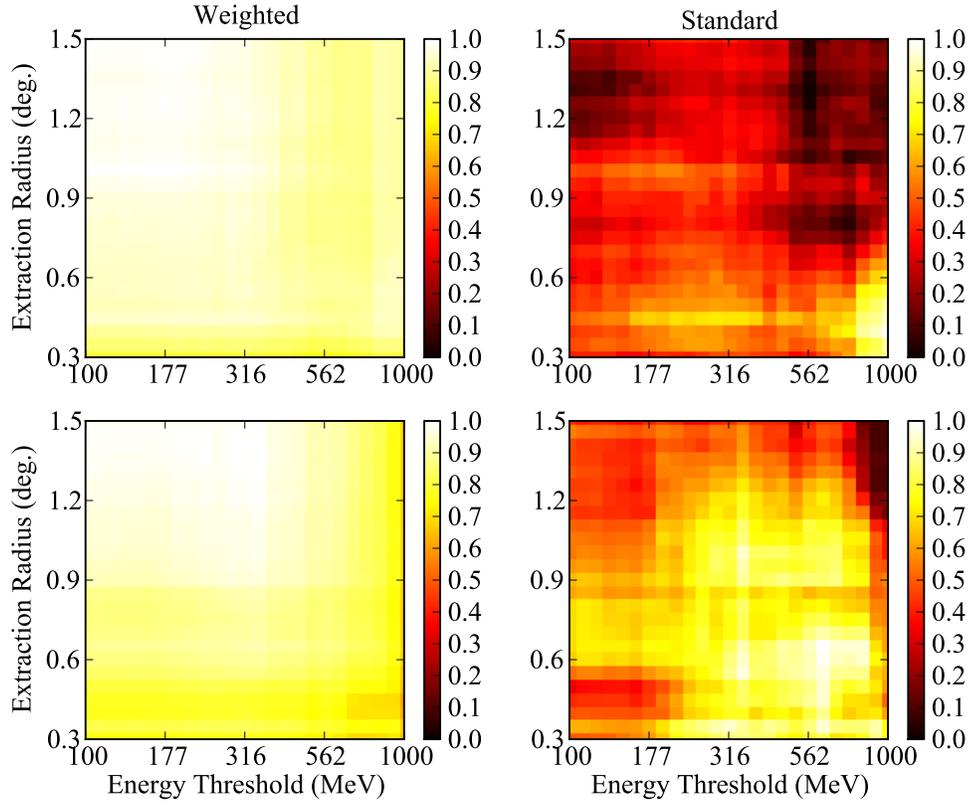}
\end{minipage}
\begingroup\renewcommand{\baselinestretch}{1.0}
\caption{A comparison of the dependence on event selection of $H_{20w}$ and $H_{20}$.  A source with flux $10^{-8}$ \fluxunits was simulated and a ``Vela-like'' light curve was added to the source photons.  The spectrum was that of Eq. \ref{eq:vela_spec}.  The test statistics were calculated over a grid of selection criteria: the y-axis gives the extraction radius (maximum angular distance from the pulsar allowed) and the x-axis indicates the threshold energy, the minimum energy allowed in the selection.  The lefthand (righthand) columns shows the results for the weighted (un-weighted) test statistic.  The two rows are two indepedent Monte Carlo realizations of the source and background.  The test statistics were converted to $\sigma$ units and independently normalized to the maximum observed value.  It is clear that the weighted statistics have a minimal dependence on the extraction criterion and, more importantly, do best when given the most data (maximum in upper lefthand corner of cut space).  On the other hand, the significance can be strongly peaked for the standard statistic and the optimal cut varies from realization to realization.}
\renewcommand{\baselinestretch}{1.5}\endgroup
\label{ch5_plot8}
\end{figure}

To make these claims concrete, we compare the weighted H test ($H_{20w}$) and the standard H test ($H_{20}$) over a grid of selections in photon position and energy.  The method and results are presented in Figure \ref{ch5_plot8}.  Similar results in a single dimension---photons with energies above 200 MeV and a grid of position selection criteria---in terms of detection threshold flux (defined in the section below) appear in Figure \ref{ch5_plot5}.  In both examples, the weighted statistics are largely insensitive to the extraction criterion.  The result can be summarized as follows: unweighted statistics depend on SNR while weighted statistics depend primarily on \emph{signal alone}, and the strategy is then to use a single, large aperture for all weighted pulsation tests.

\subsection{Performance: Type II errors (Sensitivity)}

We must also consider Type II error---false negatives.  In astrophysical terms, this is directly related to the sensitivity of the method, i.e., the number of photons that must be collected before the test statistic passes (for some percent of an ensemble) the detection criterion.

The easiest way to quantify the sensitivity is by simulation of an ensemble of pulsed sources.  As for the validation of spectral analysis, we use the Science Tool \emph{gtobssim} to simulate a point source with a diffuse background.  For the discussion below, we use a realistic pulsar spectrum, in particular
\begin{equation}
\label{eq:vela_spec}
\frac{dN}{dE} \propto (E/GeV)^{-\Gamma} \exp -E/E_c
\end{equation}
with $\Gamma=1.57$ and $E_c = 3150$GeV.  (This spectrum is typical of many pulsars and is, in fact, close to the measured spectrum for the middle-aged Vela pulsar.)  We place the source at the position of the Vela pulsar, (R.A., Decl.) = (128.8463, -45.1735).  We choose this scenario since (a) many pulsars lie in the Galactic plane, in particular along the spiral arms as projected onto the sky and (b) detection in the Galactic plane is much more difficult than at high Galactic latitudes, so improved methods are particularly helpful here.  As in Chapter \ref{ch3}, we simulate a very bright source ($\mcf = \mcf_5 = 10^{-5}$\fluxunits) and select subsets of the photons in order to achieve a target brightness.  Unlike in the spectral analysis case, we use a single realization of the diffuse background.  This is not a serious impediment since we can randomize the phase of the diffuse photons for each MC realization.  Using the same spectral models used to simulate the sources, we calculate the weights using \emph{gtsrcprob}, a Science Tool designed to implement the probability-weighting scheme.  The base weights must be modified to account for changing point source flux; we address this point below.

When the Monte Carlo events are generated by \emph{gtobssim}, the simulated event times have no inherent pulsation.  In order to convert ``unpulsed'' photons to a ``pulsed'' data set, we apply the following procedure:
\begin{enumerate}
\item Select a template for the light curve of the pulsar.  This will typically be a sum of (wrapped) Gaussian functions, and will be normalized, i.e., a probability density function for photon phase.
\item Select photons originating with the pulsed source.  This is possible because each event generated by \emph{gtobssim} is tagged with a unique identification for the generating source.
\item For each selected photon, draw (by Monte Carlo) a phase from the light curve and assign it to the photon.
\end{enumerate}
For background photons, we simply assign a phase drawn from the uniform distribution.  In this prescription, there is no sense of a time-to-phase mapping.  If we wish to use an explicit ephemeris, e.g., specifying a reference time, a period, and a period derivative, then this timing solution predicts the phase as a function of time.  We then must perform the following additional steps to make the simulated times consistent with the simulated phases:
\begin{enumerate}
\item For the simulated time, use the time-to-phase mapping to calculate the phase \emph{predicted} by the timing solution.
\item Use the timing solution to determine the pulsar period at the simulation time.
\item Adjust the simulated time by $\delta\,t = P(t)\times(\phi_{sim} - \phi_{mod})\,\mathrm{mod}\,1$, with $\phi_{sim}$ the phase drawn from the light curve and $\phi_{mod}$ the phase predicted by the timing solution, i.e., the smallest possible adjustment that will align the simulated time with the desired light curve.  Provided $P(t)$ is small enough, the change in exposure is negligible.
\end{enumerate}
After applying this procedure, one can use the timing solution to convert the adjusted simulated times to phase, and these phases will follow the light curve distribution.  These steps are not necessary for testing pulsed sensitivity and we include them here only for completeness.

\begin{figure}
\begin{minipage}{6in}
\includegraphics[width=6in]{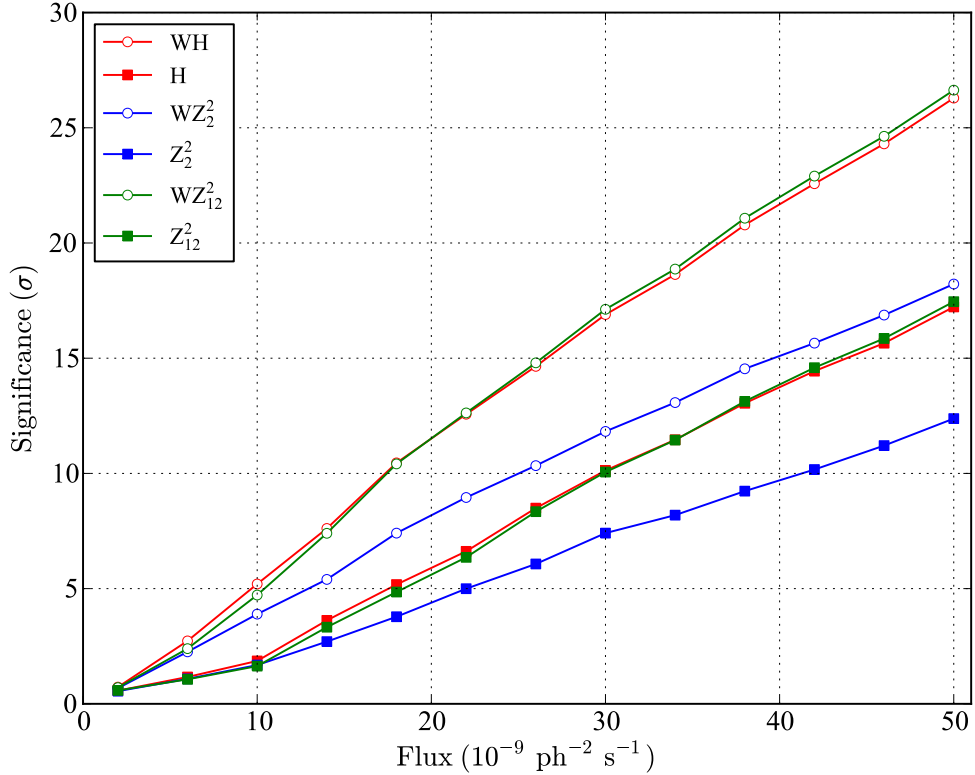}
\end{minipage}
\begingroup\renewcommand{\baselinestretch}{1.0}
\caption{The dependence of significance in $\sigma$ units as a function of flux.  The reported value at a given flux is that attained by at least $68\%$ of the ensemble of 50 MC realizations.  The trend is approximately linear, as explained in the main text, making it simple to invert the relation.  The light curve here is a single Gaussian peak with $\sigma=0.03$.}
\renewcommand{\baselinestretch}{1.5}\endgroup
\label{ch5_plot10}
\end{figure}

With these procedures, we have in hand an ensemble of pulsars at a series of intrinsic fluxes with arbitrary light curves.  To test the sensitivity, we determine the \emph{flux threshold} for each method given a particular LC morphology, i.e., the flux for which a source would be detected according to some criteria (see below) a majority of the time.  More specifically, we
\begin{enumerate}
\item At a sequence of intrinsic fluxes, evaluate the test statistic for each method for each member of the ensemble.  At each flux, we must recalculate the probability weights.  These were originally assigned with the \emph{simulation} flux values, $\mcf_{sim}$, and we now wish to described a source with target flux $\mcf_{tar}$.  The new weights are given by $w_{tar}^{-1} - 1 = (w_{sim}^{-1} - 1)\times\mcf_{sim}/\mcf_{tar}$.
\item Using the asymptotic calibration, calculate the chance probability of Type I error (the tail probability in the asymptotic distribution) and convert it to $\sigma$ units, i.e., a value of $X\sigma$ is the two-sided tail probability of a normal distribution integrated from $X$ to $\infty$.
\item Determine the flux for which $68\%$ of the ensemble deliver a $\sigma$ value above some pre-determined threshold (see below).  This is the \emph{detection} or {flux threshold}.  To determine the $68\%$ level with some level of robustness, we fit the ensemble values with a normal distribution and report the appropriate quantile value.
\end{enumerate}
To determine the flux threshold, we need to invert the calculated quantity (tail probability in $\sigma$ units) to the desired quantity (flux threshold).  Fortunately, the relationship between the two is linear.  This dependence may be slightly surprising.  Significance canonically scales as $\sqrt{N}$, the collected events, or in this case, since we integrate for exactly one year, $\sqrt{\mcf}$, the square root of the source flux.  However, when we increase the source flux without increasing the background flux, we also increase the SNR, and in general significance is proportional to the square root of SNR.  Taken together, the dependence is linear, i.e., $\sigma\propto\mcf$.  This dependence for a particular source is shown in Figure \ref{ch5_plot10}.

Above, we mentioned a pre-determined threshold for the tail probability in $\sigma$ units: test statistics above this threshold cause us to discard the null hypothesis, i.e., claim detection of a pulsed source.  In the following, we choose $4\sigma$ for as the confidence level for detection.  It provides a very small probability of Type I error, about 1 in 15800, and it is also not too different from the LAT Collaboration's internal threshold of $5\sigma$.  On the other hand, as we have seen, sample sizes of order $100$ (typical for a faint pulsar) begin to depart from their asymptotic calibrations at around this confidence level.

We are now ready to characterize the effect of adding probability-based weights to existing statistics on sensitivity.  To give a modest survey of the statistics outlined above, we present weighted and un-weighted versions of $H_{20}$, $Z^2_{12}$, and $Z^2_2$.  Recall that $H_{20}$ is an ``omnibus'' test that depends little on light curve morphology, whereas we expect $Z^2_{12}$ to perform well for light curves with sharp peaks and $Z^2_2$ to perform well for light curves with broad features.

\begin{figure}
\begin{minipage}{6in}
\includegraphics[width=6in]{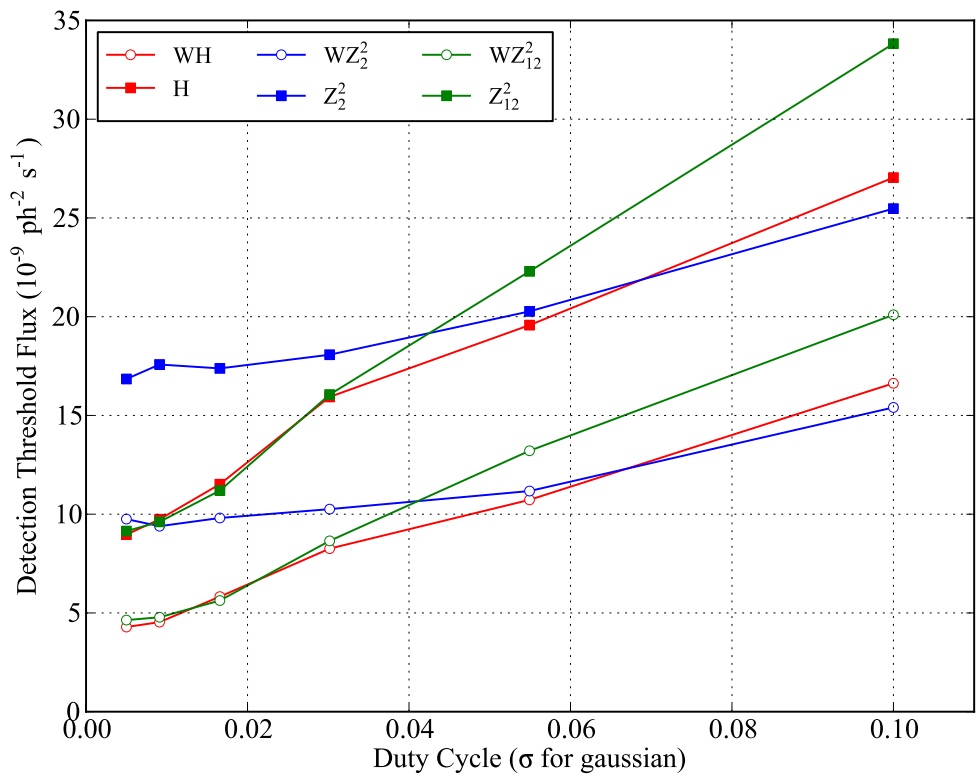}
\end{minipage}
\begingroup\renewcommand{\baselinestretch}{1.0}
\caption{The $68\%$ flux detection threshold based on a $4\sigma$ detection criterion for a single-peaked Gaussian light curve as a function of duty cycle.  Light curves with narrow (broad) peaks lie to the left (right).  The un-weighted thresholds are about twice those of the weighted thresholds, indepedent of duty cycle or statistic.}
\renewcommand{\baselinestretch}{1.5}\endgroup
\label{ch5_plot4}
\end{figure}

\begin{figure}
\begin{minipage}{6in}
\includegraphics[width=6in]{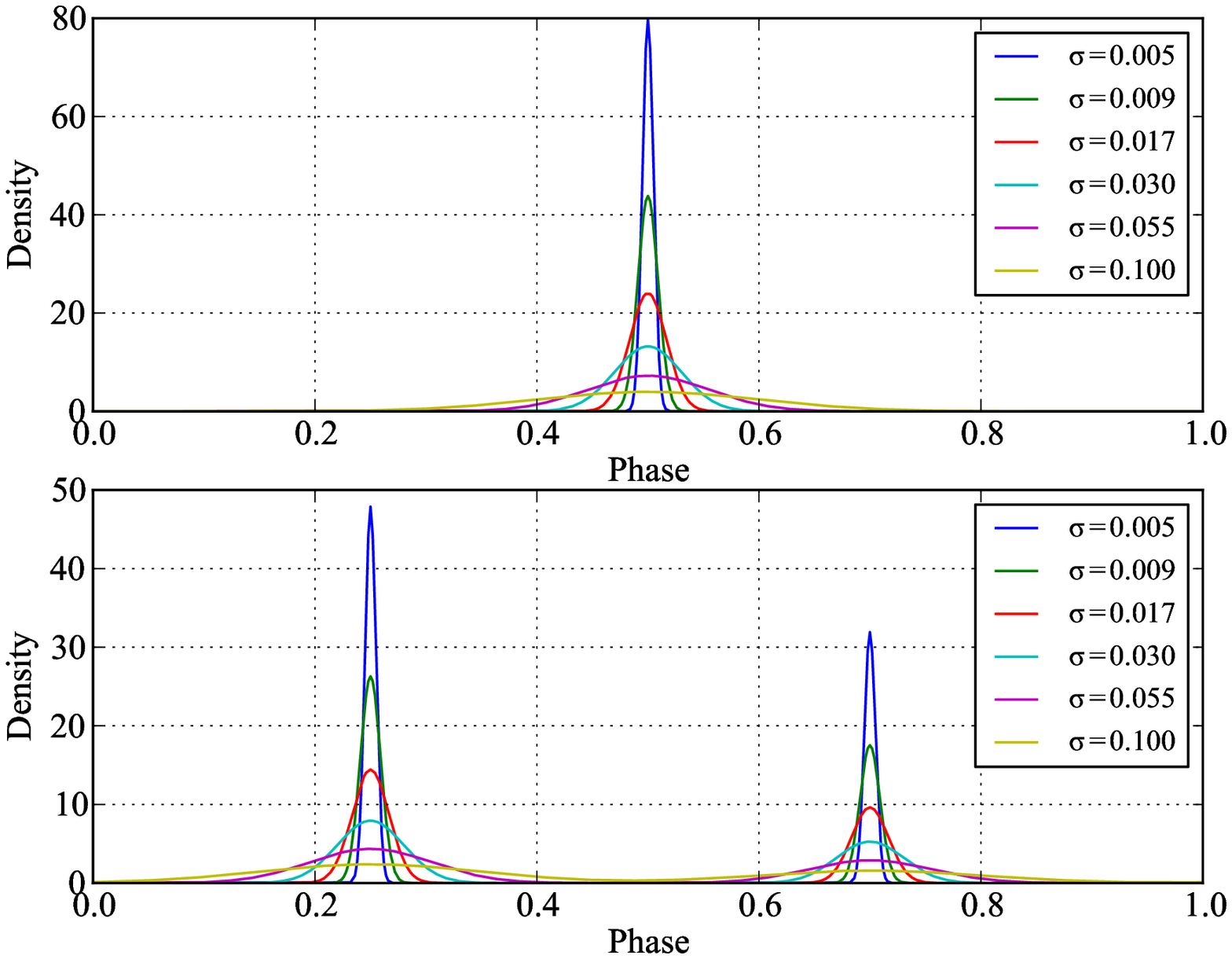}
\end{minipage}
\begingroup\renewcommand{\baselinestretch}{1.0}
\caption{The templates used for the single- and double-peaked light curves in the determination of the detection flux thresholds.  The functional form is a wrapped Gaussian, i.e., $f(\phi)= \sum\limits_{i=\infty}^{\infty} g(\phi + i)$ with $g(\phi,\mu,\sigma) = (2\pi)^{-0.5}\,\exp(-0.5\,(\phi-\mu)^2/\sigma^2)$.  In the first panel, $\mu=0.5$, while in the second panel, $\mu_1=0.25$, and $\mu_2=0.70$.  In this panel, the ratio of the peak heights is 3/2, and the peak widths are identical.}
\renewcommand{\baselinestretch}{1.5}\endgroup
\label{ch5_plot11}
\end{figure}

The primary result, in Figure \ref{ch5_plot4}, shows the flux threshold for each method as a function of ``duty cycle'', the fraction of the full phase for which there is appreciable pulsed emission.  In this case, the light curves are single Gaussian peaks with a variety of values for their $\sigma$ parameter; the templates are shown in Figure \ref{ch5_plot11}.  For a fixed flux, increasing the duty cycle decreases the peak flux, or SNR, and the primary dependence is then an inverse relation between flux threshold and duty cycle.  However, there is additional dependence from the nature of each test.  It is clear, e.g., that the $H_{20}$ and $Z^2_{12}$ perform significantly better than $Z^2_2$ for low duty cycle sources, while $Z^2_2$ maintains a slight edge for broad light curves.  The H test does a good job for all duty cycles.

More importantly, it is clear that, independent of pulsar duty cycle (unsurprisingly), the weighted statistics enjoy about a factor of two smaller threshold for detection than their unweighted counterparts.  This has nontrivial implications for the detected pulsar population.  E.g., if the pulsars follow a linear logN-logS relation with slope $\alpha$, using weighted statistics yields $\approx2^{\alpha}$ additional detections.

\subsubsection{Two-peaked light curves}
\begin{figure}
\begin{minipage}{6in}
\includegraphics[width=6in]{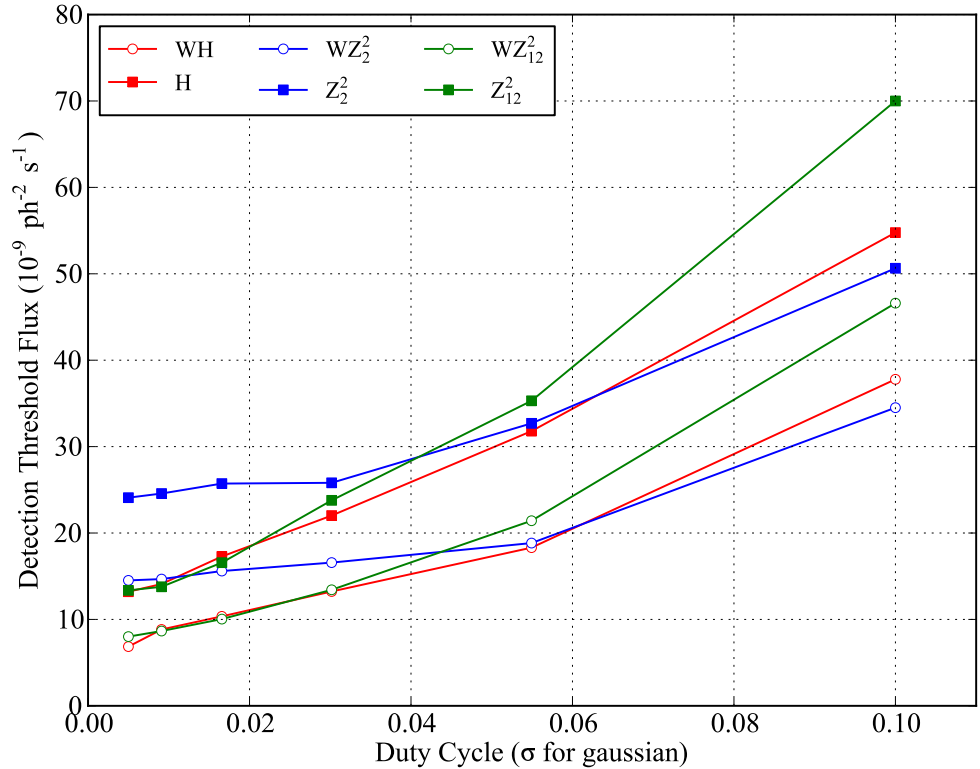}
\end{minipage}
\begingroup\renewcommand{\baselinestretch}{1.0}
\caption{The $68\%$ flux detection threshold based on a $4\sigma$ detection criterion for a double-peaked Gaussian light curve as a function of duty cycle.  The flux threshold for the weighted statistic is $1.5-2.0$ times lower than the unweighted statistics.}
\renewcommand{\baselinestretch}{1.5}\endgroup
\label{ch5_plot12}
\end{figure}

Many pulsar light curves illustrate twin peaks, often separated by about 0.4-0.5 cycles.  We repeat the analysis of the previous section using a template comprising two Gaussian peaks separated by $0.45$ in phase.  Real pulsar peaks often have unequal intensities (with an energy-dependent ratio.)  We reflect that here with a slightly-dominant leading peak.  The templates are shown in Figure \ref{ch5_plot11}.  As seen in Figure \ref{ch5_plot12}, the overall flux thresholds are unsurprisingly increased: at a fixed flux, spreading photons between multiple peaks decreases the overall SNR.  The weighted statistics maintain a comfortably decreased flux threshold.

\subsubsection{Pulsars with DC Emission Components}

\begin{figure}
\begin{minipage}{6in}
\includegraphics[width=6in]{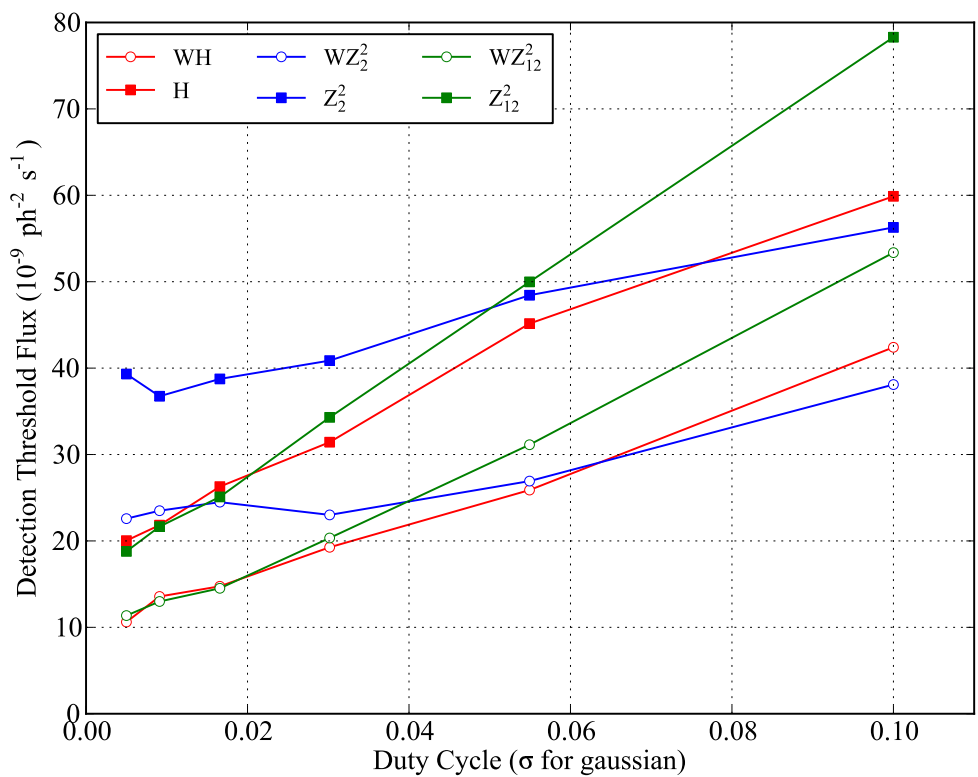}
\end{minipage}
\begingroup\renewcommand{\baselinestretch}{1.0}
\caption{The $68\%$ flux detection threshold based on a $4\sigma$ detection criterion for a single-peaked Gaussian light curve as a function of duty cycle.  Here, the pulsed fraction has been decreased to $50\%$, i.e., half of the photons from the source are uniformly distributed and half are drawn from the Gaussian light curve.  The flux threshold for the weighted statistic is $1.5-2.0$ times lower than the unweighted statistics, with some slight dependence on duty cycle and statistic.}
\renewcommand{\baselinestretch}{1.5}\endgroup
\label{ch5_plot13}
\end{figure}

Some pulsars (not respecting their nomenclature\footnote{Given the original notion of pulsar as a ``pulsating source of \emph{radio}'', with the advent of ``radio-quiet pulsars'' this ship may already have sailed.}) emit an appreciable portion of their flux as unpulsed $\gamma$ rays, e.g. PSR J1836+5925\cite{j1836p5925}.  (Models of magnetospheric emission generally allow for a sustained GeV component under certain geometries and viewing angles.)  Unpulsed emission could be a confounding factor for the weighted test, since unpulsed photons from the source will carry heavy weight but be uniformly distributed, decreasing the effective SNR.  And indeed, in Figure \ref{ch5_plot13}, where we have used a single-peaked light curve with a pulsed fraction of $1/2$, we see (a) an overall increased flux threshold on the order of two, as expected (b) a slight increase in the ratio of weighted-to-unweighted detection thresholds.  However, the ratio is $1.5$ to $2.0$, indicating the weighted statistics still offer significantly improved sensitivity for pulsars with appreciable DC emission.

\subsubsection{Effect of Uncertainties in Spectral Parameters}
\label{ch5:uncertain_spectrum}
From these demonstrations, it is clear that the probability-weighted statistics have, on average, a factor of about 2 improved sensitivity relative to the unweighted versions.  One potential objection is that, in determining the weights for this validation, we have used the known spectrum.  With real data, of course, we must first estimate the spectrum in order to determine the weights.  To assess the impact of using estimated parameters with concomitant uncertainty, we make use of the Monte Carlo ensembles developed for the validation of spectral analysis and described in Chapter \ref{ch3_post}.

For this application, we simulated 20 realizations of a point source at the position of the Vela pulsar with the same power law with exponential cutoff spectrum as employed above, Eq. \ref{eq:vela_spec}.  To this data we added phase from a single-Gaussian light curve with $\sigma=0.03$.  From Figure \ref{ch5_plot4}, we see that the detection threshold for such a configuration is estimated at $\approx8\times10^{-9}$\fluxunits.  Thus, following the procedure outlined in Chapter \ref{ch3_post} for constructing ensembles of point sources with a particular flux, we generate 20 realizations of this ``pulsar'' with a flux of $8\times10^{-9}$\fluxunits\footnote{At this flux, there are of order 100 source photons for the 1-year integration period, allowing for asymptotic calibration.}.  First, we calculate the probability weights using the known model parameters for the pulsar and the diffuse background, i.e., the ``ideal'' case.  We then perform a ML spectral fit with \ptl to estimate the spectral parameters.  Since the simulated point source is very dim relative to the background, it is inappropriate to attempt to fit a spectral model with three degrees of freedom.  We therefore fix the cutoff energy to $100$ GeV, essentially a power law spectrum.  This approach is \emph{conservative} since we are using an incorrect spectral model.  Using the best-fit values for the flux density and the photon index, we calculate a new set of probability weights.  Finally, we use a weighted $H$ test to determine the significance (a) using the ``ideal'' weights and (b) using the ``measured'' weights.

\begin{figure}
\begin{minipage}{6in}
\includegraphics[width=6in]{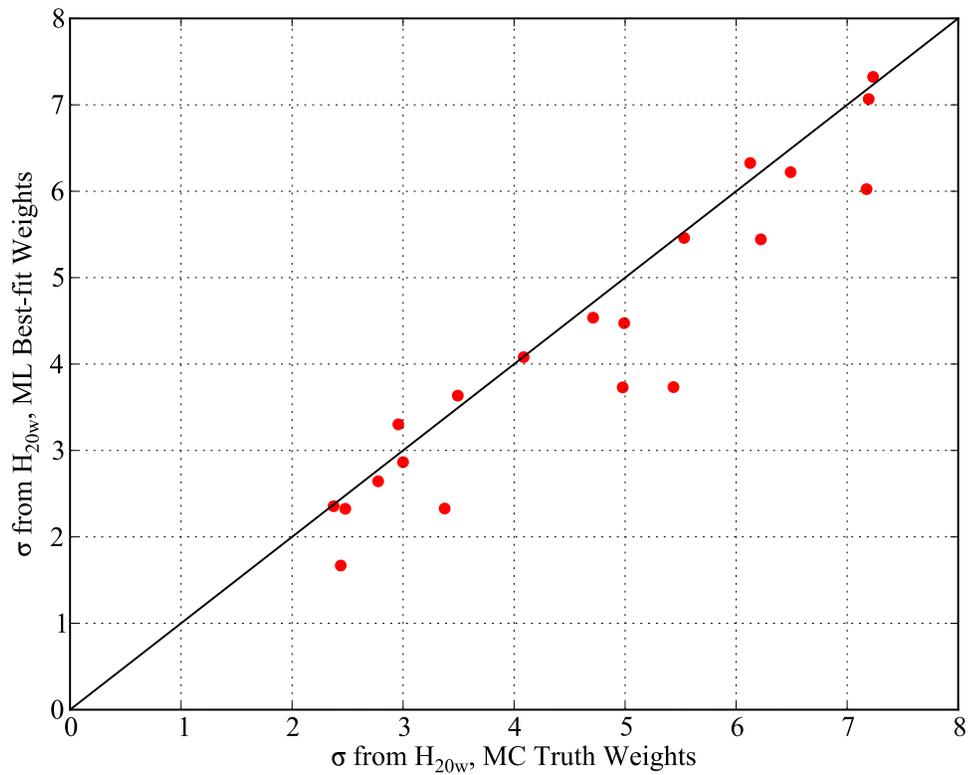}
\end{minipage}
\begingroup\renewcommand{\baselinestretch}{1.0}
\caption{The significance for each member of the ensemble described in the text calculated using weights derived from the model used to simulated the data (the Monte Carlo ``truth'') and derived from the ML model.  The ML-derived values are very similar to those obtained using the known spectrum.}
\renewcommand{\baselinestretch}{1.5}\endgroup
\label{ch5_plot14}
\end{figure}

We compare the results---in $\sigma$ units---in Figure \ref{ch5_plot14}.  In general, the detection significance obtained with the ``measured'' weights is slightly lower than that obtained with the ``ideal'' weights, although in a few cases statistical fluctuations lead to the opposite outcome.  Comparing the populations means, the the overall significance using measured weights is decreased to $92\%$ of the ideal case.  Recalling that the flux threshold depends linearly on the significance, we estimate the flux threshold is increased in the case of measured weights by $\approx10\%$.  This effect is relatively small compared to the factor of $1.5-2.0$ increase in flux threshold seen between the weighted and unweighted versions of the $H$ test.  We therefore conclude that---even accounting for uncertainties in the spectral parameters used to calculate the probability weights---weighted statistics offer a significance improvement in sensitivity.

\subsubsection{Comparison of Pulsed and Unpulsed Detection Thresholds}
The machinery established above---performing spectral fits on an ensemble to compute the weighted statistics using probabilities estimated from an ML fit---also provides for directly comparing the DC (unpulsed) source significance with the pulsed significance.  Recall from the discussion in Chapter \ref{ch4} that there is an asymptotic calibration for the test statistic obtained from the likelihood ratio for a single parameter whose null value lies on a boundary.  In that case and here the parameter was the flux of a point source which is zero in the null case.  To apply this calibration, we proceed as in Chapter \ref{ch4} and fix all parameters but the flux.  We set the cutoff energy to $100$ GeV as above, and we set the photon index to $\Gamma=2.0$.

Next, we perform a ML fit on the ensemble of sources described in the previous section to determine the best-fit value for the flux density.  The DC significance is then determined by
\begin{equation}
\sigma_{DC} = \sqrt{2\times\log\mathcal{L}_{opt}/\mathcal{L}_0},
\end{equation}
with $\mathcal{L}_{opt}$ the likelihood value obtained with the best-fit flux density and $\mathcal{L}_0$ the likelihood value obtained with the flux density set to zero.

As in the previous section, we calculate the probability weights using both the Monte Carlo truth values of the parameter and with the best-fit spectrum---in this case, with cutoff energy \emph{and} photon index fixed---to estimate a pulsed significance with the $H_{20w}$ test.

\begin{figure}
\begin{minipage}{6in}
\includegraphics[width=6in]{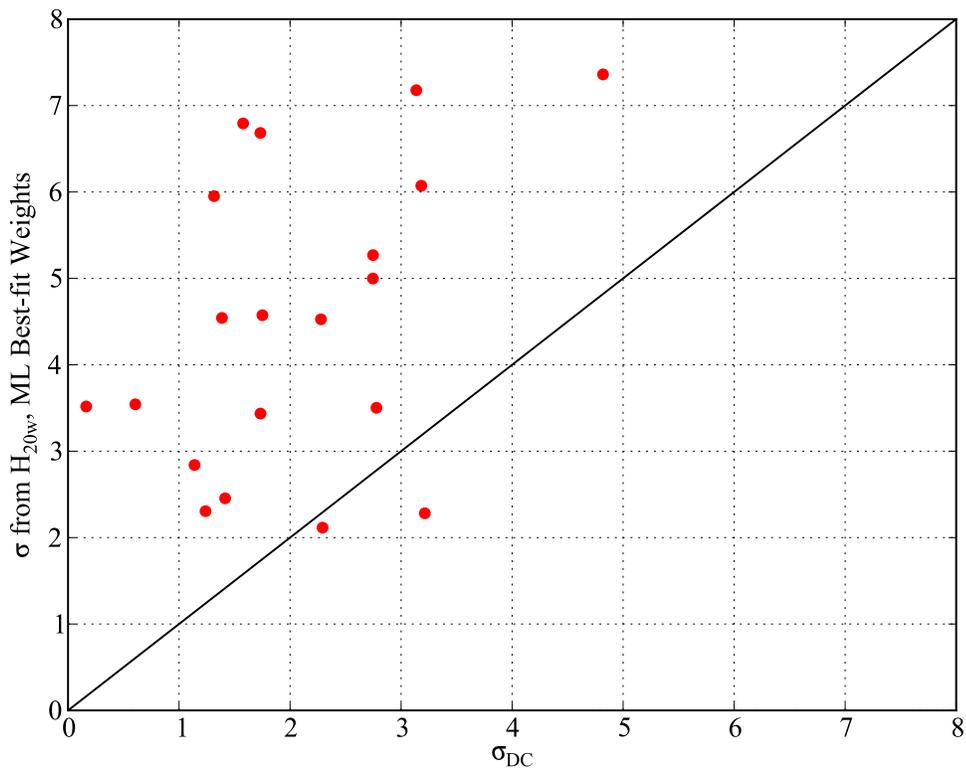}
\end{minipage}
\begingroup\renewcommand{\baselinestretch}{1.0}
\caption{A comparison of the significance of pulsed detection versus unpulsed detection.  The pulsed detection significance was calculated using the ML best-fit spectrum with the weighted $H$ test, while the unpulsed significance was derived from a likelihood ratio test as described in the text.  For nearly all sources, the pulsations are more strongly detected than the DC emission.}
\renewcommand{\baselinestretch}{1.5}\endgroup
\label{ch5_plot15}
\end{figure}

We compare the unpulsed and pulsed significances in Figure \ref{ch5_plot15}.  The vast majority of ensemble members are detected more significantly through pulsations than through unpulsed emission.  The measured significance for pulsed detections is a factor of $2.2$ greater than for DC detection.  If we assume that the detection flux threshold is linear in $\sigma_{DC}$ as it is in the pulsed significance, this means we require sources to be about twice as bright on average to detection them through DC emission rather than pulsed emission\footnote{This result, of course, depends strongly on the light curve morphology.}.  It is amusing that this is about the same factor we observe when comparing weighted and unweighted statistics.

\begin{figure}
\begin{minipage}{6in}
\includegraphics[width=6in]{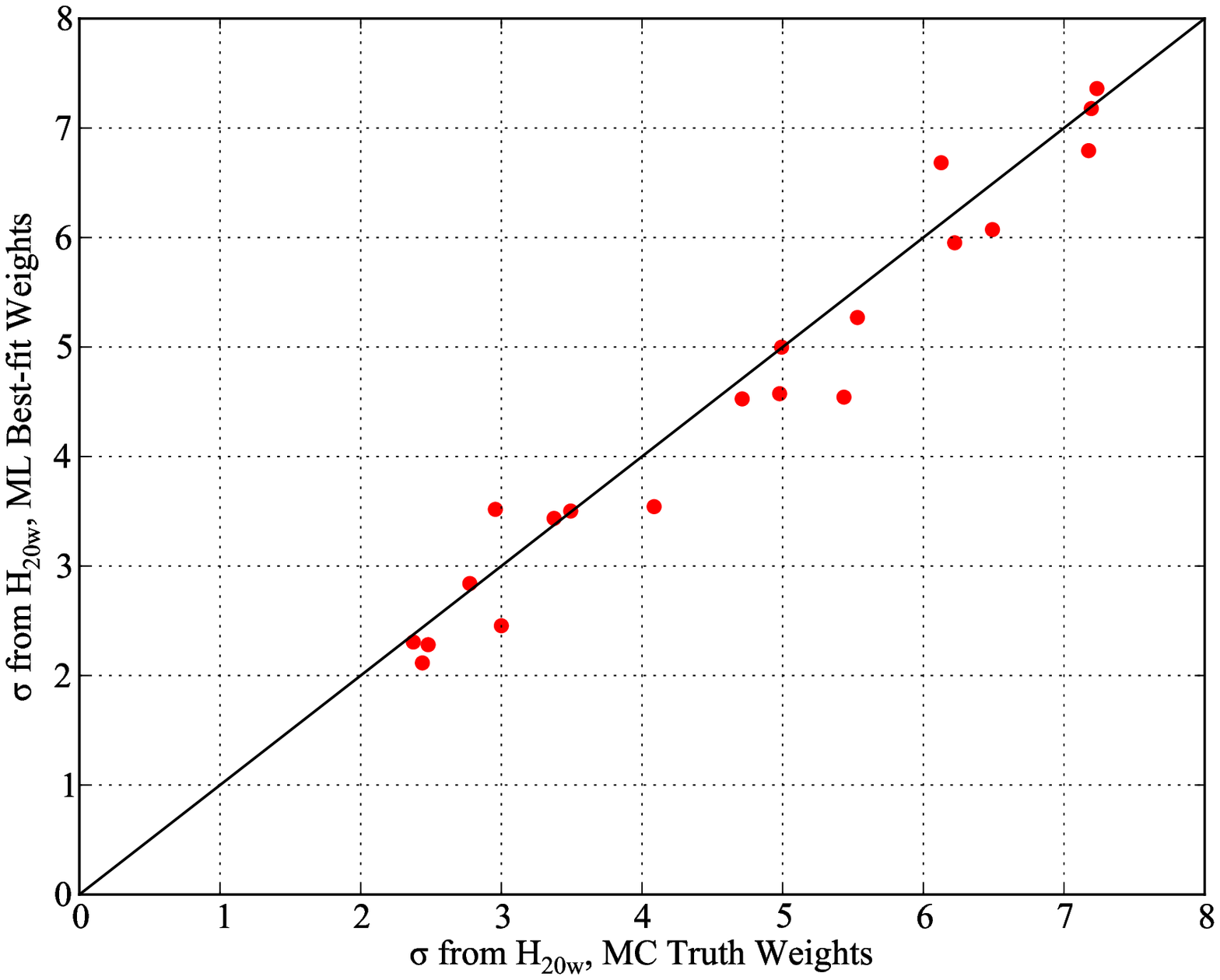}
\end{minipage}
\begingroup\renewcommand{\baselinestretch}{1.0}
\caption{As Figure \ref{ch5_plot14} but with weights derived from a ML fit with the pulsar spectrum fixed to a power law with $\Gamma=2.0$ and only the flux allowed to vary. }
\renewcommand{\baselinestretch}{1.5}\endgroup
\label{ch5_plot16}
\end{figure}

To determine how fitting the spectra with only one degree of freedom affects the performance of the weighted statistics, we compare in Figure \ref{ch5_plot16} the significance computed from the ``ideal'' weights and from the weights determined from the best-fit spectrum (recalling that the photon index and cutoff energy are fixed).  Interestingly, the ``measured'' weights deliver results comparable to those calculated in \S\ref{ch5:uncertain_spectrum}, indicating an insensitivity to the overall spectral shape.

\section{Calculating Pulsed Sensitivity}
One of the primary observables in constraining pulsar emission mechanisms is the overlap between radio-loud and $\gamma$-ray bright pulsars.  An important extension of this is placing the strongest constraints possible on the $\gamma$-ray flux from radio-loud pulsars.  In the material above, we saw that weighted statistics outperform both unweighted versions of the same statistics \emph{and} unpulsed significance tests.  Thus, pulsed upper limits are a natural candidate for constraining the $\gamma$-ray emission.

Although much work has been done on extracting analytic pulsed upper limits (see \cite{dejager_3} for an enlightening discussion), the weighted method is not particularly amenable to this approach.  Each source comes with its own set of probability weights that depends strongly on the pulsar spectrum and its position on the sky.  Thus, while we can extract analytic upper limits once we know the weight distribution, by calculating the weights we have already done most of the work!  The machinery we developed above for determining flux detection thresholds is precisely what is needed for an upper limit/sensitivity.

To determine the sensitivity of \fermi to pulsed detections, we essentially repeat the procedure outlined in the previous section, viz.
\begin{enumerate}
\item Simulate an ensemble of point sources with some spectral shape.
\item Add phase from, e.g., a single-peaked or a double-peaked light curve.
\item Select a significance level for detection, e.g., the $4\sigma$ criterion adopted in our analyses.
\item Determine the fraction of the ensemble that must exceed the detection significance to claim a detectable flux.  (We used $68\%$.)
\item Calculate the flux detection threshold as outlined above.
\end{enumerate}
This flux detection threshold is the desired sensitivity, i.e., the flux level at which we can typically ($68\%$ of the time) detect the source with significance exceeding our threshold ($4\sigma$.)  Sensitivity maps giving the pulsed flux detection threshold as a function of position on the sky can be generated by repeating this exercise over a grid of positions.

\section{Summary}

We motivated the use of unbinned, Fourier-based statistics for the detection of pulsations and we outlined some of their properties.  We showed that they could be modified to incorporate probability weights while retaining their calibration.  Appealing to gains made in using likelihood---i.e., knowledge of the IRF---we proposed that likelihood-derived probability weights would offer improvement over statistics formulated with counts alone.  We defined the time-averaged probability that a given photon comes from a particular source in terms of the ML spectral model for the source and its background.  We found that the weighted versions of the $Z^2_m$ and $H$ test had very little dependence on the particular data selection used in contradistinction to the unweighted versions (Figure \ref{ch5_plot8}).  By evaluating test statistics for ensembles of simulated pulsars, we estimated the detection flux threshold for a given statistical test and showed that the flux threshold for weighted tests was a factor of $1.5-2$ lower than the unweighted test (Figures \ref{ch5_plot4} and \ref{ch5_plot12}).  Finally, we demonstrated that these improvements depended very little on the fraction of pulsed emission (Figure \ref{ch5_plot13}) or on exact knowledge of the spectrum (Figures \ref{ch5_plot14} and \ref{ch5_plot16}).  Additionally, we found that pulsed sensitivity is about twice that of the unpulsed sensitivity (Figure \ref{ch5_plot15}) for a fully-pulsed source.

Taken together, these results represent a significant gain in the search for periodic signals from pulsars and suggest the adoption of weighted statistics---in particular, $H_{20w}$---as an omnibus test for pulsation.

%
%
\nocite{*}   
\bibliographystyle{plain}
\bibliography{uwthesis}

\begin{thebibliography}{10}

\bibitem{cta1}
A.~A. {Abdo et al.}
\newblock {The Fermi Gamma-Ray Space Telescope Discovers the Pulsar in the
  Young Galactic Supernova Remnant CTA 1}.
\newblock {\em Science}, 322:1218--, November 2008.

\bibitem{msp_pop}
A.~A. {Abdo et al.}
\newblock {A Population of Gamma-Ray Millisecond Pulsars Seen with the Fermi
  Large Area Telescope}.
\newblock {\em Science}, 325:848 (MSP Population), August 2009.

\bibitem{blind_search_16}
A.~A. {Abdo}~et al.
\newblock {Detection of 16 Gamma-Ray Pulsars Through Blind Frequency Searches
  Using the Fermi LAT}.
\newblock {\em Science}, 325:840--, August 2009.

\bibitem{fermi_limb}
A.~A. {Abdo}~et al.
\newblock {Fermi large area telescope observations of the cosmic-ray induced
  {$\gamma$}-ray emission of the Earth's atmosphere}.
\newblock {\em \prd}, 80(12):122004--+, December 2009.

\bibitem{vela1}
A.~A. {Abdo}~et al.
\newblock {Fermi Large Area Telescope Observations of the Vela Pulsar}.
\newblock {\em \apj}, 696:1084 (Vela), May 2009.

\bibitem{local_diffuse}
A.~A. {Abdo et al.}
\newblock {Fermi LAT Observation of Diffuse Gamma Rays Produced Through
  Interactions Between Local Interstellar Matter and High-energy Cosmic Rays}.
\newblock {\em \apj}, 703:1249--1256, October 2009.

\bibitem{lsi61303}
A.~A. {Abdo}~et al.
\newblock {Fermi LAT Observations of LS I $+61{\deg}303$: First Detection of an
  Orbital Modulation in GeV Gamma Rays}.
\newblock {\em \apjl}, 701:L123 (LS I +61$^{\circ}$ 303), August 2009.

\bibitem{j2021}
A.~A. {Abdo et al.}
\newblock {Pulsed Gamma-rays from PSR J2021+3651 with the Fermi Large Area
  Telescope}.
\newblock {\em \apj}, 700:1059 (PSR J2021+3651), August 2009.

\bibitem{j0030}
A.~A. {Abdo}~et al.
\newblock {Pulsed Gamma Rays from the Millisecond Pulsar J0030+0451 with the
  Fermi Large Area Telescope}.
\newblock {\em \apj}, 699:1171 (PSR J0030+0451), July 2009.

\bibitem{onorbit}
A.~A. {Abdo et al.}
\newblock {The on-orbit calibration of the Fermi Large Area Telescope}.
\newblock {\em Astroparticle Physics}, 32:193--219, October 2009.

\bibitem{starburst}
A.~A. {Abdo et al.}
\newblock {Detection of Gamma-Ray Emission from the Starburst Galaxies M82 and
  NGC 253 with the Large Area Telescope on Fermi}.
\newblock {\em \apjl}, 709:L152--L157, February 2010.

\bibitem{1fgl}
A.~A. {Abdo}~et al.
\newblock {Fermi Large Area Telescope First Source Catalog}.
\newblock {\em \apjs}, 188:405--436, June 2010.

\bibitem{j1836p5925}
A.~A. {Abdo}~et al.
\newblock {Fermi Large Area Telescope Observations of PSR J1836+5925}.
\newblock {\em \apj}, 712:1209--1218, April 2010.

\bibitem{q2_diffuse}
A.~A. {Abdo et al.}
\newblock {Fermi Observations of Cassiopeia and Cepheus: Diffuse Gamma-ray
  Emission in the Outer Galaxy}.
\newblock {\em \apj}, 710:133--149, February 2010.

\bibitem{isotropic_spec}
A.~A. {Abdo et al.}
\newblock {Spectrum of the Isotropic Diffuse Gamma-Ray Emission Derived from
  First-Year Fermi Large Area Telescope Data}.
\newblock {\em Physical Review Letters}, 104(10):101101--+, March 2010.

\bibitem{1lac}
A.~A. {Abdo et al.}
\newblock {The First Catalog of Active Galactic Nuclei Detected by the Fermi
  Large Area Telescope}.
\newblock {\em \apj}, 715:429--457, May 2010.

\bibitem{1pcat}
A.~A. {Abdo et al.}
\newblock {The First Fermi Large Area Telescope Catalog of Gamma-ray Pulsars}.
\newblock {\em \apjs}, 187:460--494, April 2010.

\bibitem{diffuse1}
F.~A. {Aharonian} and A.~M. {Atoyan}.
\newblock {Broad-band diffuse gamma ray emission of the galactic disk}.
\newblock {\em \aap}, 362:937--952, October 2000.

\bibitem{timing_noise}
Z.~{Arzoumanian}, D.~J. {Nice}, J.~H. {Taylor}, and S.~E. {Thorsett}.
\newblock {Timing behavior of 96 radio pulsars}.
\newblock {\em \apj}, 422:671--680, February 1994.

\bibitem{blind_search_technique}
W.~B. {Atwood}, M.~{Ziegler}, R.~P. {Johnson}, and B.~M. {Baughman}.
\newblock {A Time-differencing Technique for Detecting Radio-quiet Gamma-Ray
  Pulsars}.
\newblock {\em \apjl}, 652:L49--L52, November 2006.

\bibitem{tkr}
W.~B. {Atwood et al.}
\newblock {Design and initial tests of the Tracker-converter of the Gamma-ray
  Large Area Space Telescope}.
\newblock {\em Astroparticle Physics}, 28:422--434, December 2007.

\bibitem{lat_instrument}
W.~B. {Atwood}~et al.
\newblock {The Large Area Telescope on the Fermi Gamma-Ray Space Telescope
  Mission}.
\newblock {\em \apj}, 697:1071--1102, June 2009.

\bibitem{annular_gap}
{X.-N.} {Bai} and A.~{Spitkovsky}.
\newblock {Modeling of Gamma-Ray Pulsar Light Curves with Force-Free Magnetic
  Field}.
\newblock {\em ArXiv e-prints}, October 2009.

\bibitem{3c_cat}
A.~S. {Bennett}.
\newblock {The revised 3C catalogue of radio sources.}
\newblock {\em \memras}, 68:163--+, 1962.

\bibitem{beran1}
R.~J. Beran.
\newblock Asymptotic theory of a class of tests for uniformity of a circular
  distribution.
\newblock {\em The Annals of Mathematical Statistics}, 40(4):1196--1206, 1969.

\bibitem{geminga_gray_pulse}
D.~L. {Bertsch et al.}
\newblock {Pulsed high-energy gamma-radiation from Geminga (1E0630 + 178)}.
\newblock {\em \nat}, 357:306--+, May 1992.

\bibitem{cosb}
G.~F. {Bignami et al.}
\newblock {The COS-B experiment for gamma-ray astronomy}.
\newblock {\em Space Science Instrumentation}, 1:245--268, August 1975.

\bibitem{bandt}
J.~{Binney} and S.~{Tremaine}.
\newblock {\em {Galactic dynamics}}.
\newblock 1987.

\bibitem{hubble_m87}
J.~A. {Biretta}, W.~B. {Sparks}, and F.~{Macchetto}.
\newblock {Hubble Space Telescope Observations of Superluminal Motion in the
  M87 Jet}.
\newblock {\em \apj}, 520:621--626, August 1999.

\bibitem{sunmoon}
M.~Brigida.
\newblock Moon and quiet sun detection with fermi-lat observatory.
\newblock {\em Nuclear Instruments and Methods in Physics Research Section A:
  Accelerators, Spectrometers, Detectors and Associated Equipment}, In Press,
  Uncorrected Proof:--, 2010.

\bibitem{clemson}
L.~E. {Brown}, D.~D. {Clayton}, and D.~H. {Hartmann}.
\newblock {Gamma ray pulsar analysis from photon probability maps}.
\newblock In {C.~R.~Shrader, N.~Gehrels, \& B.~Dennis}, editor, {\em NASA
  Conference Publication}, volume 3137 of {\em NASA Conference Publication},
  pages 267--272, February 1992.

\bibitem{cos-b_pulse_search}
R.~{Buccheri et al.}
\newblock {Search for pulsed gamma-ray emission from radio pulsars in the COS-B
  data}.
\newblock {\em \aap}, 128:245--251, November 1983.

\bibitem{two_pulsar_paper}
F.~{Camilo et al.}
\newblock {Radio Detection of LAT PSRs J1741-2054 and J2032+4127: No Longer
  Just Gamma-ray Pulsars}.
\newblock {\em \apj}, 705:1--13, November 2009.

\bibitem{chr}
K.~S. {Cheng}, C.~{Ho}, and M.~{Ruderman}.
\newblock {Energetic radiation from rapidly spinning pulsars. I - Outer
  magnetosphere gaps. II - VELA and Crab}.
\newblock {\em \apj}, 300:500--539, January 1986.

\bibitem{chernoff_miracle}
Herman Chernoff.
\newblock On the distribution of the likelihood ratio.
\newblock {\em The Annals of Mathematical Statistics}, 25(3):573--578, 1954.

\bibitem{cr94}
J.~{Chiang} and R.~W. {Romani}.
\newblock {An outer gap model of high-energy emission from rotation-powered
  pulsars}.
\newblock {\em \apj}, 436:754--761, December 1994.

\bibitem{fft}
James~W. Cooley and John~W. Tukey.
\newblock An algorithm for the machine calculation of complex fourier series.
\newblock {\em Mathematics of Computation}, 19(90):297--301, 1965.

\bibitem{pc82}
J.~K. {Daugherty} and A.~K. {Harding}.
\newblock {Electromagnetic cascades in pulsars}.
\newblock {\em \apj}, 252:337--347, January 1982.

\bibitem{pc96}
J.~K. {Daugherty} and A.~K. {Harding}.
\newblock {Gamma-Ray Pulsars: Emission from Extended Polar CAP Cascades}.
\newblock {\em \apj}, 458:278--+, February 1996.

\bibitem{dejager_3}
O.~C. {de Jager}.
\newblock {On periodicity tests and flux limit calculations for gamma-ray
  pulsars}.
\newblock {\em \apj}, 436:239--248, November 1994.

\bibitem{dejager_2}
O.~C. {de Jager} and I.~{B{\"u}sching}.
\newblock {The H-test probability distribution revisited: Improved
  sensitivity}.
\newblock {\em ArXiv e-prints}, May 2010.

\bibitem{dejager_1}
O.~C. {de Jager}, B.~C. {Raubenheimer}, and J.~W.~H. {Swanepoel}.
\newblock {A poweful test for weak periodic signals with unknown light curve
  shape in sparse data}.
\newblock {\em \aap}, 221:180--190, August 1989.

\bibitem{deutsch}
A.~J. {Deutsch}.
\newblock {The electromagnetic field of an idealized star in rigid rotation in
  vacuo}.
\newblock {\em Annales d'Astrophysique}, 18:1--+, January 1955.

\bibitem{guppi}
R.~{DuPlain}, S.~{Ransom}, P.~{Demorest}, P.~{Brandt}, J.~{Ford}, and A.~L.
  {Shelton}.
\newblock {Launching GUPPI: the Green Bank Ultimate Pulsar Processing
  Instrument}.
\newblock In {\em Society of Photo-Optical Instrumentation Engineers (SPIE)
  Conference Series}, volume 7019 of {\em Society of Photo-Optical
  Instrumentation Engineers (SPIE) Conference Series}, August 2008.

\bibitem{dyks_rudak}
J.~{Dyks} and B.~{Rudak}.
\newblock {Two-Pole Caustic Model for High-Energy Light Curves of Pulsars}.
\newblock {\em \apj}, 598:1201--1206, December 2003.

\bibitem{magic}
C.~Baixeras et~al.
\newblock Commissioning and first tests of the magic telescope.
\newblock {\em Nuclear Instruments and Methods in Physics Research Section A:
  Accelerators, Spectrometers, Detectors and Associated Equipment},
  518(1-2):188 -- 192, 2004.
\newblock Frontier Detectors for Frontier Physics: Proceedin.

\bibitem{veritas}
T.~C.~Weekes et~al.
\newblock Veritas: the very energetic radiation imaging telescope array system.
\newblock {\em Astroparticle Physics}, 17(2):221 -- 243, 2002.

\bibitem{cygx3}
{Fermi LAT Collaboration}.
\newblock {Modulated High-Energy Gamma-Ray Emission from the Microquasar Cygnus
  X-3}.
\newblock {\em Science}, 326:1512--, December 2009.

\bibitem{sas2}
C.~E. {Fichtel et al.}
\newblock {High-energy gamma-ray results from the second small astronomy
  satellite}.
\newblock {\em \apj}, 198:163--182, May 1975.

\bibitem{bwp}
A.~S. {Fruchter}, D.~R. {Stinebring}, and J.~H. {Taylor}.
\newblock {A millisecond pulsar in an eclipsing binary}.
\newblock In {H.~{\"O}gelman \& E.~P.~J.~van den Heuvel}, editor, {\em Timing
  Neutron Stars}, pages 163--+, 1989.

\bibitem{gaisser}
T.~K. {Gaisser}.
\newblock {\em {Cosmic Rays and Particle Physics}}.
\newblock January 1991.

\bibitem{gj}
P.~{Goldreich} and W.~H. {Julian}.
\newblock {Pulsar Electrodynamics}.
\newblock {\em \apj}, 157:869--+, August 1969.

\bibitem{healpix}
K.~M. {G{\'o}rski et al.}
\newblock {HEALPix: A Framework for High-Resolution Discretization and Fast
  Analysis of Data Distributed on the Sphere}.
\newblock {\em \apj}, 622:759--771, April 2005.

\bibitem{halpern_blazar}
J.~P. {Halpern}, M.~{Eracleous}, R.~{Mukherjee}, and E.~V. {Gotthelf}.
\newblock {3EG J2016+3657: Confirming an EGRET Blazar behind the Galactic
  Plane}.
\newblock {\em \apj}, 551:1016--1023, April 2001.

\bibitem{geminga_xray_pulse}
J.~P. {Halpern} and S.~S. {Holt}.
\newblock {Discovery of soft X-ray pulsations from the gamma-ray source
  Geminga}.
\newblock {\em \nat}, 357:222--224, May 1992.

\bibitem{harding_slot}
A.~K. {Harding}, J.~V. {Stern}, J.~{Dyks}, and M.~{Frackowiak}.
\newblock {High-Altitude Emission from Pulsar Slot Gaps: The Crab Pulsar}.
\newblock {\em \apj}, 680:1378--1393, June 2008.

\bibitem{hewish68}
A.~{Hewish}, S.~J. {Bell}, J.~D.~H. {Pilkington}, P.~F. {Scott}, and R.~A.
  {Collins}.
\newblock {Observation of a Rapidly Pulsating Radio Source}.
\newblock {\em \nat}, 217:709--713, February 1968.

\bibitem{hess}
J.~A. Hinton.
\newblock The status of the hess project.
\newblock {\em New Astronomy Reviews}, 48(5-6):331 -- 337, 2004.
\newblock 2nd VERITAS Symposium on the Astrophysics of Extragalactic Sources.

\bibitem{jackson}
J.~D. {Jackson}.
\newblock {\em {Classical Electrodynamics, 3rd Edition}}.
\newblock July 1998.

\bibitem{calcal}
W.N. Johnson, J.E. Grove, B.F. Phlips, J.~Ampe, S.~Singh, and E.~Ponslet.
\newblock The construction and performance of the csi hodoscopic calorimeter
  for the glast beam test engineering module.
\newblock {\em Nuclear Science, IEEE Transactions on}, 48(4):1182 --1189, aug
  2001.

\bibitem{tune}
T.~{Kamae}, N.~{Karlsson}, T.~{Mizuno}, T.~{Abe}, and T.~{Koi}.
\newblock {Parameterization of {$\gamma$}, e$^{+/-}$, and Neutrino Spectra
  Produced by p-p Interaction in Astronomical Environments}.
\newblock {\em \apj}, 647:692--708, August 2006.

\bibitem{egret_intro}
G.~{Kanbach et al.}
\newblock {The project EGRET (Energetic Gamma-Ray Experiment Telescope) on
  NASA's Gamma-Ray Observatory (GRO)}.
\newblock {\em \ssr}, 49:69--84, 1988.

\bibitem{oso3}
W.~L. {Kraushaar}, G.~W. {Clark}, G.~P. {Garmire}, R.~{Borken}, P.~{Higbie},
  V.~{Leong}, and T.~{Thorsos}.
\newblock {High-Energy Cosmic Gamma-Ray Observations from the OSO-3 Satellite}.
\newblock {\em \apj}, 177:341--+, November 1972.

\bibitem{starburst2}
M.~D. {Lehnert} and T.~M. {Heckman}.
\newblock {The Nature of Starburst Galaxies}.
\newblock {\em \apj}, 472:546--+, December 1996.

\bibitem{longair}
M.~S. {Longair}.
\newblock {\em {High energy astrophysics. Vol.1: Particles, photons and their
  detection}}.
\newblock March 1992.

\bibitem{mattox}
J.~R. {Mattox et al.}
\newblock {The Likelihood Analysis of EGRET Data}.
\newblock {\em \apj}, 461:396--+, April 1996.

\bibitem{maura}
M.~A. {McLaughlin} and J.~M. {Cordes}.
\newblock {Gamma-Ray Pulsars: Modeling and Searches}.
\newblock {\em ArXiv Astrophysics e-prints}, October 2003.

\bibitem{acd}
A.~A. {Moiseev}, R.~C. {Hartman}, J.~F. {Ormes}, D.~J. {Thompson}, M.~J.
  {Amato}, T.~E. {Johnson}, K.~N. {Segal}, and D.~A. {Sheppard}.
\newblock {The anti-coincidence detector for the GLAST large area telescope}.
\newblock {\em Astroparticle Physics}, 27:339--358, June 2007.

\bibitem{egret_pulse_search}
H.~I. {Nel et al.}
\newblock {EGRET High-Energy Gamma-Ray Pulsar Studies. III. A Survey}.
\newblock {\em \apj}, 465:898--+, July 1996.

\bibitem{nocedal99}
Jorge Nocedal and Stephen~J. Wright.
\newblock {\em Numerical Optimization}.
\newblock Springer, August 2000.

\bibitem{hans}
M.~E. {{\"O}zel} and H.~A. {Mayer-Ha{\ss}elwander}.
\newblock {A method to improve the visibility of time-variable gamma-ray
  sourcesin structured background.}
\newblock {\em \aap}, 125:130--135, August 1983.

\bibitem{ramanamurthy}
P.~V. {Ramanamurthy}, C.~E. {Fichtel}, D.~A. {Kniffen}, P.~{Sreekumar}, and
  D.~J. {Thompson}.
\newblock {Possible Evidence for Pulsed Emission of High-Energy Gamma Rays by
  PSR B0656+14}.
\newblock {\em \apj}, 458:755--+, February 1996.

\bibitem{r96}
R.~W. {Romani}.
\newblock {Gamma-Ray Pulsars: Radiation Processes in the Outer Magnetosphere}.
\newblock {\em \apj}, 470:469--+, October 1996.

\bibitem{ry95}
R.~W. {Romani} and {I.-A.} {Yadigaroglu}.
\newblock {Gamma-ray pulsars: Emission zones and viewing geometries}.
\newblock {\em \apj}, 438:314--321, January 1995.

\bibitem{blind_search_8}
P.~M. {Saz Parkinson et al.}
\newblock {Eight gamma-ray pulsars discovered in blind frequency searches of
  Fermi LAT data}.
\newblock {\em ArXiv e-prints}, June 2010.

\bibitem{be_bound}
I.~G. Shevtsova.
\newblock Sharpening of the upper bound of the absolute constant in the
  berry--esseen inequality.
\newblock {\em Theory of Probability and its Applications}, 51(3):549--553,
  2007.

\bibitem{blazar_jets}
M.~{Sikora}, M.~C. {Begelman}, and M.~J. {Rees}.
\newblock {Comptonization of diffuse ambient radiation by a relativistic jet:
  The source of gamma rays from blazars?}
\newblock {\em \apj}, 421:153--162, January 1994.

\bibitem{smith_timing}
D.~A. {Smith et al.}
\newblock {Pulsar timing for the Fermi gamma-ray space telescope}.
\newblock {\em \aap}, 492:923--931, December 2008.

\bibitem{spitkovsky06}
A.~{Spitkovsky}.
\newblock {Time-dependent Force-free Pulsar Magnetospheres: Axisymmetric and
  Oblique Rotators}.
\newblock {\em \apjl}, 648:L51--L54, September 2006.

\bibitem{agile}
M.~{Tavani et al.}
\newblock {The AGILE Mission}.
\newblock {\em \aap}, 502:995--1013, August 2009.

\bibitem{sas2_anticenter}
D.~J. {Thompson}, C.~E. {Fichtel}, R.~C. {Hartman}, D.~A. {Kniffen}, and R.~C.
  {Lamb}.
\newblock {Final SAS-2 gamma-ray results on sources in the galactic anticenter
  region}.
\newblock {\em \apj}, 213:252--262, April 1977.

\bibitem{sas2_vela}
D.~J. {Thompson}, C.~E. {Fichtel}, D.~A. {Kniffen}, and H.~B. {Ogelman}.
\newblock {SAS-2 high-energy gamma-ray observations of the VELA pulsar}.
\newblock {\em \apjl}, 200:L79--L82, September 1975.

\bibitem{egret_calib}
D.~J. {Thompson et al.}
\newblock {Calibration of the Energetic Gamma-Ray Experiment Telescope (EGRET)
  for the Compton Gamma-Ray Observatory}.
\newblock {\em \apjs}, 86:629--656, June 1993.

\bibitem{wilks}
S.~S. Wilks.
\newblock The large-sample distribution of the likelihood ratio for testing
  composite hypotheses.
\newblock {\em The Annals of Mathematical Statistics}, 9(1):60--62, 1938.

\bibitem{integral}
C.~{Winkler et al.}
\newblock {The INTEGRAL mission}.
\newblock {\em \aap}, 411:L1--L6, November 2003.

\bibitem{ccsn_review}
S.~{Woosley} and T.~{Janka}.
\newblock {The physics of core-collapse supernovae}.
\newblock {\em Nature Physics}, 1:147--154, December 2005.

\bibitem{pdg}
W.-M. et~al. {Yao}.
\newblock {Review of Particle Physics}.
\newblock {\em {Journal of Physics G}}, 33:1+, 2006.

\bibitem{xray_msp}
L.~{Zhang} and K.~S. {Cheng}.
\newblock {X-ray and gamma-ray emission from millisecond pulsars}.
\newblock {\em \aap}, 398:639--646, February 2003.

\end{thebibliography}
%
%
\appendix
\raggedbottom\sloppy
 
 
\chapter{The Asymptotic Null Distribution of $H_m$}
\label{appB}

Recall the $H_m$ statistic is defined as
\begin{equation}
H_m = \mathrm{max}[Z^2_i - c(i-1)]\equiv\mathrm{max}[X_i],\, 1\leq i \leq m.
\end{equation}
In its original formulation, $m=20$ and $c=4$.  The addition of the constant term $c(i-1)$ suppresses contributions from the higher harmonics in the null case.  (If $c = 0$, the $m$th value is always the maximum.)

From the above definition, $H_m$ is an (extreme) order statistic of the $X_i$, a collection of $m$ dependent, non-identically distributed RVs.  While the distribution of such a statistic is often difficult or impossible to obtain in a useful form, in this case, the relatively simple mutual dependence of the variables (they satisfy the Markov property), together with the form of the conditional distributions (exponential), admits a concise, closed form solution for the asymptotic null distribution of $H_m$.

\section{The Joint Probability Density Function of $\vec{X}$}

Let $X_i \equiv Z^2_i - c(i-1)$.  (For convenience, we interpret $Z^2_0$ as $0$, giving $X_0 = c$.)  If we assume the asymptotic distribution for $Z^2_m$, then $X_{i+1}$ can be obtained from $X_{i}$ by adding a $\chi^2_2$ distributed variable and subtracting $c$.  That is,
\begin{equation}
\label{conditional}
f_{X_{i+1}|X_i}(x_{i+1} | x_i) = \chi^2_2(x_{i+1} - x_{i} + c).
\end{equation}

Now, let $\vec{X}_m$ be a random vector in $\mathbb{R}^m$, such that $H_m$ is the maximum element of $\Vec{X}_m$.  We construct the joint pdf for $\vec{X}$ as a product of conditional distributions:
\begin{align*}
f_{\vec{X}_m}(\vec{x}_m) = &f_{X_m|\vec{X}_{m-1}}(x_m |\vec{x}_{m-1})\times f_{X_{m-1}|\vec{X}_{m-2}}(x_{m-1} | \vec{x}_{m-2})\times\cdots\
\\ \times &f_{X_2|X_1}(x_2 | x_1)\times f_{X_1}(x_1).
\end{align*}
From Eq. \ref{conditional}, this reduces to
\begin{equation}
f_{\vec{X}_m}(\vec{x}_m) = \prod_{i=1}^m \chi^2_2(x_i - x_{i-1} + c),
\end{equation}
demonstrating the Markov property.  From this form, it is easy to show that the marginal distribution for $X_i$ is $\chi^2_{2i}$ by reducing the integral over the remaining $m-1$ terms to a series of convolutions.  The normalization may also be verified in this way. 

Inserting the explicit form for $\chi^2_2(x)=\frac{1}{2}\exp(-\frac{x}{2})\,\theta(x)$, where $\theta(x)$ is the Heaviside step function restricing support to positive arguments, yields a significant simplification:
\begin{equation}
\label{explicit}
f_{\vec{X}_m}(\vec{x}_m) = \frac{\alpha^{m-1}}{2}\times\left[\prod_{i=1}^m \theta(x_i - x_{i-1} + c)\right]\exp\left(-\frac{x_m}{2}\right),
\end{equation}
where we have defined for convenience $\alpha\equiv \frac{1}{2}\exp\left(-\frac{c}{2}\right)$.

\section{The Cumulative Distribution Function of $H_m$}

$H_m$ is just the maximum element of the vector $\vec{X}_m$.  Thus, the probability to observe a value less than or equal to $h_m$ is simply the integral of the $f_{\vec{X}}(\vec{x})$ over all values of $\vec{x}$ with all elements of $\vec{x}$ less than or equal to $h_m$.  That is,
\begin{equation*}
F_{H_m}(h_m) = \prod_{i=1}^{m}\left(\int_{-\infty}^{h_m} dx_i\right) f_{\vec{X}}(\vec{X}).
\end{equation*}
Inserting the form obtained in Eq. \ref{explicit} yields

\begin{equation}
\label{f1}
F_{H_m}(h_m) = \frac{\alpha^{m-1}}{2} \prod_{i=1}^{m}\left[\int_{-\infty}^{h_m} dx_i \theta\left(x_i - x_{i-1} + c\right)\right]\exp\left(-\frac{x_m}{2}\right).
\end{equation}

The two important features of the joint pdf---the simple relation between the dependent variables and the simple form of the conditional distributions---yields an integral expression for $F_{H_m}$ in which the integrand consists of an exponential in a single variable.  The main difficulty lies in determining the support of the integrand specified by the product of step functions.

\subsection{Reduction of $F_{H_m}(h_m)$}
We can develop the integral in Eq. \ref{f1} recursively.  First, we make a change of variables in the rightmost integral: $u_m \equiv x_m - x_{m-1} + c$.  This integral is then

$$\int_{0}^{h - x_{m-1} + c}du_m\, \exp\left(\frac{-u_m -x_{m-1}+c}{2}\right) = \alpha^{-1}\left[\exp\left(\frac{-x_{m-1}}{2}\right) - \exp\left(-\frac{h+c}{2}\right)\right].$$

The lefthand term, togther with the remaining integrals, is the cumulative distribution function for an $m-1$ harmonic H-test, $F_{H_{m-1}}$.  The righthand is the integral of unity with nontrivial limits of integration.  That is,
\begin{equation}
\label{f2}
F_{H_m}(h_m) = F_{H_{m-1}}(h_{m}) - \alpha^{m-1}\times\exp\left(-\frac{h_m}{2}\right) I_{m-1},
\end{equation}
where
$$I_n(h) = \prod_{i=1}^{n}\left[\int_{-\infty}^{h} dx_i\, \theta\left(x_i - x_{i-1} + c\right)\right].$$
Fully reducing $F_{H_m}$ yields a power series in $\alpha$:
\begin{equation}
\label{f3}
F_{H_m}(h_m) = 1 - \exp\left(-\frac{h_m}{2}\right)\times\sum_{n=0}^{m-1} \alpha^n I_{n}(h_m).
\end{equation}
Thus, the only remaining task is to evaluate $I_n$.

\subsection{Evaluation of $I_n$}

We begin with a change of variables to eliminate the step functions.  Let $u_i\equiv x_i -\sum_{j=1}^{i-1} (x_j - c)$.  Then
\begin{equation*}
I_n(h) = \prod_{i=1}^{n}\left[\int_{0}^{B_i} du_i \right].
\end{equation*}
Here, $B_i = h + (i-1)c - \sum_{j=1}^{i-1} x_j$, and we note that $B_n = B_{n-1} + c - u_{n-1}$.  We can evaluate the $n$ integrals recursively.  With each integration, we make the change of integration variable to $q_i \equiv B_{i} + (n-i+1)c - u_i$.  For instance, evaluating the rightmost integral, we have
\begin{align*}
I_n(h) &= \prod_{i=1}^{n-1}\left(\int_{0}^{B_i} du_i \right) B_n
\\	   &= \prod_{i=1}^{n-2}\left(\int_{0}^{B_i} du_i \right) \int_0^{B_{n-1}} du_{n-1}\, B_{n-1} + c - u_{n-1}
\\     &= \prod_{i=1}^{n-2}\left(\int_{0}^{B_i} du_i \right) \int_{2c}^{B_{n-1}+2c} dq_{n-1}\, (q_{n-1} - c)
\\     &= \left[\prod_{i=1}^{n-2}\left(\int_{0}^{B_i} du_i \right) \int_{2c}^{B_{n-1}+2c} dq_{n-1}\, q_{n-1}\right] - cI_{n-1}.
\end{align*}
This form is typical as one continues to integrate.  The change of variable always produces a monomial in the integration variable.  The upper boundary then produces a term in the integration variable of the next integral while the lower boundary produces a monomial of $c$.  This separation allows a recursive development for $I_n$, and combining the recursive terms yields
\begin{equation}
\label{ifinal}
I_n(h) = \frac{(h+nc)^n}{n!} - \sum_{j=1}^{n}I_{n-j-1}\ \frac{(jc)^{j}}{j!}.
\end{equation}
We note that $I_0(h)=1$ and $I_1(h)=h$.  The use of logarithms facilitates the numerical evaluation of this expressions for large $h$ and $n$.

\subsection{Monte Carlo Validation}
\begin{figure}
\begin{minipage}{6in}
\includegraphics[width=6in]{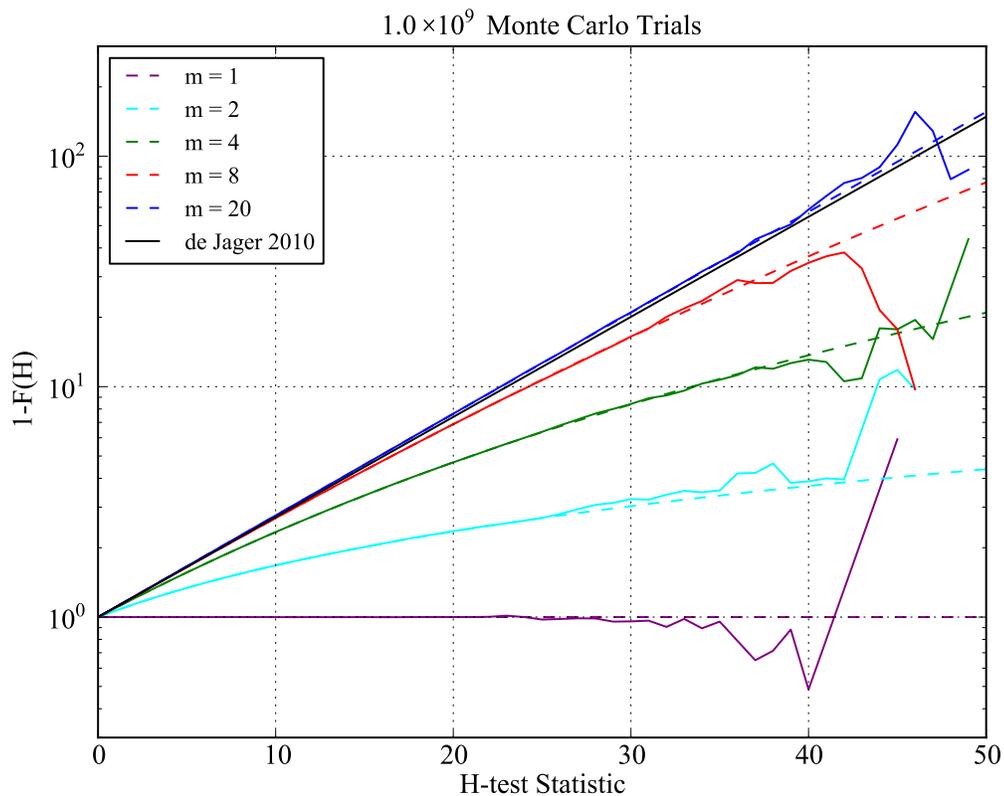}
\end{minipage}
\begingroup\renewcommand{\baselinestretch}{1.0}
\caption{We show the survival function ($1-F(H)$) of the asymptotic distribution for H and for a sample distribution ($N=10^{9}$) drawn from the null distribution by simulation.  To reduce the scale, we divide the survival function for a $\chi^2$ variable with two degrees of freedom, viz. $1-F(x)=\exp(-0.5x)$.  For each maximum harmonic, the sample distributions agree with the asymptotic calibration.}
\renewcommand{\baselinestretch}{1.5}\endgroup
\label{ch5_plot6}
\end{figure}

\begin{figure}
\begin{minipage}{6in}
\includegraphics[width=6in]{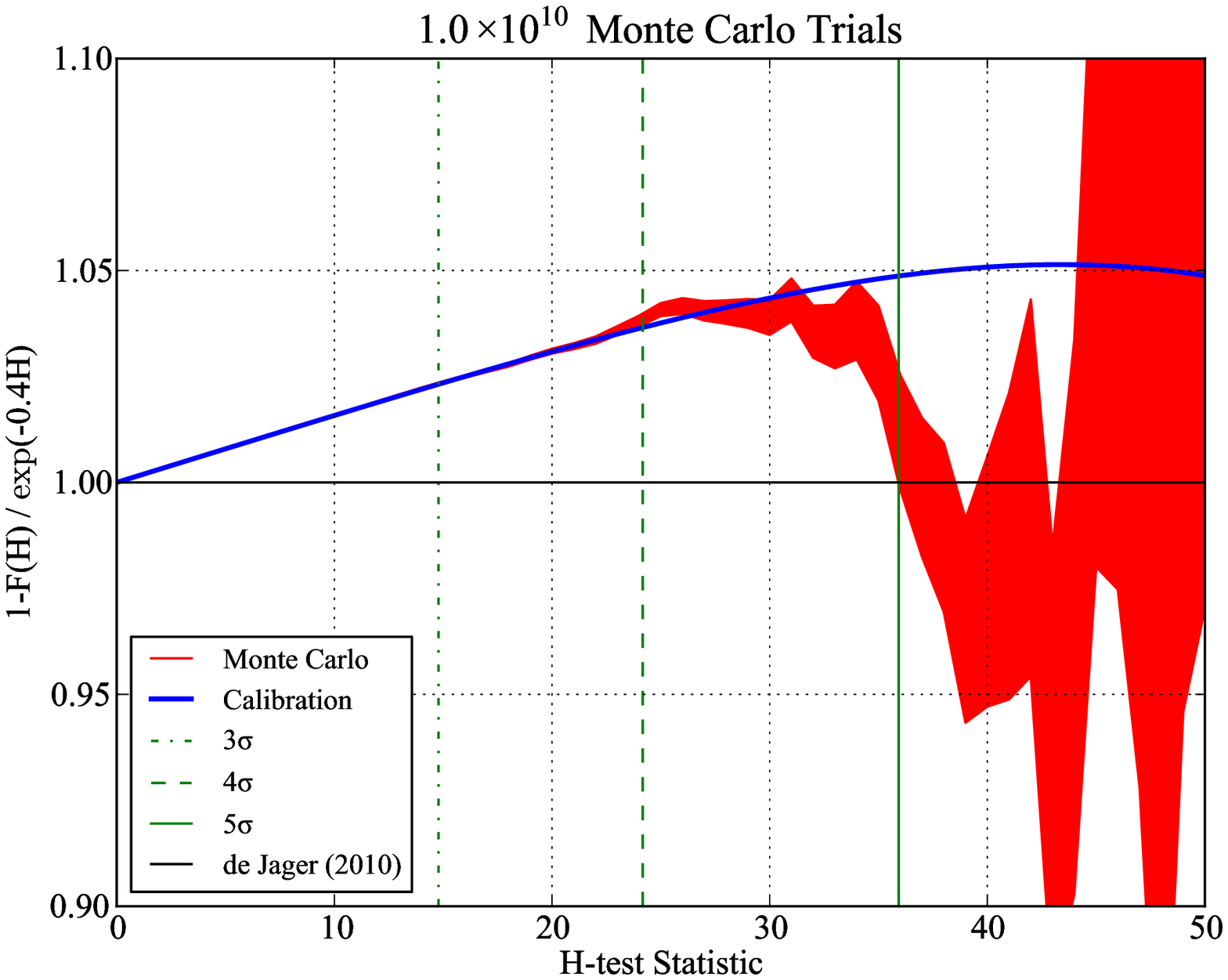}
\end{minipage}
\begingroup\renewcommand{\baselinestretch}{1.0}
\caption{We show the survival function ($1-F(H)$) of the asymptotic distribution for H and for a sample distribution ($N=10^{10}$) drawn from the null distribution by simulation.  To reduce the scale, we divide the survival function by the ``first-order'' approximation $1-F(H)\approx\exp(-0.4H)$\cite{dejager_2}.  This approximation gives excellent results ($<5\%$ deviation from the true distribution) to $H=50$ and above.  We have marked the values $H$ corresponding to (two-sided) significance levels of $3\sigma$, $4\sigma$, and $5\sigma$ for reference.}
\renewcommand{\baselinestretch}{1.5}\endgroup
\label{ch5_plot7}
\end{figure}

To check our results, we performed Monte Carlo simulations of the $H_{20}$ statistic in the asymptotic null case.  Specifically, for each realization of $H_{20}$, we drew $20$ realizations from a $\chi^2$ distribution with one degree of freedom and determined $H$ accordingly (with $c=4$.)  In Figure \ref{ch5_plot6}, we show the results of $10^9$ Monte Carlo trials for a variety of maximum harmonics, viz. $H_1$, $H_2$, $H_4$, $H_8$, and $H_{20}$.  The results are in good agreement with the asymptotic distribution derived here.  Next, we focus on $H_{20}$ with $10^{10}$ trials, shown in Figure \ref{ch5_plot7}.  These results are also in good agreement.

\subsection{Behavior at Large H}
\begin{figure}
\begin{minipage}{6in}
\includegraphics[width=6in]{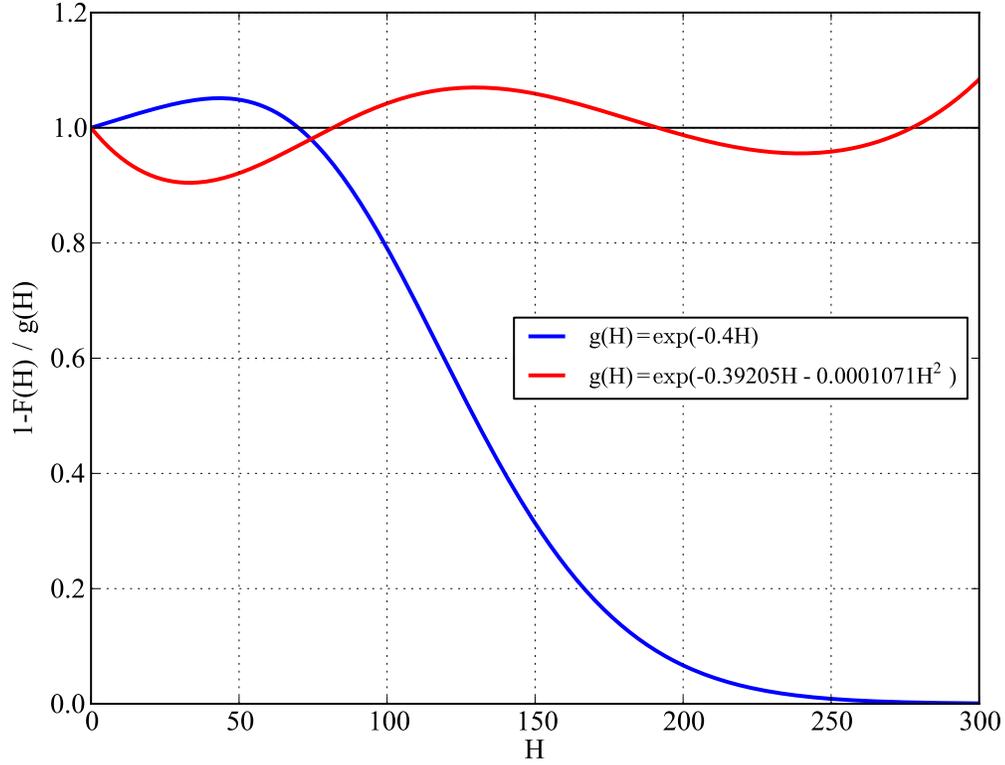}
\end{minipage}
\begingroup\renewcommand{\baselinestretch}{1.0}
\caption{The ratio of the true asymptotic distribution for $H_{20}$ to the approximate expression $\exp(-0.4H_{20})$.  At $H_{20}\approx40$, the survival function of the asymptotic distribution begins to decrease more rapidly than exponential approximation, indicating the importance of a quadratic term.  We show such a term, determined by least squares, with the red curve, which provides an accurate estimation of the survival function over a broad range.}
\renewcommand{\baselinestretch}{1.5}\endgroup
\label{ch5_plot17}
\end{figure}

Finally, we consider the behavior of the null distribution at large values of $H_{20}$.  We have seen that $\exp({-0.4H})$ is an excellent approximation for the true survival function for any ``practical'' application\footnote{Finite sample size effects are almost certainly larger than the discprepancy between the approximation and the asymptotic distribution once $H_{20}$ is of order 50.}.  Nonetheless, it is interesting to consider very large values of $H$ to investigate the trend of the true distribution relative to the simple exponential approximation.  We show the ratio of the two expression in Figure \ref{ch5_plot17}.  The exponential first-order in $H$ becomes a poor approximation after $H\approx100$ (again, sufficiently large for any practical need!).  By performing a least squares fit to minimize the difference between the asymptotic distribution and an approximation of the form $\exp(-aH - bH^2)$, we obtain an expression accurate to $10\%$ over the range $0\leq H\leq 300$,
\begin{equation}
1-F(H)\approx\exp(-0.39205H - 0.0001071H^2).
\end{equation}

\chapter{Pulsation Searches With Non-uniform Null Distributions}
\label{appC}

As discussed above in Chapter \ref{ch5}\footnote{We adopt the notation of Chapter \ref{ch5} here.}, when the null distribution of the phases departs from uniformity, the $\psi_k$ can retain independence even for large sample sizes.  Below, we outline two approaches to partially overcome this deficiency.

First, if the departure from non-uniformity is in some sense small, we can proceed as before.  We must modify the definition of the $\psi_k$ to account for the non-uniformity so that they are once again asymptotically normally distributed.  We begin by explicitly calculating the null distribution $F_{\Phi}(\phi)$, based on the IRF and S/C pointing history.  Essentially, $f_{\Phi}(\phi) \approx \epsilon(\phi)$, i.e., the null distribution is roughly the exposure calculated as a function of phase.  The approximation comes in choosing a shape for the (typically unknown) source spectrum used as a weight when summing the exposure over energy, and in the form of time-dependent background leakage.  Provided we are able to adequately characterize the IRF and model the time-dependent background, we can calculate the null distribution $F_\Phi(\phi)$ and thus the moments of the $G_k$.  Then,
\begin{equation}
\sqrt{n}(\psi_k - \mu_{G_k})/\sigma_{G_k}) \sim \mathcal{N},
\end{equation}
as before.  The moments are somewhat cumbersome to calculate as the distributions of $G_k$ are now not monotonic.  For an arbitrary phase distribution $F_{\Phi}(\phi)$, the distribution of $c_k$ is given by
\begin{equation}
\label{eqn:gdist}
F_{C_k}(x) = \sum_{i=1}^k \int_{(2i-1+\delta(x))/2k}^{(2i-1-\delta(x))/2k}d\phi\, f_{\Phi}(\phi)
\end{equation}
with $\delta=1 - \arccos\,(x)/\pi$.  The distribution for $S_k$ may be obtained from Eq. \ref{eqn:gdist} by letting $i\rightarrow i+1/4$.  These distribution functions are in turn used to calculate the required moments $\mu$ and $\sigma$.

Now, provided that the departure from uniformity is mild, we expect the same arguments about the asymptotic dependence of $\psi_k$ to apply, at least approximately, and a $Z^2_m$ statistic constructed using these new $\psi_k$ will follow, again approximately, a $\chi^2_m$ distribution.  However, the risk of overestimating the significance of a detection due departure from the $\chi^2_m$ calibration leads us to prefer a second method.

\begin{figure}
\begin{minipage}{6in}
\includegraphics[width=6in]{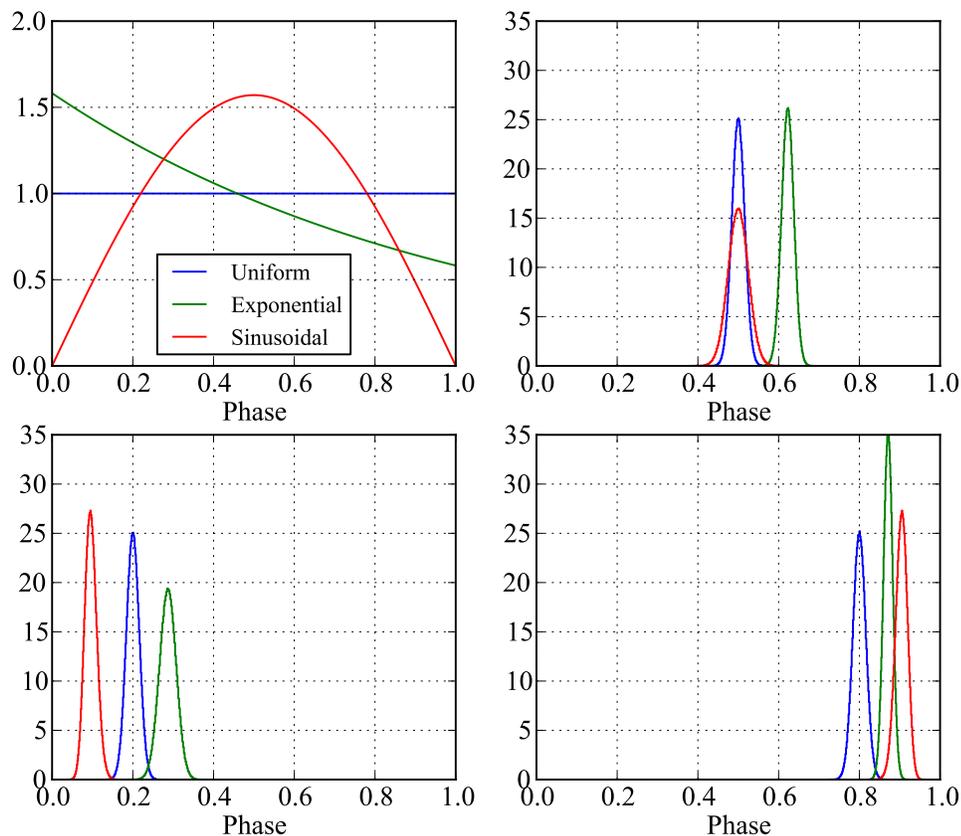}
\end{minipage}
\begingroup\renewcommand{\baselinestretch}{1.0}
\caption{The resulting ``signal'' after processing an input with $F_{\Phi}(\phi_i)$ as outlined in the text.  We simulate $10^6$ realizations from a von Mises distribution with a sharp peak and then generate histograms of $F_{\Phi}(\phi_i)$ for each of three distributions.  The uniform distribution is trivial and represents the unprocessed signal.  A sinusoidal distribution might arise, e.g., if the period of interest is very close to twice the orbital period of the S/C.  An exponential distribution is difficult to produce physically but provides insight into asymmetric distributions.  In the upper left panel, we plot $f_{\Phi}(\phi)$ for each of the exposure scenarios.  In the remaining three panels, we show the processed signal for three different absolute phases of the signal.  Clockwise from upper right, the signal is centered $\phi=0.5$, $0.8$, and $0.2$.}
\renewcommand{\baselinestretch}{1.5}\endgroup
\label{appc_plot1}
\end{figure}

In a second approach, we again calculate the null distribution $F_{\Phi}(\phi)$ explicitly from the exposure.  However, we now switch our fundamental set of rvs from $\{\phi_i\}$ to $\{F_{\Phi}(\phi_i)\}$, i.e., the cdf evaluated at the observed phases.  If we have calculated $F_{\Phi}(\phi)$ correctly, then the $\{F_{\Phi}(\phi_i)\}$ are guaranteed to follow a uniform distribution in the null case.  Further, $F_{\Phi}(\phi_i)\in[0,1)$, so we can use the $\{F_{\Phi}(\phi_i)\}$ directly in the (weighted and unweighted) tests we have already developed.  The drawback to this approach is a distortion of the input signal by ``processing'' by $F_{\Phi}(\phi_i)$.  Even worse, this processing depends on absolute phase, i.e., we lose the invariance under translations on the circle that has been a property of the statistics we have outlined\footnote{To be clear---the test statistic itself is still independent of absolute phase, but the processed signal $F_{\Phi}(\phi)$ \emph{does} depend on the absolute phase}.  However, we should not be surprised by this.  The departure from uniform exposure inevitably increases our sensitivity to some signals and decreases it for others.  A few examples appear in Figure \ref{appc_plot1} where we have taken some simple analytic distributions---a uniform distribution, a decaying exponential\footnote{An exponential distibution is a somewhat pathological example since it is discontinuous.}, and a sinusoid---for the exposure and processed a sharp signal---represented by a von Mises distribution.  We see that, e.g., when the signal is aligned with the peak of the sinusoidal exposure, the processed signal is broadened, whereas if it is near the minimum of the exposure, the signal is sharpened.  (This gain is offset by the appreciable decrease in the number of collected photons.)

\end{document}